\documentclass[11pt,a4paper,tightenlines,preprint,floatfix,nofootinbib,aps,prc,superscriptaddress,showkeys,showpacs]{revtex4}

\usepackage[dvips]{graphicx}
\usepackage{amsmath,amssymb,amsopn,bm,slashed}
\usepackage{color}
\usepackage{epstopdf}
\usepackage{grffile}

\usepackage[bookmarks,pdfhighlight=/O,colorlinks=false,pdfstartview=FitH]{hyperref}

\DeclareMathOperator{\im}{Im}
\DeclareMathOperator{\re}{Re}

\newcommand{\dd}{\text{d}}
\newcommand{\ee}{\text{e}}
\newcommand{\ii}{\text{i}}

\DeclareMathOperator{\Tr}{\ensuremath{\mathrm{Tr}}}
\newcommand{\beq}{\begin{equation}}
\newcommand{\eeq}{\end{equation}}
\newcommand{\bce}{\begin{center}}
\newcommand{\ece}{\end{center}}

\newcommand{\GeV}{\ensuremath{\mathrm{GeV}}}
\newcommand{\fm}{\ensuremath{\mathrm{fm}}}
\newcommand{\corres}{\ensuremath{\mathrel{\widehat{=}}}}

\begin{document}

\title{Off-equilibrium photon production during the chiral phase transition}

\author{Frank Michler}
\email{michler@th.physik.uni-frankfurt.de}
\affiliation{Institut f{\"u}r Theoretische Physik, Goethe-Universit{\"a}t
Frankfurt, Max-von-Laue-Stra{\ss}e 1, D-60438 Frankfurt, Germany}

\author{Hendrik van Hees}
\email{hees@fias.uni-frankfurt.de}
\affiliation{Institut f{\"u}r Theoretische Physik, Goethe-Universit{\"a}t
Frankfurt, Max-von-Laue-Stra{\ss}e 1, D-60438 Frankfurt, Germany}
\affiliation{Frankfurt Institute for Advanced Studies (FIAS),
Ruth-Moufang-Stra{\ss}e 1, D-60438 Frankfurt, Germany}

\author{Dennis D. Dietrich}
\email{dietrich@th.physik.uni-frankfurt.de}
\affiliation{Institut f{\"u}r Theoretische Physik, Goethe-Universit{\"a}t
Frankfurt, Max-von-Laue-Stra{\ss}e 1, D-60438 Frankfurt, Germany}

\author{Stefan Leupold}
\email{stefan.leupold@physics.uu.se} 
\affiliation{Institutionen f{\"o}r fysik och
  astronomi, Uppsala Universitet, Box 516, 75120 Uppsala, Sweden}

\author{Carsten Greiner}
\email{carsten.greiner@th.physik.uni-frankfurt.de}
\affiliation{Institut f{\"u}r Theoretische Physik, Goethe-Universit{\"a}t
Frankfurt, Max-von-Laue-Stra{\ss}e 1, D-60438 Frankfurt, Germany}

\date{\today}

\begin{abstract} 
  In the early stage of ultrarelativistic heavy-ion collisions chiral
  symmetry is restored temporarily. During this so-called chiral phase
  transition, the quark masses change from their constituent to their
  bare values. This mass shift leads to the spontaneous non-perturbative
  creation of quark-antiquark pairs, which effectively contributes
  to the formation of the quark-gluon plasma. We investigate the photon
  production induced by this creation process. We provide an approach
  that eliminates possible unphysical contributions from the vacuum
  polarization and renders the resulting photon spectra integrable in
  the ultraviolet domain. The off-equilibrium photon numbers are of
  quadratic order in the perturbative coupling constants while a thermal
  production is only of quartic order.  Quantitatively, we find, however,
  that for the most physical mass-shift scenarios and for photon momenta
  larger than 1 GeV the off-equilibrium processes contribute less
  photons than the thermal processes.
\end{abstract}


\maketitle

\section{Introduction}  
\label{sec_intro}

Ultrarelativistic heavy-ion collision experiments allow for studying
strongly interacting matter under extreme conditions. Such experiments
are currently performed at the Relativistic Heavy Ion Collider (RHIC) at
the Brookhaven National Laboratory (BNL) and at the Large Hadron
Collider (LHC) at the European Organization for Nuclear Research
(CERN). Furthermore, they will be carried out at the future Facility for
Antiproton and Ion Research (FAIR) at the Helmholtz Center for Heavy Ion
Research (GSI) and at the future Nuclotron-based Ion
Collider Facility (NICA) at the Joint Institute for Nuclear Research
(JINR). One main objective of these experiments is the creation and
exploration of the so-called quark-gluon plasma (QGP), a state of matter
of deconfined quarks and gluons. The two basic
  features of the strong interaction, namely confinement
  \cite{BS:1992} and asymptotic freedom \cite{GW:1973,Pol:1973} predict
that this state is created at high densities and temperatures
\cite{Shuryak:1978ij,Shuryak:1977ut,Yag:2005,Muller:2006ee,Friman:2011zz},
which occur during ultrarelativistic heavy-ion collisions.

The lifetime of the QGP created during a heavy-ion collision is expected
to be of the order of up to $5-10 \;\fm/c$ \cite{Yag:2005}. After that
it transforms into a gas of hadrons. Thus, experiments cannot access the
QGP directly, which makes the determination of the properties of this
state difficult. Therefore, it is important to find theoretical
signatures that provide a distinction between a hadron gas and a
QGP. Furthermore, one has to identify experimental observables from
which one can draw conclusions on theses theoretical signatures
\cite{Yag:2005,Muller:2006ee,Friman:2011zz}. One important category of
these observables are direct photons as electromagnetic probes. As they
only interact electromagnetically, their mean free path is much larger
than the spatial extension of the QGP. Therefore, they leave the QGP
almost undisturbed once they have been produced and thus provide direct
insight into the early stage of the collision.

One important aspect in this context is that the quark-gluon plasma, as
it occurs in a heavy-ion collision, is not a static medium. It first
thermalizes over a finite timescale, then expands and cools down before
it hadronizes finally. This non-equilibrium dynamics has always been a
major motivation for investigations in non-equilibrium quantum field
theory
\cite{Schw61,Bakshi:1962dv,Bakshi:1963bn,Kel64,Cra68,Danielewicz:1982kk,Chou:1984es,Landsman:1986uw,Greiner:1998vd,Berges:2001fi,Nahrgang:2011mg,Nahrgang:2011mv}. Besides
the role of possible memory effects during the time evolution
\cite{P1984305,Greiner:1994xm,Kohler:1995zz,Xu:1999aq,Juchem:2004cs,Juchem:2003bi,Schenke:2005ry,Schenke:2006uh,Michler:2009dy},
it is of particular interest how the finite lifetime of the quark-gluon
plasma itself affects the resulting photon spectra.

The first investigations on this topic were done by Boyanovski et al.\
\cite{Wang:2000pv,Wang:2001xh}. The authors first specified the density
matrix at some initial time, $t_{0}$, for a thermalized quark-gluon
plasma not containing any photon, i.e.,
\begin{equation}
 \label{eq:1:initial_condition}
 \hat{\rho}(t_{0}) = \frac{\ee^{-\beta\hat{H}_{\text{QCD}}}}{\Tr \ee^{-\beta\hat{H}_{\text{QCD}}}} \ ,
\end{equation}
with $\beta=1/T$. Afterwards, the authors propagated the system from the
initial time, $t_{0}$, to a later time, $t$, under the influence of the
electromagnetic interaction and determined the photon number at this
point of time. One main result was the prediction of contributions of
first-order QED processes, which are forbidden kinematically in thermal
equilibrium. Furthermore, the photon spectra resulting from these
processes flattened into a power law decay for photon energies $\omega_{\vec{k}}>1.5\;
\GeV$ ($\omega_{\vec{k}}=|\vec{k}|$ with $\vec{k}$ denoting the three-momentum of 
the photon) and thus dominated over higher equilibrium contributions such as
gluon Compton scattering and quark pair annihilation with associated 
gluon production in that domain.

On the other hand, the investigations in \cite{Wang:2000pv,Wang:2001xh}
were also accompanied by the problem that the photon spectra decayed too
slowly for being integrable in the ultraviolet (UV) domain. In this
domain, these spectra behaved as $1/\omega^{3}_{\vec{k}}$ , which means that the total
number density and the total energy density of the emitted photons were
logarithmically and linearly divergent, respectively. Furthermore, the
authors did not include the process where a quark-antiquark pair
together with a photon is spontaneously created out of the vacuum. Such
a process is conceivable, as the temporal change of the background can 
provide energy for particle creation.

An inclusion of this process in a subsequent work
\cite{Boyanovsky:2003qm} revealed the even more serious problem that the
contribution from the vacuum polarization, which is included in this
process, is divergent for any given photon energy, $\omega_{\vec{k}}$. The authors
argued that this contribution is unphysical and can be eliminated
(renormalized) by rescaling the photon field operators with the vacuum
wavefunction renormalization, $\sqrt{Z}$, but they did not provide a
detailed calculation from which this could be inferred. Furthermore, one
still encounters the problem with the UV behavior of the remaining
contributions.

Later on, the topic was also picked up by Fraga et al.\
\cite{Fraga:2003sn,Fraga:2004ur} where the ansatz used in
\cite{Wang:2000pv,Wang:2001xh,Boyanovsky:2003qm} was considered as
doubtful, as it came along with the mentioned problems. In particular,
the concerns raised in \cite{Fraga:2003sn,Fraga:2004ur} were the
following:
\begin{itemize}
\item In \cite{Wang:2000pv,Wang:2001xh,Boyanovsky:2003qm} the time at which the photons
  are observed has been kept finite. Either this corresponds to
  measuring photons that are not free asymptotic states or it
  corresponds to suddenly turning off the electromagnetic interaction at
  this point in time. Both cases are questionable.
\item A system of quarks and gluons, which undergoes electromagnetic
  interactions, necessarily contains photons. Hence, taking an initial
  state without any photons and without the Hamiltonian for
  electromagnetism corresponds to switching on the electromagnetic
  interaction at the initial time, which is questionable as well.  It
  was shown in \cite{Arleo:2004gn} that the ansatz used in
  \cite{Wang:2000pv,Wang:2001xh,Boyanovsky:2003qm} is indeed equivalent
  to such a scenario.
\item The divergent contribution from the vacuum polarization is
  unphysical and thus needs to be renormalized. Nevertheless, the
  renormalization procedure presented in \cite{Boyanovsky:2003qm} is not
  coherent since no derivation of the photon yield with rescaled field
  operators has been presented in \cite{Boyanovsky:2003qm}.
\end{itemize}
The authors of \cite{Fraga:2003sn,Fraga:2004ur} did, however, not
provide an alternative approach for how to handle the mentioned problems
in a consistent manner. Solely in \cite{Arleo:2004gn} it was indicated
that the question of finite-lifetime effects could be addressed within
the 2PI (two-particle irreducible) approach even though the conservation
of gauge invariance remains challenging. Later on Boyanovsky et al.
insisted on their approach \cite{Boyanovsky:2003rw} and objected to the
arguments by \cite{Fraga:2003sn,Fraga:2004ur} as follows.
\begin{itemize}
\item Non-equilibrium quantum field theory is an initial-value problem.
  This means that the density matrix of the considered system is first
  specified at some initial time and then propagated to a later time by
  the time-evolution operator. For that reason, the Hamiltonian is not modified
  by introducing a time-dependent artificial coupling as it would be the
  case for a `switching on' and a later `switching off' of the
  electromagnetic interaction.
\item The quark-gluon plasma, as it occurs in a heavy-ion collision, has
  a lifetime of only a few $\fm/c$.  Therefore, taking the time to
  infinity is unphysical as this limit requires the inclusion of
  non-perturbative phase-transition effects on the photon production.
\item The renormalization technique of \cite{Boyanovsky:2003rw} provides
  a rescaling of the photon field operators such that the photon number
  operator actually counts asymptotic photon states with amplitude one.
\end{itemize}

This debate has actually been one of our motivations to study the
mentioned problems. In the first attempt, we have modeled the finite
lifetime of the (thermalized) quark-gluon plasma during a heavy-ion
collision by introducing time-dependent occupation numbers in the photon
self-energy \cite{Michler:2009hi}. This ansatz allowed us to renormalize
the divergent contribution from the vacuum polarization consistently.
Furthermore, if the occupation numbers are switched on and off again to
mimic the time evolution of a quark-gluon plasma during a heavy-ion
collision, it also renders the resulting photon spectra integrable in the 
ultraviolet domain if one
takes into account that both the creation and the hadronization of the
quark-gluon plasma take place over a finite interval of time. The photon
spectra, however, remain non-integrable in the UV domain if they are considered at a point
of time, $t$, where the plasma still exists. Thus, the problem with the
UV behavior is not under control for the general case.

It is conceivable that these shortcomings result from a violation of the
Ward-Takahashi identities within the model descriptions
\cite{Wang:2000pv,Wang:2001xh,Boyanovsky:2003qm,Michler:2009hi} on
photon production from an evolving QGP. Therefore, in the present work,
we consider a conceptually different scenario, where the production of
quark pairs and photons results from a change in the quark
  mass. In contrast to \cite{Michler:2009hi}, such a scenario has the
crucial advantage that it allows for a first-principle description by
introducing a Yukawa-like source term in the QED Lagrangian. The source
term couples the fermion field to a scalar, time-dependent background
field, $\phi(t)$. Thereby, the fermions effectively achieve a
time-dependent mass, which is compatible with the Ward-Takahashi
identities.

During the chiral phase transition in the very early stage of a
heavy-ion collision the quark mass drops from its constituent value,
$m_{c}$, to its bare value, $m_{b}$. It has been shown in
\cite{Greiner:1995ac,Greiner:1996wz} that this change leads to the
spontaneous and non-perturbative production of quark-antiquark
  pairs, which in turn contributes effectively to the creation of a
quark-gluon plasma. We investigate the photon emission arising from this
creation process. In this context, the emitted photons do not only serve
as a signature for the finite thermalization time of the QGP itself but,
in particular, also for the nature of the chiral phase transition. The
quarks can obtain energy by the coupling to the time-dependent source
field. Therefore, photons can be produced in first-order QED processes,
which would be kinematically forbidden in a static thermal
equilibrium. We restrict ourselves to these first-order QED processes
but maintain the coupling to the source field up to all orders.
Similar investigations have been performed in \cite{Blaschke:2011is} on electron-positron annihilation into a single photon 
in the presence of a strong laser field, with the pair creation induced by a time-dependent electromagnetic background field \cite{Schmidt:1998vi}.

In this context, there is a crucial difference to the approaches in
\cite{Wang:2000pv,Wang:2001xh,Boyanovsky:2003qm,Michler:2009hi}: There
the photon numbers have been considered at finite times, $t$. In the
present work we will show, however, that the photon numbers have to be
extracted in the limit $t \to \infty$, i.e.\ for free asymptotic states,
which are the only observable ones because they reach the
detectors. For this purpose, we specify our initial
  state at $t_{0}\rightarrow-\infty$ and introduce an adiabatic
  switching of the electromagnetic interaction, i.e.,
\begin{equation}
 \hat{H}_{\text{EM}}(t) \rightarrow f_{\varepsilon}(t)\hat{H}_{\text{EM}}(t) \ , \quad 
                                    \text{with} \quad f_{\varepsilon}(t)=\ee^{-\varepsilon|t|} \quad \text{and} \quad \varepsilon>0 \ .
\end{equation}
The photon numbers are then considered in the limit $t\rightarrow\infty$
and we let $\varepsilon\rightarrow0$ at the end of our calculation.
Hence, we pursue an in/out description as suggested in
\cite{Fraga:2003sn,Fraga:2004ur}. In this paper we shall demonstrate
that this procedure eliminates possible unphysical contributions from
the vacuum polarization and, furthermore, renders the resulting photon
spectra integrable in the UV domain if it is taken into account that the mass change takes
place over a finite time interval, $\tau$. In
  particular, we also show that keeping the exact sequence of limits,
  i.e., taking first $t\rightarrow\infty$ and then
  $\varepsilon\rightarrow0$, is indeed essential and that interchanging
  them leads to an inadequate definition of the photon number. In
  particular, we point out that it effectively comes along with a
  violation of the Ward-Takahashi identities. This again
    underlines the fact that a definition of physically meaningful
    photon numbers at finite time is problematic.

This paper is structured as follows. In Sec.~\ref{sec:gaug-inv-phot}, we
present our approach to photon production arising from the chiral mass
shift. We calculate the photon yield up to first order in
$\alpha_{\text{e}}$ but maintain the coupling to the external background
field up to all orders. For this purpose, we construct our
interaction picture in such a way that it already incorporates the
underlying interaction term of the external Yukawa field with the
quarks in the unperturbed Hamiltonian of our system. The photon yield
from first-order QED processes is then obtained by a standard
perturbative calculation. We shall also prove explicitly that the
coupling of the fermion field to the scalar background field does
not violate the Ward-Takahashi identities.

Before we turn to our subsequent investigations concerning photon
production, we provide in Sec.~\ref{sec:pair-production} an insertion on
pair production arising from the chiral mass shift. There we will
basically generalize the results of \cite{Greiner:1995ac,Greiner:1996wz}
on the quark and antiquark occupation numbers from asymptotic to
finite times. One crucial result is that at any finite time, $t$, the
total energy density of the fermionic sector features a logarithmic
divergence. This is even the case for a mass parametrization, which is
arbitrarily often continuously differentiable. On the other hand, the
truly observable fermion occupation number, i.e.\ the one at $t \to
+\infty$, is not divergent. These features might indicate that the
resulting spectrum of photons radiated by the produced quarks inherits a
divergence, since for the photons one sums over the whole history. We
will see that this is in general not the case for the observable photon
spectrum.

In Sec.~\ref{sec:photon-production}, we present our numerical results on
chiral photon production. There we compare again different mass
parameterizations, $m(t)$. For the case of an instantaneous mass shift,
the loop integral entering the photon self-energy features a linear
divergence, caused by the scaling behavior of the quark-antiquark
occupation numbers with respect to the fermion momentum, $\vec{p}$. When
regulating this divergence by a numerical cutoff in the loop integral,
the resulting photon spectra decay $\propto 1/\omega^{3}_{\vec{k}}$ in the ultraviolet
domain and thus feature the same pathology as in
\cite{Wang:2000pv,Wang:2001xh}. When we turn from an instantaneous mass
shift to a mass shift over a {\em finite} time interval, $\tau$, which
corresponds to a more realistic scenario, the mentioned divergence in
the loop integral is cured and, furthermore, the resulting photon
spectra become integrable in the ultraviolet domain.  So the logarithmic
divergence in the fermionic energy density at finite times does not
manifest itself in form of a similar pathology in the the total energy
density of the photonic sector.

In Sec.~\ref{sec:conclusions} we close with a summary and an outlook to
future investigations. Technical details are relegated to six appendices.

\section{Gauge-invariant model for chiral photon production}
\label{sec:gaug-inv-phot}
Our starting point is the Hamiltonian in the Schr{\"o}dinger picture,
$\hat{H}_{\text{S}}(t)$, governing the time evolution of the system
\begin{subequations}
 \label{eq:hamiltonian}
 \begin{eqnarray}
   \hat{H}_{\text{S}}(t) & = & \hat{H}_{0}+\hat{H}_{g}(t)+\hat{H}_{\text{EM}} \ , \\
   \hat{H}_{0} & = & \int \dd^{3}x\mbox{
   }\hat{\bar{\psi}}_{\text{S}}(\vec{x})\left(-\ii
     \vec{\gamma}\cdot\vec{\nabla}+m_{c}\right) 
   \hat{\psi}_{\text{S}}(\vec{x})
   +\frac{1}{2}\int\dd^{3}x\left(\hat{\vec{E}}^{2}_{\text{S}}(\vec{x})+\hat{\vec{B}}^{2}_{\text{S}}(\vec{x})\right)
   \ , \\
   \hat{H}_{g}(t) & = & g\phi(t)\int \dd^{3}x\mbox{
   }\hat{\bar{\psi}}_{\text{S}}(\vec{x})\hat{\psi}_{\text{S}}(\vec{x}) \
   , \\ 
   \hat{H}_{\text{EM}}   & = & \int \dd^{3}x\mbox{
   }\hat{j}_{\mu,\text{S}}(\vec{x})\hat{A}^{\mu}_{\text{S}}(\vec{x}) \ . 
 \end{eqnarray}
\end{subequations}
Here $\hat{H}_{0}$ contains the kinetic part for the fermions and the
photons, $\hat{H}_{g}(t)$ the coupling of the fermions to the external
source field, $\phi(t)$, and $\hat{H}_{\text{EM}}$ the electromagnetic
interaction between the fermions and the photons. The current operator,
$\hat{j}_{\mu,\text{S}}(\vec{x})$, is given by
\begin{equation}
 \label{eq:2:current_schroed}
 \hat{j}_{\mu,\text{S}}(\vec{x})=e\hat{\bar{\psi}}_{\text{S}}(\vec{x})\gamma_{\mu}\hat{\psi}_{\text{S}}(\vec{x}) \ ,
\end{equation}
with $e$ and $\gamma_{\mu}$ denoting the electromagnetic coupling and
the Dirac matrices, respectively. The subscript `S' indicates that the
operators are taken in the Schr{\"o}dinger picture. In this picture, the
time evolution of the system is described by the density matrix,
$\hat{\rho}_{\text{S}}(t)$, which reads
\begin{equation}
 \label{eq:2:density}
 \hat{\rho}_{\text{S}}(t) = \sum_{n} p_{n}(t)\left|n\right\rangle\left\langle n\right| \ .
\end{equation}
Here $\left\lbrace\left|n\right\rangle\right\rbrace$ can be any
orthonormal and complete set of state vectors, i.e., it has the
properties
$$
 \sum_{n}\left|n\right\rangle\left\langle n\right|=\hat{I} \quad\text{and}\quad \left\langle n|m\right\rangle=\delta_{nm} \ ,
$$
and $p_{n}(t)$ denotes the probability that the system is found in the
state $\left|n\right\rangle$ at a given time, $t$. The density matrix
(\ref{eq:2:density}) obeys the equation of motion
\begin{equation}
 \ii \partial_{t}\hat{\rho}_{\text{S}}(t) = \left[\hat{H}_{\text{S}}(t),\hat{\rho}_{\text{S}}(t)\right] \ .
\end{equation}
This equation is formally solved by
\begin{subequations}
 \begin{eqnarray}
   \hat{\rho}_{\text{S}}(t) & = &
   \hat{U}_{\text{S}}(t,t_{0}) \hat{\rho}_{\text{S}}(t_{0})
   \hat{U}^{\dagger}_{\text{S}}(t,t_{0}) \ ,  \\
   \hat{U}_{\text{S}}(t,t_{0}) & = &
   T\left\lbrace\exp\left[-\ii\int_{t_0}^{t}\dd
       t'\hat{H}_{\text{S}}(t')\right]\right\rbrace \ .
 \end{eqnarray}
\end{subequations}
Here $t_{0}$ is the initial time and $T$ denotes time
ordering. $\hat{U}_{\text{S}}(t,t_{0})$ is the so-called time-evolution
operator which solves the equation of motion
\begin{equation}
 \ii \partial_{t}\hat{U}_{\text{S}}(t,t_{0}) = \hat{H}_{\text{S}}(t)\hat{U}_{\text{S}}(t,t_{0}) \ .
\end{equation}
The expectation value of an observable characterized by the operator
$\hat{O}_{\text{S}}$ is given by
\begin{eqnarray}
  \left \langle \hat{O}_{\text{S}}\right \rangle & = & \mbox{Tr}\left\lbrace\hat{\rho}_{\text{S}}(t)\hat{O}_{\text{S}}\right\rbrace \nonumber \\
  & = & \mbox{Tr}\left\lbrace
    \hat{U}_{\text{S}}(t,t_{0})\hat{\rho}_{\text{S}}(t_{0})
    \hat{U}^{\dagger}_{\text{S}}(t,t_{0})\hat{O}_{\text{S}}
  \right\rbrace \nonumber \\
  & = & \mbox{Tr}\left\lbrace
    \hat{\rho}_{\text{S}}(t_{0})\hat{U}^{\dagger}_{\text{S}}(t,t_{0})
    \hat{O}_{\text{S}}\hat{U}_{\text{S}}(t,t_{0})
  \right\rbrace \nonumber \\
  & = & \mbox{Tr}\left\lbrace
    \hat{\rho}_{\text{S}}(t_{0})\hat{O}_{\text{H}}(t)
  \right\rbrace \nonumber \ .
\end{eqnarray}
In the last step, we have introduced the operator in the Heisenberg
picture,
\begin{eqnarray}
 \label{eq:2:heisenberg}
 \hat{O}_{\text{H}}(t) = \hat{U}^{\dagger}_{\text{S}}(t,t_{0})\hat{O}_{\text{S}}\hat{U}_{\text{S}}(t,t_{0}) \ .
\end{eqnarray}
Since the scalar background field, $\phi(t)$, is assumed to be classical
and only time-dependent, the fermions effectively obtain a time
dependent mass,
\begin{equation}
 \label{eq:2:timedep_mass}
 m(t)=m_{c}+g\phi(t) \ .
\end{equation}
As we shall demonstrate below (see Eq. (\ref{eq:2:wti})) and again in
greater detail in appendix \ref{sec:appb}, this is compatible with the
Ward-Takahashi identities. It has been shown in
\cite{Greiner:1995ac,Greiner:1996wz} that the change of the quark mass
from its constituent value, $m_{c}$, to its bare value, $m_{b}$, during
the chiral phase transition in the very early stage of a heavy-ion
collision leads to the spontaneous pair creation of quarks and
antiquarks. We now investigate the photon emission arising from this
creation process. We assume that our system does not contain any quarks,
antiquarks, or photons initially. The initial density matrix,
$\hat{\rho}_{\text{S}}(t_{0})$, is hence given by
\begin{equation}
 \label{eq:2:initial_density}
 \hat{\rho}_{\text{S}}(t_{0})=\left|0_{q\bar{q}}\right\rangle \left\langle 0_{q\bar{q}}\right|\otimes
                              \left|0_{\gamma}\right \rangle \left \langle 0_{\gamma}\right| \ .
\end{equation}
Here $\left|0_{q\bar{q}}\right\rangle$ and
$\left|0_{\gamma}\right\rangle$ denote the vacuum states of the
fermionic and the photonic sector, respectively. Since the fermion mass
changes in time, it is important to point out that
$\left|0_{q\bar{q}}\right\rangle$ is defined with regard to the initial,
constituent mass, $m_{c}$. The initial time, $t_{0}$, is chosen from the
domain where the quark mass is still at this value, i.e., $t_{0}\le
t^{'}_{0}$ with $t^{'}_{0}$ denoting the time at which the change of the
quark mass begins. In the case of parameterizations, $m(t)$, for which
the time derivative, $\dot{m}(t)$, has a non-compact support, it is
sufficient to ensure that $g\phi(t)\ll m_{c}$ for $t\le t^{'}_{0}$.

\begin{figure}[htb]
 \begin{center}
  \includegraphics[height=3.0cm]{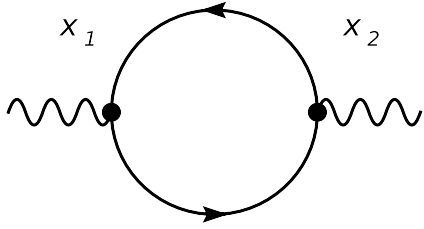}
  \caption{The photon self-energy, $\ii \Pi^{<}_{\mu\nu}(x_{1},x_{2})$,
    is given by the one-loop approximation with fermion propagators
    dressed by the background field, $\phi(t)$.}
  \label{fig:2:chiral_one_loop}
 \end{center}
\end{figure}

As the electromagnetic coupling is small, we pursue a calculation at the
first order in $\alpha_{e}$ but keep all orders in $g$. For this purpose,
we construct an interaction picture in a way that it incorporates
$\hat{H}_{g}(t)$,
\begin{subequations}
 \begin{eqnarray}
   \hat{O}_{\text{J}}(t)           & = & \hat{U}^{g,\dagger}_{0}(t,t_{0})\hat{O}_{\text{S}}\hat{U}^{g}_{0}(t,t_{0}) \ , \label{eq:2:intpic_op} \\
   \hat{U}^{g}_{0}(t,t_{0}) & = & T\left\lbrace\exp\left[-\ii\int_{t_0}^{t}\dd t'\hat{H}^{g}_{0}(t')\right]\right\rbrace \ , \\
   \hat{H}^{g}_{0}(t)       & = & \hat{H}_{0}+\hat{H}_{g}(t) \ .
 \end{eqnarray}
\end{subequations}
Here we have introduced the subscript `J' in order to distinguish our
interaction picture from the standard one in which only the kinetic part
of the Hamiltonian $\hat{H}^{g}_{0}(t)$ is included in the
time-evolution operator. This standard interaction picture, denoted by
`I', will also come into play later. It can be shown that in our
interaction picture the operators obey the analogous equations of motion
as in the standard one
\begin{eqnarray}
 \ii \partial_{t}\hat{O}_{\text{J}}(t) & = & \left[\hat{O}_{\text{J}}(t),\hat{H}_{0,\text{J}}(t)\right] \ ,
\end{eqnarray}
where $\hat{H}_{0,\text{J}}(t)$ is given by
\begin{eqnarray}
 \hat{H}_{0,\text{J}}(t) & = & \hat{U}^{g,\dagger}_{0}(t,t_0)\hat{H}^{g}_{0}(t)\hat{U}^{g}_{0}(t,t_0) \ .
\end{eqnarray}
Furthermore, we can construct an interaction-picture time-evolution operator, $\hat{U}_{\text{J}}(t,t_0)$, as
\begin{subequations}
 \begin{eqnarray}
  \label{eq:2:intpic_evol}
  \hat{U}_{\text{J}}(t,t_0) & = & T\left\lbrace\exp\left[-\ii\int_{t_0}^{t}\dd t'\hat{H}_{\text{J}}(t')\right]\right\rbrace \ , \\
  \hat{H}_{\text{J}}(t)     & = & \hat{U}^{g,\dagger}_{0}(t,t_0)\hat{H}_{\text{EM}}\hat{U}^{g}_{0}(t,t_0) \ .
 \end{eqnarray}
\end{subequations}
In analogy to the standard case, it fulfills the equations of motion,
\begin{equation}
 \ii \partial_{t}\hat{U}_{\text{J}}(t,t_0) = \hat{U}_{\text{J}}(t,t_0)\hat{H}_{\text{J}}(t) \ ,
\end{equation}
as well as the identity,
\begin{equation}
 \label{eq:2:ident_int}        
 \hat{U}_{\text{S}}(t,t_0)=\hat{U}^{g}_{0}(t,t_0)\hat{U}_{\text{J}}(t,t_0) \ .
\end{equation}
Accordingly, the Heisenberg-picture operator representation
(\ref{eq:2:heisenberg}) is related to (\ref{eq:2:intpic_op}) by
\begin{equation}
 \hat{O}_{\text{H}}(t) = \hat{U}^{\dagger}_{\text{J}}(t,t_0)\hat{O}_{\text{J}}(t)\hat{U}_{\text{J}}(t,t_0) \ .
\end{equation}
Moreover, (\ref{eq:2:intpic_op}) can be related to the standard
interaction-picture representation as
\begin{equation} 
 \label{eq:2:intpic_link}
 \hat{O}_{\text{J}}(t) = \hat{U}^{\dagger}_{\text{I}}(t,t_0)\hat{O}_{\text{I}}(t)\hat{U}_{\text{I}}(t,t_0) \ .
\end{equation}
Here we have introduced
\begin{subequations}
 \begin{eqnarray}
   \hat{U}_{\text{I}}(t,t_0) & = & T\left\lbrace\exp\left[-\ii\int_{t_0}^{t}\dd t'\hat{H}^{g}_{\text{I}}(t')\right]\right\rbrace \ , \\
   \hat{H}^{g}_{\text{I}}(t) & = & g\phi(t)\int\dd^{3}x\mbox{ }\hat{\bar{\psi}}_{\text{I}}(x)\hat{\psi}_{\text{I}}(x) \ , \\
   \hat{O}_{\text{I}}(t)     & = & \hat{U}^{\dagger}_{0}(t,t_{0})\hat{O}_{\text{S}}\hat{U}_{0}(t,t_{0}) \ , \\
   \hat{U}_{0}(t,t_0)        & = & \ee^{-\ii\hat{H}_{0}(t-t_0)} \ ,
 \end{eqnarray}
\end{subequations}
and taken into account that
\begin{equation}
 \label{eq:2:ident_intstd}
 \hat{U}^{g}_{0}(t,t_{0}) = \hat{U}_{0}(t,t_{0})\hat{U}_{\text{I}}(t,t_{0}) \ .
\end{equation}
Before we can start with our calculations on photon production, we still
have to determine the form of the fermion- and photon-field operators
within the interaction-picture representation, `J'. For this purpose, we
take into account that in the standard interaction picture, `I', the
photon-field operator is given by
\begin{subequations}
 \begin{eqnarray}
  \label{eq:2:phot_int}
   \hat{A}^{\mu}_{\text{I}}(x) & = & \sum_{\lambda}\int\frac{\dd^{3}k}{(2\pi)^{3}}\frac{1}{\sqrt{2\omega_{\vec{k}}}}
   \left[
     \varepsilon^{\mu}(\vec{k},\lambda)\hat{a}_{\text{I}}(\vec{k},\lambda,t)\ee^{\ii \vec{k}\cdot\vec{x}}+
     \varepsilon^{\mu,*}(\vec{k},\lambda)\hat{a}^{\dagger}_{\text{I}}(\vec{k},\lambda,t)
     \ee^{-\ii\vec{k}\cdot\vec{x}} \right] \ , \label{eq:2:chiral_phot_decom} \\
   \hat{a}_{\text{I}}(\vec{k},\lambda,t) & = & \hat{a}_{\text{S}}(\vec{k},\lambda)\ee^{-\ii\omega_{\vec{k}}(t-t_{0})} \ , \\
   \hat{a}^{\dagger}_{\text{I}}(\vec{k},\lambda,t) & = &
   \hat{a}^{\dagger}_{\text{S}}(\vec{k},\lambda)\ee^{\ii\omega_{\vec{k}}(t-t_{0})} \ . 
 \end{eqnarray}
\end{subequations}
Here $\omega_{\vec{k}}$ is the free photon energy given by 
\begin{equation}
  \label{eq:2:photen}
  \omega_{\vec{k}}=|\vec{k}| \ .
\end{equation}
The creation and annihilation operators, $\hat{a}^{\dagger}_{\text{S}}(\vec{k},\lambda)$ and $\hat{a}_{\text{S}}(\vec{k},\lambda)$, fulfill the commutation relation,
\begin{equation}
  \left[\hat{a}_{\text{S}}(\vec{k},\lambda),\hat{a}^{\dagger}_{\text{S}}(\vec{k}',\lambda')\right] = (2\pi)^{3}\delta_{\lambda\lambda'}
  \delta^{(3)}\left(\vec{k}-\vec{k}'\right) \ ,
\end{equation}
with all other commutators vanishing. Since $\hat{U}_{\text{I}}(t,t_0)$
contains only fermion-field operators and hence commutes with both
$\hat{a}_{\text{I}}(\vec{k},\lambda,t)$ and
$\hat{A}^{\mu}_{\text{I}}(\vec{x},t)$, it immediately follows from
(\ref{eq:2:intpic_link}) that
\begin{subequations}
 \label{eq:2:op_phot}
 \begin{eqnarray}
   \hat{a}_{\text{J}}(\vec{k},\lambda,t) & = &
   \hat{a}_{\text{I}}(\vec{k},\lambda,t) \ , \label{eq:2:op_phot_a} \\ 
   \hat{A}^{\mu}_{\text{J}}(\vec{x},t)  & = & \hat{A}^{\mu}_{\text{I}}(\vec{x},t) \ .  \label{eq:2:op_phot_b}
 \end{eqnarray}
\end{subequations}
The fermion-field operator, $\hat{\psi}_{\text{J}}(x)$, obeys the
equation of motion,
\begin{eqnarray}
  \ii \partial_{t}\hat{\psi}_{\text{J}}(x) & = &
  \left[\hat{\psi}_{\text{J}}(x),\hat{H}_{0,\text{J}}(t)\right]
  \nonumber \\
  & = & \left[-\ii \gamma_{0}\vec{\gamma} \cdot \vec{\nabla} +
    \gamma_{0}m(t)\right]\hat{\psi}_{\text{J}}(x) \ ,
\end{eqnarray}
with $m(t)$ given by (\ref{eq:2:timedep_mass}). In a more compact form,
this can be rewritten as
\begin{equation}
 \label{eq:2:dirac_chiral}
  \left[\ii \gamma^{\mu}\partial_{\mu}-m(t)\right]\hat{\psi}_{\text{J}}(x)=0 \ ,
\end{equation}
which is just the Dirac equation with a time-dependent, scalar
mass. Since both momentum and spin are conserved, it is convenient to
expand $\hat{\psi}_{\text{J}}(x)$ in terms of positive and negative
energy eigenfunctions with momentum, $\vec{p}$, and spin, $s$,
\begin{equation}
 \label{eq:2:expansion}
 \hat{\psi}_{\text{J}}(x) = \sum_{s}\int\frac{\dd^{3}p}{(2\pi)^{3}}\left[\hat{b}_{\vec{p},s}\psi_{\vec{p},s,\uparrow}(x)+
                                                                \hat{d}^{\dagger}_{-\vec{p},s}\psi_{\vec{p},s,\downarrow}(x)\right] \ .
\end{equation}
The creation and annihilation operators are constructed such that they
obey the anticommutation relations
\begin{subequations}
 \label{eq:2:commutation_fermions}
 \begin{eqnarray}
  \left\lbrace\hat{b}^{\dagger}_{\vec{p},s},\hat{b}_{\vec{q},r}\right\rbrace & = & (2\pi)^{3}\delta_{rs}\delta^{(3)}(\vec{p}-\vec{q}) \ , \\
  \left\lbrace\hat{d}^{\dagger}_{\vec{p},s},\hat{d}_{\vec{q},r}\right\rbrace & = & (2\pi)^{3}\delta_{rs}\delta^{(3)}(\vec{p}-\vec{q}) \ ,
 \end{eqnarray}  
\end{subequations}
with all other anticommutators vanishing and that both
$\hat{b}_{\vec{p},s}$ and $\hat{d}_{-\vec{p},s}$ annihilate the initial
fermionic vacuum state,
\begin{subequations}
 \begin{align}
   \hat{b}_{\vec{p},s}\left|0_{q\bar{q}}\right \rangle & =
   \hat{d}_{-\vec{p},s}\left|0_{q\bar{q}}\right \rangle = 0 \
   , \\
   \left \langle 0_{q\bar{q}}\right|\hat{b}^{\dagger}_{\vec{p},s} & =
   \left \langle 0_{q\bar{q}}\right|\hat{d}^{\dagger}_{-\vec{p},s} = 0 \
   .
 \end{align}
\end{subequations}
The positive- and negative-energy state wavefunctions
$\psi_{\vec{p},s,\uparrow\downarrow}(x)$ fulfill the Dirac equation
(\ref{eq:2:dirac_chiral}), i.e.,
\begin{equation}
 \label{eq:2:dirac_chiral_wf}
  \left [\ii \gamma^{\mu}\partial_{\mu}-m(t)\right ]\psi_{\vec{p},s,\uparrow\downarrow}(x)=0 
\end{equation}
with the initial conditions
\begin{subequations}
 \begin{eqnarray}
  \label{eq:2:ini_con}
  \psi_{\vec{p},s,\uparrow}(x)  \rightarrow & \psi^{c}_{\vec{p},s,\uparrow}(x)   = u_{c}(\vec{p},s)\ee^{-\ii(E^{c}_{\vec{p}}t-\vec{p}\cdot\vec{x})} & \quad\mbox{for  } t\le t^{'}_{0} \ , \\
  \psi_{\vec{p},s,\downarrow}(x) \rightarrow & \psi^{c}_{\vec{p},s,\downarrow}(x) = v_{c}(\vec{p},s)\ee^{\ii (E^{c}_{\vec{p}}t+\vec{p}\cdot\vec{x})} & \quad\mbox{for  } t\le t^{'}_{0} \ .
 \end{eqnarray}
\end{subequations}
Choosing the Dirac representation for $\gamma^{\mu}$, i.e.,
$$
 \gamma^{0} = \begin{pmatrix}
               I & 0 \\
               0 & -I  
              \end{pmatrix} \ , \
 \gamma^{i} = \begin{pmatrix}
               0           & \sigma^{i} \\
               -\sigma^{i} & 0  
              \end{pmatrix} \ ,
$$
with $\sigma^{i}$ denoting the Pauli matrices, the spinors
$u_{c}(\vec{p},s)$ and $v_{c}(\vec{p},s)$ read
\begin{subequations}
 \begin{eqnarray}
   u_{c}(\vec{p},s) & = & \begin{pmatrix}
                           \cos\varphi^{c}_{\vec{p}}\mbox{ }\chi_{s} \\
                           \sin\varphi^{c}_{\vec{p}}\mbox{ }\frac{\vec{\sigma}\cdot\vec{p}}{p}\chi_{s}
                          \end{pmatrix} \ , \\
   v_{c}(\vec{p},s) & = & \begin{pmatrix}
                           \sin\varphi^{c}_{\vec{p}}\mbox{ }\chi_{s} \\
                           -\cos\varphi^{c}_{\vec{p}}\mbox{ }\frac{\vec{\sigma}\cdot\vec{p}}{p}\chi_{s}
                          \end{pmatrix} \ .
 \end{eqnarray}
\end{subequations}
$\cos\varphi^{c}_{\vec{p}}$ and $\sin\varphi^{c}_{\vec{p}}$ are given by
\begin{subequations}
 \label{eq:2:sincos}
 \begin{eqnarray}
  \cos\varphi^{c}_{\vec{p}} & = & \sqrt{\frac{E^{c}_{\vec{p}}+m_{c}}{2E^{c}_{\vec{p}}}} \ , \\
  \sin\varphi^{c}_{\vec{p}} & = & \sqrt{\frac{E^{c}_{\vec{p}}-m_{c}}{2E^{c}_{\vec{p}}}} \ .
 \end{eqnarray}
\end{subequations}
$E^{c}_{\vec{p}}$ is the onshell particle energy for given momentum,
$\vec{p}$. The superscript `c' denotes that it is defined with respect
to the initial constituent-quark mass, $m_{c}$. The Weyl spinors,
$\chi_{s}$, can be chosen as any orthonormal set of two-component vectors
fulfilling the completeness relation
$$
\sum_{s}\chi_{s}\bar{\chi}_{s} = I \ .
$$
In order to solve (\ref{eq:2:dirac_chiral_wf}), we parametrize
$\psi_{\vec{p},s,\uparrow\downarrow}(x)$ by
\begin{equation}
 \label{eq:2:wave_param}
 \psi_{\vec{p},s,\uparrow\downarrow}(x) = \begin{pmatrix}
                                           \alpha_{\vec{p},\uparrow\downarrow}(t)\chi_{s} \\
                                           \beta_{\vec{p},\uparrow\downarrow}(t)\frac{\vec{\sigma}\cdot\vec{p}}{p}\chi_{s}
                                          \end{pmatrix}\ee^{\ii \vec{p}\cdot\vec{x}} \ ,
\end{equation}
which leads to the following equations of motion for
$\alpha_{\vec{p},\uparrow\downarrow}(t)$ and
$\beta_{\vec{p},\uparrow\downarrow}(t)$
\begin{subequations}
 \label{eq:2:eom_param}
 \begin{eqnarray}
  \ii \partial_{t}\alpha_{\vec{p},\uparrow\downarrow}(t) 
   & = & p\beta_{\vec{p},\uparrow\downarrow}(t)+m(t)\alpha_{\vec{p},\uparrow\downarrow}(t) \ , \\
  \ii \partial_{t}\beta_{\vec{p},\uparrow\downarrow}(t) 
   & = & p\alpha_{\vec{p},\uparrow\downarrow}(t)-m(t)\beta_{\vec{p},\uparrow\downarrow}(t) \ .
 \end{eqnarray}
\end{subequations}
The initial conditions read
\begin{subequations}
 \label{eq:2:icon_param}
 \begin{eqnarray}
   \alpha_{\vec{p},\uparrow}(t)   \rightarrow & \cos\varphi^{c}_{\vec{p}}\mbox{ }\ee^{-\ii E^{c}_{\vec{p}}t} & \quad\mbox{for } t\le t^{'}_{0}  \ , \\
   \alpha_{\vec{p},\downarrow}(t) \rightarrow & \sin\varphi^{c}_{\vec{p}}\mbox{ }\ee^{+iE^{c}_{\vec{p}}t} & \quad\mbox{for } t\le t^{'}_{0}  \ , \\
   \beta_{\vec{p},\uparrow}(t)    \rightarrow & \sin\varphi^{c}_{\vec{p}}\mbox{ }\ee^{-\ii E^{c}_{\vec{p}}t} & \quad\mbox{for } t\le t^{'}_{0}  \ ,\\
   \beta_{\vec{p},\downarrow}(t)  \rightarrow & -\cos\varphi^{c}_{\vec{p}}\mbox{ }\ee^{+iE^{c}_{\vec{p}}t} & \quad\mbox{for } t\le t^{'}_{0}  \ .
 \end{eqnarray}
\end{subequations}
It can be shown that the wavefunction parameters,
$\alpha_{\vec{p},\uparrow\downarrow}(t)$ and
$\beta_{\vec{p},\uparrow\downarrow}(t)$, fulfill the normalization
condition
\begin{equation}
 \label{eq:2:param_normalization}
 \left|\alpha_{\vec{p},\uparrow\downarrow}(t)\right|^{2}+\left|\beta_{\vec{p},\uparrow\downarrow}(t)\right|^{2}=1 \ ,
\end{equation}
as well as the relations
\begin{subequations}
 \label{eq:2:param_link}
 \begin{eqnarray}
  \alpha_{\vec{p},\downarrow}(t) & = & \beta^{*}_{\vec{p},\uparrow}(t) \ , \\
  \beta_{\vec{p},\downarrow}(t)  & = & -\alpha^{*}_{\vec{p},\uparrow}(t) \ .
 \end{eqnarray}
\end{subequations}
Relations (\ref{eq:2:param_normalization}) and (\ref{eq:2:param_link})
essentially show that $\psi_{\vec{p},s,\uparrow\downarrow}(x)$ indeed
form an orthonormal basis for given momentum, $\vec{p}$, and spin, $s$,
as they obviously imply
\begin{subequations}
 \begin{eqnarray}
  \psi^{\dagger}_{\vec{p},s,\uparrow}(x)\psi_{\vec{p},s,\uparrow}(x)     =
  \psi^{\dagger}_{\vec{p},s,\downarrow}(x)\psi_{\vec{p},s,\downarrow}(x) = 1 \ , \\
  \psi^{\dagger}_{\vec{p},s,\uparrow}(x)\psi_{\vec{p},s,\downarrow}(x)   =
  \psi^{\dagger}_{\vec{p},s,\downarrow}(x)\psi_{\vec{p},s,\uparrow}(x)   = 0 \ .
 \end{eqnarray}
\end{subequations}
Now we turn to the description of photon emission induced by the chiral
mass shift. Since the quark mass is time-dependent only and our
system is initially given by the vacuum state
(\ref{eq:2:initial_density}), it is spatially homogeneous. For such a
system, the photon number at a given time, $t$, reads
\begin{equation}
\begin{split}
 \label{eq:2:a_phot_heis}
 \frac{\dd^{6}n_{\gamma}(t)}{\dd^{3}x\dd^{3}k} &=
 \frac{1}{(2\pi)^{3}V}\sum_{\lambda=\perp}\left \langle \hat{n}_{\text{H}}(\vec{k},\lambda,t)\right \rangle
 \\
 &= \frac{1}{(2\pi)^{3}V}
 \sum_{\lambda=\perp} \left \langle  \hat{a}^{\dagger}_{\text{H}}(\vec{k},\lambda,t)
   \hat{a}_{\text{H}}(\vec{k},\lambda,t) \right \rangle \ .
\end{split}
\end{equation}
The sum runs over all physical (transverse) polarizations. Before we can
continue with the evaluation of (\ref{eq:2:a_phot_heis}), we first have
to clarify under which circumstances
$\hat{a}_{\text{H}}(\vec{k},\lambda,t)$ together with its Hermitian
conjugate actually allows for an interpretation as a single-photon
operator. For this purpose, we recall the plane-wave decomposition
(\ref{eq:2:phot_int}) which together with (\ref{eq:2:op_phot_b}) implies
\begin{eqnarray}
 \label{eq:phot_heis}
 \hat{A}^{\mu}_{\text{H}}(x) & = & \hat{U}^{\dagger}_{\text{J}}(t,t_0)\hat{A}^{\mu}_{\text{J}}(x)
                                   \hat{U}_{\text{J}}(t,t_0) \nonumber \\
                             & = & \sum_{\lambda}\int\frac{\dd^{3}k}{(2\pi)^{3}}\frac{1}{\sqrt{2\omega_{\vec{k}}}}
                                    \left[
                                     \varepsilon^{\mu}(\vec{k},\lambda)\hat{a}_{\text{H}}(\vec{k},\lambda,t)\ee^{\ii \vec{k}\cdot\vec{x}}+
                                     \varepsilon^{\mu,*}(\vec{k},\lambda)\hat{a}^{\dagger}_{\text{H}}(\vec{k},\lambda,t)\ee^{-\ii \vec{k}\cdot\vec{x}}
                                    \right] \ ,
\end{eqnarray}
and accordingly
\begin{equation}
 \label{eq:phot_mode_heis}
 \hat{a}_{\text{H}}(\vec{k},\lambda,t) = \hat{U}^{\dagger}_{\text{J}}(t,t_0)\hat{a}_{\text{J}}(\vec{k},\lambda,t)
                                         \hat{U}_{\text{J}}(t,t_0) \ .
\end{equation}
The interpretation of $\hat{a}_{\text{J}}(\vec{k},\lambda,t)$ from
(\ref{eq:2:op_phot_a}) as a single-photon operator is evident since
(\ref{eq:2:op_phot_b}) describes a free electromagnetic field. On the
other hand, (\ref{eq:phot_heis}) describes an interacting
electromagnetic field. Hence, the same interpretation for
(\ref{eq:phot_mode_heis}) is not justified in general. It is, however,
possible in the limit $t\rightarrow\pm\infty$ for free asymptotic
fields. Such fields are obtained by introducing an adiabatic switching
of the electromagnetic interaction,
\begin{equation}
 \label{eq:2:switch}
 \hat{H}_{\text{EM}}(t) \rightarrow f_{\varepsilon}(t)\hat{H}_{\text{EM}}(t) \ , \quad 
                                    \text{with} \quad f_{\varepsilon}(t)=\ee^{-\varepsilon|t|} \quad \text{and} \quad \varepsilon>0 \ .
\end{equation}
According to the Gell-Mann and Low theorem \cite{FW:2002}, such a
switching can also be applied to construct the eigenstates of an
interacting theory out of those for a non-interacting theory. Upon the
introduction of $f_{\varepsilon}(t)$, the interaction-picture time-
evolution operator turns into
\begin{equation}
 \hat{U}_{\text{J}}(t,t_0)\rightarrow\hat{U}^{\varepsilon}_{\text{J}}(t,t_0)
  = T\left\lbrace\exp\left[-\ii\int_{t_0}^{t}\dd t'f_{\varepsilon}(t')\hat{H}_{\text{J}}(t')\right]\right\rbrace \ .
\end{equation}
In order obtain physically well defined results for the photon numbers,
we hence have to specify our initial state at $t_{0}\rightarrow-\infty$
and consider eq. (\ref{eq:phot_heis}) in the limit $t\rightarrow\infty$
for free asymptotic states. Such an approach corresponds to an in/out
description as suggested in \cite{Fraga:2003sn,Fraga:2004ur}. Since the
introduction of $f_{\varepsilon}(t)$ per se is an artificial procedure,
we have to take $\varepsilon\rightarrow0$ at the end of our
calculation. As a first step, we expand
$\hat{a}_{\text{H}}(\vec{k},\lambda,t)$ to first order in the
electromagnetic coupling, $e$:
\begin{equation}
\begin{split}
  \label{eq:2:phot_expand}
  \hat{a}_{\text{H}}(\vec{k},\lambda,t) & =
  \hat{U}^{\varepsilon,\dagger}_{\text{J}}(t,-\infty)\hat{a}_{\text{J}}(\vec{k},\lambda,t)\hat{U}^{\varepsilon}_{\text{J}}(t,-\infty)
   \\
  & \approx \hat{a}_{\text{J}}(\vec{k},\lambda,t)+\ii\int_{-\infty}^{t}\dd
  t'f_{\varepsilon}(t')\left[\hat{H}_{\text{J}}(t'),
    \hat{a}_{\text{J}}(\vec{k},\lambda,t)\right] \\
  & = \hat{a}_{\text{I}}(\vec{k},\lambda,t)-\ii
  \frac{\varepsilon^{\mu}(\vec{k},\lambda)}{\sqrt{2\omega_{\vec{k}}}} \int_{-\infty}^{t}\dd
  t'\int \dd^{3}x f_{\varepsilon}(t')\hat{j}_{\mu,\text{J}}(\vec{x},t')\ee^{\ii\left[\omega_{\vec{k}}t'-\vec{k}\cdot\vec{x}\right]}
   \ ,
\end{split}
\end{equation}
where we have made use of relations (\ref{eq:2:op_phot}) and
$[\hat{a}_{\text{I}}(\vec{k},\lambda,t),\hat{j}_{\mu,\text{J}}(\vec{x},t')]=0$.
Upon insertion of (\ref{eq:2:phot_expand}) into
(\ref{eq:2:a_phot_heis}), we obtain for the photon yield
\begin{equation}
 \label{eq:2:yield_1st}
 2\omega_{\vec{k}}\frac{\dd^{6}n^{\varepsilon}_{\gamma}(t)}{\dd^{3}x\dd^{3}k} = \frac{\gamma^{\mu\nu}(k)}{(2\pi)^{3}V}\int\underline{\dd^{4}x_{1}}\int\underline{\dd^{4}x_{2}}
 \mbox{ }\ii \Pi^{<}_{\nu\mu}(x_{1},x_{2})\ee^{\ii k(x_1-x_2)} \ ,
\end{equation}
with the underline denoting that
$$ 
\int\underline{\dd^{4}x} = \int\dd^{3}x\int_{-\infty}^{t}\dd tf_{\varepsilon}(t) \ .
$$
We have introduced the photon polarization tensor
\begin{equation}
 \label{eq:2:chiral_polten}
 \gamma^{\mu\nu}(k) =
 \sum_{\lambda=\perp}\varepsilon^{\mu,*}(\vec{k},\lambda)
 \varepsilon^{\nu}(\vec{k},\lambda) = \begin{cases} 
   -\eta^{\mu\nu}-\frac{k^{\mu}k^{\nu}}{\vec{k}^{2}} &
   \quad\mbox{for}\quad\mu,\nu\in\left\lbrace1,2,3\right\rbrace \\ 0 &
   \quad \mbox{otherwise}
 \end{cases} \ ,
\end{equation}
with $\eta^{\mu\nu}=\text{diag}\left\lbrace1,-1,-1,-1\right\rbrace$ as
well as the current-current correlator
\begin{equation}
 \label{eq:2:pse_chiral}
 \ii \Pi^{<}_{\nu\mu}(x_{1},x_{2}) =
 \left \langle 
   \hat{j}^{\dagger}_{\mu,\text{J}}(x_{2})\hat{j}_{\nu,\text{J}}(x_{1})\right \rangle
 \ ,
\end{equation}
which describes the photon self-energy to first order in $\alpha_{e}$.
The average is taken with respect to the initial density matrix,
$\hat{\rho}(t_{0}\rightarrow-\infty)$, which is simply given by the
vacuum expression (\ref{eq:2:initial_density}). Hence, we have
\begin{equation}
 \label{eq:2:pse_ini_vac}
 \ii \Pi^{<}_{\nu\mu}(x_{1},x_{2}) =
 \left \langle 0_{q\bar{q}}\right|\hat{j}^{\dagger}_{\mu,\text{J}}(x_{2})\hat{j}_{\nu,\text{J}}(x_{1})\left|0_{q\bar{q}}\right \rangle
 \ .
\end{equation}
In the interaction picture representation, 'J', the current operator
(\ref{eq:2:current_schroed}) reads
\begin{equation}
 \label{eq:2:current_int}
 \hat{j}_{\mu,\text{J}}(x)=e\hat{\bar{\psi}}_{\text{J}}(x)\gamma_{\mu}\hat{\psi}_{\text{J}}(x) \ .
\end{equation}
Upon insertion of (\ref{eq:2:current_int}) into (\ref{eq:2:pse_ini_vac})
and performing a Wick decomposition, we obtain
\begin{equation}
 \label{eq:2:pse_eval}
 \ii \Pi^{<}_{\nu\mu}(x_{1},x_{2})
 = e^2\mbox{Tr}\left\lbrace\gamma_{\mu}S^{<}_{F}(x_{1},x_{2})\gamma_{\nu}S^{>}_{F}(x_{2},x_{1})\right\rbrace \ .
\end{equation}
This expression corresponds to the one-loop approximation, which is
depicted in Fig. \ref{fig:2:chiral_one_loop}.
The propagators entering expression (\ref{eq:2:pse_eval}) are given by
\begin{subequations}
 \label{eq:2:propagators}
 \begin{eqnarray}
   S^{>}_{F}(x_{1},x_{2}) & = & -\ii \left \langle \hat{\psi}_{\text{J}}(x_{1})\hat{\bar{\psi}}_{\text{J}}(x_{2})\right\rangle \nonumber \\
   & = & -\ii \sum_{s}\int\frac{\dd^{3}p}{(2\pi)^{3}}\psi_{\vec{p},s,\uparrow}(x_{1})\otimes
     \bar{\psi}_{\vec{p},s,\uparrow}(x_{2}) \ , \label{eq:2:propagators_more} \\
   S^{<}_{F}(x_{1},x_{2}) & = & \ii
   \left \langle \hat{\bar{\psi}}_{\text{J}}(x_{2})\hat{\psi}_{\text{J}}(x_{1})\right \rangle
   \nonumber \\ 
   & = & \ii \sum_{s}\int\frac{\dd^{3}p}{(2\pi)^{3}}\psi_{\vec{p},s,\downarrow}(x_{1})\otimes
     \bar{\psi}_{\vec{p},s,\downarrow}(x_{2}) \ . \label{eq:2:propagators_less}
 \end{eqnarray}
\end{subequations}
Since the spatial dependence of $\psi_{\vec{p},s,\uparrow\downarrow}(x)$
is included entirely in the factor $\ee^{\ii \vec{p}\cdot\vec{x}}$, it
is convenient to formally separate it off via
\begin{equation}
 \label{eq:2:separation}
 \psi_{\vec{p},s,\uparrow\downarrow}(x) = \psi^{'}_{\vec{p},s,\uparrow\downarrow}(t)\ee^{\ii \vec{p}\cdot\vec{x}} \ .
\end{equation}
Then $\psi^{'}_{\vec{p},s,\uparrow\downarrow}(t)$ fulfills
\begin{equation}
 \label{eq:2:dirac_mod}
 \left [\ii \gamma^{0}\partial_{t}+\gamma^{i}p_{i}-m(t)\right ]\psi^{'}_{\vec{p},s,\uparrow\downarrow}(t)=0 \ .
\end{equation}
With the help of (\ref{eq:2:separation}), we can rewrite
(\ref{eq:2:propagators}) as
\begin{subequations}
 \begin{eqnarray}
   S^{>}_{F}(x_{1},x_{2}) & = & \int\frac{\dd^{3}p}{(2\pi)^{3}}S^{>}_{F}(\vec{p},t_{1},t_{2})\ee^{\ii \vec{p}\cdot(\vec{x}_{1}-\vec{x}_{2})} \ , \\
   S^{<}_{F}(x_{1},x_{2}) & = & \int\frac{\dd^{3}p}{(2\pi)^{3}}S^{<}_{F}(\vec{p},t_{1},t_{2})\ee^{\ii \vec{p}\cdot(\vec{x}_{1}-\vec{x}_{2})} \ ,
 \end{eqnarray}
\end{subequations}
with the propagators in mixed time-momentum representation given by
\begin{subequations}
 \label{eq:2:propagators_momentum}
 \begin{eqnarray}
   S^{>}_{F}(\vec{p},t_{1},t_{2}) & = & -\ii \sum_{s}\psi^{'}_{\vec{p},s,\uparrow}(t_{1})\bar{\psi}^{'}_{\vec{p},s,\uparrow}(t_{2}) \ , \\
   S^{<}_{F}(\vec{p},t_{1},t_{2}) & = & \ii\sum_{s}\psi^{'}_{\vec{p},s,\downarrow}(t_{1})\bar{\psi}^{'}_{\vec{p},s,\downarrow}(t_{2}) \ .
 \end{eqnarray}
\end{subequations}
By virtue of eq. (\ref{eq:2:propagators_momentum}) we can express the
photon self-energy in mixed time-momentum representation by
\begin{subequations}
 \label{eq:2:pse_momentum_chiral}
 \begin{eqnarray}
   \ii \Pi^{<}_{\mu\nu}(x_{1},x_{2})         & = & \int\frac{\dd^{3}p}{(2\pi)^{3}}\mbox{ }\ii \Pi^{<}_{\mu\nu}(\vec{p},t_{1},t_{2})
   \ee^{\ii \vec{p}\cdot(\vec{x}_{1}-\vec{x}_{2})} \label{eq:2:pse_momentum_chiral_a} \ , \\
   \ii \Pi^{<}_{\mu\nu}(\vec{p},t_{1},t_{2}) & = & e^2\int\frac{\dd^{3}q}{(2\pi)^{3}}\mbox{Tr}
   \left\lbrace\gamma_{\mu}S^{<}(\vec{p}+\vec{q},t_{1},t_{2})
     \gamma_{\nu}S^{>}(\vec{q},t_{2},t_{1})\right\rbrace \label{eq:2:pse_momentum_chiral_b} \ .
 \end{eqnarray}
\end{subequations}
Finally, this enables us to rewrite the expression for the photon yield
(\ref{eq:2:yield_1st}) as
\begin{equation}
 \label{eq:2:photon_yield_chiral_eps}
 2\omega_{\vec{k}}\frac{\dd^{6}n^{\varepsilon}_{\gamma}(t)}{\dd^{3}x\dd^{3}k} = \frac{1}{(2\pi)^{3}}\int_{-\infty}^{t}\dd t_{1}\int_{-\infty}^{t}\dd t_{2}
 \mbox{ }f_{\varepsilon}(t_{1})f_{\varepsilon}(t_{2})\ii\Pi^{<}_{T}(\vec{k},t_{1},t_{2})\ee^{\ii\omega_{\vec{k}}(t_{1}-t_{2})} \ , 
\end{equation}
where we have introduced
$$
 \ii\Pi^{<}_{T}(\vec{k},t_{1},t_{2}) = \gamma^{\mu\nu}(k)\ii \Pi^{<}_{\nu\mu}(\vec{k},t_{1},t_{2}) \ .
$$
We show in Appendix~\ref{sec:appa} that
(\ref{eq:2:photon_yield_chiral_eps}) can be written as the (space-time
integrated) absolute square of a first-order QED transition amplitude
and is thus positive (semi-) definite. Therefore, it cannot adopt unphysical
negative values.

As in \cite{Michler:2009hi}, the photon self-energy entering
(\ref{eq:2:photon_yield_chiral_eps}) is given by the one-loop
approximation (\ref{eq:2:pse_momentum_chiral_b}). The crucial
difference, however, is that the underlying scenario has been addressed
within a first-principle description. In particular, the propagators
entering (\ref{eq:2:pse_momentum_chiral_b}) are determined by the
equations of motion,
\begin{subequations}
 \label{eq:2:eom_propm}
 \begin{eqnarray}
   \left(\ii \gamma^{0}\partial_{t_{1}}+\gamma^{i}p_{i}-m(t_{1})\right)S^{\lessgtr}_{F}(\vec{p},t_{1},t_{2}) & = & 0 \ , \\
   \left(\ii \gamma^{0}\partial_{t_{2}}-\gamma^{i}p_{i}+m(t_{2})\right)S^{\lessgtr}_{F}(\vec{p},t_{1},t_{2}) & = & 0 \ .
 \end{eqnarray}
\end{subequations}
From (\ref{eq:2:eom_propm}) it follows that our description fulfills the
Ward-Takahashi identities for the photon self-energy,
\begin{equation}
 \label{eq:2:wti}
 \partial_{t_{1}}\ii \Pi^{<}_{0\mu}(\vec{k},t_{1},t_{2})-\ii k^{j}\ii \Pi^{<}_{j\mu}(\vec{k},t_{1},t_{2}) = 0 \ ,
\end{equation}
and is hence consistent with $U(1)$ gauge invariance. A more explicit
verification of (\ref{eq:2:wti}) with (\ref{eq:2:pse_momentum_chiral_b})
expressed in terms of the wavefunction parameters
(\ref{eq:2:wave_param}) is given in Appendix~\ref{sec:appb}. Moreover,
the observable photon yield is extracted from
(\ref{eq:2:photon_yield_chiral_eps}) for free asymptotic states by
successively taking the limits $t\rightarrow\infty$ and then
$\varepsilon\rightarrow0$ after the time integrations have been
performed, i.e.,
\begin{equation}
 \label{eq:2:photon_yield_chiral_aspt}
  2\omega_{\vec{k}}\frac{\dd^{6}n_{\gamma}}{\dd^{3}x\dd^{3}k} = 
   \lim_{\varepsilon\rightarrow0}\frac{1}{(2\pi)^{3}}\int_{-\infty}^{\infty}\dd t_{1}\int_{-\infty}^{\infty}\dd t_{2}
 f_{\varepsilon}(t_{1})f_{\varepsilon}(t_{2})\ii\Pi^{<}_{T}(\vec{k},t_{1},t_{2})\ee^{\ii\omega_{\vec{k}}(t_{1}-t_{2})} \ .
\end{equation}

Taking into account that (\ref{eq:2:chiral_polten}) has a purely
spacelike structure, it follows from (\ref{eq:b:chiral_pseij}) that $\ii
\Pi^{<}_{T}(\vec{k},t_{1},t_{2})$ reads in terms of wavefunction parameters
\begin{equation}
\begin{split}
 \label{eq:2:pse_contract_chiral}
 \ii \Pi^{<}_{T}(\vec{k},t_{1},t_{2}) = 4e^2\int\frac{\dd^{3}p}{(2\pi)^{3}} &
 \left[
   \alpha^{*}_{\vec{p},\uparrow}(t_{1})\beta_{\vec{p}+\vec{k},\downarrow}(t_{1})
   \beta^{*}_{\vec{p}+\vec{k},\downarrow}(t_{2})\alpha_{\vec{p},\uparrow}(t_{2})
 \right.  \\
 &
 +\beta^{*}_{\vec{p},\uparrow}(t_{1})\alpha_{\vec{p}+\vec{k},\downarrow}(t_{1})
 \alpha^{*}_{\vec{p}+\vec{k},\downarrow}(t_{2})\beta_{\vec{p},\uparrow}(t_{2})
  \\
 -\frac{x(px+\omega_{\vec{k}})}{|\vec{p}+\vec{k}|} & \left(
   \alpha^{*}_{\vec{p},\uparrow}(t_{1})\beta_{\vec{p}+\vec{k},\downarrow}(t_{1})
   \alpha^{*}_{\vec{p}+\vec{k},\downarrow}(t_{2})\beta_{\vec{p},\uparrow}(t_{2})
 \right.  \\
 & \left.+\left.
     \beta^{*}_{\vec{p},\uparrow}(t_{1})\alpha_{\vec{p}+\vec{k},\downarrow}(t_{1})
     \beta^{*}_{\vec{p}+\vec{k},\downarrow}(t_{2})\alpha_{\vec{p},\uparrow}(t_{2})
   \right)\right] \ ,
\end{split}
\end{equation}
with $p$ and $x$ denoting the absolute value of the fermion momentum, $\vec{p}$, and the cosine of the angle between $\vec{p}$ and
$\vec{k}$, respectively. It follows from (\ref{eq:2:icon_param}) that
(\ref{eq:2:pse_contract_chiral}) reduces to the vacuum polarization
if both time arguments, $t_{1}$ and $t_{2}$, are taken from the domain
where the fermion mass is still at its initial value, $m_{c}$,
\begin{eqnarray}
  \label{eq:2:chiral_vacpol}
  \ii \Pi^{<}_{T,0}(\vec{k},t_{1},t_{2}) & = &   \ii \Pi^{<}_{T,0}(\vec{k},t_{1}-t_{2}) \nonumber \\
  & = &   2e^2\int\frac{\dd^{3}p}{(2\pi)^{3}}
  \left\lbrace
    1+\frac{px(px+\omega_{\vec{k}})+m^{2}_{c}}{E^{c}_{\vec{p}}E^{c}_{\vec{p}+\vec{k}}}
  \right\rbrace
  \ee^{\ii \left(E^{c}_{\vec{p}+\vec{k}}+E^{c}_{\vec{p}}\right)(t_{1}-t_{2})} \ .                                      
\end{eqnarray}
Due to the chiral mass shift, (\ref{eq:2:pse_contract_chiral}) will
acquire an additional non-stationary contribution,
\begin{equation}
 \label{eq:2:pse_decomp}
 \ii \Pi^{<}_{T}(\vec{k},t_{1},t_{2}) = \ii \Pi^{<}_{T,0}(\vec{k},t_{1}-t_{2})+ \ii
 \Delta \Pi^{<}_{T}(\vec{k},t_{1},t_{2}) \ ,
\end{equation}
depending on both time arguments separately. In appendix \ref{sec:appa}
we show that the contribution from the vacuum polarization
(\ref{eq:2:chiral_vacpol}) vanishes when taking the successive limits
$t\rightarrow\infty$ and $\varepsilon\rightarrow0$ so that only
contributions from mass-shift effects characterized by
$\ii\Delta\Pi^{<}_{T}(\vec{k},t_{1},t_{2})$ remain. Thereby, we also
point out that keeping \emph{this} order of limits is indeed
crucial to eliminate the vacuum contribution and that the latter shows
up again if the limits are interchanged. Moreover, we
  demonstrate in appendix \ref{sec:appf} that adhering to the correct
  order of limits is also essential to obtain physically reasonable
  results from the mass-shift effects.  Together with
(\ref{eq:2:pse_contract_chiral}), expression
(\ref{eq:2:photon_yield_chiral_aspt}) describes photon production
induced by the chiral mass shift at first order in $\alpha_{e}$ but to
all orders in $g$.

\section{Pair production from dynamical mass shifts}
\label{sec:pair-production}
Before we turn to the numerical investigations on photon production
arising from the chiral mass shift, we first provide an insertion on
quark-pair production. It has been shown in
\cite{Greiner:1995ac,Greiner:1996wz} that the asymptotic quark/antiquark
occupation numbers are highly sensitive to the order of
differentiability of the considered mass parametrization, $m(t)$. We
are now going to extend these investigations to the time dependence of
the quark and antiquark occupation numbers for different mass functions,
$m(t)$. We consider pair production arising from the chiral mass shift
only. The starting point for our considerations is hence the fermionic
part of the Hamiltonian in the interaction picture, J,
\begin{align}
 \label{eq:3:hamilton_density}
 \hat{H}_{\text{J}}(t) & = \int \dd^{3}x \;
 \hat{\bar{\psi}}_{\text{J}}(x) \left[-\ii
   \vec{\gamma}\cdot\vec{\nabla} + m(t) \right]
 \hat{\psi}_{\text{J}}(x) \ .
\end{align}
To simplify the notation, the subscript `J' is dropped from now on. With
the help of (\ref{eq:2:expansion}) and (\ref{eq:2:wave_param}) we can
rewrite (\ref{eq:3:hamilton_density}) as
\begin{align}
 \label{eq:3:hamiltonian_short}
 \hat{H}(t) = \sum_{s}\int\frac{\dd^{3}p}{(2\pi)^{3}}
               \left\lbrace\Omega(t)\left[
                                     \hat{b}^{\dagger}_{\vec{p},s}\hat{b}_{\vec{p},s}-
                                     \hat{d}_{-\vec{p},r}\hat{d}^{\dagger}_{-\vec{p},s}
                                     \right]+
                \Lambda(t)\hat{b}^{\dagger}_{\vec{p},s}\hat{d}^{\dagger}_{-\vec{p},s}+
                \Lambda^{*}(t)\hat{d}_{-\vec{p},s}\hat{b}_{\vec{p},s}
               \right\rbrace \ ,
\end{align}
where we have introduced
\begin{subequations}
 \label{eq:3:hamilton_matrix}
 \begin{align}
\begin{split}
   \Omega(t) & = \bar{\psi}_{\vec{p},s,\uparrow}(x)\left [-\ii\vec{\gamma}\cdot\vec{\nabla}+m(t)\right ]
                 \psi_{\vec{p},s,\uparrow}(x)  \\
             & = -\bar{\psi}_{\vec{p},s,\downarrow}(x)\left [-\ii\vec{\gamma}\cdot\vec{\nabla}+m(t)\right ]
                  \psi_{\vec{p},s,\downarrow}(x)  \\
             & = p\left [\alpha^{*}_{\vec{p},\uparrow,}(t)\beta_{\vec{p},\uparrow}(t)+
                 \beta^{*}_{\vec{p},\uparrow}(t)\alpha_{\vec{p},\uparrow}(t)\right ]+
                 m(t)\left [\left|\alpha_{\vec{p},\uparrow}(t)\right|^{2}-
                 \left|\beta_{\vec{p},\uparrow,}(t)\right|^{2}\right
               ] \label{eq:3:hamilton_matrix_a} \ ,
\end{split}
 \\
\begin{split}
   \Lambda(t) & =
   \bar{\psi}_{\vec{p},s,\uparrow}(x)\left [-\ii\vec{\gamma}\cdot\vec{\nabla}+m(t)\right ]
   \psi_{\vec{p},s,\downarrow}(x)  \\
   & =
   p\left [\alpha^{*}_{\vec{p},\uparrow}(t)\beta_{\vec{p},\downarrow}(t)+
     \beta^{*}_{\vec{p},\uparrow}(t)\alpha_{\vec{p},\downarrow}(t)\right ]+
   m(t)\left [\alpha^{*}_{\vec{p},\uparrow}(t)\alpha_{\vec{p},\downarrow}(t)-
     \beta^{*}_{\vec{p},\uparrow,}(t)\beta_{\vec{p},\downarrow}(t)\right ] \label{eq:3:hamilton_matrix_b}
   \ .
\end{split}
 \end{align}
\end{subequations}
The second equality in (\ref{eq:3:hamilton_matrix_a}) follows
immediately from (\ref{eq:2:param_link}). Introducing
\begin{equation} 
  \hat{A}(t) = \begin{pmatrix}
                \Omega(t)      & \Lambda(t)   \\
                \Lambda^{*}(t) & -\Omega(t) 
               \end{pmatrix}
 \text{,}
\end{equation}
we can rewrite (\ref{eq:3:hamiltonian_short}) in an even more compact
form,
\begin{equation}
 \label{eq:3:hamiltonian_matrix}
 \hat{H}(t) = \sum_{s}\int\frac{\dd^{3}p}{(2\pi)^{3}} 
              \begin{pmatrix}
               \hat{b}_{\vec{p},s} \\
               \hat{d}^{\dagger}_{-\vec{p},s}
              \end{pmatrix}^{\dagger}
              \hat{A}(t)
              \begin{pmatrix}
               \hat{b}_{\vec{p},s} \\
               \hat{d}^{\dagger}_{-\vec{p},s}
              \end{pmatrix} \ .
\end{equation}
Now the particle number density is extracted from
(\ref{eq:3:hamiltonian_matrix}) by diagonalizing this expression with
respect to $\hat{b}_{\vec{p},s}$ and $\hat{d}_{-\vec{p},s}$ via a
Bogolyubov transformation \cite{FW:2002},
\begin{equation}
 \label{eq:3:Bogolyubov}
   \begin{pmatrix}
     \hat{\tilde{b}}_{\vec{p},s}(t) \\
     \hat{\tilde{d}}^{\dagger}_{-\vec{p},s}(t)
   \end{pmatrix}
 = \begin{pmatrix}
    \xi_{\vec{p},s}(t)       & \eta_{\vec{p},s}(t)    \\
    -\eta^{*}_{\vec{p},s}(t) & \xi^{*}_{\vec{p},s}(t)
    \end{pmatrix}
    \begin{pmatrix}
      \hat{b}_{\vec{p},s} \\
      \hat{d}^{\dagger}_{-\vec{p},s}
    \end{pmatrix}
 \equiv  \hat{C}(t)
          \begin{pmatrix}
          \hat{b}_{\vec{p},s} \\
          \hat{d}^{\dagger}_{-\vec{p},s}
         \end{pmatrix} \ .
\end{equation}
In order to maintain the anticommutation relations
(\ref{eq:2:commutation_fermions}) under (\ref{eq:3:Bogolyubov}), the
Bogolyubov coefficients have to satisfy the relation
\begin{equation}
  \label{eq:3:normalization_param}
  \left|\xi_{\vec{p},s}(t)\right|^{2}+\left|\eta_{\vec{p},s}(t)\right|^{2} = 1 \ .
\end{equation}
Furthermore, they have to fulfill the initial conditions,
\begin{subequations}
  \label{eq:icon_bog}
  \begin{eqnarray}
  \xi_{\vec{p},s}(t_{0})  & = & 1 \ , \label{eq:3:icon_bog_xi} \\
  \eta_{\vec{p},s}(t_{0}) & = & 0 \ . \label{eq:3:icon_bog_eta}
 \end{eqnarray}
\end{subequations}
The Bogolyubov particle number density for given momentum, $\vec{p}$, and spin,
$s$, is then defined as \cite{FW:2002}
\begin{equation}
\begin{split}
 \label{eq:3:bogolyubov_numbers}
 \frac{\dd^{6}n_{q\bar{q}}(t)}{\dd^{3}x\dd^{3}p} & =
 \frac{1}{(2\pi)^{3}V}\sum_{s} \left \langle
   \hat{\tilde{b}}^{\dagger}_{\vec{p},s}(t)
   \hat{\tilde{b}}_{\vec{p},s}(t)
 \right \rangle \\
 & = \frac{1}{(2\pi)^{3}V}\sum_{s} \left \langle
   \hat{\tilde{d}}^{\dagger}_{-\vec{p},s}(t)
   \hat{\tilde{d}}_{-\vec{p},s}(t)
 \right \rangle \\
 & = \frac{1}{(2\pi)^{3}}\sum_{s}\left|\eta_{\vec{p},s}(t)\right|^{2}.
\end{split}
\end{equation}
By virtue of (\ref{eq:3:Bogolyubov}), we can obviously rewrite
(\ref{eq:3:hamiltonian_short}) as
\begin{equation}
\begin{split}
 \label{eq:3:hamiltonian_bogolyubov}
  \hat{H}(t) & =  \sum_{s}\int\frac{\dd^{3}p}{(2\pi)^{3}}
                   \begin{pmatrix}
                    \hat{\tilde{b}}_{\vec{p},s}(t) \\
                    \hat{\tilde{d}}^{\dagger}_{-\vec{p},s}(t)
                   \end{pmatrix}^{\dagger}
                   \hat{C}(t)\hat{A}(t)\hat{C}^{\dagger}(t)
                   \begin{pmatrix}
                     \hat{\tilde{b}}_{\vec{p},s}(t) \\
                     \hat{\tilde{d}}^{\dagger}_{-\vec{p},s}(t)
                   \end{pmatrix} \\
             & =  \sum_{s}\int\frac{\dd^{3}p}{(2\pi)^{3}}
             \begin{pmatrix}
                    \hat{\tilde{b}}_{\vec{p},s}(t) \\
                    \hat{\tilde{d}}^{\dagger}_{-\vec{p},s}(t)
                   \end{pmatrix}^{\dagger}
                   \hat{A}_{B}(t)
                   \begin{pmatrix}
                    \hat{\tilde{b}}_{\vec{p},s}(t) \\
                    \hat{\tilde{d}}^{\dagger}_{-\vec{p},s}(t)
                   \end{pmatrix} \ .
\end{split}
\end{equation}
The adjoint matrix $\hat{C}^{\dagger}(t)$ denotes the inverse
transformation of (\ref{eq:3:Bogolyubov}) and reads
\begin{equation} 
  \label{eq:3:Bogolyubov_inverse} 
  \hat{C}^{\dagger}(t) = \begin{pmatrix}
    \xi^{*}_{\vec{p},s}(t)  & -\eta_{\vec{p},s}(t)   \\
    \eta^{*}_{\vec{p},s}(t) &  \xi_{\vec{p},s}(t)
                         \end{pmatrix} \ .
\end{equation}
For $\hat{A}_{B}(t)$ to be diagonal, $\xi_{\vec{p},s}(t)$ and
$\eta_{\vec{p},s}(t)$ have to be determined such that
\begin{equation}
 v_{+}(t) = \begin{pmatrix}
             \xi^{*}_{\vec{p},s}(t) \\
             \eta^{*}_{\vec{p},s}(t)
            \end{pmatrix}
            \quad\mbox{,}\quad
            v_{-}(t) = \begin{pmatrix}
              -\eta_{\vec{p},s}(t) \\
              \xi_{\vec{p},s}(t)
            \end{pmatrix} \ ,
\end{equation}
are the (orthonormal) eigenvectors of $\hat{A}(t)$ for the respective
eigenvalues, $\lambda_{\pm}(t)$. These are obtained as
\begin{equation}
\begin{split}
  & \quad\text{det}\left[\hat{A}(t)-\lambda_{\pm}(t)\hat{I}\right] = 0 \\
  \Leftrightarrow & \quad\lambda_{\pm}(t) = \pm\sqrt{\Omega^{2}(t)+|\Lambda(t)|^{2}} \ .
\end{split}
\end{equation}
With the help of (\ref{eq:2:param_normalization}) and
(\ref{eq:2:param_link}), they are further evaluated to
\begin{eqnarray}
 \quad\lambda_{\pm}(t) = E_{\vec{p}}(t) \ .
\end{eqnarray}
Here we have introduced the dispersion relation
\begin{equation}
 \label{eq:3:dispersion}
 E_{\vec{p}}(t) = \sqrt{p^{2}+m^{2}(t)} \ .
\end{equation}
In terms of the transformed operators (\ref{eq:3:Bogolyubov}), the
Hamiltonian (\ref{eq:3:hamiltonian_short}) then reads
\begin{eqnarray}
 \label{eq:3:hamiltonian_transform}
 \hat{H}(t) & =           & \sum_{s}\int\frac{\dd^{3}p}{(2\pi)^{3}}E_{\vec{p}}(t)
                             \left[
                              \hat{\tilde{b}}^{\dagger}_{\vec{p},s}(t)\hat{\tilde{b}}_{\vec{p},s}(t)-
                              \hat{\tilde{d}}_{-\vec{p},s}(t)\hat{\tilde{d}}^{\dagger}_{-\vec{p},s}(t)
                             \right] \nonumber \\
            & \rightarrow & \sum_{s}\int\frac{\dd^{3}p}{(2\pi)^{3}}E_{\vec{p}}(t)
                             \left[
                              \hat{\tilde{b}}^{\dagger}_{\vec{p},s}(t)\hat{\tilde{b}}_{\vec{p},s}(t)+
                              \hat{\tilde{d}}^{\dagger}_{-\vec{p},s}(t)\hat{\tilde{d}}_{-\vec{p},s}(t)
                             \right] \ .
\end{eqnarray}
In the second step, $\hat{H}(t)$ has been normal ordered with respect to
(\ref{eq:3:Bogolyubov}) in order to avoid an infinitely negative vacuum
energy. It follows immediately from (\ref{eq:3:hamiltonian_transform})
and (\ref{eq:3:Bogolyubov}) that the energy density is given by
\begin{eqnarray}
 \label{eq:3:energy_density}
 \frac{\dd^{3}E_{q\bar{q}}(t)}{\dd^{3}x} 
  & = & \frac{1}{V}\left\langle 0_{q\bar{q}}\right|\hat{H}(t)\left|0_{q\bar{q}}\right\rangle \nonumber \\
  & = & 2\sum_{s}\int\frac{\dd^{3}p}{(2\pi)^{3}}E_{\vec{p}}(t)\left|\eta_{\vec{p},s}(t)\right|^{2} \ .
\end{eqnarray}
We see that it corresponds to (\ref{eq:3:bogolyubov_numbers}) integrated
over the momentum modes, $\vec{p}$, which justifies the definition of
(\ref{eq:3:bogolyubov_numbers}) as the particle-number density at a
given time, $t$. The additional factor of $2$ in
(\ref{eq:3:energy_density}) arises from the fact that this expression
describes the energy density carried by quarks and antiquarks together
whereas (\ref{eq:3:bogolyubov_numbers}) corresponds to the number of
quarks which is equal to the number of antiquarks. Together with
\begin{equation}
 \left[\hat{A}(t)-\lambda_{\pm}(t)\hat{I}\right]v_{\pm}(t) = \pm E_{\vec{p}}(t)v_{\pm}(t) \ ,
\end{equation}
we obtain the following linear system of equations for
$\xi_{\vec{p},s}(t)$ and $\eta_{\vec{p},s}(t)$,
\begin{subequations}
 \label{eq:bogolubov_param_sys}
 \begin{eqnarray}
  \Omega(t)\xi_{\vec{p},s}(t)+\Lambda^{*}(t)\eta_{\vec{p},s}(t) & = & E_{\vec{p}}(t)\xi_{\vec{p},s}(t) \ , \\
  \Lambda(t)\xi_{\vec{p},s}(t)-\Omega(t)\eta_{\vec{p},s}(t)     & = & E_{\vec{p}}(t)\eta_{\vec{p},s}(t) \ .
 \end{eqnarray}
\end{subequations}
In appendix \ref{sec:appc} it is shown that this system is solved by
\begin{subequations}
 \label{eq:3:bogolyubov_param}
 \begin{eqnarray}
   \xi_{\vec{p},s}(t)  & = & \ee^{\ii E_{\vec{p}}(t)t}\left[
     \sqrt{\frac{E_{\vec{p}}(t)+m(t)}{2E_{\vec{p}}(t)}}\alpha_{\vec{p},\uparrow}(t)+
     \sqrt{\frac{E_{\vec{p}}(t)-m(t)}{2E_{\vec{p}}(t)}}\beta_{\vec{p},\uparrow}(t)
   \right] \ , \label{eq:3:bogolyubov_param_xi} \\
   \eta_{\vec{p},s}(t) & = & \ee^{\ii E_{\vec{p}}(t)t}\left[
     \sqrt{\frac{E_{\vec{p}}(t)+m(t)}{2E_{\vec{p}}(t)}}\alpha_{\vec{p},\downarrow}(t)+
     \sqrt{\frac{E_{\vec{p}}(t)-m(t)}{2E_{\vec{p}}(t)}}\beta_{\vec{p},\downarrow}(t)
   \right] \ . \label{eq:3:bogolyubov_param_eta}
 \end{eqnarray}
\end{subequations}
The phase factor, $\ee^{\ii E_{\vec{p}}(t)t}$, has been introduced to
satisfy the initial condition (\ref{eq:3:icon_bog_xi}). Furthermore, it
allows us to rewrite $\hat{\psi}(x)$ in terms of positive- and
negative-energy wavefunctions of the respective momentary mass, $m(t)$,
i.e.,
\begin{equation}
 \label{eq:3:reexpansion}
 \hat{\psi}(x) = \sum_{s}\int\frac{\dd^{3}p}{(2\pi)^3}\left[
                                                     \hat{\tilde{b}}_{\vec{p},s}(t)\tilde{\psi}_{\vec{p},s,\uparrow}(x)+
                                                     \hat{\tilde{d}}^{\dagger}_{-\vec{p},s}(t)\tilde{\psi}_{\vec{p},s\downarrow}(x)
                                                    \right] \ ,
\end{equation}
with $\tilde{\psi}_{\vec{p},s\uparrow\downarrow}(x)$ given by
\begin{subequations}
 \begin{eqnarray}
   \tilde{\psi}_{\vec{p},s\uparrow}(x)     & = & \begin{pmatrix}
     \cos\varphi_{\vec{p}}(t)\mbox{ }\chi_{s} \\
     \sin\varphi_{\vec{p}}(t)\mbox{ }\frac{\vec{\sigma}\cdot\vec{p}}{p}\chi_{s}
                                                \end{pmatrix}
                                                \ee^{-\ii E_{\vec{p}}(t)t}\ee^{\ii \vec{p}\cdot\vec{x}} \ , \\
   \tilde{\psi}_{\vec{p},s\downarrow}(x) & = & \begin{pmatrix}
                                                 \sin\varphi_{\vec{p}}(t)\mbox{ }\chi_{s} \\
                                                 -\cos\varphi_{\vec{p}}(t)\mbox{ }\frac{\vec{\sigma}\cdot\vec{p}}{p}\chi_{s}
                                                \end{pmatrix}
                                                \ee^{+iE_{\vec{p}}(t)t}\ee^{\ii \vec{p}\cdot\vec{x}} \ .
 \end{eqnarray}
\end{subequations}
In analogy to (\ref{eq:2:sincos}), we have introduced
\begin{subequations}
 \begin{eqnarray}
   \cos\varphi_{\vec{p}}(t) & = & \sqrt{\frac{E_{\vec{p}}(t)+m(t)}{2E_{\vec{p}}(t)}} \ , \\
   \sin\varphi_{\vec{p}}(t) & = & \sqrt{\frac{E_{\vec{p}}(t)-m(t)}{2E_{\vec{p}}(t)}} \ .
 \end{eqnarray}
\end{subequations}
The Bogolyubov transformation (\ref{eq:3:bogolyubov_numbers}) hence corresponds to a reexpansion of the 
fermion-field operators in terms of the instantaneous eigenstates of the Hamilton-density operator
\begin{equation}
  \hat{h}_{D}(t) = -\ii\gamma_{0}\vec{\gamma}\cdot\vec{\nabla}+\gamma_{0}m(t) \ , 
\end{equation}
which is demonstrated in greater detail in appendix \ref{sec:appc}. The same procedure 
has been applied in \cite{Filatov:2007ha}. The crucial difference to our approach is that the 
authors us an expansion in the form of (\ref{eq:3:reexpansion}) to derive the equations of motion 
for the field operators (\ref{eq:3:Bogolyubov}) and eventually a kinetic equation for 
the Bogolyubov particle number density (\ref{eq:3:bogolyubov_numbers}) from the Dirac equation with a 
time-dependent mass. To the contrary, we extract the Bogolyubov parameters and hence the 
particle number density by translating (\ref{eq:3:Bogolyubov}) into relations between 
$\psi_{\vec{p},s,\uparrow\downarrow}(x)$ and $\tilde{\psi}_{\vec{p},s,\uparrow\downarrow}(x)$ and 
projecting the Bogolyubov parameters out of the latter (also see Eqs. (\ref{eq:c:wavef_trans})-(\ref{eq:c:projection}) in appendix \ref{sec:appc}).

With the help of (\ref{eq:2:param_link}), both $\xi_{\vec{p},s}(t)$ and
$\eta_{\vec{p},s}(t)$ can be expressed alternatively in terms of
(complex conjugated) negative- and positive-energy wavefunction
parameters, respectively. The Bogolyubov particle number density
(\ref{eq:3:bogolyubov_numbers}) thus reads
\begin{equation}
 \label{eq:3:bogolyubov_final}
  \frac{\dd^{6}n_{q\bar{q}}(t)}{\dd^{3}x\dd^{3}p} = 
   \frac{1}{(2\pi)^{3}}
   \left\lbrace
   1+\frac{2p
    \mathrm{Re} \left [
      \alpha^{*}_{\vec{p},\downarrow}(t)\beta_{\vec{p},\downarrow}(t)
    \right] +m(t) \left [
      \left|\alpha_{\vec{p},\downarrow}(t)\right|^{2}-
      \left|\beta_{\vec{p},\downarrow}(t)\right|^{2} \right
    ]}{E_{\vec{p}}(t)}
 \right\rbrace \ .
\end{equation}
We see that it can be expressed entirely in terms of the negative-energy
wavefunction parameters, $\alpha_{\vec{p},\downarrow}(t)$ and
$\beta_{\vec{p},\downarrow}(t)$. This is a result one would also expect
intuitively since $\psi_{\vec{p},s,\downarrow}(x)$ describes initially a
negative-energy state but then also acquires a positive-energy component
from the mass shift measured by (\ref{eq:3:bogolyubov_final}). This
expression is, indeed, just the absolute square of the projection of
$\psi_{\vec{p},s,\downarrow}(x)$ on
$\tilde{\psi}_{\vec{p},s,\uparrow}(x)$ summed over the spin index,
$s$. For completeness we mention that (\ref{eq:3:bogolyubov_final}) can
also be obtained by projecting the respective propagator
(\ref{eq:2:propagators_less}) on the positive-energy wavefunction of
the respective current mass, $m(t)$,
\begin{eqnarray}
  \frac{\dd^{6}n_{q\bar{q}}(t)}{\dd^{3}x\dd^{3}p} & = & -\frac{\ii}{(2\pi)^{3}V}\left.\sum_{s}\int \dd^{3}x_{1}\int \dd^{3}x_{2}
                                     \tilde{\psi}^{\dagger}_{\vec{p},s,\uparrow}(x_{1})
                                     S^{<}_{F}(x_1,x_2)\gamma_{0}\psi_{\vec{p},s,\uparrow}(x_2)
                                     \right|_{t_{1}=t_{2}=t} \ ,
\end{eqnarray}
which has been used in \cite{Greiner:1995ac,Greiner:1996wz} in the
asymptotic limit $t\rightarrow\infty$. Thus
(\ref{eq:3:bogolyubov_final}) generalizes the result therein to finite
times, $t$.

\begin{figure}[htb]
 \begin{center}
  \includegraphics[height=6.0cm]{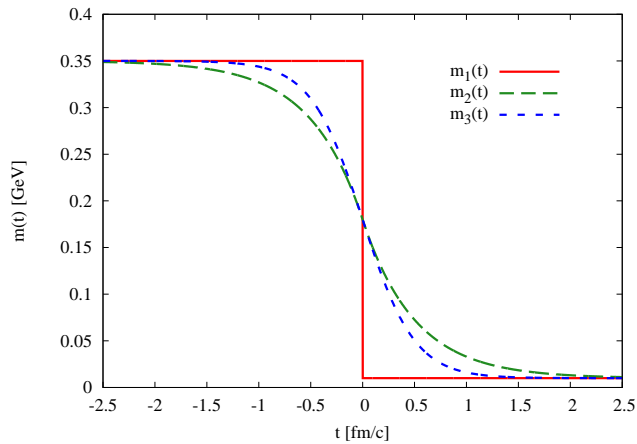}
  \caption{During the chiral phase transition, the quark/antiquark mass
    changes from its constituent value, $m_{c}$, to its bare value,
    $m_{b}$.}
  \label{fig:3:chiral_masschange}
 \end{center}
\end{figure}

For our investigations on the time dependence of
(\ref{eq:3:bogolyubov_final}), we model the change of the fermion mass
from its initial constituent value, $m_c$, to its final bare value,
$m_{b}$, by three different mass parameterizations,
\begin{subequations}
 \label{eq:3:chiral_massparam}
 \begin{eqnarray}
   m_{1}(t) & = & \frac{m_{c}+m_{b}}{2}-\frac{m_{c}-m_{b}}{2}\mbox{ }\mbox{sign}(t) \ , \label{eq:3:chiral_massparam_1} \\
   m_{2}(t) & = & \frac{m_{c}+m_{b}}{2}-\frac{m_{c}-m_{b}}{2}\mbox{ }\mbox{sign}(t)\left(1-\ee^{-2|t|/\tau}\right) \ , \label{eq:3:chiral_massparam_2} \\
   m_{3}(t) & = & \frac{m_{c}+m_{b}}{2}-\frac{m_{c}-m_{b}}{2}\mbox{ }\mbox{tanh}\left(\frac{2t}{\tau}\right) \ , \label{eq:3:chiral_massparam_3}
 \end{eqnarray}
\end{subequations}
with $\mbox{sign}(t)$ given by
$$
 \mbox{sign}(t) = \begin{cases}
                    1  & t>0 \\
                    -1 & t<0 \\
                    0  & t=0
                  \end{cases} \ .
$$
The mass parameterizations
(\ref{eq:3:chiral_massparam_1})-(\ref{eq:3:chiral_massparam_3}) are
depicted in Fig.~\ref{fig:3:chiral_masschange}. Analogously to
\cite{Greiner:1995ac,Greiner:1996wz}, we have chosen $m_{c}=0.35 \;
\GeV$ and $m_{b}=0.01 \; \GeV$ and assumed a transition time of
$\tau=1.0 \; \fm/c$.

It has been shown in \cite{Greiner:1995ac,Greiner:1996wz} that the
asymptotic occupation numbers are very sensitive to the order of
differentiability of the considered mass parametrization, $m(t)$. For
the case of an instantaneous mass shift described by
(\ref{eq:3:chiral_massparam_1}), the equation of motion
(\ref{eq:2:dirac_chiral_wf}) for the positive- and negative-energy wave
function is solved analytically with the ansatz
\begin{subequations}
 \label{eq:3:ansatz_inst}
 \begin{eqnarray}
   \psi_{\vec{p},s,\uparrow}(x)   & = & \begin{cases}
     \psi^{c}_{\vec{p},s,\uparrow}(x)                   & \mbox{for}\quad t<0 \ , \\
     \alpha_{\vec{p},s}\psi^{b}_{\vec{p},s,\uparrow}(x)+
     \beta_{\vec{p},s}\psi^{b}_{\vec{p},s,\downarrow}(x)  & \mbox{for}\quad t\ge0 \ ,
   \end{cases} \\
   \psi_{\vec{p},s,\downarrow}(x) & = & \begin{cases}
     \psi^{c}_{\vec{p},s,\downarrow}(x)                   & \mbox{for}\quad t<0 \ , \\
     \gamma_{\vec{p},s}\psi^{b}_{\vec{p},s,\downarrow}(x)+
     \delta_{\vec{p},s}\psi^{b}_{\vec{p},s,\uparrow}(x) &
     \mbox{for}\quad t\ge0 \ .
   \end{cases}
 \end{eqnarray}
\end{subequations}
For $t<0$, $\psi_{\vec{p},s,\uparrow}(x)$ and
$\psi_{\vec{p},s,\downarrow}(x)$ describe positive- and negative-energy
states of mass $m_{c}$, respectively, whereas they turn into
superpositions of positive- and negative-energy states of mass $m_{b}$
for $t\ge0$. From the continuity condition
\begin{equation}
 \label{eq:3:continuity}
 \psi_{\vec{p},s,\uparrow\downarrow}(\vec{x},0^{-})=\psi_{\vec{p},s,\uparrow\downarrow}(\vec{x},0^{+}) \ ,
\end{equation}
we obtain
\begin{subequations}
 \begin{eqnarray}
  \label{eq:3:coeff_link}
  \alpha_{\vec{p},s} = & \gamma_{\vec{p},s} & = \cos\varphi^{b}_{\vec{p}}\cos\varphi^{c}_{\vec{p}}+\sin\varphi^{b}_{\vec{p}}\sin\varphi^{c}_{\vec{p}} \ , \\
  \beta_{\vec{p},s}  = & -\delta_{\vec{p},s} & = \sin\varphi^{b}_{\vec{p}}\cos\varphi^{c}_{\vec{p}}-\cos\varphi^{b}_{\vec{p}}\sin\varphi^{c}_{\vec{p}} \ .
 \end{eqnarray}
\end{subequations}
As the coefficients $\alpha_{\vec{p},s}$ and $\beta_{\vec{p},s}$ do not
explicitly depend on the spin, $s$, this index will be omitted from now
on. The occupation numbers thus read
\begin{equation}
\begin{split}
 \label{eq:3:chiral_partinst}
 \frac{\dd^{6}n_{q\bar{q}}(t)}{\dd^{3}x\dd^{3}p} &= \frac{2\beta^{2}_{\vec{p}}}{(2\pi)^{3}} \\
 &= \frac{1}{(2\pi)^{3}}\left[1-\frac{p^{2}+m_{b}m_{c}}{E^{b}_{\vec{p}}E^{c}_{\vec{p}}}\right]
\end{split}
\end{equation}
for $t>0$, whereas they vanish for $t<0$. For $p\gg m_{b},m_{c}$, the
expression (\ref{eq:3:chiral_partinst}) can be approximated as
\begin{equation}
 \label{eq:3:chiral_partinst_approx}
 \frac{\dd^{6}n_{q\bar{q}}(t)}{\dd^{3}x\dd^{3}p} \simeq \frac{(m_{c}-m_{b})^{2}}{(2\pi)^{3}2p^{2}}+\mathcal{O}(1/p^{4}) \ ,
\end{equation}
which means that the total particle number density and the total energy
density of the fermionic sector are linearly and quadratically
divergent, respectively.

\begin{figure}[htb]
 \begin{center}
  \includegraphics[height=6.0cm]{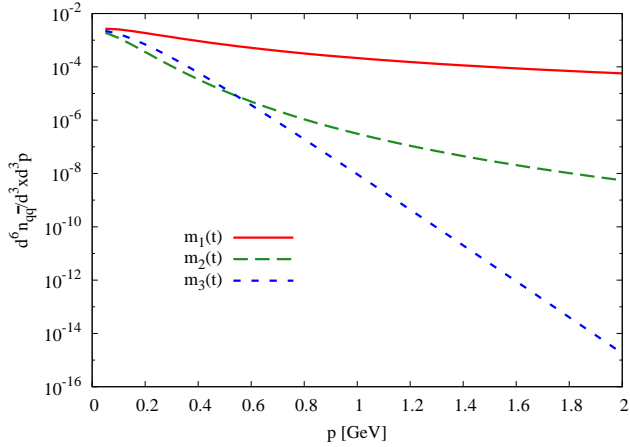}
  \caption{Asymptotic particle spectra for the different 
           mass parameterizations given in Eqs. (\ref{eq:3:chiral_massparam}). The 
           decay behavior is highly sensitive to the order of differentiability of $m(t)$. Both for 
           $m_{2}(t)$ and $m_{3}(t)$, we have chosen $\tau=1.0$ fm/c.}
  \label{fig:3:chiral_asymptcomp}
 \end{center}
\end{figure}

This artifact can be removed if the mass shift is assumed to take place
over a finite time interval, $\tau$. In this case, the occupation
numbers are obtained by solving (\ref{eq:2:eom_param}) numerically for
the negative-energy wavefunction parameters,
$\alpha_{\vec{p}¸\downarrow,}(t)$ and $\beta_{\vec{p}¸\downarrow}(t)$,
which are then inserted into
(\ref{eq:3:bogolyubov_final}). Fig. \ref{fig:3:chiral_asymptcomp}
compares the asymptotic particle spectra for the different mass
parameterizations, $m_{i}(t)$.

\begin{figure}[htb]
 \center
 \includegraphics[height=5.0cm]{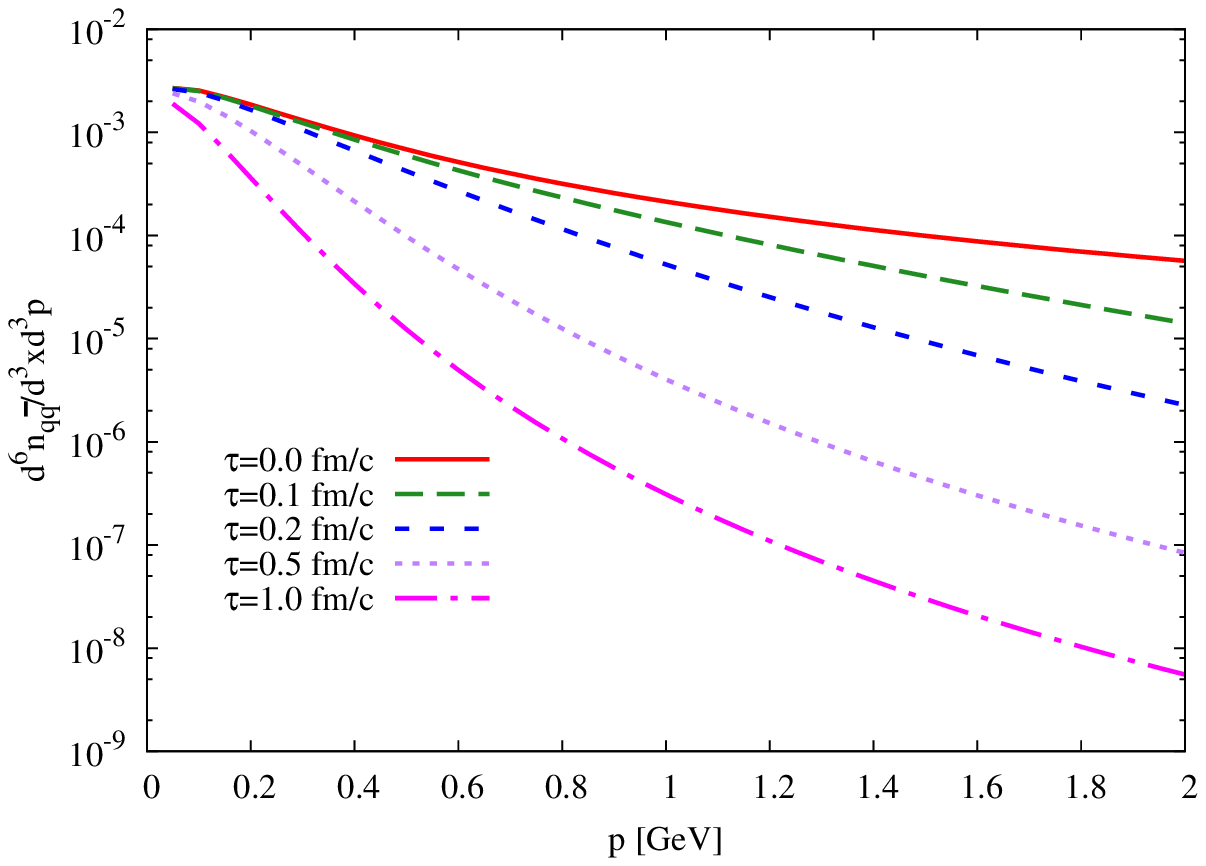}
 \includegraphics[height=5.0cm]{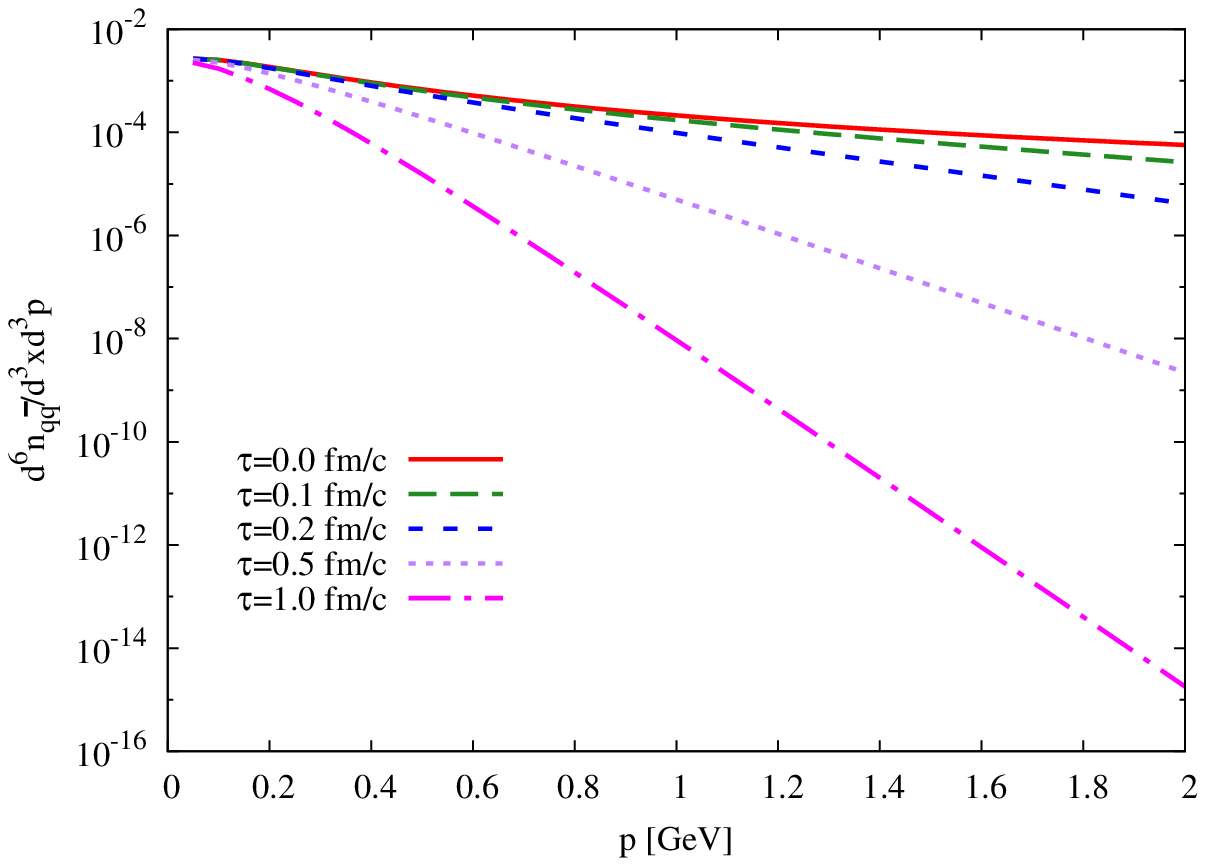}
 \caption{Asymptotic particle spectra for $m_{2}(t)$ (left panel) and $m_{3}(t)$ (right panel) 
          for different transition times, $\tau$. In each case, the suppression at large $p$ with respect to an instantaneous mass shift is the stronger the 
          more slowly ($\tau$ increasing) the mass shift is assumed to take place. As expected, both 
          parameterizations reproduce expression (\ref{eq:3:chiral_partinst}) in the limit $\tau\rightarrow 0$.}
 \label{fig:3:chiral_tau0}
\end{figure}

Analogously to \cite{Greiner:1995ac,Greiner:1996wz}, we find that if we
turn from $m_{1}(t)$ to $m_{2}(t)$, which is continuously differentiable
once, the occupation numbers decay $\propto 1/p^{6}$ and are hence
suppressed relative to the case with the instantaneous
transition. Moreover, if we turn from $m_{2}(t)$ to $m_{3}(t)$, which is
continuously differentiable infinitely many times, the occupation
numbers are further suppressed to an exponential decay. In the limit
$\tau\rightarrow0$, both $m_{2}(t)$ and $m_{3}(t)$ reproduce expression
(\ref{eq:3:chiral_partinst}), which is depicted in
Fig. \ref{fig:3:chiral_tau0}.

\begin{figure}[htb]
 \center
 \includegraphics[height=5.0cm]{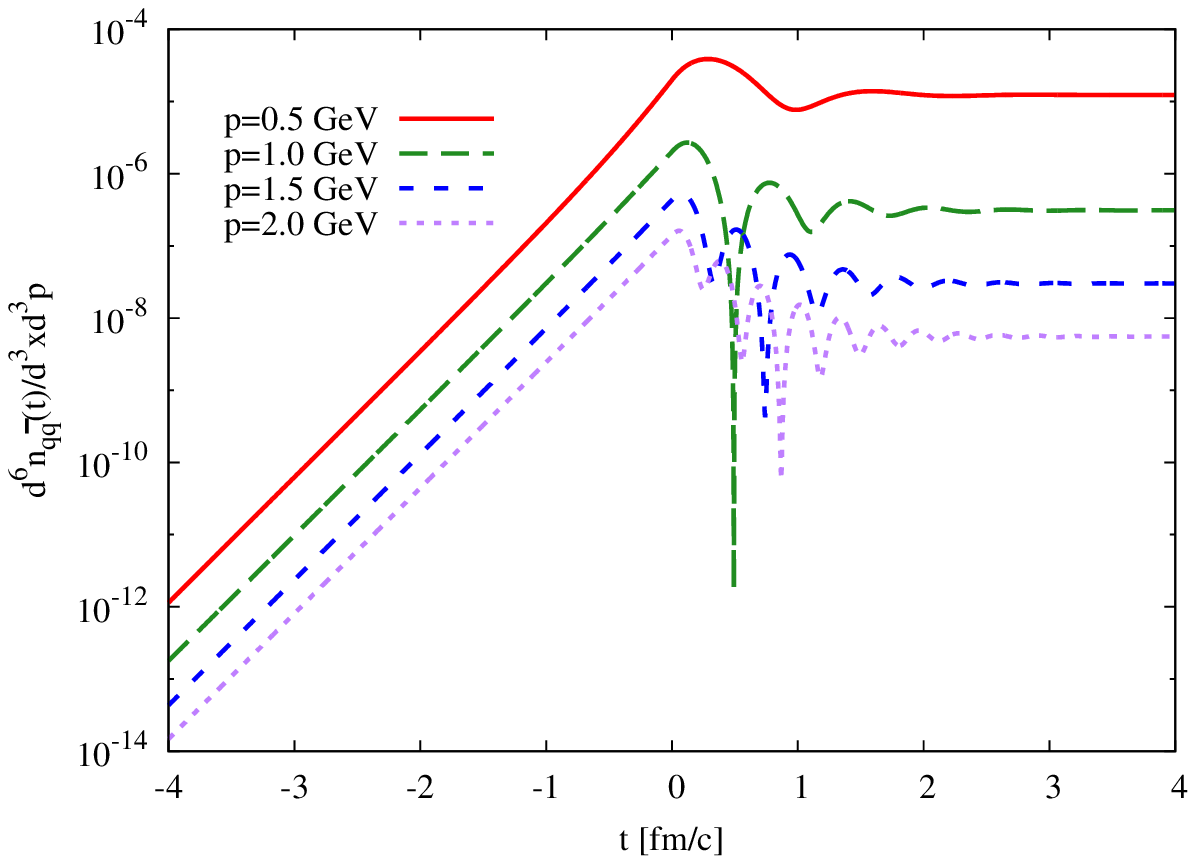}
 \includegraphics[height=5.0cm]{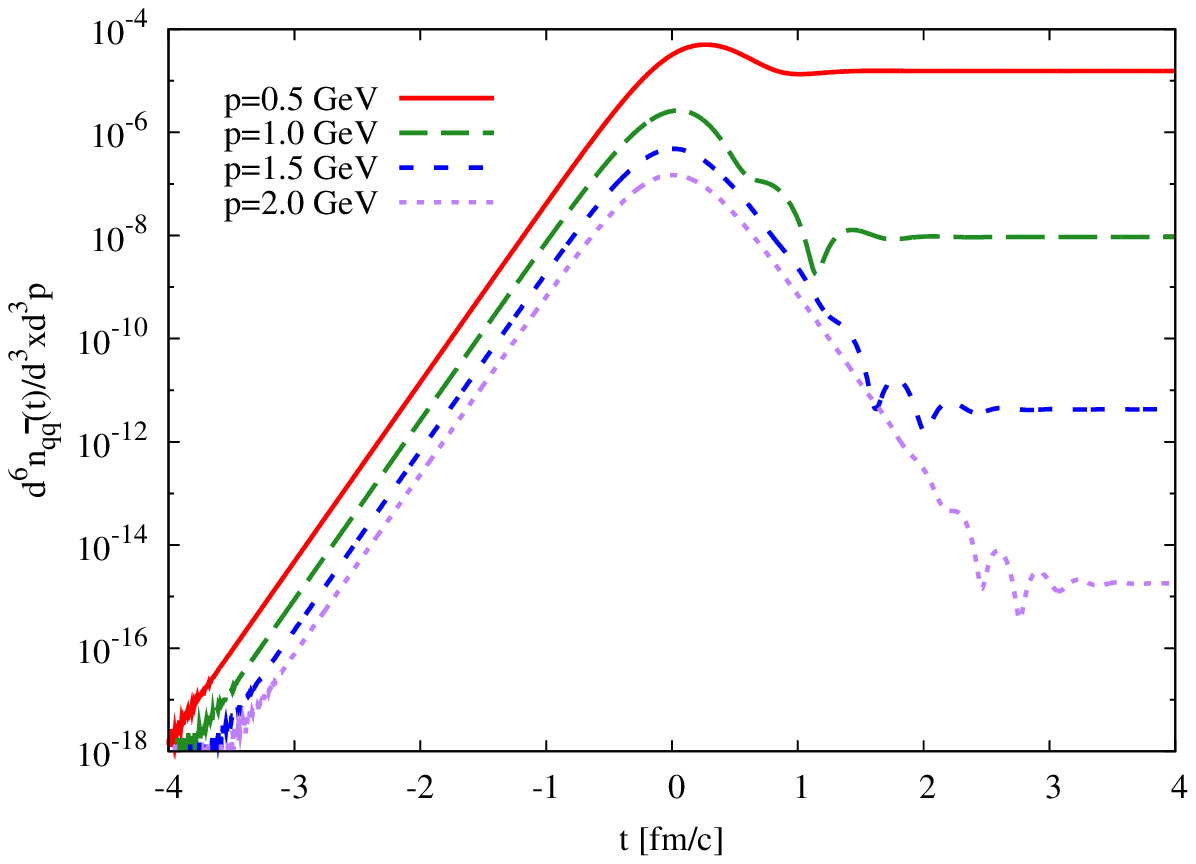}
 \caption{Time dependence of occupation numbers for $m_{2}(t)$ (left
   panel) and $m_{3}(t)$ (right panel) for $\tau=1.0$ fm/c. For large values of $p$, they
   exhibit an 'overshoot' over their asymptotic value around $t=0$, which
   is particularly distinctive for $m_{3}(t)$.}
 \label{fig:3:chiral_timedep}
\end{figure}

\begin{figure}[htb]
 \center
 \includegraphics[height=5.0cm]{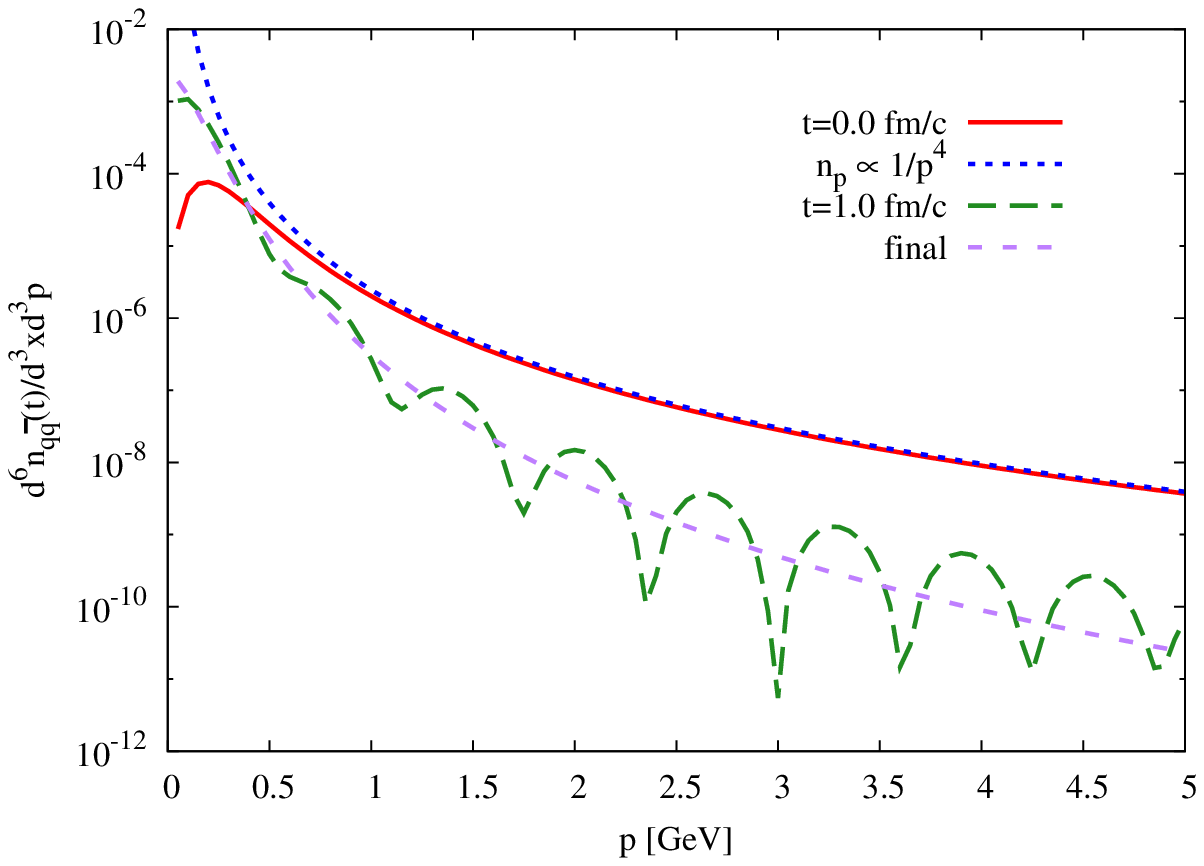}
 \includegraphics[height=5.0cm]{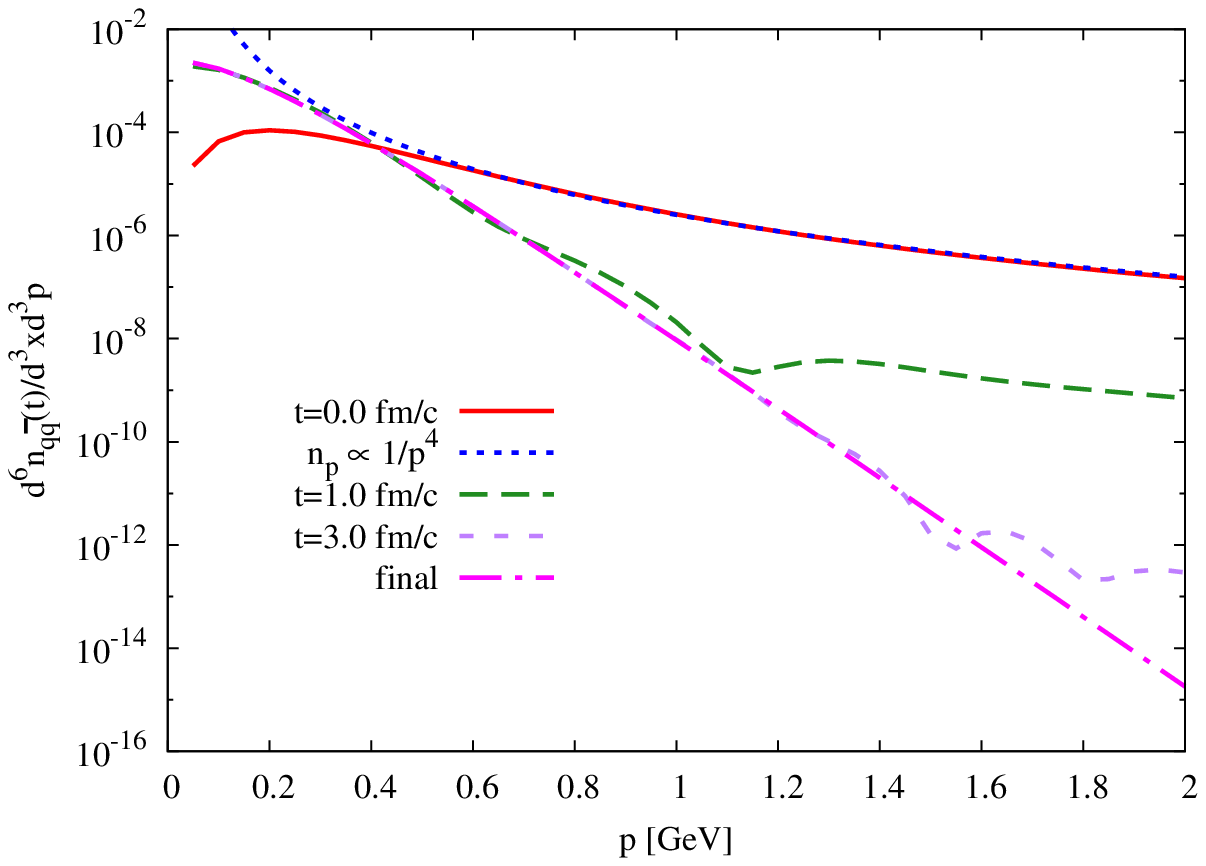}
 \caption{Particle spectra for $m_{2}(t)$ (left panel) and $m_{3}(t)$
   (right panel) at different times, $t$, for a transition time 
   of $\tau=1.0$ fm/c. For both parameterizations, the
   characteristic decay behavior only emerges for $t\rightarrow\infty$,
   whereas one encounters a decay behavior $\propto 1/p^{4}$ for large
   $p$ around $t=0$.}
 \label{fig:3:chiral_timedepspec}
\end{figure}

We shall briefly explain how the sensitivity of the asymptotic
occupation numbers on the mass parametrization, $m_{i}(t)$, comes
about. For this purpose, we consider the Bogolyubov parameters in terms
of negative-energy wavefunction parameters,
\begin{subequations}
 \label{eq:3:chiral_bog}
 \begin{eqnarray}
   \xi^{*}_{\vec{p},s}(t) & = & \sqrt{\frac{E_{\vec{p}}(t)-m(t)}{2E_{\vec{p}}(t)}}\alpha^{*}_{\vec{p},\downarrow}(t)-
   \sqrt{\frac{E_{\vec{p}}(t)+m(t)}{2E_{\vec{p}}(t)}}\beta^{*}_{\vec{p},\downarrow}(t)
   \ , \label{eq:3:chiral_bog_xi} \\
   \eta_{\vec{p},s}(t)    & = & \sqrt{\frac{E_{\vec{p}}(t)+m(t)}{2E_{\vec{p}}(t)}}\alpha_{\vec{p},\downarrow}(t)+
   \sqrt{\frac{E_{\vec{p}}(t)-m(t)}{2E_{\vec{p}}(t)}}\beta_{\vec{p},\downarrow}(t)
   \ , \label{eq:3:chiral_bog_eta}
 \end{eqnarray}
\end{subequations}
with (\ref{eq:3:chiral_bog_xi}) following from
(\ref{eq:3:bogolyubov_param_xi}) and (\ref{eq:2:param_link}). We do not
take into account the phase factor, $\ee^{\ii E_{\vec{p}}(t)t}$, as it
drops out when taking the absolute square of (\ref{eq:3:chiral_bog_eta})
to obtain the Bogolyubov particle number density
(\ref{eq:3:bogolyubov_numbers}). It follows from (\ref{eq:2:eom_param})
that $\xi^{*}_{\vec{p},s}(t)$ and $\eta_{\vec{p},s}(t)$ then obey the
equations of motion,
\begin{subequations}
 \begin{eqnarray}
   \ii \partial_{t}\xi^{*}_{\vec{p},s}(t)  & = & -\ii \frac{p\dot{m}(t)}{2E^{2}_{\vec{p}}(t)}\eta_{\vec{p},s}(t)-
   E_{\vec{p}}(t)\xi^{*}_{\vec{p},s}(t) \ , \\
   \ii \partial_{t}\eta_{\vec{p},s}(t)     & = & \ii
   \frac{p\dot{m}(t)}{2E^{2}_{\vec{p}}(t)}\xi^{*}_{\vec{p},s}(t)+ E_{\vec{p}}(t)\eta_{\vec{p},s}(t) \ .
 \end{eqnarray}
\end{subequations}
In the limit $p\gg m(t)$, these equations of motion are approximately
solved by
\begin{equation}
 \begin{split}
   \xi^{*}_{\vec{p},s}(t) & =  \ee^{\ii pt} \ ,\\
   \eta_{\vec{p},s}(t) & = \frac{\ee^{-\ii pt}}{2p}\int_{t_{0}}^{t}\dd
   t'\dot{m}(t')\ee^{2 \ii pt'} \ .
 \end{split}
\end{equation}
Hence, the Bogolyubov particle number density in that domain reads
\begin{equation}
 \label{eq:3:occup_approx}
 \frac{\dd^{6}n_{q\bar{q}}(t)}{\dd^{3}x\dd^{3}p} \simeq
 \frac{(m_{c}-m_{b})^{2}}{(2\pi)^{3}2p^{2}}\left|\int_{-\infty}^{t}\dd t'\chi(t')\ee^{2 \ii pt'}\right|^{2}
 \ ,
\end{equation}
where we have formally introduced $\chi(t)$ by means of the relation
$\dot{m}(t)=\chi(t)(m_{c}-m_{b})$ and taken $t_{0}\rightarrow-\infty$
since $\dot{m}(t)\rightarrow0$ for $t\rightarrow\pm\infty$. In
particular, in the asymptotic limit, $t\rightarrow\infty$, we have for
the quark/antiquark occupation numbers,
\begin{equation}
 \label{eq:3:occup_approx_asympt}
 \left.\frac{\dd^{6}n_{q\bar{q}}(t)}{\dd^{3}x\dd^{3}p}\right|_{t\rightarrow\infty} \simeq \frac{(m_{c}-m_{b})^{2}}{(2\pi)^{3}2p^{2}}\left|\int_{-\infty}^{\infty}\dd t'\chi(t')\ee^{2 \ii pt'}\right|^{2} \ .
\end{equation}
So for a mass shift over a finite time interval, $\tau$,
(\ref{eq:3:chiral_partinst_approx}) is effectively modulated with the
absolute square of the Fourier transform of $\chi(t)$ from $t$ to
$2p$. So the particle numbers for $p\gg m_{b},m_{c}$ are suppressed by
an additional factor of $1/p^{2}$ each time the order of
differentiability of $m(t)$ is increased by one. In 
particular, for $m(t)=m_{2}(t)$ we have
\begin{equation}
 \left.\frac{\dd^{6}n_{q\bar{q}}(t)}{\dd^{3}x\dd^{3}p}\right|_{t\rightarrow\infty} \simeq \frac{(m_{c}-m_{b})^{2}}{(2\pi)^{3}2p^{2}}\frac{1}{(1+p^{2}\tau^{2})^{2}} \ .
\end{equation}
As a next step, we extend the investigations in
\cite{Greiner:1995ac,Greiner:1996wz} to the time dependence of the
occupation numbers. Fig.~\ref{fig:3:chiral_timedep} shows the time
evolution of (\ref{eq:3:bogolyubov_final}) for different momentum modes.
Here we see that for hard momentum modes, the occupation numbers exhibit
a strong `overshoot' over their asymptotic values by several orders of
magnitude in the region of strong mass gradients. This means that the
particle spectra exhibit their decay behavior characteristic for the
order of differentiability of $m(t)$ only in the limit
$t\rightarrow\infty$, which is depicted in
Fig. \ref{fig:3:chiral_timedepspec}.

\begin{figure}[htb]
 \center
 \includegraphics[height=5.0cm]{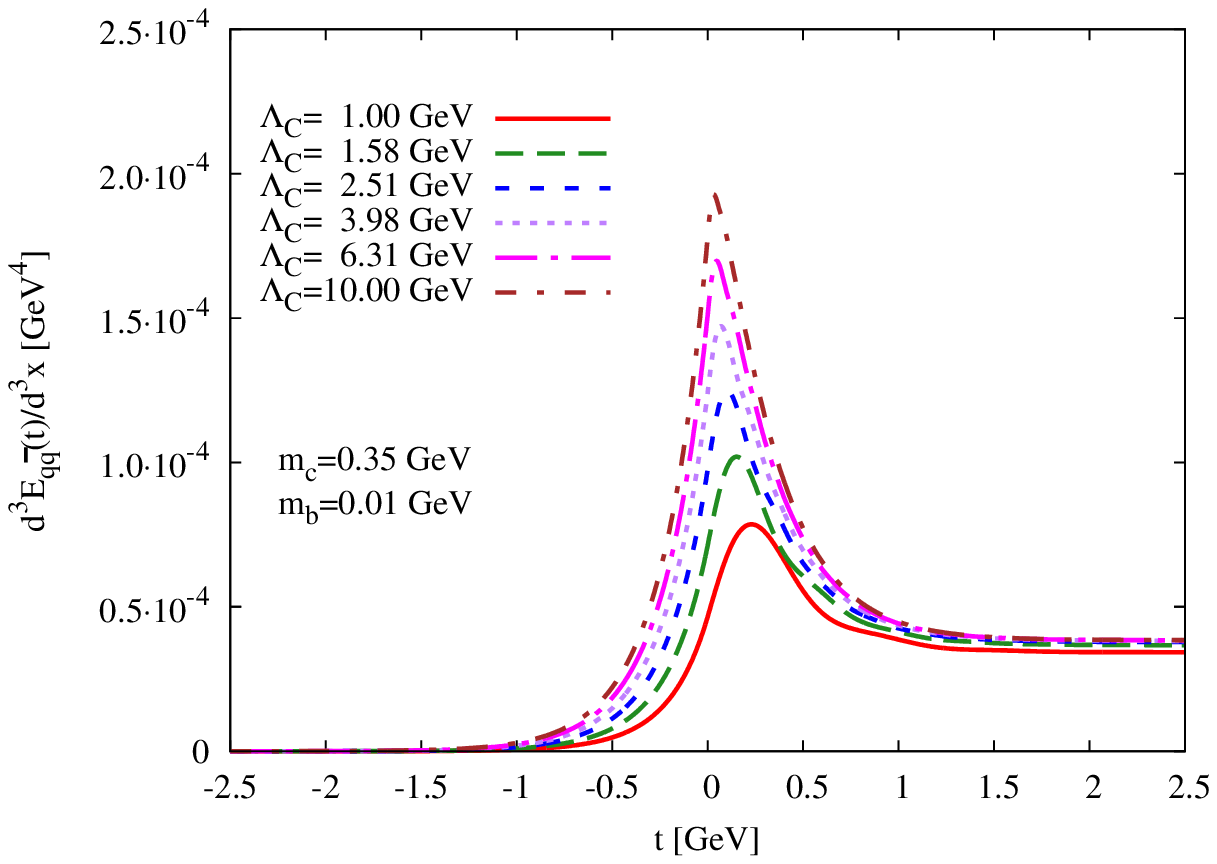}
 \includegraphics[height=5.0cm]{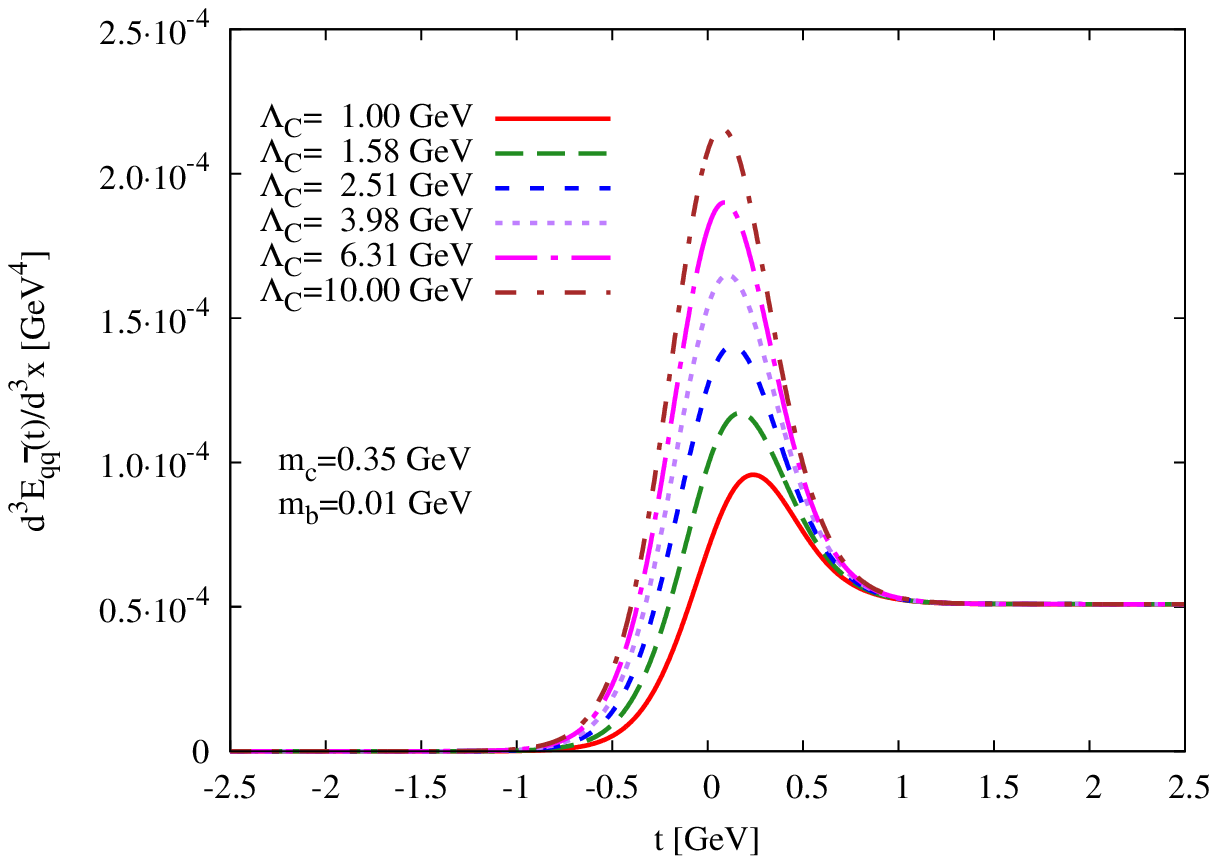}
 \caption{Time evolution of the total energy density for $m(t)=m_{2}(t)$ (left
   panel) and $m(t)=m_{3}(t)$ (right panel) for $\tau=1.0$ fm/c. It features a logarithmic
   divergence in the regions of strong mass gradients, which disappears again as soon as 
   the quark/antiquark mass has reached its final bare value, $m_{b}$.}
 \label{fig:3:chiral_logdiv}
\end{figure}

At finite times, however, the particle spectra decay as $1/p^{4}$ both
for $m_{2}(t)$ and for $m_{3}(t)$ for $p\gg m_{b},m_{c}$. This can be
understood by taking into account that at intermediate times the time
integral in (\ref{eq:3:occup_approx}) runs from $-\infty$ to $t$, so
that one effectively carries out a Fourier transform over a
discontinuous function.  Hence, expression (\ref{eq:3:bogolyubov_final})
picks up an additional factor of $1/p^{2}$ compared to the case of an
instantaneous mass shift. This implies that at finite times, $t$, the
total particle number density is finite whereas the total energy density
features a logarithmic divergence. This is depicted in
Fig. \ref{fig:3:chiral_logdiv} showing its time evolution with the loop
integral entering (\ref{eq:3:energy_density}) being regulated by a
cutoff at different values of $p=\Lambda_{C}$.
The considered values of $\Lambda_{C}$ follow an exponentially
increasing sequence given by
$$
 \Lambda_{C}(i) = \Lambda_{1}\ee^{0.2\ln{\frac{\Lambda_{2}}{\Lambda_{1}}}i} \ ,
$$
with $\Lambda_{1}=1$ GeV, $\Lambda_{2}=10$ GeV, and $i$ going from $0$
to $5$.  This choice is such that in the region of strong mass
gradients, the total energy density always increases by a constant
amount for consecutive $\Lambda_{C}(i)$, which reflects the logarithmic
divergence of energy density (\ref{eq:3:energy_density}). This
divergence, however, only shows up in regions of strong mass gradients
and disappears again as soon as the fermion mass has reached its final
value. Among other things, this can also be inferred from
Fig. \ref{fig:3:chiral_logdiv}.

For $m(t)=m_{2}(t)$ and $t>0$, the particle spectrum does not decay
strictly monotonously for $p\gg m_{b},m_{c}$, but instead exhibits an
oscillatory behavior. This can be understood by taking into account that
\begin{align}
 \left|\int_{-\infty}^{t}\dd t'\chi(t')e^{2\ii pt'}\right|^{2}
  = & \frac{e^{-2t/\tau}}{4(1+p^{2}\tau^{2})}
      +\frac{1-e^{-2t/\tau}\left(\cos{2pt}-p\tau\sin{2pt}\right)}{(1+p^{2}\tau^{2})^{2}}
\end{align}
for $t\ge0$ from which the oscillatory behavior of
(\ref{eq:3:occup_approx}) in $p$ at positive times becomes apparent. For
$t\rightarrow\infty$, the oscillating terms disappear so that the
asymptotic particle spectrum shows a strictly monotonous decay again.

\begin{figure}[htb]
 \begin{center}
  \includegraphics[height=6.0cm]{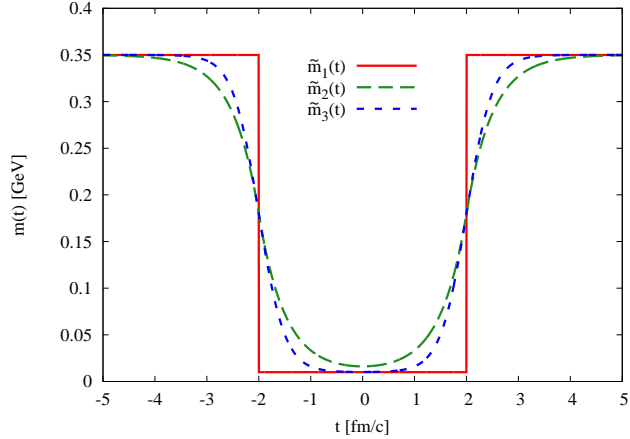}
  \caption{The mass is first changed from $m_{c}$ to $m_{b}$ and then
    back to $m_{c}$ again to simulate the temporary restoration of
    chiral symmetry.}
  \label{fig:3:chiral_masschange_back}
 \end{center}
\end{figure}

For the sake of completeness, we also consider the case where the
fermion mass first undergoes a change from its constituent value,
$m_{c}$, to its bare value, $m_{b}$, and then again back to its
constituent value, $m_{c}$, after a certain period of time, $\tau_{L}$,
to simulate the temporary restoration of chiral symmetry during a
heavy-ion collision. Again we consider different mass parameterizations,
\begin{subequations}
 \label{eq:3:chiral_massparam_back}
 \begin{align}
  \tilde{m}_{1}(t) = \frac{m_{c}+m_{b}}{2}-\frac{m_{c}-m_{b}}{2} & \text{sign}\left(\frac{\tau_{L}^{2}}{4}-t^{2}\right) \ , 
  \label{eq:3:chiral_massparam_back1} \\
  \tilde{m}_{2}(t) = \frac{m_{c}+m_{b}}{2}-\frac{m_{c}-m_{b}}{2} &
  \mbox{sign}\left(t+\frac{\tau_{L}}{2}\right)
  \left(1-\ee^{-|2t+\tau_{L}|/\tau}\right) \nonumber \\
  \times & \mbox{sign}\left(\frac{\tau_{L}}{2}-t\right)
  \left(1-\ee^{-|2t-\tau_{L}|/\tau}\right) \ ,
  \label{eq:3:chiral_massparam_back2} \\
  \tilde{m}_{3}(t) = \frac{m_{c}+m_{b}}{2}-\frac{m_{c}-m_{b}}{2} &
  \mbox{tanh}\left(\frac{2t+\tau_{L}}{\tau}\right)
  \mbox{tanh}\left(\frac{\tau_{L}-2t}{\tau}\right) \ .
  \label{eq:3:chiral_massparam_back3}
 \end{align}
\end{subequations}
Here $\tau_{L}$ denotes the lifetime of the quark-gluon plasma during
which the chiral symmetry is restored. For our numerical investigations,
we choose $\tau_{L}=4$ fm/c for which
(\ref{eq:3:chiral_massparam_back1})-(\ref{eq:3:chiral_massparam_back3})
are depicted in Fig. \ref{fig:3:chiral_masschange_back}.

\begin{figure}[htb]
 \begin{center}
  \includegraphics[height=6.0cm]{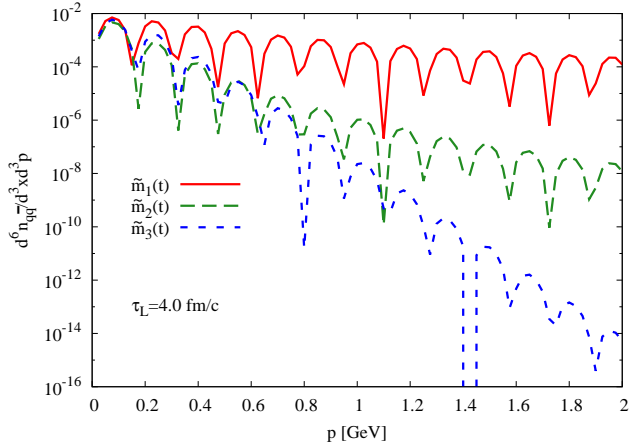}
  \caption{Asymptotic particle spectra for the different mass
    parameterizations in Eqs. (\ref{eq:3:chiral_massparam_back}).  Analogously
    to the first scenario where the mass is not restored to its initial value, 
    the decay behavior is highly sensitive to
    the order of differentiability of $\tilde{m}(t)$. Both for
    $\tilde{m}_{2}(t)$ and for $\tilde{m}_{3}(t)$ we have chosen $\tau=1.0
    \; \fm/c$.}
  \label{fig:3:chiral_asymptcomp_back}
 \end{center}
\end{figure}

For the case of $\tilde{m}_{1}(t)$, where both mass shifts are
assumed to take place instantaneously, the Dirac equation
(\ref{eq:2:dirac_chiral_wf}) is solved with an ansatz similar to
(\ref{eq:3:ansatz_inst}):
\begin{subequations}
 \begin{eqnarray}
   \psi_{\vec{p},s,\uparrow}(x)    & = & \begin{cases}
     \psi^{c}_{\vec{p},s,\uparrow}(x)                            & \mbox{for}\quad t<-\tau_{L}/2 \ , \\
     \alpha_{\vec{p}}\psi^{b}_{\vec{p},s,\uparrow}(x)+
     \beta_{\vec{p}}\psi^{b}_{\vec{p},s,\downarrow}(x)           & \mbox{for}\quad -\tau_{L}/2\le t\le\tau_{L}/2 \ , \\
     \tilde{\alpha}_{\vec{p}}\psi^{c}_{\vec{p},s,\uparrow}(x)+
     \tilde{\beta}_{\vec{p}}\psi^{c}_{\vec{p},s,\downarrow}(x)   & \mbox{for}\quad t>\tau_{L}/2 \ ,
                                        \end{cases} \\
   \psi_{\vec{p},s,\downarrow}(x) & = & \begin{cases}
                                         \psi^{c}_{\vec{p},s,\downarrow}(x)                          & \mbox{for}\quad t<-\tau_{L}/2 \ , \\
                                         \gamma_{\vec{p}}\psi^{b}_{\vec{p},s,\downarrow}(x)+
                                         \delta_{\vec{p}}\psi^{b}_{\vec{p},s,\uparrow}(x)            & \mbox{for}\quad -\tau_{L}/2\le t\le\tau_{L}/2 \ , \\
                                         \tilde{\gamma}_{\vec{p}}\psi^{c}_{\vec{p},s,\downarrow}(x)+
                                         \tilde{\delta}_{\vec{p}}\psi^{c}_{\vec{p},s,\uparrow}(x)    & \mbox{for}\quad t>\tau_{L}/2 \ .
                                        \end{cases} 
\end{eqnarray}
\end{subequations}
The successive application of the continuity conditions
$\psi_{\vec{p},s,\uparrow\downarrow}(\vec{x},-\tau/2^{-})=\psi_{\vec{p},s,\uparrow\downarrow}(\vec{x},-\tau/2^{+})$
and
$\psi_{\vec{p},s,\uparrow\downarrow}(\vec{x},\tau/2^{-})=\psi_{\vec{p},s,\uparrow\downarrow}(\vec{x},\tau/2^{+})$
leads to
\begin{subequations}
 \begin{eqnarray}
   \alpha_{\vec{p}}         & = & \gamma^{*}_{\vec{p}} \nonumber \\
   & = & \ee^{\ii (E^{c}_{\vec{p}}-E^{b}_{\vec{p}})\tau_{L}/2}
   \left(
     \cos\varphi^{b}_{\vec{p}}\cos\varphi^{c}_{\vec{p}}+
     \sin\varphi^{b}_{\vec{p}}\sin\varphi^{c}_{\vec{p}}
   \right) \ , \\
   \beta_{\vec{p}}          & = & -\delta^{*}_{\vec{p}} \nonumber \\
   & = & \ee^{\ii (E^{c}_{\vec{p}}+E^{b}_{\vec{p}})\tau_{L}/2}
   \left(
     \sin\varphi^{b}_{\vec{p}}\cos\varphi^{c}_{\vec{p}}-
     \cos\varphi^{b}_{\vec{p}}\sin\varphi^{c}_{\vec{p}}
   \right) \ , \\
   \tilde{\alpha}_{\vec{p}} & = & \tilde{\gamma}_{\vec{p}}^{*} \nonumber \\
   & = & \alpha^{2}_{\vec{p}}+\beta^{2}_{\vec{p}} \ , \\
   \tilde{\beta}_{\vec{p}}  & = & -\tilde{\delta}_{\vec{p}}^{*} \nonumber \\
   & = & 2 \ii \mbox{Im}\left\lbrace\alpha^{*}_{\vec{p}}\beta_{\vec{p}}\right\rbrace \ .
 \end{eqnarray}
\end{subequations}
Hence, for $-\tau_{L}/2\le t\le\tau_{L}/2$, the occupations numbers are
given by (\ref{eq:3:chiral_partinst}) whereas for $t>\tau_{L}/2$ we have
\begin{eqnarray}
 \label{eq:3:chiral_partinst_back}
 \frac{\dd^{6}n_{q\bar{q}}(t)}{\dd^{3}x\dd^{3}p} & = & \frac{2}{(2\pi)^{3}}\left|\tilde{\beta}_{\vec{p}}\right|^{2} \nonumber \\
 & = & \frac{2}{(2\pi)^{3}}\left[1-\left(\frac{p^{2}+m_{b}m_{c}}{E^{b}_{\vec{p}}E^{c}_{\vec{p}}}\right)^{2}\right]
 \sin^{2}(E^{b}_{\vec{p}}\tau_{L}) \nonumber \\
 & \simeq & 2\frac{(m_{c}-m_{b})^{2}}{(2\pi)^{3}p^{2}}\sin^{2}(p\tau_{L})\quad\mbox{for}\quad p\gg m_{b},m_{c}.
\end{eqnarray}
Consequently, for given momentum, $\vec{p}$, the final occupation number is highly
sensitive to $\tau_{L}$ and becomes maximal if the condition
$$
E^{b}_{\vec{p}}\tau_{L} = \frac{2n+1}{2}\pi\quad\text{with}\quad n\in\mathbb{N}_{0} \ ,
$$
is satisfied. Comparing the asymptotic particle number density for different
mass parameterizations $\tilde{m}(t)$, we observe the same sensitivity
on the respective order of differentiability for large $p$ as we did in
the first scenario. This is depicted in
Fig. \ref{fig:3:chiral_asymptcomp_back} and follows immediately from
expression (\ref{eq:3:occup_approx}).

\begin{figure}[htb]
 \center
 \includegraphics[height=5.0cm]{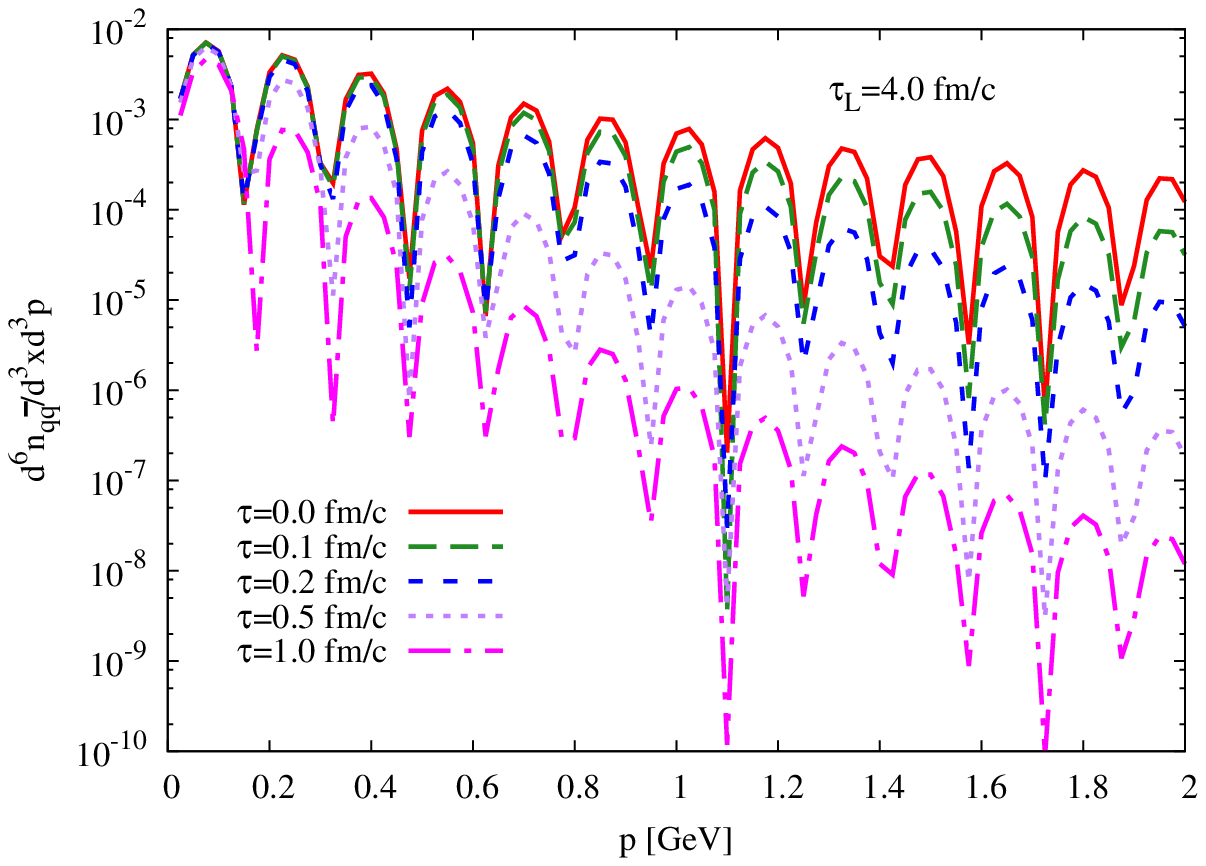}
 \includegraphics[height=5.0cm]{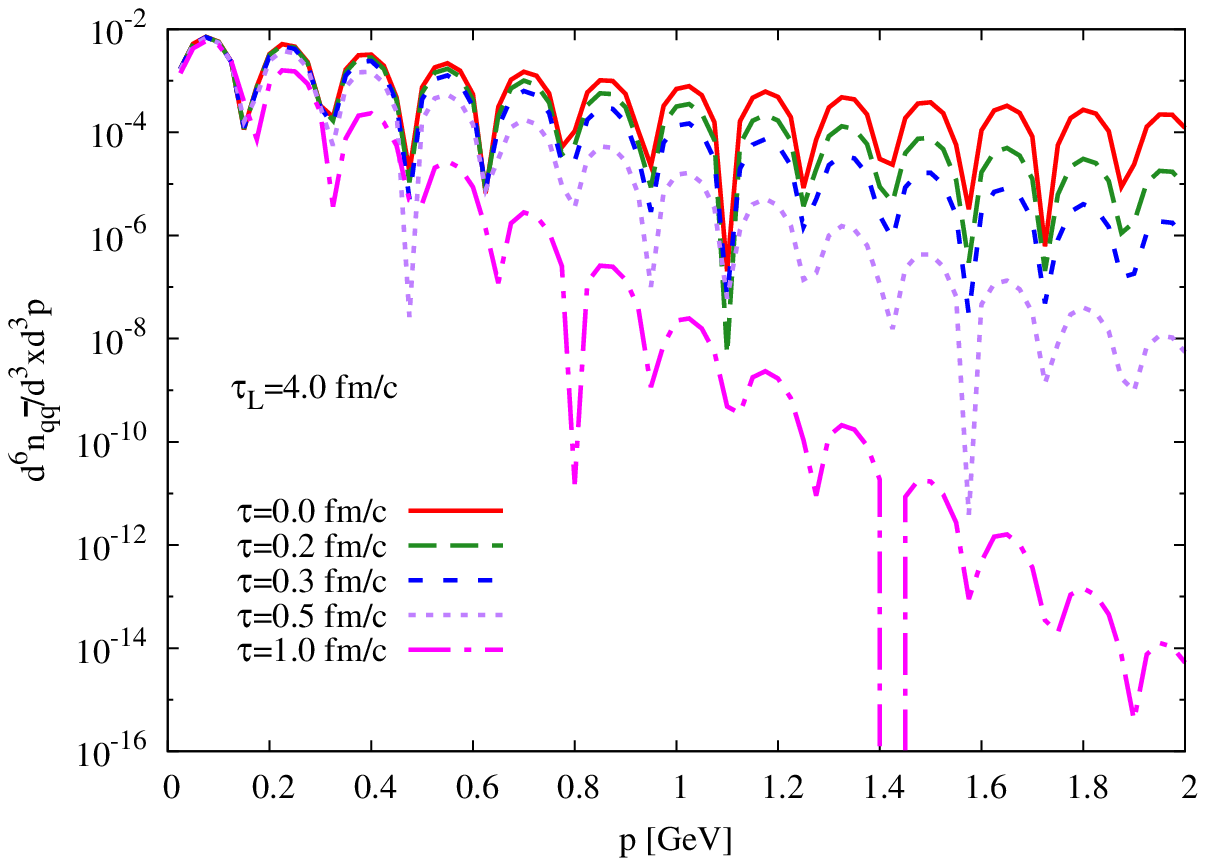}
 \caption{Asymptotic particle spectra for $\tilde{m}_{2}(t)$ (left panel) and
   $\tilde{m}_{3}(t)$ (right panel)  for different values of
   $\tau$. Analogously to the first scenario, the suppression at large $p$ compared to the 
   instantaneous case is the stronger the more slowly ($\tau$ increasing) both 
   mass changes are assumed to take place. As it must be, expression (\ref{eq:3:chiral_partinst_back}) 
   is reproduced in the limit $\tau\rightarrow0$ for both parameterizations.}
 \label{fig:3:chiral_tau0_back}
\end{figure}

Similarly to the first case, both (\ref{eq:3:chiral_massparam_back2})
and (\ref{eq:3:chiral_massparam_back3}) reproduce the particle spectrum
of (\ref{eq:3:chiral_massparam_back1}) in the limit $\tau\rightarrow0$,
which is shown in figure \ref{fig:3:chiral_tau0_back}.

\begin{figure}[htb]
 \center
 \includegraphics[height=5.0cm]{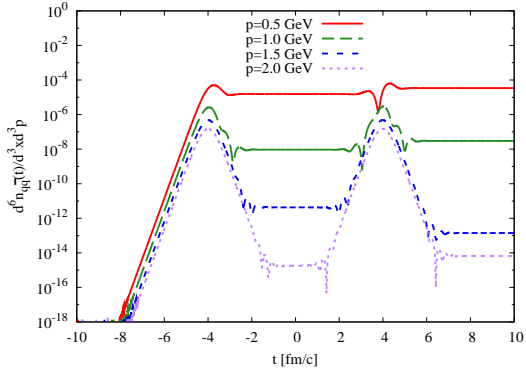}
 \caption{Time evolution of the particle number density for $\tilde{m}_{3}(t)$ with
   $\tau=1.0 \; \fm/c$ and $\tau_{L}=8.0 \; \fm/c$. They are again changing
   by several orders of magnitude when the mass is being restored to its
   initial constituent value, $m_{c}$.}
 \label{fig:3:chiral_timedep_back}
\end{figure}

Fig.~\ref{fig:3:chiral_timedep_back} shows the time evolution of the
particle number density for $\tilde{m}_{3}(t)$ with $\tau=1 \; \fm/c$ and
$\tau_{L}=8 \; \fm/c$. As it should be, the time dependence in the
second scenario is the same as in the first one until we start changing
the fermion mass back to its initial constituent value, $m_{c}$. Hence,
the occupation numbers first saturate at the same value as they did
within the first scenario. As soon as a second mass gradient shows up,
they are again changing by several orders of magnitude before they
saturate at their final asymptotic value, which is usually different
from the first one.

Such a behavior follows immediately from expression
(\ref{eq:3:occup_approx}). At times when the fermion mass is at its bare
value, $m_{b}$, the integral entering it effectively represents a full
Fourier transform over an at least continuous function while we again
encounter a Fourier transform over a discontinuous function again when
the fermion mass is being changed back to its constituent value,
$m_{c}$.

\begin{figure}[htb]
 \begin{center}
  \includegraphics[height=6.0cm]{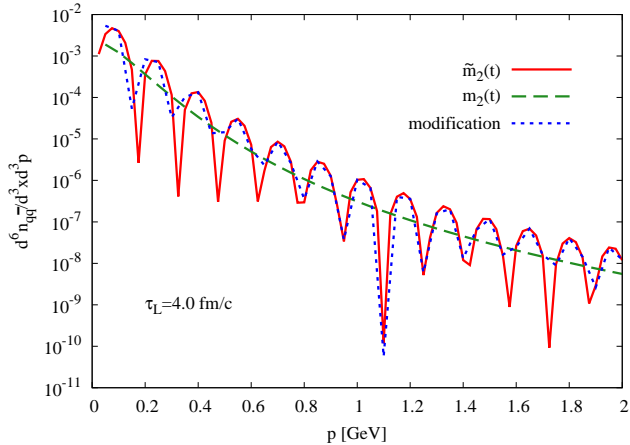}
  \caption{Comparison of asymptotic particle spectra for $\tilde{m}_{2}(t)$ and $m_{2}(t)$ for $\tau=1.0$ fm/c.
    For large values of $p$, the asymptotic occupation numbers of $\tilde{m}_{2}(t)$ essentially differ from those of $m_{2}(t)$ 
    by a modification factor of $4\sin^{2}{p\tau_{L}}$.}
  \label{fig:3:chiral_asymptcomp_scen}
 \end{center}
\end{figure}

For completeness, we still have to investigate in more detail how the
occupation numbers are modified if the quark mass is changed back to its
initial value, $m_{c}$. For this purpose, we take into account that our
mass parametrization (\ref{eq:3:chiral_massparam}) and
(\ref{eq:3:chiral_massparam_back}) can be written in the general form
\begin{subequations}
 \begin{eqnarray}
   m(t)         & = & \frac{m_{c}+m_{b}}{2}-\frac{m_{c}-m_{b}}{2}f(t) \ , \\
   \tilde{m}(t) & = & \frac{m_{c}+m_{b}}{2}-\frac{m_{c}-m_{b}}{2}f\left(t+\frac{\tau_{L}}{2}\right)f\left(\frac{\tau_{L}}{2}-t\right) \ ,
 \end{eqnarray}
\end{subequations}
with $f(t)$ increasing monotonously from $-1$ to $1$ and fulfilling the
condition $f(-t)=-f(t)$ (odd under time inversion). Hence, we have
\begin{equation}
 \label{eq:3:chi_back}
 \tilde{\chi}(t) = f\left(\frac{\tau_{L}}{2}-t\right)\chi\left(\frac{\tau_{L}}{2}+t\right)- 
                   \chi\left(\frac{\tau_{L}}{2}-t\right)f\left(\frac{\tau_{L}}{2}+t\right) \ .
\end{equation}
The Fourier transform of (\ref{eq:3:chi_back}) from $t$ to $2p$ is given
by
\begin{equation}
  \int_{-\infty}^{\infty}\dd t'\tilde{\chi}(t')\ee^{2 \ii pt'} = 2 \ii \im \left\lbrace 
    \ee^{-\ii p\tau_{L}}\int_{-\infty}^{\infty}\dd t'
    f\left(\tau_{L}-t'\right)\chi(t')\ee^{2 \ii pt'}
  \right\rbrace \ .
\end{equation}
Thus, unlike the case where the fermion mass is changed instantaneously,
the occupation numbers for $p\gg m_{b},m_{c}$ in the second scenario are
generally different from those in the first scenario multiplied by a
factor of $4\sin^{2}(p\tau_{L})$. Nevertheless, if $\tau_{L}$ is
significantly larger than $\tau$, we have $f(\tau_{L}-t)\approx1$ for
those times, $t$, where $\chi(t)$ is significantly different from
zero. We can hence approximate
\begin{equation}
  \int_{-\infty}^{\infty}\dd t'\tilde{\chi}(t')\ee^{2 \ii pt'} \approx -2 \ii \sin p_{L}\tau\int_{-\infty}^{\infty}\dd t'\chi(t')\ee^{2 \ii pt'},
\end{equation}
where we have also taken into account that $\chi(-t)=\chi(t)$. The
latter follows immediately from $f(-t)=-f(t)$. Hence, for
$\tau\ll\tau_{L}$, the asymptotic occupation numbers for hard
quark-antiquark pairs are given by
\begin{equation}
  \left.\frac{\dd^{6}n_{q\bar{q}}(t)}{\dd^{3}x\dd^{3}p}\right|_{t\rightarrow\infty} = 2\frac{(m_{b}-m_{c})^{2}}{(2\pi)^{3}p^{2}}\left|\int_{-\infty}^{\infty}\dd t'\chi(t')\ee^{2 \ii pt'}\right|^{2}\sin^{2}(p\tau_{L}) \ ,
\end{equation}
which just corresponds to (\ref{eq:3:occup_approx}) modified by a factor
of $4\sin^{2}(p\tau_{L})$. This is illustrated in
Fig.~\ref{fig:3:chiral_asymptcomp_scen} for the asymptotic particle
spectrum of $\tilde{m}_{2}(t)$, where the dotted line represents the
spectrum for $m_{2}(t)$ multiplied by $4\sin^{2}(p\tau_{L})$.

To summarize, we have found that the occupation numbers in the 
asymptotic limit show a strong dependence on the `smoothness' 
of the considered mass parametrization, $m(t)$. For the case 
of an instantaneous mass shift, they scale $\propto (m_{c}-m_{b})^{2}/p^{2}$ 
for $p\gg m_{c},m_{b}$ which means that the total number density is 
linearly divergent. This artifact is removed if the mass shift takes place 
over a finite time interval, $\tau$. In particular, the quark/antiquark 
occupation numbers at $p\gg m_{c},m_{b}$ and $p\gg 1/\tau$ scale
$\propto (m_{c}-m_{b})^{2}/p^{6}\tau^{4}$ for $m_{2}(t)$ (continuously 
differentiable once) and are suppressed even further to an exponential decay for 
$m_{3}(t)$ (continuously differentiable infinitely many times). 

Our investigations have shown that the pathological scaling behavior for 
an instantaneous mass shift essentially results from high-momentum Fourier 
components. These components are suppressed for mass shifts over a finite time 
interval such that the scaling behavior of the quark/antiquark occupation 
numbers becomes physically reasonable. A very similar effect occurs when considering back-to-back 
particle-antiparticle correlations in high energy nuclear collisions for mass 
parameterizations with different order of differentiability \cite{Vourdas:1988cd,Razumov:1994xu,Asakawa:1996xx,Asakawa:1998cx,Dudek:2010nr,Knoll:2010hd} .

Furthermore, if the quark mass is changed back from its bare value, $m_{b}$, 
to its constituent value, $m_{c}$, the lifetime of the chirally restored phase enters the
asymptotic occupations numbers at large $p\gg m_{b},m_{c}$ in the form
of a factor of $4\sin^{2}(p\tau_{L})$ if $\tau\ll\tau_{L}$. From 
the semiclassical point of view, the oscillations in the particle spectra 
emerging from this factor correspond to a multiple scattering of 
the quarks/antiquarks at both mass gradients.

At intermediate times, however, the quark/antiquark occupation numbers decay $\propto 1/p^{4}$ for large $p$ 
in regions of strong mass gradients. This, in turn, leads to a transient logarithmic
divergence in the total energy density of the quarks and antiquarks. At first sight 
such a divergence might be disturbing. We stress, however, that only
the asymptotic energy density, i.e.\ for $t \to +\infty$, constitutes an
observable in the sense of S-matrix theory, as the interpretation of (\ref{eq:3:bogolyubov_numbers}) as quark/antiquark occupation 
numbers is only justified for asymptotic times where $\dot{m}(t)=0$. The reason is that the dispersion 
relation (\ref{eq:3:dispersion}) then actually characterizes free and thus detectable particles wheres it only describes 
quasiparticles for $\dot{m}(t)\neq0$ \cite{Filatov:2007ha}. Moreover, the asymptotic value does
not show any divergence as long as the mass shift is smooth enough,
which is a physically reasonable condition. Yet, there is one more twist
in this argument when we proceed to the photon production.  Photons are
produced (and, in principle, leave the system) at any instant of time.
Thus one might suspect that the tremendously large intermediate fermion
numbers leave their imprint on the number of emitted photons. It is
conceivable that the asymptotic number of photons, which is an
accumulation over the whole time history, becomes large or even diverges
just because there have been very many fermionic emitters at
intermediate times. We will see in the following that this is not the
case. It is important to understand that the disappearance of the large
fermion numbers is not a damping effect, but a quantum mechanical
interference effect. Collisions between the fermions, which would provide
a loss rate, i.e.\ damping, are not included in our approach. On the
other hand, the full quantum effects are retained, which can lead to
interference patterns that are unintuitive from a classical point of
view. This applies in particular to quantities which are not 
observable anyhow like, for instance, the number of fermions at finite times. In
the same way the asymptotic photon number can turn out to be reasonably
small in spite of the fact that the emitted photons seem to pile up
during the whole history of the process.

\section{Photon production}
\label{sec:photon-production}

We now turn to our investigations on photon production. As it has been
shown in the previous section, the asymptotic quark/antiquark occupation
numbers exhibit a strong sensitivity to the order of differentiability
of the time dependence of the mass, $m(t)$. In particular, they are
rendered integrable in the ultraviolet domain, if the mass shift is assumed to take place over a
finite time interval, $\tau$. We now investigate whether the resulting
photon spectra exhibit a similar sensitivity and if the model of chiral
photon production is accordingly suitable to describe finite-lifetime
effects on the photon emission from a quark-gluon plasma. As one lesson
from section~\ref{sec:pair-production} we recall that only the
asymptotic particle numbers constitute observables \cite{Filatov:2007ha}, while
quantities defined by the analogous expressions with
  interpolating fields have no definite interpretation as particle
  numbers. Thus, in the following we concentrate on the asymptotic
photon numbers (\ref{eq:2:photon_yield_chiral_aspt}).

\subsection{Instantaneous mass shift}
First, we consider photon production for an instantaneous mass shift at
$t=0$ as this special case still allows for an almost complete
analytical treatment. In particular, the individual contributions to the
photon yield allow for an interpretation as first-order QED processes
and their interference among each other.

On the other hand, we show that the assumption of an instantaneous mass
shift comes along with essentially three unphysical artifacts. In
section \ref{subsubsec:photon-proudction-evaluate}, we calculate the
individual contributions to the photon yield
(\ref{eq:2:photon_yield_chiral_aspt}). Thereby, we will demonstrate that
the overall loop integral entering (\ref{eq:2:photon_yield_chiral_aspt})
features a linear divergence caused by the decay behavior of the
quark/antiquark occupation numbers (\ref{eq:3:chiral_partinst}) for
$p\gg m_{b},m_{c}$. We regulate this divergence by cutting off the loop
integral at $p=\Lambda_{C}$.

We will see in section \ref{subsubsec:photon-proudction-asymptotic} that
then the resulting asymptotic photon spectra decay as $1/\omega^{3}_{\vec{k}}$, with
$\omega_{\vec{k}}$ denoting the photon energy (\ref{eq:2:photen}), in the ultraviolet domain for given
value of $\Lambda_{C}$. The total number density and the total energy
density of the emitted photons are hence logarithmically and linearly
divergent, respectively.

Finally, as we mention at the end of section
\ref{subsubsec:photon-proudction-asymptotic} and discuss in greater
detail in appendix \ref{sec:appe}, another problem appears in the limit
$m_{b}\rightarrow 0$. In that limit, the loop integral over the
contributions describing quark and antiquark bremsstrahlung and
quark-antiquark-pair annihilation into a photon feature a collinear and
an anticollinear singularity, respectively.

\subsubsection{Evaluation of contributions to the photon yield}
\label{subsubsec:photon-proudction-evaluate}
To evaluate the contributions to the photon yield for an instantaneous
mass shift, we first undo the contraction
$$
\ii\Pi^{<}_{T}(\vec{k},t_{1},t_{2}) = \gamma^{\mu\nu}(k)\Pi^{<}_{\nu\mu}(\vec{k},t_{1},t_{2}) \ ,
$$
and express the photon self-energy in terms of positive and negative
energy wavefunctions, i.e.,
\begin{align}
 \label{eq:4:pse_wavef}
 \ii\Pi^{<}_{\mu\nu}(\vec{k},t_{1},t_{2}) = & e^{2}\sum_{r,s}\int\frac{\dd^{3}p}{(2\pi)^{3}}
 \left [\bar{\psi}^{'}_{\vec{p},r,\uparrow}(t_{1})\gamma_{\mu}
   \psi^{'}_{\vec{p}+\vec{k},s,\downarrow}(t_{1}) \right ]\cdot
 \left [\bar{\psi}^{'}_{\vec{p}+\vec{k},s,\downarrow}(t_{2})\gamma_{\nu}
   \psi^{'}_{\vec{p},r,\uparrow}(t_{2}) \right ] \nonumber \\
  = & e^{2}\sum_{r,s}\int\frac{\dd^{3}p}{(2\pi)^{3}}
 \left [\bar{\psi}^{'}_{\vec{p},r,\uparrow}(t_{1})\gamma_{\mu}
   \psi^{'}_{\vec{p}+\vec{k},s,\downarrow}(t_{1}) \right ]\cdot
 \left [\bar{\psi}^{'}_{\vec{p},r,\uparrow}(t_{2})\gamma_{\nu}
   \psi^{'}_{\vec{p}+\vec{k},s,\downarrow}(t_{2}) \right ]^{*} \ .
\end{align}
Upon insertion of ({\ref{eq:4:pse_wavef}) into
  (\ref{eq:2:photon_yield_chiral_aspt}) and interchanging both time
  integrations with the loop integral, we can rewrite
  (\ref{eq:2:photon_yield_chiral_aspt}) as
\begin{equation}
 \label{eq;4:photonum_inst}
 2\omega_{\vec{k}}\frac{\dd^{6}n_{\gamma}}{\dd^{3}x \dd^{3}k} = \lim_{\varepsilon\rightarrow0}\frac{\gamma^{\mu\nu}(k)}{(2\pi)^{3}}
 \sum_{r,s}\int\frac{\dd^{3}p}{(2\pi)^{3}}I^{\varepsilon,*}_{\mu}(\vec{p},\vec{k},r,s)
 I^{\varepsilon}_{\nu}(\vec{p},\vec{k},r,s) \ .
\end{equation}
Here we have introduced 
\begin{equation}
 \label{eq:4:amp_inst}
 I^{\varepsilon}_{\mu}(\vec{p},\vec{k},r,s) = e\int_{-\infty}^{\infty}\underline{\dd t}\mbox{ }\bar{\psi}^{'}_{\vec{p},r,\uparrow}(t)
                                              \gamma_{\mu}\psi^{'}_{\vec{p}+\vec{k},s,\downarrow}(t)\ee^{\ii\omega_{\vec{k}}t} \ ,
\end{equation}
with the underline denoting that the time integral is regulated by the
convergence factor $f_{\varepsilon}(t)=\ee^{-\varepsilon|t|}$. Moreover,
with the help of (\ref{eq:3:ansatz_inst}) and (\ref{eq:3:coeff_link}),
we obtain
\begin{equation}
  \begin{split}
 \label{eq:4:amp_inst_eval}
 I^{\varepsilon}_{\mu}(\vec{p},\vec{k},r,s) = e & \left[
   \bar{u}_{c}(\vec{p},r)\gamma_{\mu}v_{c}(\vec{p}+\vec{k},s)
   \frac{1}{\varepsilon+\ii\omega^{c}_{1}(\vec{p},\vec{k})}
 \right.  \\
 & +\alpha_{\vec{p}}\alpha_{\vec{p}+\vec{k}}
 \bar{u}_{b}(\vec{p},r)\gamma_{\mu}v_{b}(\vec{p}+\vec{k},s)
 \frac{1}{\varepsilon-\ii\omega^{b}_{1}(\vec{p},\vec{k})}  \\
 & -\alpha_{\vec{p}}\beta_{\vec{p}+\vec{k}}
 \bar{u}_{b}(\vec{p},r)\gamma_{\mu}u_{b}(\vec{p}+\vec{k},s)
 \frac{1}{\varepsilon+\ii\omega^{b}_{2}(\vec{p},\vec{k})}  \\
 & +\beta_{\vec{p}}\alpha_{\vec{p}+\vec{k}}
 \bar{v}_{b}(\vec{p},r)\gamma_{\mu}v_{b}(\vec{p}+\vec{k},s)
 \frac{1}{\varepsilon-\ii\omega^{b}_{3}(\vec{p},\vec{k})}  \\
 & \left. -\beta_{\vec{p}}\beta_{\vec{p}+\vec{k}}
   \bar{v}_{b}(\vec{p},r)\gamma_{\mu}u_{b}(\vec{p}+\vec{k},s)
   \frac{1}{\varepsilon+\ii\omega^{b}_{4}(\vec{p},\vec{k})} \right] \ .
\end{split}
\end{equation}
To keep the notation short, we have introduced the frequencies
\begin{subequations}
 \label{eq:4:freq}
 \begin{eqnarray}
  \omega^{b,c}_{1}(\vec{p},\vec{k}) & = & E^{b,c}_{\vec{p}+\vec{k}}+E^{b,c}_{\vec{p}}+\omega_{\vec{k}} \ , \label{eq:4:freq_1} \\
  \omega^{b,c}_{2}(\vec{p},\vec{k}) & = & E^{b,c}_{\vec{p}+\vec{k}}-E^{b,c}_{\vec{p}}-\omega_{\vec{k}} \ , \label{eq:4:freq_2} \\
  \omega^{b,c}_{3}(\vec{p},\vec{k}) & = & E^{b,c}_{\vec{p}+\vec{k}}-E^{b,c}_{\vec{p}}+\omega_{\vec{k}} \ , \label{eq:4:freq_3} \\
  \omega^{b,c}_{4}(\vec{p},\vec{k}) & = & E^{b,c}_{\vec{p}+\vec{k}}+E^{b,c}_{\vec{p}}-\omega_{\vec{k}} \ . \label{eq:4:freq_4}
 \end{eqnarray}
\end{subequations}
Since (\ref{eq:4:freq_1})-(\ref{eq:4:freq_4}) are either positive or
negative definite for both $m_{c}$ and $m_{b}$ taking the limit 
$\varepsilon\rightarrow0$ leads to
\begin{equation}
\begin{split}
 \label{eq:amp_inst}
 \lim_{\varepsilon\rightarrow0}I^{\varepsilon}_{\mu}(\vec{p},\vec{k},r,s)
 = I_{\mu}(\vec{p},\vec{k},r,s) = -\ii e & \left[
   \bar{u}_{c}(\vec{p},r)\gamma_{\mu}v_{c}(\vec{p}+\vec{k},s)
   \frac{1}{\omega^{c}_{1}(\vec{p},\vec{k})}
 \right.  \\
 & -\alpha_{\vec{p}}\alpha_{\vec{p}+\vec{k}}
 \bar{u}_{b}(\vec{p},r)\gamma_{\mu}v_{b}(\vec{p}+\vec{k},s)
 \frac{1}{\omega^{b}_{1}(\vec{p},\vec{k})}  \\
 & -\alpha_{\vec{p}}\beta_{\vec{p}+\vec{k}}
 \bar{u}_{b}(\vec{p},r)\gamma_{\mu}u_{b}(\vec{p}+\vec{k},s)
 \frac{1}{\omega^{b}_{2}(\vec{p},\vec{k})}  \\
 & -\beta_{\vec{p}}\alpha_{\vec{p}+\vec{k}}
 \bar{v}_{b}(\vec{p},r)\gamma_{\mu}v_{b}(\vec{p}+\vec{k},s)
 \frac{1}{\omega^{b}_{3}(\vec{p},\vec{k})}  \\
 & \left. -\beta_{\vec{p}}\beta_{\vec{p}+\vec{k}}
   \bar{v}_{b}(\vec{p},r)\gamma_{\mu}u_{b}(\vec{p}+\vec{k},s)
   \frac{1}{\omega^{b}_{4}(\vec{p},\vec{k})} \right] \ .
\end{split}
\end{equation}
With the help of
\begin{subequations}
 \begin{eqnarray}
  u_{c}(\vec{p},s) & = & \alpha_{\vec{p}}u_{b}(\vec{p},s)+\beta_{\vec{p}}v_{b}(\vec{p},s) \ , \\
  v_{c}(\vec{p},s) & = & \alpha_{\vec{p}}v_{b}(\vec{p},s)-\beta_{\vec{p}}u_{b}(\vec{p},s) \ ,
 \end{eqnarray}
\end{subequations}
we can rewrite (\ref{eq:amp_inst}) in the following more compact form:
\begin{equation}
\begin{split}
 \label{eq:4:amp_inst_comp}
 I_{\mu}(\vec{p},\vec{k},r,s) = \ii e & \left[
   \alpha_{\vec{p}}\alpha_{\vec{p}+\vec{k}}
   \bar{u}_{b}(\vec{p},r)\gamma_{\mu}v_{b}(\vec{p}+\vec{k},s)
   \left(\frac{1}{\omega^{b}_{1}(\vec{p},\vec{k})}-\frac{1}{\omega^{c}_{1}(\vec{p},\vec{k})}\right)
 \right.  \\
 & +\alpha_{\vec{p}}\beta_{\vec{p}+\vec{k}}
 \bar{u}_{b}(\vec{p},r)\gamma_{\mu}u_{b}(\vec{p}+\vec{k},s)
 \left(\frac{1}{\omega^{b}_{2}(\vec{p},\vec{k})}+\frac{1}{\omega^{c}_{1}(\vec{p},\vec{k})}\right)
  \\
 & +\beta_{\vec{p}}\alpha_{\vec{p}+\vec{k}}
 \bar{v}_{b}(\vec{p},r)\gamma_{\mu}v_{b}(\vec{p}+\vec{k},s)
 \left(\frac{1}{\omega^{b}_{3}(\vec{p},\vec{k})}-\frac{1}{\omega^{c}_{1}(\vec{p},\vec{k})}\right)
  \\
 & \left.+ \beta_{\vec{p}}\alpha_{\vec{p}+\vec{k}}
   \bar{v}_{b}(\vec{p},r)\gamma_{\mu}u_{b}(\vec{p}+\vec{k},s)
   \left(\frac{1}{\omega^{b}_{4}(\vec{p},\vec{k})}+\frac{1}{\omega^{c}_{1}(\vec{p},\vec{k})}\right)
 \right] \ .
\end{split}
\end{equation}
Since we have $\alpha_{\vec{p}}\rightarrow1$ and $\beta_{p}\rightarrow0$
for $m_{b}\rightarrow m_{c}$, expression (\ref{eq:4:amp_inst_comp})
vanishes in this case and we will have no photon production, as it
should be. Taking a closer look at the spinor structure of the
particular contributions to (\ref{eq:4:amp_inst_comp}) allows us to
interpret them as first-order QED-transition amplitudes. It is hence
convenient to split up (\ref{eq:4:amp_inst_comp}) as
\begin{subequations}
 \label{eq:4:amp_inst_split}
 \begin{eqnarray}
  I_{\mu}(\vec{p},\vec{k},r,s)     & = & \sum_{i=1}^{4}I^{i}_{\mu}(\vec{p},\vec{k},r,s) \ , \\
  I^{1}_{\mu}(\vec{p},\vec{k},r,s) & = & \ii e \alpha_{\vec{p}}\alpha_{\vec{p}+\vec{k}}
                                         \bar{u}_{b}(\vec{p},r)\gamma_{\mu}v_{b}(\vec{p}+\vec{k},s)
                                         \left(\frac{1}{\omega^{b}_{1}(\vec{p},\vec{k})}-\frac{1}{\omega^{c}_{1}(\vec{p},\vec{k})}\right) \ , \\
  I^{2}_{\mu}(\vec{p},\vec{k},r,s) & = & \ii e \alpha_{\vec{p}}\beta_{\vec{p}+\vec{k}}
                                         \bar{u}_{b}(\vec{p},r)\gamma_{\mu}u_{b}(\vec{p}+\vec{k},s)
                                         \left(\frac{1}{\omega^{b}_{2}(\vec{p},\vec{k})}+\frac{1}{\omega^{c}_{1}(\vec{p},\vec{k})}\right) \ , \\
  I^{3}_{\mu}(\vec{p},\vec{k},r,s) & = & \ii e \beta_{\vec{p}}\alpha_{\vec{p}+\vec{k}}
                                         \bar{v}_{b}(\vec{p},r)\gamma_{\mu}v_{b}(\vec{p}+\vec{k},s)
                                         \left(\frac{1}{\omega^{b}_{3}(\vec{p},\vec{k})}-\frac{1}{\omega^{c}_{1}(\vec{p},\vec{k})}\right) \ , \\
  I^{4}_{\mu}(\vec{p},\vec{k},r,s) & = & \ii e \beta_{\vec{p}}\beta_{\vec{p}+\vec{k}}
                                         \bar{v}_{b}(\vec{p},r)\gamma_{\mu}u_{b}(\vec{p}+\vec{k},s)
                                         \left(\frac{1}{\omega^{b}_{4}(\vec{p},\vec{k})}+\frac{1}{\omega^{c}_{1}(\vec{p},\vec{k})}\right) \ ,
 \end{eqnarray}
\end{subequations}
with the individual contributions describing the spontaneous creation of
a quark-antiquark pair together with a photon ($i=1$), quark bremsstrahlung
($i=2$), antiquark bremsstrahlung ($i=3$) and quark-antiquark pair
annihilation into a photon ($i=4$). With the help of
(\ref{eq:4:amp_inst_split}), we can rewrite (\ref{eq;4:photonum_inst})
as
\begin{equation}
\begin{split}
 \label{eq:4:photnum_inst_split}
 2\omega_{\vec{k}}\frac{\dd^{6}n_{\gamma}}{\dd^{3}x \dd^{3}k} =
 \frac{\gamma^{\mu\nu}(k)}{(2\pi)^{3}}\sum_{r,s}\int
 \frac{\dd^{3}p}{(2\pi)^{3}} & \left\lbrace
   \sum_{i}I^{i,*}_{\mu}(\vec{p},\vec{k},r,s)I^{i}_{\nu}(\vec{p},\vec{k},r,s)
 \right. \\
 & +\left.  2\mathrm{Re}
   \left[\sum_{i<j}I^{i,*}_{\mu}(\vec{p},\vec{k},r,s)
     I^{j}_{\nu}(\vec{p},\vec{k},r,s)\right] \right\rbrace \ .
\end{split}
\end{equation}
The first term in (\ref{eq:4:photnum_inst_split}) describes the direct
contributions from first-order QED processes whereas the second one
describes the interference among them. For further considerations, we
introduce the shorthand notation
\begin{equation}
 \label{eq:4:shorthand}
 I_{ij}(\vec{p},\vec{k},t) = \gamma^{\mu\nu}(k)\re\left\lbrace\sum_{r,s}I^{i,*}_{\mu}(\vec{p},\vec{k},r,s)
                             I^{j}_{\nu}(\vec{p},\vec{k},r,s)\right\rbrace \ ,
\end{equation}
in which (\ref{eq:4:photnum_inst_split}) reads
\begin{equation}
 \label{eq:4:yield_fin_re}
 2\omega_{\vec{k}}\frac{\dd^{6}n_{\gamma}}{\dd^{3}x\dd^{3}k} =
 \frac{1}{(2\pi)^{3}} \int\frac{\dd^{3}p}{(2\pi)^{3}} 
 \left \lbrace
   \sum_{i=1}^{4}I_{ii}(\vec{p},\vec{k})+
   2\sum_{i<j}I_{ij}(\vec{p},\vec{k})
 \right\rbrace \ .
\end{equation}
The evaluation of the individual contributions to
(\ref{eq:4:photnum_inst_split}) is a lengthy but straightforward
procedure and demonstrated exemplarily in appendix \ref{sec:appc}. The
direct contributions from first-order QED processes read
\begin{subequations}
 \label{eq:4:contr_dir}
 \begin{eqnarray}
  I_{11}(\vec{p},\vec{k})         & = & 2e^{2}\alpha^{2}_{\vec{p}+\vec{k}}\alpha^{2}_{\vec{p}}
                                        \left(1+\frac{px(px+\omega_{\vec{k}})+m^{2}_{b}}{E^{b}_{\vec{p}+\vec{k}}E^{b}_{\vec{p}}}\right)
                                        \left(\frac{1}{\omega^{b}_{1}(\vec{p},\vec{k})}-\frac{1}{\omega^{c}_{1}(\vec{p},\vec{k})}\right)^{2}
                                        \label{eq:4:contr_11} \ , \\
  \tilde{I}_{22}(\vec{p},\vec{k}) & = & 4e^{2}\beta^{2}_{\vec{p}+\vec{k}}\alpha^{2}_{\vec{p}}
                                        \left(1-\frac{px(px+\omega_{\vec{k}})+m^{2}_{b}}{E^{b}_{\vec{p}+\vec{k}}E^{b}_{\vec{p}}}\right)
                                        \left(\frac{1}{\omega^{b}_{2}(\vec{p},\vec{k})}+\frac{1}{\omega^{c}_{1}(\vec{p},\vec{k})}\right)^{2}
                                        \label{eq:4:contr_22} \ , \\
  I_{44}(\vec{p},\vec{k})         & = & 2e^{2}\beta^{2}_{\vec{p}+\vec{k}}\beta^{2}_{\vec{p}}
                                        \left(1+\frac{px(px+\omega_{\vec{k}})+m^{2}_{b}}{E^{b}_{\vec{p}+\vec{k}}E^{b}_{\vec{p}}}\right)
                                        \left(\frac{1}{\omega^{b}_{4}(\vec{p},\vec{k})}+\frac{1}{\omega^{c}_{1}(\vec{p},\vec{k})}\right)^{2}
                                        \label{eq:4:contr_44} \ .
 \end{eqnarray}
\end{subequations}
As in section \ref{sec:gaug-inv-phot}, $x$ denotes the cosine of the
polar angle between $\vec{p}$ and $\vec{k}$. Moreover, for the
interference contributions we obtain
\begin{subequations}
 \label{eq:4:contr_int}
 \begin{align}
  \tilde{I}_{12}(\vec{p},\vec{k}) = & -4e^{2}\alpha_{\vec{p}+\vec{k}}\beta_{\vec{p}+\vec{k}}\alpha^{2}_{\vec{p}}
                                      \frac{m_{b}|\vec{p}+\vec{k}|}{E^{b}_{p}E^{b}_{\vec{p}+\vec{k}}}
                                      \left(1-\frac{px(px+\omega_{\vec{k}})}{|\vec{p}+\vec{k}|^{2}}\right) \nonumber \\
                                    & \cdot\left(\frac{1}{\omega^{b}_{1}(\vec{p},\vec{k})}-\frac{1}{\omega^{c}_{1}(\vec{p},\vec{k})}\right)
                                      \left(\frac{1}{\omega^{b}_{2}(\vec{p},\vec{k})}+\frac{1}{\omega^{c}_{1}(\vec{p},\vec{k})}\right) 
                                      \label{eq:4:contr_12} \ , \\
  I_{14}(\vec{p},\vec{k}) = & -2e^{2}\alpha_{\vec{p}+\vec{k}}\beta_{\vec{p}+\vec{k}}\alpha_{\vec{p}}\beta_{\vec{p}}
                              \frac{p|\vec{p}+\vec{k}|}{E^{b}_{p}E^{b}_{\vec{p}+\vec{k}}}
                              \left(1+\frac{x(px+\omega_{\vec{k}})(E^{b}_{\vec{p}+\vec{k}}E^{b}_{p}+m^{2}_{b})}{p|\vec{p}+\vec{k}|^{2}}\right) \nonumber \\
                            & \cdot\left(\frac{1}{\omega^{b}_{1}(\vec{p},\vec{k})}-\frac{1}{\omega^{1}_{c}(\vec{p},\vec{k})}\right)
                              \left(\frac{1}{\omega^{b}_{4}(\vec{p},\vec{k})}+\frac{1}{\omega^{1}_{c}(\vec{p},\vec{k})}\right)  
                              \label{eq:4:contr_14} \ , \\
  I_{23}(\vec{p},\vec{k}) = & -2e^{2}\alpha_{\vec{p}+\vec{k}}\beta_{\vec{p}+\vec{k}}\alpha_{\vec{p}}\beta_{\vec{p}}
                              \frac{p|\vec{p}+\vec{k}|}{E^{b}_{p}E^{b}_{\vec{p}+\vec{k}}}
                              \left(1-\frac{x(px+\omega_{\vec{k}})(E^{b}_{\vec{p}+\vec{k}}E^{b}_{p}-m^{2}_{b})}{p|\vec{p}+\vec{k}|^{2}}\right) \nonumber \\
                            & \cdot\left(\frac{1}{\omega^{b}_{2}(\vec{p},\vec{k})}+\frac{1}{\omega^{c}_{1}(\vec{p},\vec{k})}\right) 
                              \left(\frac{1}{\omega^{b}_{3}(\vec{p},\vec{k})}-\frac{1}{\omega^{c}_{1}(\vec{p},\vec{k})}\right)
                              \label{eq:4:contr_23} \ , \\
  \tilde{I}_{24}(\vec{p},\vec{k}) = & -4e^{2}\alpha_{\vec{p}+\vec{k}}\beta_{\vec{p}+\vec{k}}\beta^{2}_{\vec{p}}
                                      \frac{m_{b}|\vec{p}+\vec{k}|}{E^{b}_{p}E^{b}_{\vec{p}+\vec{k}}}
                                      \left(1-\frac{px(px+\omega_{\vec{k}})}{|\vec{p}+\vec{k}|^{2}}\right) \nonumber \\
                                    & \cdot\left(\frac{1}{\omega^{b}_{3}(\vec{p},\vec{k})}-\frac{1}{\omega^{c}_{1}(\vec{p},\vec{k})}\right)
                                      \left(\frac{1}{\omega^{b}_{4}(\vec{p},\vec{k})}+\frac{1}{\omega^{c}_{1}(\vec{p},\vec{k})}\right)
                                      \label{eq:4:contr_34} \ .
 \end{align}
\end{subequations}
We have taken into account that the still to be carried out
loop integrals over $\dd^{3}p$ yield the same contribution
for $I_{22}(\vec{p},\vec{k},t)$ and $I_{33}(\vec{p},\vec{k},t)$, for
$I_{12}(\vec{p},\vec{k},t)$ and $I_{13}(\vec{p},\vec{k},t)$ and for
$I_{24}(\vec{p},\vec{k},t)$ and $I_{34}(\vec{p},\vec{k},t)$. Hence, 
in each case, these contributions can be taken together to one single 
contribution, i.e.,
\begin{subequations}
 \label{eq:4:sum_up}
 \begin{align}
   \tilde{I}_{22}(\vec{p},\vec{k},t) & \corres I_{22}(\vec{p},\vec{k},t)+I_{33}(\vec{p},\vec{k},t) \corres 2I_{22}(\vec{p},\vec{k},t) \ , \\
   \tilde{I}_{12}(\vec{p},\vec{k},t) & \corres I_{12}(\vec{p},\vec{k},t)+I_{13}(\vec{p},\vec{k},t) \corres 2I_{12}(\vec{p},\vec{k},t) \ , \\
   \tilde{I}_{24}(\vec{p},\vec{k},t) & \corres I_{24}(\vec{p},\vec{k},t)+I_{34}(\vec{p},\vec{k},t) \corres 2I_{24}(\vec{p},\vec{k},t) \ .
 \end{align}
\end{subequations}
The $\corres$-sign denotes that the equalities hold with respect to
the integration over $\dd^{3}p$. As the next step, we investigate the
asymptotic behavior of the different $I_{ij}(\vec{p},\vec{k})$ for
$p\rightarrow\infty$ to determine whether the integration over the loop
momentum is finite,
\begin{subequations}
 \label{eq:4:asymptotics}
 \begin{align}
  I_{11}(\vec{p},\vec{k})         = & e^{2}(1+x^{2})\frac{(m^{2}_{b}-m^{2}_{c})^{2}}{8p^{6}}+\mathcal{O}\left(\frac{1}{p^{7}}\right) \ , \\
  \tilde{I}_{22}(\vec{p},\vec{k}) = & \frac{e^{2}(m_{b}-m_{c})^{2}(1-x^2)(1+x)^{2}}{\omega^{2}_{\vec{k}}\left[p^{2}(1-x^{2})+m^{2}_b\right]}
                                      \left(1+\frac{m_{b}+m_{c}+2\omega_{\vec{k}}}{p}\right) \nonumber \\
                                    & -\frac{e^{2}(m_{b}-m_{c})^{2}(1-x^2)(1+x)}{p\omega_{\vec{k}}\left[p^{2}(1-x^{2})+m^{2}_b\right]}
                                      +\mathcal{O}\left(\frac{1}{p^{4}}\right) \ , \\
  I_{44}(\vec{p},\vec{k})         = & e^{2}(1+x^{2})\frac{(m_{c}-m_{b})^{4}}{8p^{6}}+\mathcal{O}\left(\frac{1}{p^{7}}\right) \ , \\
  \tilde{I}_{12}(\vec{p},\vec{k}) = & -\frac{e^{2}(m_{b}-m_{c})(m_{b}^2-m_{c}^2)m_b(1-x^{2})(1+x)}
                                            {2\omega_{\vec{k}}p^{3}\left[p^{2}(1-x^{2})+m^{2}_b\right]}+\mathcal{O}\left(\frac{1}{p^{6}}\right) \ , \\
  I_{14}(\vec{p},\vec{k})         = & -e^{2}(1+x^{2})\frac{(m^{2}_{b}-m^{2}_{c})(m_{b}-m_{c})^{2}}{8p^{6}}+\mathcal{O}\left(\frac{1}{p^{7}}\right) \ , \\
  I_{23}(\vec{p},\vec{k})         = & \frac{e^{2}(m_{b}-m_{c})^{2}(1-x^{2})}{2\omega_{\vec{k}}^{2}\left[p^{2}(1-x^{2})+m^{2}_b\right]} 
                                      \left(1+\frac{m_{b}+m_{c}+\omega_{\vec{k}}(1+2x)}{2p}\right)+\mathcal{O}\left(\frac{1}{p^{4}}\right) \ , \\
  \tilde{I}_{24}(\vec{p},\vec{k}) = & \frac{e^{2}(m_c-m_b)^{3}(1-x^{2})(1-x)}{2\omega_{\vec{k}}p^{3}\left[p^{2}(1-x^{2})+m^{2}_{b}\right]}+ 
                                      \mathcal{O}\left(\frac{1}{p^{6}}\right) \ .
 \end{align}
\end{subequations}
As the integration measure, $\dd^{3}p$, still contributes another factor
of $p^{2}$ to the integrand, the latter has to be of the order of
$1/p^{4}$ for the loop integral to be finite. We see, however, that the
contributions describing quark ($i=2$) or antiquark ($i=3$)
bremsstrahlung and the interference between these two processes feature
terms decaying as $1/p^{2}$ and $1/p^{3}$ in each case and that these
terms do not cancel each other.  Thus, the overall integrand behaves as
$1/p^{2}$ for large $p$, which means that the loop integral is linearly
divergent.

In order to handle this divergence, we note that
$\tilde{I}_{22}(\vec{p},\vec{k})$ and $I_{23}(\vec{p},\vec{k})$, from
which this divergence arises, scale with the Bogolyubov particle number
$\propto 1/p^{2}$ for large $p$. This behavior is an artifact from the
instantaneous mass shift \cite{Greiner:1995ac,Greiner:1996wz}
(cf. section ~\ref{sec:pair-production}) and can be regulated if the
mass shift is assumed to take place over a finite time interval,
$\tau$. This will be confirmed below in section
~\ref{sec:photons_smooth}.  Hence, from the conceptual point of view,
the linear divergence in the loop integral does not require a
renormalization and is here regulated by cutting the loop integral at
$p=\Lambda_{C}$.

\subsubsection{Asymptotic photon spectra}
\label{subsubsec:photon-proudction-asymptotic}

Fig.~\ref{fig:4:chiral_inst_comp} shows the resulting photon spectra for
different values of $\Lambda_{C}$. As in
\cite{Greiner:1995ac,Greiner:1996wz}, we have chosen $m_{c}=0.35 \;
\GeV$ and $m_{b}=0.01 \; \GeV$. One can see that the photon spectrum
drops as $1/\omega^{3}_{\vec{k}}$ in the ultraviolet domain such that the total
number density and the total energy density of the emitted photons are logarithmically and linearly
divergent, respectively, for given $\Lambda_{C}$. We will investigate
below if this is also an artifact of the instantaneous mass shift.
\begin{figure}[htb]
 \begin{center}
   \includegraphics[height=5.0cm]{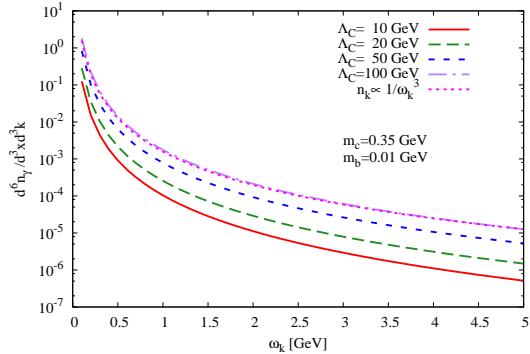}
   \caption{Photon spectra for an instantaneous mass shift for different values of $\Lambda_{C}$. 
            In each case, they decay as $1/\omega^{3}_{\vec{k}}$ in the ultraviolet domain.}
  \label{fig:4:chiral_inst_comp}
 \end{center}
\end{figure}

Furthermore, we investigate the dependence of the photon yield on the
values of the constituent mass, $m_{c}$, and the bare mass, $m_{b}$,
which is depicted in Fig. \ref{fig:4:chiral_massdep}.
\begin{figure}[htb]
 \center
 \includegraphics[height=5.0cm]{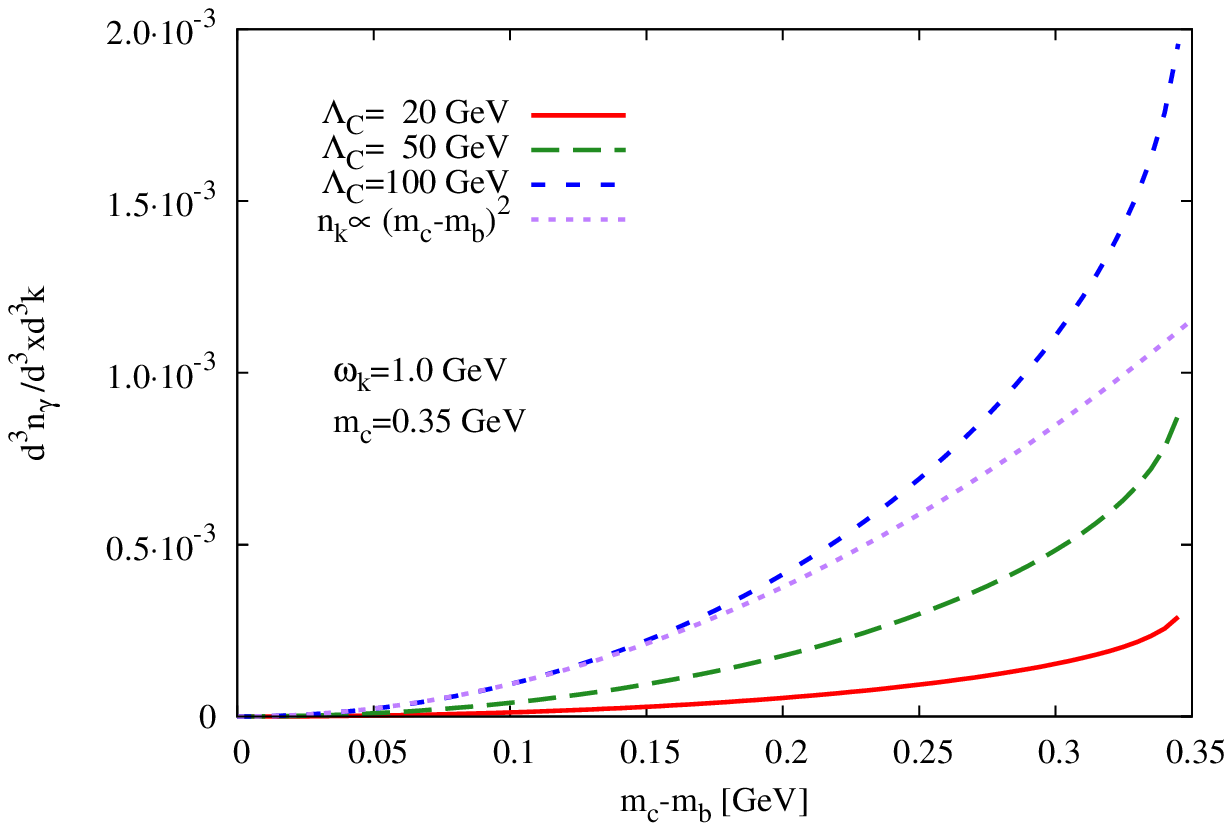}
 \includegraphics[height=5.0cm]{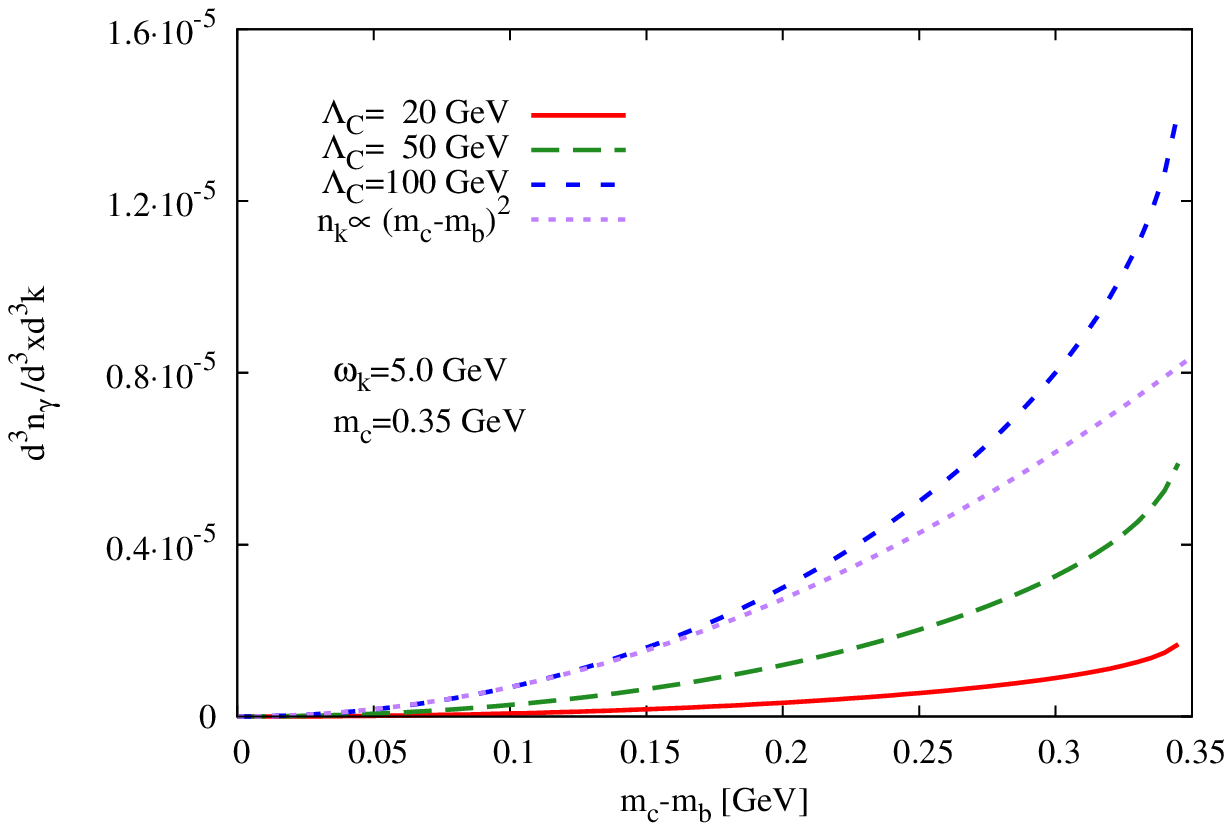}
 \caption{Dependence of the photon yield on the magnitude of 
          the instantaneous mass shift for $\omega_{\vec{k}}=1.0 \; \GeV$ (left panel) 
          and $\omega_{\vec{k}}=5.0 \; \GeV$ (right panel). For small mass shifts, the 
          yield scales $\propto(m_{c}-m_{b})^{2}$, which is particularly visible for 
          large values of $\Lambda_{C}$.}
 \label{fig:4:chiral_massdep}
\end{figure}
For small mass shifts, the photon yield scales with $(m_{c}-m_{b})^{2}$,
which is particularly visible for large $\Lambda_{C}$. This can be
understood by taking into account that the dominant contributions to
(\ref{eq:4:photnum_inst_split}) are given by (\ref{eq:4:contr_22}) and
(\ref{eq:4:contr_23}), which for large $p$ both scale with occupation
numbers $\propto (m_{c}-m_{b})^{2}/p^{2}$. For large mass shifts, i.e., for
$m_{b}\rightarrow0$ at fixed $m_{c}$, the photon yield starts to deviate
from this quadratic scaling. In that limit, the photon yield diverges
due to a collinear and an anti-collinear singularity in the loop
integral. This issue will also be discussed in greater detail in
appendix \ref{sec:appe}. From the phenomenological point of view,
however, it is not a serious problem since the quark masses stay finite
even in the chirally restored phase.

\subsubsection{Summary of results}
So far our investigations on chiral photon production have shown that
the scenario of an instantaneous mass shift essentially comes along with
three unphysical artifacts.

Firstly, the loop integral entering
(\ref{eq:2:photon_yield_chiral_aspt}) features a linear divergence. In
particular, this divergence arises from the contributions describing
quark and antiquark bremsstrahlung and the interference between these
two processes. It is caused by the quark and antiquark occupation
numbers scaling $\propto (m_{c}-m_{b})^{2}/2p^{2}$ for $p\gg
m_{b},m_{c}$.  The latter is an artifact of the instantaneous change of
the quark mass, and the mentioned divergence has been regulated by a
cutoff at $p=\Lambda_{C}$.

Furthermore, we have seen that then the asymptotic photon spectra decay
as $1/\omega^{3}_{\vec{k}}$ for any fixed value of $\Lambda_{C}$, which means
that the total number density and the total energy density of the
produced photons are logarithmically and linearly divergent,
respectively.

Finally, the asymptotic photon yield diverges for
$m_{b}\rightarrow0$. We have demonstrated that this divergence is due to
a collinear and an anticollinear singularity in the loop integral over
the contributions (\ref{eq:4:contr_22}) and (\ref{eq:4:contr_44})
describing quark and antiquark bremsstrahlung and quark-antiquark
annihilation into a photon, respectively. This is, however, a less
serious problem than the two previous ones as it can be circumvented by
leaving the bare mass, $m_{b}$, finite. The latter is justified from the
phenomenological point of view since the quarks masses stay finite even
in the chirally restored phase.

Hence, as the next step, we have to determine if and to which extent
these problems are regulated if the chiral mass shift is assumed to take
place over a finite time interval, $\tau$, which corresponds to a
physically more realistic scenario.

\subsection{Mass shift over a finite time interval}
\label{sec:photons_smooth}

\subsubsection{Calculation of photon numbers}
For the general case of a mass shift over a finite time interval,
$\tau$, both the time evolution of the fermionic wavefunctions,
$\psi_{\vec{p},s,\uparrow\downarrow}(x)$, and the time integrals
entering (\ref{eq:2:photon_yield_chiral_eps}) require a numerical
treatment. Hence, we have to find a way to extract the physical photon
numbers from (\ref{eq:2:photon_yield_chiral_eps}), i.e., the
contributions which persist after taking the successive limits
$t\rightarrow\infty$ and $\varepsilon\rightarrow0$. For this purpose, we
consider the photon self-energy in terms of positive- and
negative-energy wavefunctions again,
\begin{equation}
\begin{split}
 \label{eq:4:pse_wavef_smth}
 \ii\Pi^{<}_{\mu\nu}(\vec{k},t_{1},t_{2}) & =  e^{2}\sum_{r,s}\int \frac{\dd^{3}p}{(2\pi)^{3}}
 \left [\bar{\psi}^{'}_{\vec{p},r,\uparrow}(t_{1})\gamma_{\mu}
   \psi^{'}_{\vec{p}+\vec{k},s,\downarrow}(t_{1}) \right ]\cdot
 \left [ \bar{\psi}^{'}_{\vec{p}+\vec{k},s,\downarrow}(t_{2})\gamma_{\nu}
   \psi^{'}_{\vec{p},r,\uparrow}(t_{2}) \right] \\
 & =  \sum_{r,s}\int\frac{d^{3}p}{(2\pi)^{3}}
 j_{\mu}(\vec{p},\vec{k},t_{1})j^{*}_{\nu}(\vec{p},\vec{k},t_{2}) \ ,
\end{split}
\end{equation}
where we have introduced the effective current
\begin{equation}
 \label{eq:4:current}
 j_{\mu}(\vec{p},\vec{k},r,s,t) = e\bar{\psi}^{'}_{\vec{p},r,\uparrow}(t)\gamma_{\mu}\psi^{'}_{\vec{p}+\vec{k},s,\downarrow}(t) \ .
\end{equation}
Next, we rewrite
\begin{align}
 j_{\mu}(\vec{p},\vec{k},r,s,t) 
  & = e\left[
        \bar{\psi}^{'}_{\vec{p},r,\uparrow}(t)\gamma_{\mu}\psi^{'}_{\vec{p}+\vec{k},s,\downarrow}(t)-
        \bar{\psi}^{c,'}_{\vec{p},r,\uparrow}(t)\gamma_{\mu}\psi^{c,'}_{\vec{p}+\vec{k},s,\downarrow}(t)
       \right]+
      e\bar{\psi}^{c,'}_{\vec{p},r,\uparrow}(t)\gamma_{\mu}\psi^{c,'}_{\vec{p}+\vec{k},s,\downarrow}(t) \nonumber \\
 & = j^{\text{MST}}_{\mu}(\vec{p},\vec{k},r,s,t)+j^{0}_{\mu}(\vec{p},\vec{k},r,s,t) \ ,
\end{align}
In analogy to $\psi^{'}_{\vec{p},r,\uparrow\downarrow}(t)$, the
expressions $\psi^{c,'}_{\vec{p},r,\uparrow\downarrow}(t)$ are defined
according to the relation
$$
\psi^{c}_{\vec{p},r,\uparrow\downarrow}(x)=\psi^{c,'}_{\vec{p},r,\uparrow\downarrow}(t)\ee^{\ii\vec{p}\cdot\vec{x}} \ .
$$
Moreover, we have introduced
\begin{subequations}
 \label{eq:4:current_split}
 \begin{eqnarray}
  j^{\text{MST}}_{\mu}(\vec{p},\vec{k},r,s,t) & = & e\left[
                                                      \bar{\psi}^{'}_{\vec{p},r,\uparrow}(t)\gamma_{\mu}\psi^{'}_{\vec{p}+\vec{k},s,\downarrow}(t)-
                                                      \bar{\psi}^{c,'}_{\vec{p},r,\uparrow}(t)\gamma_{\mu}\psi^{c,'}_{\vec{p}+\vec{k},s,\downarrow}(t)
                                                     \right] \ , \label{eq:4:current_a} \\
  j^{0}_{\mu}(\vec{p},\vec{k},r,s,t)          & = & e\bar{\psi}^{c,'}_{\vec{p},r,\uparrow}(t)\gamma_{\mu}\psi^{c,'}_{\vec{p}+\vec{k},s,\downarrow}(t) \ , \label{eq:4:current_b}
 \end{eqnarray}
\end{subequations}
in the second step. Eq. (\ref{eq:4:current_a}) vanishes if the quark
mass is kept at its initial constituent value, $m_{c}$, for all times,
$t$, but (\ref{eq:4:current_b}) does not. As a consequence, these
expressions can be considered as mass-shift (MST) and vacuum
contribution to (\ref{eq:4:current}). With the help of (\ref{eq:4:current_split}), we
can decompose the photon self-energy according to
\begin{equation}
 \ii\Pi^{<}_{\mu\nu}(\vec{k},t_{1},t_{2}) = \ii\Pi^{<,0}_{\mu\nu}(\vec{k},t_{1},t_{2})+\ii\Pi^{<,\text{MST}}_{\mu\nu}(\vec{k},t_{1},t_{2})+
                                            \ii\Pi^{<,\text{INT}}_{\mu\nu}(\vec{k},t_{1},t_{2}) \ ,
\end{equation}
with the individual contributions given by
\begin{subequations}
 \label{eq:4:pse_split}
 \begin{align}
  \ii\Pi^{<,0}_{\mu\nu}(\vec{k},t_{1},t_{2})          = \sum_{r,s}\int\frac{\dd^{3}p}{(2\pi)^{3}}
                                                        & j^{0}_{\mu}(\vec{p},\vec{k},r,s,t_{1})j^{0,*}_{\nu}(\vec{p},\vec{k},r,s,t_{2}) \ , \label{eq:4:pse_vac} \\
  \ii\Pi^{<,\text{MST}}_{\mu\nu}(\vec{k},t_{1},t_{2}) = \sum_{r,s}\int\frac{\dd^{3}p}{(2\pi)^{3}}
                                                        & j^{\text{MST}}_{\mu}(\vec{p},\vec{k},r,s,t_{1})j^{\text{MST},*}_{\nu}(\vec{p},\vec{k},r,s,t_{2}) \ , \label{eq:4:pse_mst} \\
  \ii\Pi^{<,\text{INT}}_{\mu\nu}(\vec{k},t_{1},t_{2}) = \sum_{r,s}\int\frac{\dd^{3}p}{(2\pi)^{3}} 
                                                         & \left[
                                                            j^{\text{MST}}_{\mu}(\vec{p},\vec{k},r,s,t_{1})j^{0,*}_{\nu}(\vec{p},\vec{k},r,s,t_{2})
                                                           \right. \nonumber \\
                                                         & \left.
                                                            +j^{0}_{\mu}(\vec{p},\vec{k},r,s,t_{1})j^{\text{MST},*}_{\nu}(\vec{p},\vec{k},r,s,t_{2})
                                                           \right] \label{eq:4:pse_int} \ .
 \end{align}
\end{subequations}
Expressions (\ref{eq:4:pse_vac}) and (\ref{eq:4:pse_mst}) describe the
contributions from the vacuum polarization (VAC) and the mass shift (MST) to the
overall photon self-energy, respectively, whereas (\ref{eq:4:pse_int})
characterizes their interference among each other (INT) . Hence, it is
convenient to split up (\ref{eq:2:photon_yield_chiral_eps}) accordingly
\begin{equation}
 2\omega_{\vec{k}}\frac{\dd^{6}n^{\varepsilon}_{\gamma}(t)}{\dd^{3}x\dd^{3}k} 
   = \left. 2\omega_{\vec{k}}\frac{\dd^{6}n^{\varepsilon}_{\gamma}(t)}{\dd^{3}x\dd^{3}k}\right|_{\text{VAC}}+
     \left. 2\omega_{\vec{k}}\frac{\dd^{6}n^{\varepsilon}_{\gamma}(t)}{\dd^{3}x\dd^{3}k}\right|_{\text{MST}}+
     \left. 2\omega_{\vec{k}}\frac{\dd^{6}n^{\varepsilon}_{\gamma}(t)}{\dd^{3}x\dd^{3}k}\right|_{\text{INT}} \ . \\
\end{equation}
The individual contributions read
\begin{subequations}
 \label{eq:4:photnum_split}
 \begin{eqnarray}
  \left. 2\omega_{\vec{k}}\frac{\dd^{6}n^{\varepsilon}_{\gamma}(t)}{\dd^{3}x\dd^{3}k}\right|_{\text{VAC}}
     & = & \frac{\gamma^{\mu\nu}(k)}{(2\pi)^{3}}\int_{-\infty}^{t}dt_{1}\int_{-\infty}^{t}dt_{2}
           \mbox{ }\ii\Pi^{<,0}_{\nu\mu}(\vec{k},t_{1},t_{2})\ee^{\ii\omega_{\vec{k}}(t_{1}-t_{2})} \ , \label{eq:4:photnum_vac} \\
  \left. 2\omega_{\vec{k}}\frac{\dd^{6}n^{\varepsilon}_{\gamma}(t)}{\dd^{3}x\dd^{3}k}\right|_{\text{MST}}
     & = & \frac{\gamma^{\mu\nu}(k)}{(2\pi)^{3}}\int_{-\infty}^{t}dt_{1}\int_{-\infty}^{t}dt_{2}
           \mbox{ }\ii\Pi^{<,\text{MST}}_{\nu\mu}(\vec{k},t_{1},t_{2})\ee^{\ii\omega_{\vec{k}}(t_{1}-t_{2})} \ , \label{eq:4:photnum_mst} \\
  \left. 2\omega_{\vec{k}}\frac{\dd^{6}n^{\varepsilon}_{\gamma}(t)}{\dd^{3}x\dd^{3}k}\right|_{\text{INT}}
     & = & \frac{\gamma^{\mu\nu}(k)}{(2\pi)^{3}}\int_{-\infty}^{t}dt_{1}\int_{-\infty}^{t}dt_{2}
           \mbox{ }\ii\Pi^{<,\text{INT}}_{\nu\mu}(\vec{k},t_{1},t_{2})\ee^{\ii\omega_{\vec{k}}(t_{1}-t_{2})} \ . \label{eq:4:photnum_int}
 \end{eqnarray}
\end{subequations}
In appendix \ref{sec:appa} it is shown that the contribution from the
vacuum polarization (\ref{eq:4:pse_vac}) vanishes under the successive
limits $t\rightarrow\infty$ and then $\varepsilon\rightarrow0$. We now
demonstrate that the contribution from the interference term
(\ref{eq:4:pse_int}) is also eliminated by this procedure so that only
the mass-shift contribution (\ref{eq:4:photnum_mst}) remains and thus
describes the physical photon number. For this purpose, we first rewrite
the asymptotic contributions from (\ref{eq:4:pse_mst}) and
(\ref{eq:4:pse_int}) at finite $\varepsilon$ by interchanging the time
integrals with the loop integral over $\dd^{3}p$. This leads to
\begin{subequations}
 \begin{align}
  \left. 2\omega_{\vec{k}}\frac{\dd^{6}n^{\varepsilon}_{\gamma}}{\dd^{3}x\dd^{3}k}\right|_{\text{MST}} 
    = \frac{\gamma^{\mu\nu}(k)}{(2\pi)^{3}}\sum_{r,s}\int\frac{\dd^{3}p}{(2\pi)^{3}} & 
      I^{\varepsilon,*}_{\mu}(\vec{p},\vec{k},r,s)I^{\varepsilon}_{\nu}(\vec{p},\vec{k},r,s) \ , \\
  \left. 2\omega_{\vec{k}}\frac{\dd^{6}n^{\varepsilon}_{\gamma}}{\dd^{3}x\dd^{3}k}\right|_{\text{INT}}
    = \frac{\gamma^{\mu\nu}(k)}{(2\pi)^{3}}\sum_{r,s}\int\frac{\dd^{3}p}{(2\pi)^{3}} &
       \left[
        I^{\varepsilon,*}_{\mu}(\vec{p},\vec{k},r,s)J^{\varepsilon}_{\nu}(\vec{p},\vec{k},r,s)
       \right. \nonumber \\
        & +\left.
            J^{\varepsilon,*}_{\mu}(\vec{p},\vec{k},r,s)I^{\varepsilon}_{\nu}(\vec{p},\vec{k},r,s)
           \right] \ .
 \end{align}
\end{subequations}
In order to keep the notation short, we have introduced
\begin{subequations}
 \label{eq:4:intaspt}
 \begin{eqnarray}
  I^{\varepsilon}_{\mu}(\vec{p},\vec{k},r,s) & = & \int_{-\infty}^{\infty}\dd t f_{\varepsilon}(t)j^{\text{MST}}_{\mu}(\vec{p},\vec{k},r,s,t)\ee^{\ii\omega_{\vec{k}}t} \ , \label{eq:4:intaspt_mst} \\
  J^{\varepsilon}_{\mu}(\vec{p},\vec{k},r,s) & = & \int_{-\infty}^{\infty}\dd t f_{\varepsilon}(t)j^{0}_{\mu}(\vec{p},\vec{k},r,s,t)\ee^{\ii\omega_{\vec{k}}t} \ , \label{eq:4:intaspt_vac}
 \end{eqnarray}
\end{subequations}
The time integral in (\ref{eq:4:intaspt_vac}) evaluates to
\begin{equation}
 \label{eq:4:intaspt_vac_eval}
 J^{\varepsilon}_{\mu}(\vec{p},\vec{k},r,s) = e\bar{u}_{c}(\vec{p},s)\gamma_{\mu}v_{c}(\vec{p}+\vec{k},s)\frac{2\varepsilon}{\varepsilon^{2}+\omega^{c,2}_{1}(\vec{p},\vec{k})} \ ,
\end{equation}
with $\omega^{c}_{1}(\vec{p},\vec{k})$ given by (\ref{eq:4:freq_1}). To
handle the time integral entering (\ref{eq:4:intaspt_mst}), we first
split
\begin{equation}
 \label{eq:4:intaspt_split}
 I^{\varepsilon}_{\mu}(\vec{p},\vec{k},r,s) 
  = \int_{-\infty}^{T}\dd t f_{\varepsilon}(t)j^{\text{MST}}_{\mu}(\vec{p},\vec{k},r,s,t)\ee^{\ii\omega_{\vec{k}}t}+
    \int_{T}^{\infty}\dd t f_{\varepsilon}(t)j^{\text{MST}}_{\mu}(\vec{p},\vec{k},r,s,t)\ee^{\ii\omega_{\vec{k}}t} \ ,
\end{equation}
where $T\gg\tau$. Next we take into account that for $t\ge T$, the
fermionic wavefunctions have essentially turned into superpositions of
positive- and negative-energy states with respect to the final bare
mass, $m_{b}$, i.e.,
\begin{subequations}
 \label{eq:4:expand_smth}
 \begin{eqnarray}
  \psi^{'}_{\vec{p},s,\uparrow}(t)   & = & \tilde{\alpha}_{\vec{p}}\psi^{b,'}_{\vec{p},s,\uparrow}(t)+
                                           \tilde{\beta}_{\vec{p}}\psi^{b,'}_{\vec{p},s,\downarrow}(t) \ , \\
  \psi^{'}_{\vec{p},s,\downarrow}(t) & = & \tilde{\gamma}_{\vec{p}}\psi^{b,'}_{\vec{p},s,\downarrow}(t)+
                                           \tilde{\delta}_{\vec{p}}\psi^{b,'}_{\vec{p},s,\uparrow}(t) \ , 
 \end{eqnarray}
\end{subequations}
with the coefficients \emph{not} depending on time. We have introduced
the $\tilde{\cdot}$ notation in order to highlight that the expansion
coefficients are generally different from those for an instantaneous
mass shift given by (\ref{eq:3:coeff_link}). With the help of
(\ref{eq:4:expand_smth}), expression (\ref{eq:4:intaspt_split}) is
further evaluated to
\begin{equation}
\label{eq:4:intaspt_mst_eval}
\begin{split}
 I^{\varepsilon}_{\mu}(\vec{p},\vec{k},r,s)
   = & \int_{-\infty}^{T}\dd t f_{\varepsilon}(t)j^{\text{MST}}_{\mu}(\vec{p},\vec{k},r,s,t)\ee^{\ii\omega_{\vec{k}}t}  \\
   + & e\left\lbrace
          \tilde{\alpha}^{*}_{\vec{p}}\tilde{\gamma}_{\vec{p}+\vec{k}}\bar{u}_{b}(\vec{p},r)\gamma_{\mu}v_{b}(\vec{p}+\vec{k},s)
          \frac{\ee^{-\left[\varepsilon-\ii\omega^{b}_{1}(\vec{p},\vec{k})\right]T}}{\varepsilon-\ii\omega^{b}_{1}(\vec{p},\vec{k})}
         \right.  \\
     & +\tilde{\alpha}^{*}_{\vec{p}}\tilde{\delta}_{\vec{p}+\vec{k}}\bar{u}_{b}(\vec{p},r)\gamma_{\mu}u_{b}(\vec{p}+\vec{k},s)
        \frac{\ee^{-\left[\varepsilon+\ii\omega^{b}_{2}(\vec{p},\vec{k})\right]T}}{\varepsilon+\ii\omega^{b}_{2}(\vec{p},\vec{k})}  \\
     & +\tilde{\beta}^{*}_{\vec{p}}\tilde{\gamma}_{\vec{p}+\vec{k}}\bar{v}_{b}(\vec{p},r)\gamma_{\mu}v_{b}(\vec{p}+\vec{k},s)
        \frac{\ee^{-\left[\varepsilon-\ii\omega^{b}_{3}(\vec{p},\vec{k})\right]T}}{\varepsilon-\ii\omega^{b}_{3}(\vec{p},\vec{k})}  \\
     & +\tilde{\beta}^{*}_{\vec{p}}\tilde{\delta}_{\vec{p}+\vec{k}}\bar{v}_{b}(\vec{p},r)\gamma_{\mu}u_{b}(\vec{p}+\vec{k},s)
         \frac{\ee^{-\left[\varepsilon+\ii\omega^{b}_{4}(\vec{p},\vec{k})\right]T}}{\varepsilon+\ii\omega^{b}_{4}(\vec{p},\vec{k})}  \\
     & -\left.
         \bar{u}_{c}(\vec{p},r)\gamma_{\mu}v_{c}(\vec{p}+\vec{k},s)
         \frac{\ee^{-\left[\varepsilon-\ii\omega^{c}_{1}(\vec{p},\vec{k})\right]T}}{\varepsilon-\ii\omega^{c}_{1}(\vec{p},\vec{k})}
        \right\rbrace \ .
\end{split}
\end{equation}
Since the frequencies (\ref{eq:4:freq_1})-(\ref{eq:4:freq_4}) are either
positive or negative definite, taking the limit $\varepsilon\rightarrow
0$ leads to
\begin{equation}
\begin{split}
 \label{eq:4:intaspt_zero}
 I^{\varepsilon}_{\mu}(\vec{p},\vec{k},r,s) \rightarrow I_{\mu}(\vec{p},\vec{k},r,s)
  = & \int_{-\infty}^{T}\dd t\mbox{ }j^{\text{MST}}_{\mu}(\vec{p},\vec{k},r,s,t)\ee^{\ii\omega_{\vec{k}}t}  \\
  + & \ii e\left[
            \tilde{\alpha}^{*}_{\vec{p}}\tilde{\gamma}_{\vec{p}+\vec{k}}\bar{u}_{b}(\vec{p},r)\gamma_{\mu}v_{b}(\vec{p}+\vec{k},s)
            \frac{\ee^{\ii\omega^{b}_{1}(\vec{p},\vec{k})}}{\omega^{b}_{1}(\vec{p},\vec{k})}
           \right.  \\
    & -\tilde{\alpha}^{*}_{\vec{p}}\tilde{\delta}_{\vec{p}+\vec{k}}\bar{u}_{b}(\vec{p},r)\gamma_{\mu}u_{b}(\vec{p}+\vec{k},s)
      \frac{\ee^{-\ii\omega^{b}_{2}(\vec{p},\vec{k})}}{\omega^{b}_{2}(\vec{p},\vec{k})}  \\
    & +\tilde{\beta}^{*}_{\vec{p}}\tilde{\gamma}_{\vec{p}+\vec{k}}\bar{v}_{b}(\vec{p},r)\gamma_{\mu}v_{b}(\vec{p}+\vec{k},s)
      \frac{\ee^{\ii\omega^{b}_{3}(\vec{p},\vec{k})}}{\omega^{b}_{3}(\vec{p},\vec{k})}  \\
    & -\tilde{\beta}^{*}_{\vec{p}}\tilde{\delta}_{\vec{p}+\vec{k}}\bar{v}_{b}(\vec{p},r)\gamma_{\mu}u_{b}(\vec{p}+\vec{k},s)
      \frac{\ee^{-\ii\omega^{b}_{4}(\vec{p},\vec{k})}}{\omega^{b}_{4}(\vec{p},\vec{k})}  \\
    & -\left.
       \bar{u}_{c}(\vec{p},r)\gamma_{\mu}v_{c}(\vec{p}+\vec{k},s)\frac{\ee^{\ii\omega^{c}_{1}(\vec{p},\vec{k})}}{\omega^{c}_{1}(\vec{p},\vec{k})}
       \right] \ .
\end{split}
\end{equation}
We are allowed to interchange the limiting process with the remaining
time integral since $j^{\text{MST}}_{\mu}(\vec{p},\vec{k},r,s,t)\ee^{\ii\omega_{\vec{k}}t}$ 
vanishes for $t\rightarrow-\infty$ and is thus integrable by
itself on the time interval $(-\infty;T]$. As we also have from
(\ref{eq:4:intaspt_vac_eval})
\begin{equation}
 J^{\varepsilon}_{\mu}(\vec{p},\vec{k},r,s) = e\bar{u}_{c}(\vec{p},r)\gamma_{\mu}v_{c}(\vec{p}+\vec{k},s)
                                              \frac{2\varepsilon}{\omega^{c,2}_{1}(\vec{p},\vec{k})}+\mathcal{O}(\varepsilon^{2}) \ , \
                                              \mbox{for} \ \varepsilon\rightarrow0 \ ,
\end{equation}
the interference contribution vanishes in that limit. Furthermore, the
mass-shift contribution, which as a consequence of the above describes
the actual photon number, turns into
\begin{equation}
 \label{eq:4:photnum_phys}
 2\omega_{\vec{k}}\frac{\dd^{6}n_{\gamma}}{\dd^{3}x\dd^{3}k} = \frac{\gamma^{\mu\nu}(k)}{(2\pi)^{3}}\sum_{r,s}\int\frac{\dd^{3}p}{(2\pi)^{3}}
                                          I^{*}_{\mu}(\vec{p},\vec{k},r,s)I_{\nu}(\vec{p},\vec{k},r,s) \ .
\end{equation}
It follows from (\ref{eq:4:photnum_phys}) and (\ref{eq:4:intaspt_zero})
that solving the equations of motion (\ref{eq:2:eom_param}) numerically
on the time interval $[-T;T]$ essentially provides all the information
required to evaluate $I_{\mu}(\vec{p},\vec{k},r,s)$ and the asymptotic
photon numbers (\ref{eq:4:photnum_phys}). We have
$j^{\text{MST}}_{\mu}(\vec{p},\vec{k},r,s,t)\approx0$ for $t\le -T$
since $T\gg\tau$.  We thus can approximate
$$
\int_{-\infty}^{T}\dd t\mbox{ }j^{\text{MST}}_{\mu}(\vec{p},\vec{k},r,s,t)\ee^{\ii\omega_{\vec{k}}t} 
 \approx\int_{-T}^{T}\dd t\mbox{ }j^{\text{MST}}_{\mu}(\vec{p},\vec{k},r,s,t)\ee^{\ii\omega_{\vec{k}}t} \ .
$$
Hence, the numerical solution of (\ref{eq:2:eom_param}) on $[-T;T]$
allows us to evaluate the time integral entering
(\ref{eq:4:photnum_phys}) with sufficiently high accuracy. Based on this
solution, the asymptotic expansion coefficients can be projected out
from (\ref{eq:4:expand_smth}) at $t=T$.

So far, we have restricted ourselves to the scenario where the quark
mass is only changed from its constituent value, $m_{c}$, to its
bare value, $m_{b}$. When considering the second scenario, where the
quark mass is first changed from $m_{c}$ to $m_{b}$ and then back to
$m_{c}$, the asymptotic photon number is, however, determined mostly in
the same way. The only difference is that $T$ then has to be chosen such
that $T\gg\tau+\frac{\tau_{L}}{2}$ with $\tau_{L}$ denoting the lifetime
of the chirally restored phase. Moreover, one has to take into account
that the fermionic wavefunctions have turned into superpositions of
positive- and negative-energy states of mass $m_{c}$ instead of mass
$m_{b}$ for $t\ge T$, i.e.,
\begin{subequations}
 \label{eq:4:expand_smth_back}
 \begin{eqnarray}
  \psi^{'}_{\vec{p},s,\uparrow}(t)   & = & \tilde{\alpha}_{\vec{p}}\psi^{c,'}_{\vec{p},s,\uparrow}(t)+
                                           \tilde{\beta}_{\vec{p}}\psi^{c,'}_{\vec{p},s,\downarrow}(t) \ , \\
  \psi^{'}_{\vec{p},s,\downarrow}(t) & = & \tilde{\gamma}_{\vec{p}}\psi^{c,'}_{\vec{p},s,\downarrow}(t)+
                                           \tilde{\delta}_{\vec{p}}\psi^{c,'}_{\vec{p},s,\uparrow}(t) \ .
 \end{eqnarray}
\end{subequations}
Accordingly, expression (\ref{eq:4:intaspt_zero}) is replaced by
\begin{equation}
\label{eq:4:intaspt_zero_back}
\begin{split}
 I_{\mu}(\vec{p},\vec{k},r,s)
  = & \int_{-\infty}^{T}\dd t\mbox{ }j^{\text{MST}}_{\mu}(\vec{p},\vec{k},r,s,t)\ee^{\ii\omega_{\vec{k}}t}  \\
  + & \ii e\left[
            \left(\tilde{\alpha}^{*}_{\vec{p}}\tilde{\gamma}_{\vec{p}+\vec{k}}-1\right)
            \bar{u}_{c}(\vec{p},r)\gamma_{\mu}v_{c}(\vec{p}+\vec{k},s)
            \frac{\ee^{\ii\omega^{c}_{1}(\vec{p},\vec{k})}}{\omega^{c}_{1}(\vec{p},\vec{k})}
           \right.  \\
    & -\tilde{\alpha}^{*}_{\vec{p}}\tilde{\delta}_{\vec{p}+\vec{k}}
      \bar{u}_{c}(\vec{p},r)\gamma_{\mu}u_{c}(\vec{p}+\vec{k},s)
      \frac{\ee^{-\ii\omega^{c}_{2}(\vec{p},\vec{k})}}{\omega^{c}_{2}(\vec{p},\vec{k})}  \\
    & +\tilde{\beta}^{*}_{\vec{p}}\tilde{\gamma}_{\vec{p}+\vec{k}}
      \bar{v}_{c}(\vec{p},r)\gamma_{\mu}v_{c}(\vec{p}+\vec{k},s)
      \frac{\ee^{\ii\omega^{c}_{3}(\vec{p},\vec{k})}}{\omega^{c}_{3}(\vec{p},\vec{k})}  \\
    & \left.
       -\tilde{\beta}^{*}_{\vec{p}}\tilde{\delta}_{\vec{p}+\vec{k}}\bar{v}_{c}(\vec{p},r)\gamma_{\mu}u_{c}(\vec{p}+\vec{k},s)
       \frac{\ee^{-\ii\omega^{c}_{4}(\vec{p},\vec{k})}}{\omega^{c}_{4}(\vec{p},\vec{k})} 
      \right] \ .
\end{split}
\end{equation}
For completeness, we mention that with the help of
$$
\gamma^{\mu\nu}(k)=\sum_{\lambda}\varepsilon^{\mu,*}(\vec{k},\lambda)\varepsilon^{\nu}(\vec{k},\lambda) \ , 
$$
expression (\ref{eq:4:photnum_phys}) can be brought into the following
alternative absolute-square representation
\begin{equation}
   2\omega_{\vec{k}}\frac{\dd^{6}n_{\gamma}}{\dd^{3}x\dd^{3}k} = \frac{1}{(2\pi)^{3}}\sum_{\lambda,r,s}\int\frac{\dd^{3}p}{(2\pi)^{3}}
                                            \left|\varepsilon^{\mu}(\vec{k},\lambda)I_{\mu}(\vec{p},\vec{k},r,s)\right|^{2} \ .
\end{equation}
Thus, the photon number is positive (semi-) definite and cannot acquire
unphysical negative values. Furthermore, it vanishes if no mass shift
takes place at all since we then have
$j^{\text{MST}}_{\mu}(\vec{p},\vec{k},r,s,t)\equiv0$.

\subsubsection{Numerical investigations and results}
First of all, we have to determine whether the linear divergence in the
loop integral entering expression (\ref{eq:2:photon_yield_chiral_aspt})
for the photon yield is cured if the mass shift is assumed to take place
over a finite time interval, $\tau$. For this purpose, we consider the
cutoff dependence of the asymptotic photon number for different photon
energies, $\omega_{\vec{k}}$, and different mass parameterizations, $m_{i}(t)$, which is
depicted in Fig.~\ref{fig:4:convergence}. As mass parameters, we have
again chosen $m_{c}=0.35 \;\GeV$ and $m_{b}=0.01 \;\GeV$.
\begin{figure}[htb]
 \begin{center}
  \includegraphics[height=5.0cm]{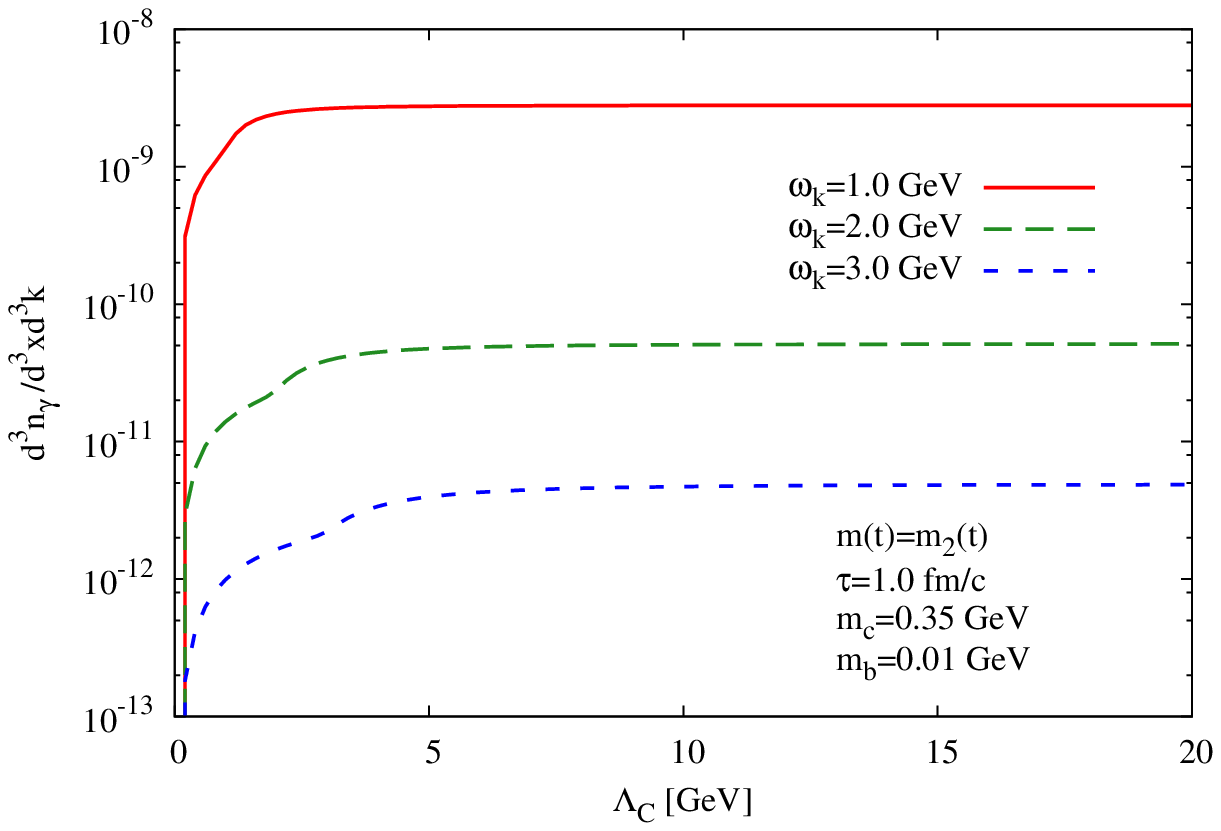}
  \includegraphics[height=5.0cm]{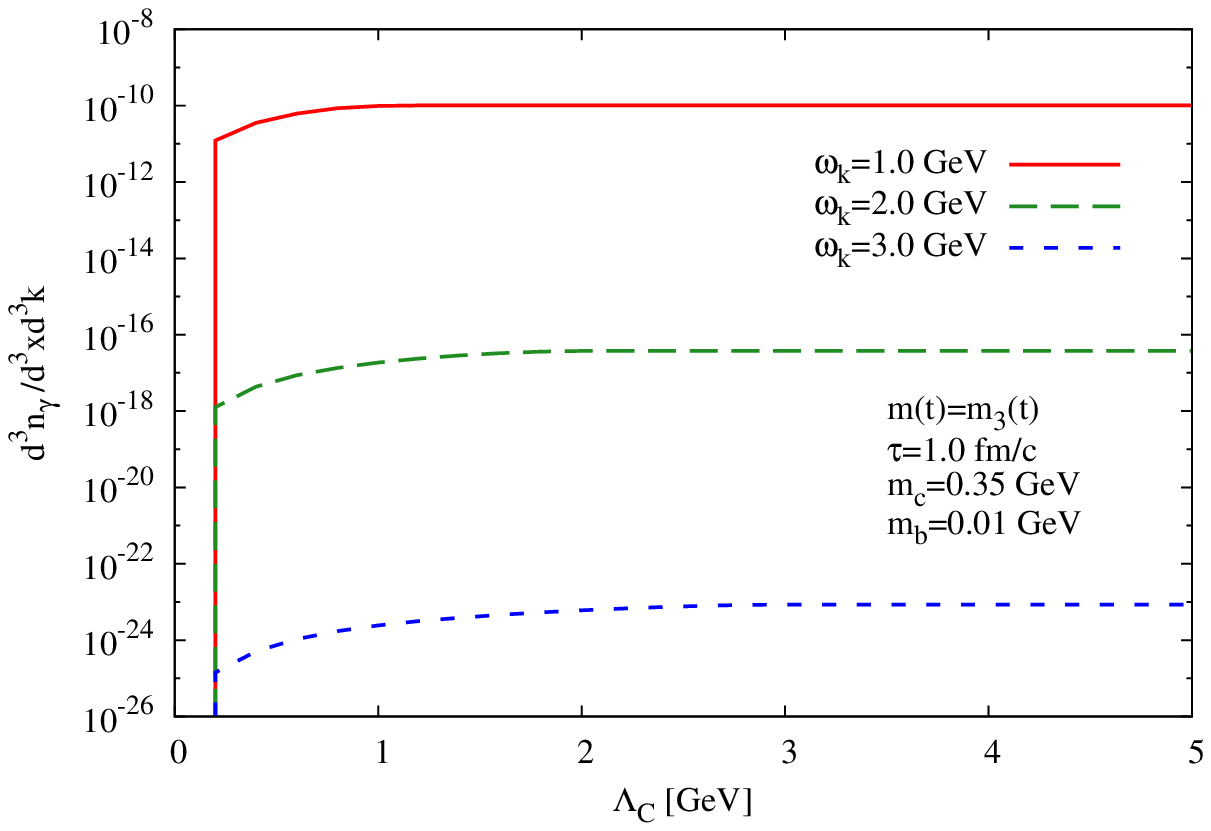}
  \caption{Cutoff dependence of asymptotic photon numbers for $m_{2}(t)$
    (left panel) and $m_{3}(t)$ (right panel).  In both cases, we have
    chosen $\tau=1.0 \; \fm/c$. The parameterizations $m_{i}(t)$ are shown 
    in Fig. \ref{fig:3:chiral_masschange} and defined in (\ref{eq:3:chiral_massparam}). 
    The loop integral is rendered finite and its saturation behavior 
     crucially depends on the order of differentiability of the considered mass 
     parametrization.}
  \label{fig:4:convergence}
 \end{center}
\end{figure}

We see that the linear divergence, which has shown up in the loop
integral for an instantaneous mass shift, is absent for both
parameterizations $m_{2}(t)$ and $m_{3}(t)$. In particular, the order of
differentiability of the considered mass parametrization, $m_{i}(t)$,
is crucial for the saturation behavior of the loop integral. For
$m_{2}(t)$ being continuously differentiable once, the loop integral
saturates at $\Lambda_{C}\simeq10 \; \GeV$ whereas it exhibits a much
faster saturation already at $\Lambda_{C}\simeq2.0-3.0 \; \GeV$ for
$m_{3}(t)$ being continuously differentiable infinitely many
times. Since the latter parametrization describes the most physical
scenario, chiral photon production can be considered as a low-momentum
phenomenon.

As the loop integral is finite for a mass shift over a finite time
interval, $\tau$, we can now turn to the UV behavior of the resulting
photon spectra. Fig.~\ref{fig:4:photspec_comp} compares the resulting
photon spectra for the different mass parameterizations. For $m_{2}(t)$
and $m_{3}(t)$, a transition time of $\tau=1.0 \; \fm/c$ has been assumed.
\begin{figure}[htb]
 \begin{center}
  \includegraphics[height=5.0cm]{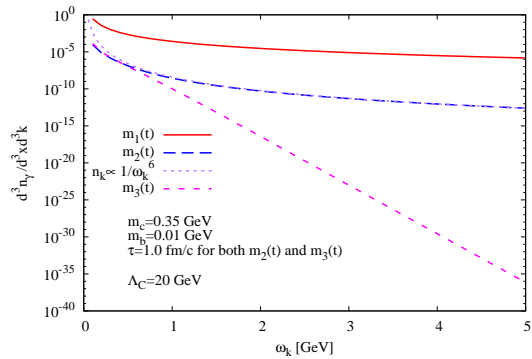}
  \caption{Asymptotic photon spectra for the different mass parameterizations 
           given by Eqs. (\ref{eq:3:chiral_massparam}). If one turns from an instantaneous mass shift ($m_{1}(t)$) to a mass shift 
           over a finite time interval ($m_{2,3}(t)$), the photon spectra are 
           rendered integrable in the ultraviolet domain. Furthermore, their 
           decay behavior there is highly sensitive to the order 
           of differentiability of the considered mass parametrization, $m_{i}(t)$.}
  \label{fig:4:photspec_comp}
 \end{center}
\end{figure}
Analogously to the particle spectra investigated in section
\ref{sec:pair-production}, we see that the asymptotic photon spectra
exhibit a strong sensitivity to the order of differentiability, i.e.,
the `smoothness' of the considered mass parametrization, $m_{i}(t)$. In
particular, the decay behavior in the ultraviolet domain is suppressed
from $\propto 1/\omega^{3}_{\vec{k}}$ to $\propto 1/\omega^{6}_{\vec{k}}$ if we turn from $m_{1}(t)$
(discontinuous parametrization) to $m_{2}(t)$ (parametrization being
continuously differentiable once). The logarithmic and linear UV
divergences in the total photon-number density and the total energy
density, respectively, are thus cured. Furthermore, if we consider the
photon spectra for $m_{3}(t)$, which is continuously differentiable
infinitely many times and hence describes the most physical scenario,
the photon numbers in the UV domain are suppressed even further to an
exponential decay.

As one can infer from Fig.~\ref{fig:4:photspec_tauzero}, the decay
behavior of the photon spectra is highly sensitive to the considered
transition time, $\tau$, for both $m_{2}(t)$ and $m_{3}(t)$. In each
case, the suppression of the photon numbers compared to the
instantaneous case is the stronger the more slowly the mass shift is
assumed to take place. As expected, both parameterizations reproduce the
photon spectra for an instantaneous mass shift in the limit
$\tau\rightarrow0$.
\begin{figure}[htb]
 \begin{center}
  \includegraphics[height=5.0cm]{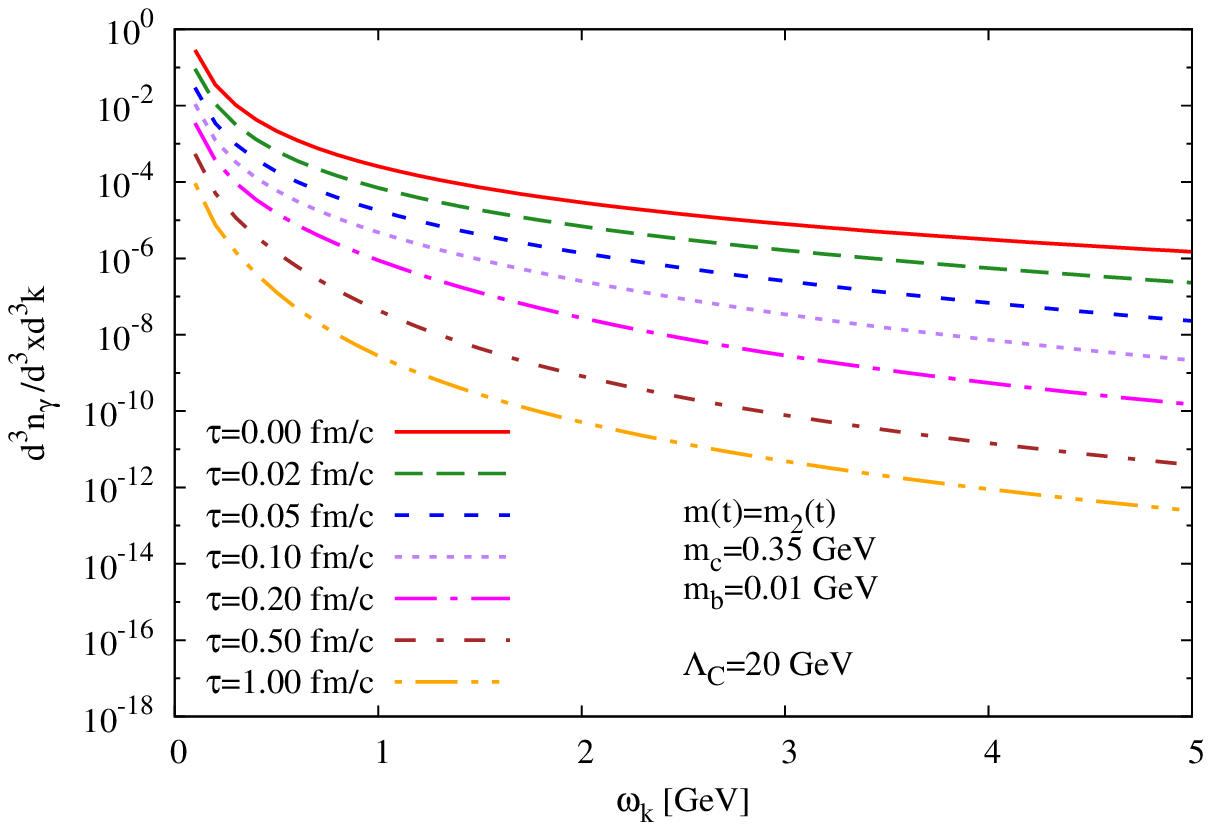}
  \includegraphics[height=5.0cm]{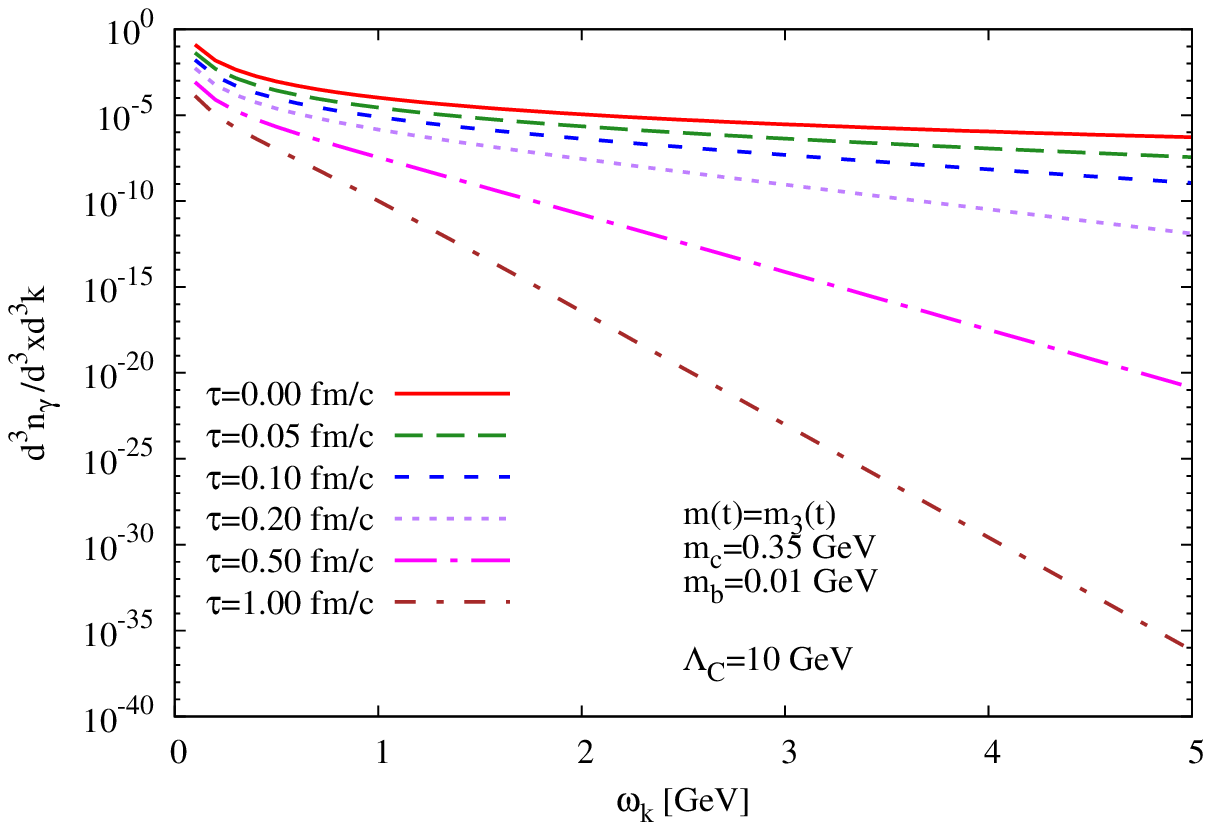}
  \caption{Asymptotic photon spectra for different transition times, $\tau$, for 
           $m_{2}(t)$ (left panel) and $m_{3}(t)$ (right panel). The suppression of the photon numbers 
           with respect to the instantaneous case is the stronger the more slowly ($\tau$ increasing) 
           the mass shift is assumed to take place.}
  \label{fig:4:photspec_tauzero}
 \end{center}
\end{figure}

It should be mentioned that because of finite machine precision, it was
first difficult to resolve the photon numbers numerically for $m_{3}(t)$
and $\tau=1.0$ fm/c in the domain $\omega_{\vec{k}}\ge 2.5$ GeV. For that reason, the
photon numbers in that domain have been extrapolated from those for
$1.5\le \omega_{\vec{k}}\le 2.5$ GeV by performing a linear regression of the
logarithms of the photon numbers.

We shall briefly point out why we have to go to comparatively small
transition times of $\tau\simeq 0.02 \;\fm/c$ for the photon numbers to
be again of the same order of magnitude as for an instantaneous mass
shift. The main reason is the different convergence behavior of the loop
integral for different mass parameterizations. For the case of an
instantaneous mass shift it features a linear divergence, which means
that all momentum modes with $p\gg m_{c},m_{b}$ and $p\gg\omega_{\vec{k}}$ contribute
more or less equally to (\ref{eq:2:photon_yield_chiral_aspt}). For the
case of a mass shift over a finite time interval, $\tau$, however, the
loop integral is UV finite so that the contributions from the different
momentum modes are suppressed with increasing $p$. This implies that the
suppression of (\ref{eq:2:photon_yield_chiral_aspt}) for given $\tau$
with respect to the instantaneous case, $m_{1}(t)$, is the stronger the
larger the value of $\Lambda_{C}$ is chosen. Therefore, the larger
$\Lambda_{C}$ is chosen the smaller $\tau$ has to be taken in order to
approximately reproduce the photon yield for $m_{1}(t)$. This can also
be inferred from Fig. \ref{fig:4:cutoff_zero} displaying the cutoff
dependence of the photon yield for $\omega_{\vec{k}}=1.0 \; \GeV$ and different
transition times, $\tau$.
\begin{figure}[htb]
 \begin{center}
  \includegraphics[height=5.0cm]{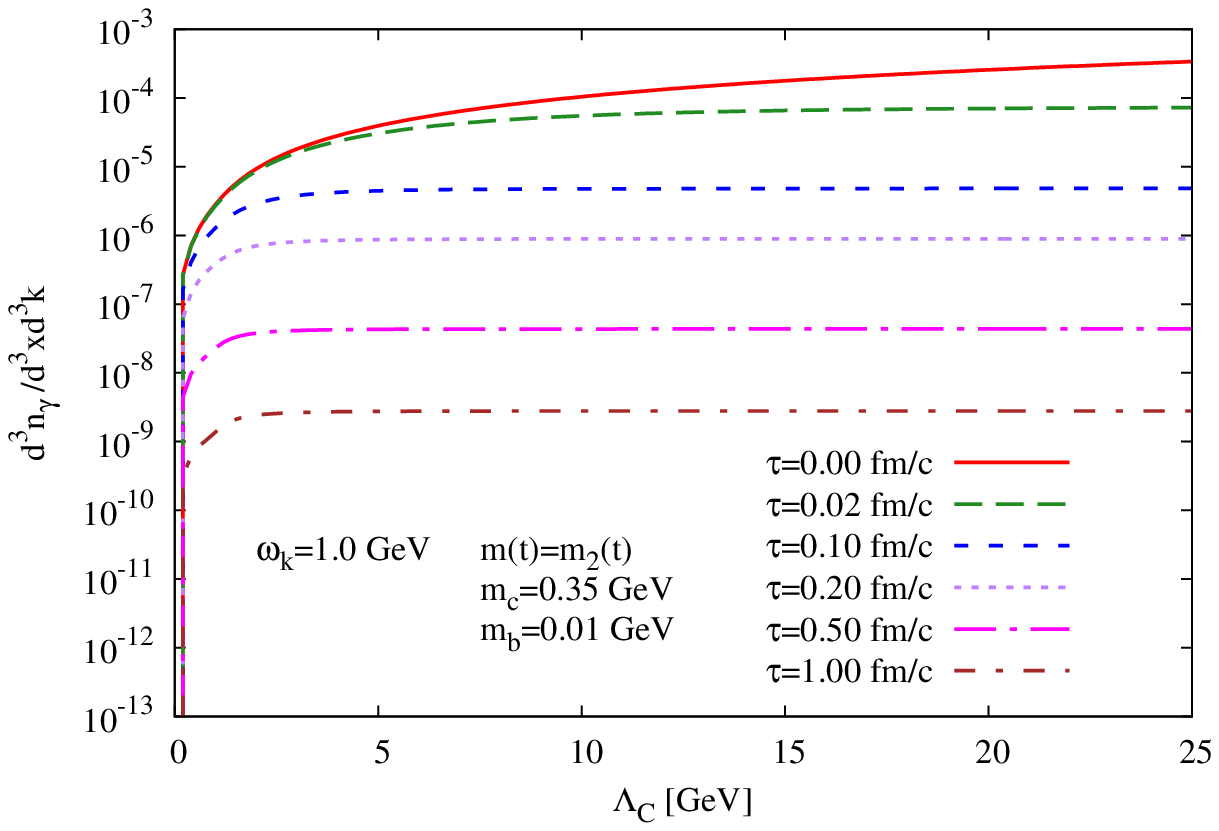}
  \includegraphics[height=5.0cm]{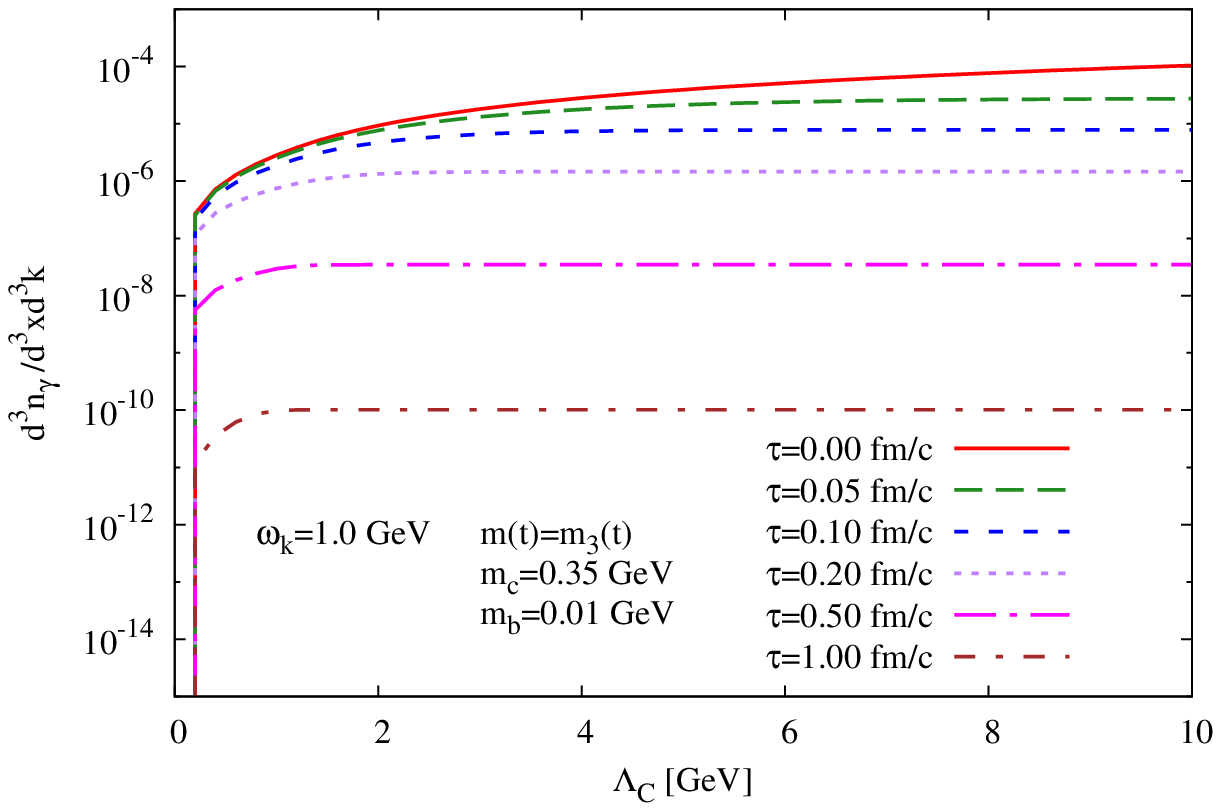}
  \caption{Cutoff dependence of the photon yield for $m_{2}(t)$ (left panel) and
    $m_{3}(t)$ (right panel) at different values of $\tau$. For given finite $\tau$ the suppression
    compared to the instantaneous case is the stronger the
    larger $\Lambda_{C}$ is chosen.}
  \label{fig:4:cutoff_zero}
 \end{center}
\end{figure}

For completeness, we also investigate the dependence of the resulting
photon spectra on the magnitude of the mass shift, which is depicted in
Fig.~\ref{fig:4:massdep}.
\begin{figure}[htb]
 \begin{center}
  \includegraphics[height=5.0cm]{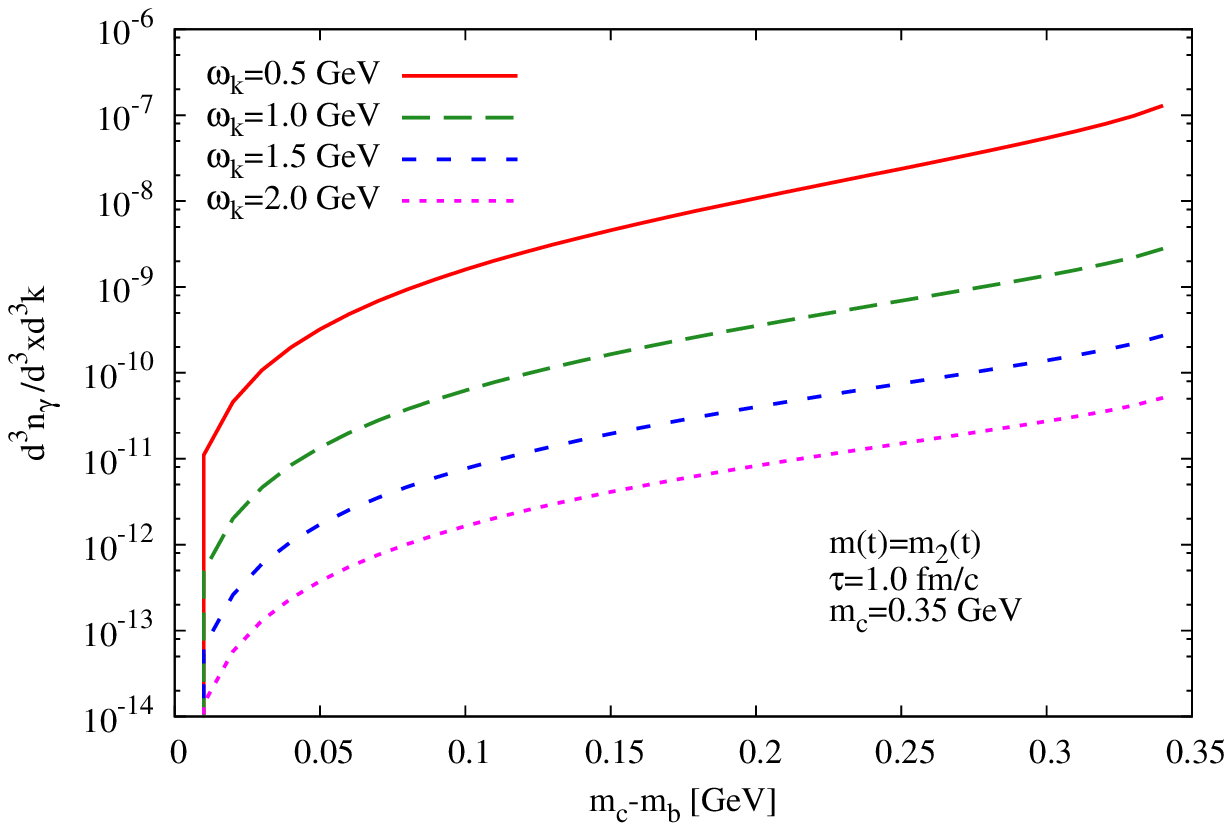}
  \includegraphics[height=5.0cm]{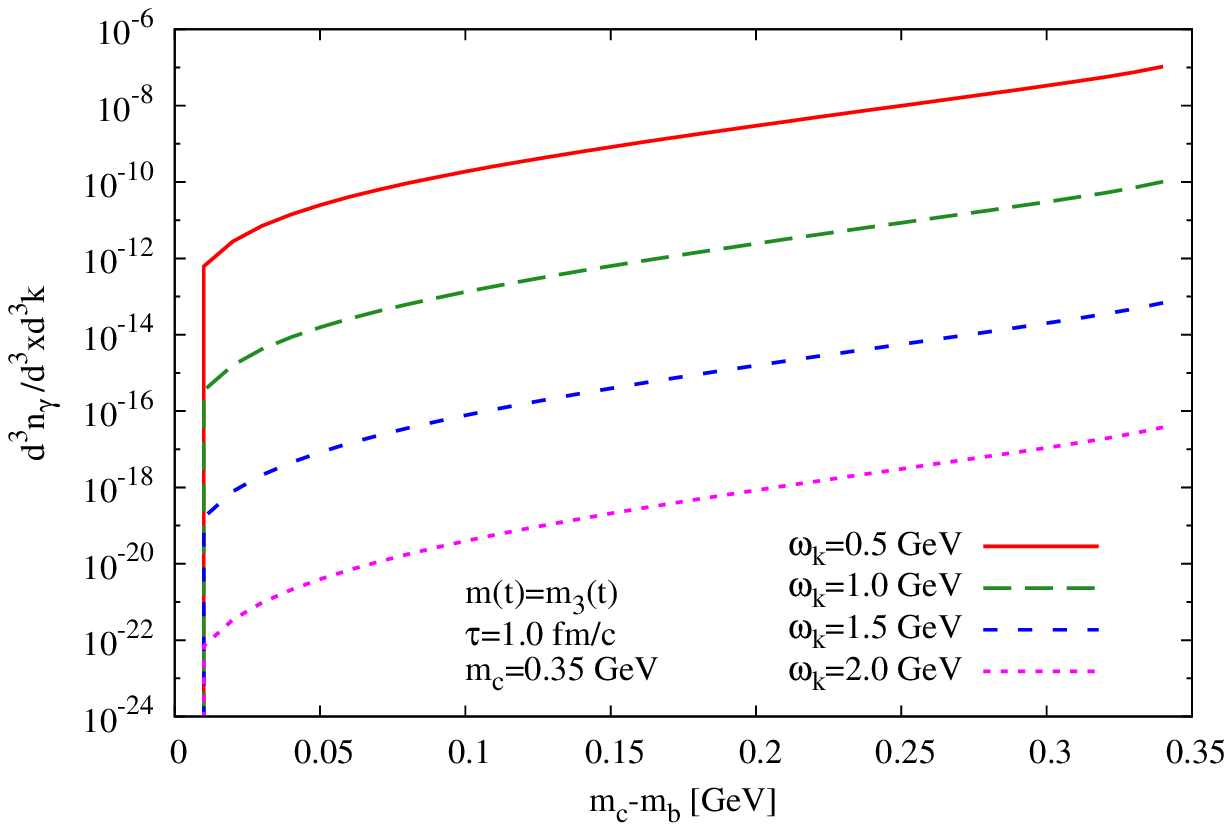}
  \caption{Dependence of the photon numbers on the considered bare mass,
    $m_{b}$, for $m_{2}(t)$ (left panel) and $m_{3}(t)$ (right
    panel). As in the instantaneous case, the photon numbers increase with the
    magnitude of the mass shift and the behavior in the limit
    $m_{b}\rightarrow0$ indicates a divergence therein.}
  \label{fig:4:massdep}
 \end{center}
\end{figure}
Similarly to the instantaneous case, the photon yield arising from the
chiral mass shift increases with the magnitude of the
latter. The change in curvature which appears for
  $m_{b}\rightarrow0$ indicates a possible divergence in this limit
  which, in analogy to the instantaneous case, could arise from a
collinear and/or anticollinear singularity in the loop integral entering
(\ref{eq:2:photon_yield_chiral_aspt}). This still requires further
investigation.

As for the asymptotic quark/antiquark occupation numbers, we also
consider the scenario where the fermion mass is first changed from
$m_{c}$ to $m_{b}$ and then back to $m_{c}$ to take into account the
finite lifetime of the chirally restored
phase. Fig. \ref{fig:4:photspec_inst_back} shows the photon spectra for
different values of $\Lambda_{C}$ for both mass shifts taking place
instantaneously which is described by $\tilde{m}_{1}(t)$. We have
assumed a lifetime of $\tau_{L}=4.0$ fm/c for the chirally restored
phase.
\begin{figure}[htb]
 \begin{center}
   \includegraphics[height=5.0cm]{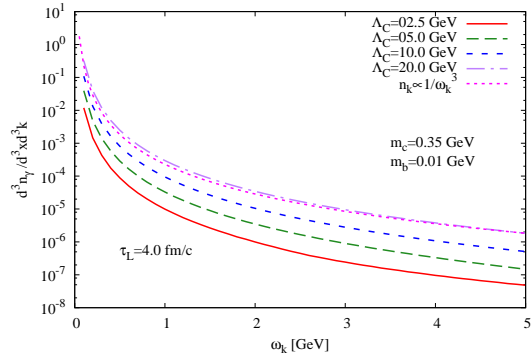}
   \caption{Asymptotic photon spectra for both mass shifts taking place instantaneously.
            The loop integral in (\ref{eq:2:photon_yield_chiral_aspt}) is again linearly divergent. Furthermore, 
            the photon spectra decay as $1/\omega_{\vec{k}}^{3}$ if this divergence is regulated via 
            a cutoff at $p=\Lambda_{C}$.}
  \label{fig:4:photspec_inst_back}
 \end{center}
\end{figure}

As for the first scenario, the loop integral entering (\ref{eq:2:photon_yield_chiral_aspt}) 
exhibits a linear divergence. If this divergence is regulated via a cutoff at $p=\Lambda_{C}$, 
the resulting photon spectrum again decays $\propto 1/\omega^{3}_{\vec{k}}$ in the ultraviolet 
domain and is hence not integrable. As one would expect from the first scenario, 
however, these pathologies are again artifacts from the (unphysical) instantaneous 
mass shifts and are resolved if both mass shifts are assumed to 
take place over a finite time interval, $\tau$.
\begin{figure}[htb]
 \begin{center}
  \includegraphics[height=5.0cm]{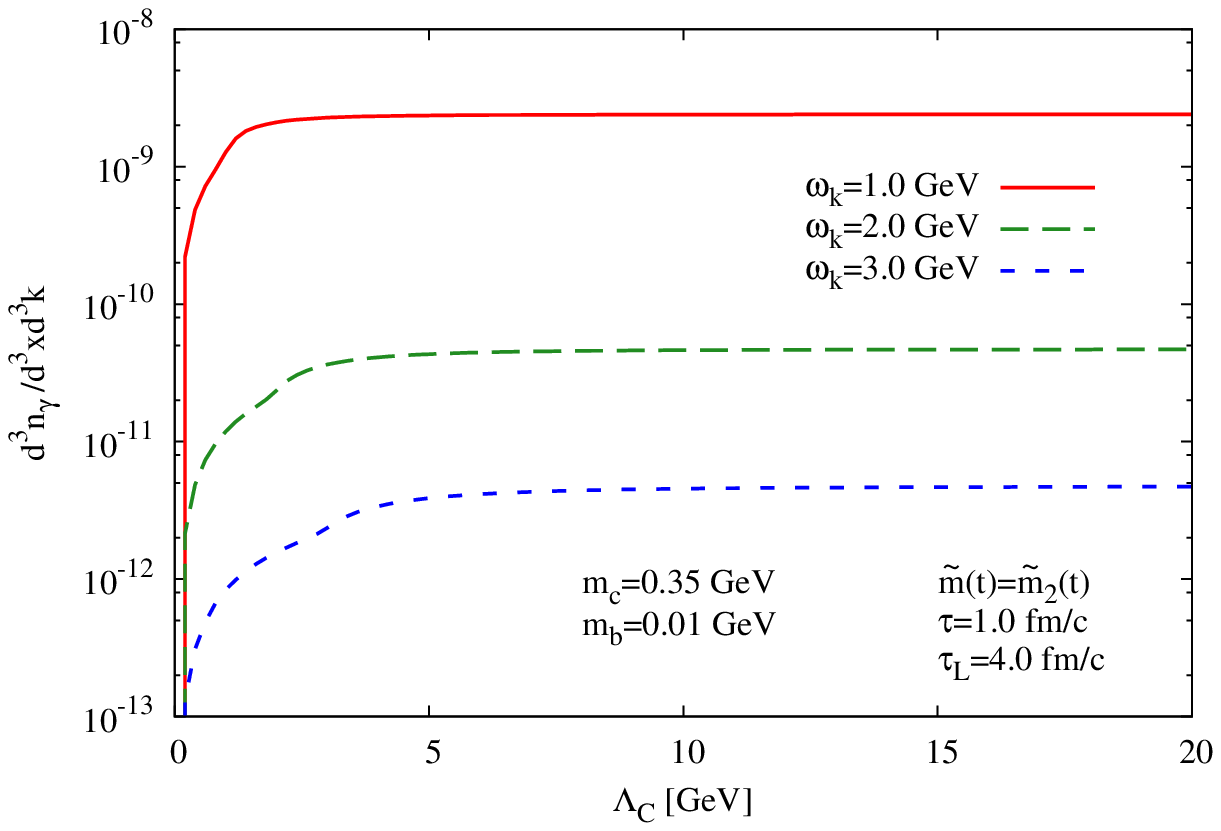}\hspace*{0.5cm}
  \includegraphics[height=5.0cm]{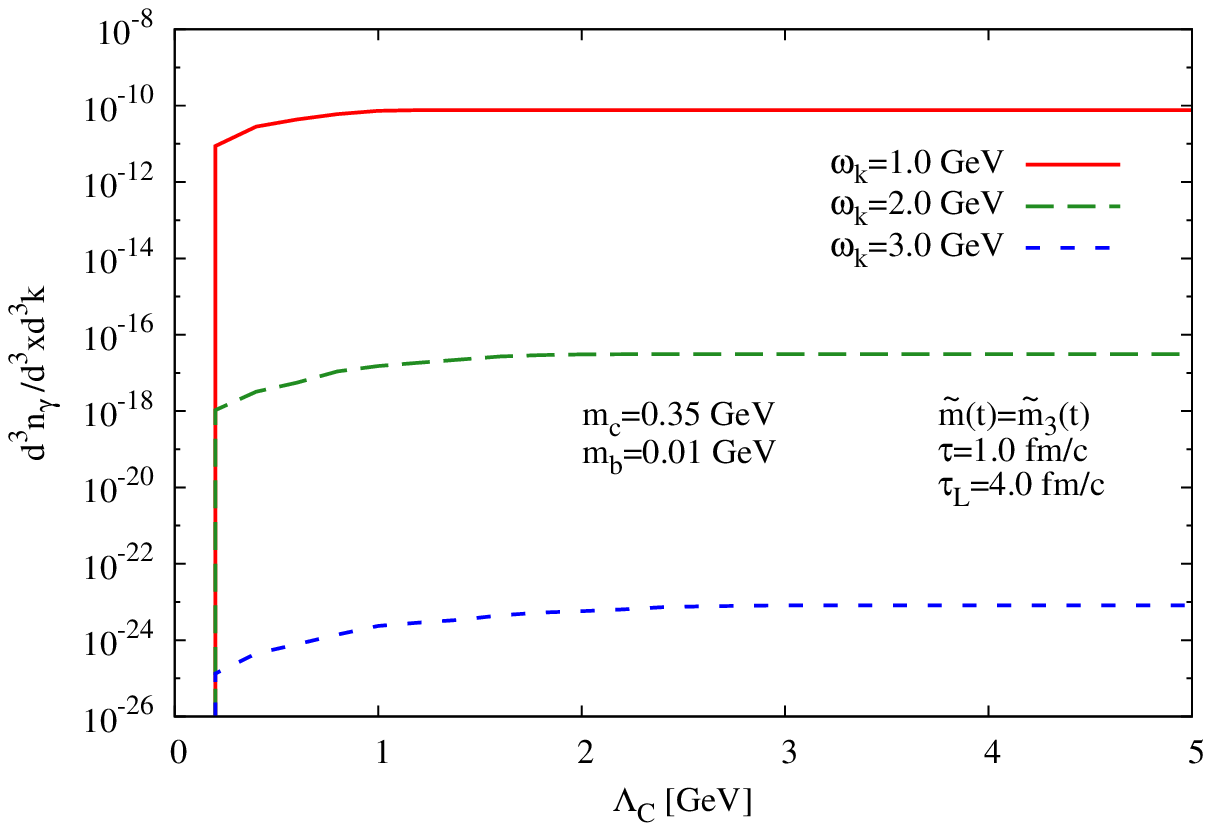}
  \caption{Cutoff dependence of asymptotic photon numbers for $\tilde{m}_{2}(t)$
    (left panel) and $\tilde{m}_{3}(t)$ (right panel).  In both cases, we have
    chosen $\tau=1.0 \; \fm/c$ and $\tau_{L}=4.0 \; \fm/c$. The parameterizations $\tilde{m}_{i}(t)$ are shown 
    in Fig. \ref{fig:3:chiral_masschange_back} and defined in (\ref{eq:3:chiral_massparam_back}). 
    Like in the first scenario, the loop integral is rendered finite and its saturation behavior 
    crucially depends on the order of differentiability of the considered mass parametrization.}
  \label{fig:4:convergence_back}
 \end{center}
\end{figure}

In particular, one can infer from Fig. \ref{fig:4:convergence_back} that
the loop integral again saturates around $\Lambda_{C}=10 \; \GeV$ and
$\Lambda_{C}=2$-$3\;\GeV$ for $\tilde{m}_{2}(t)$ and $\tilde{m}_{3}(t)$,
respectively. Moreover, Fig. \ref{fig:4:photspec_back_comp} shows that
the resulting photon spectra exhibit the same sensitivity to the order
of differentiability of the considered mass parametrization,
$\tilde{m}_{i}(t)$, just like in the previous case. The photon spectra decay
$\propto 1/\omega^{6}_{\vec{k}}$ for $\tilde{m}_{2}(t)$ (continuously differentiable
once) in the ultraviolet domain and are suppressed further to an
exponential decay for $\tilde{m}_{3}(t)$ (continuously differentiable
infinitely many times).
\begin{figure}[htb]
 \begin{center}
  \includegraphics[height=5.0cm]{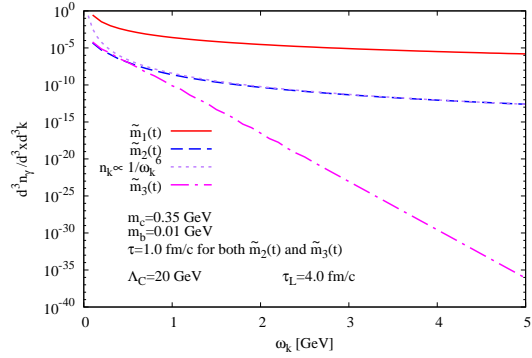}
  \caption{Asymptotic photon spectra for the different mass parameterizations 
           given by Eqs. (\ref{eq:3:chiral_massparam_back}). As expected, the photon spectra are 
           again rendered integrable in the ultraviolet domain if 
           both mass changes take place over a finite time interval, $\tau$. 
           Furthermore, they show the same sensitivity to the order of differentiability 
           of the considered mass parametrization, $\tilde{m}_{i}(t)$, as for a single mass change.}
  \label{fig:4:photspec_back_comp}
 \end{center}
\end{figure}

As it must be, the photon spectra show the same sensitivity to the
change duration, $\tau$, as for the first scenario, i.e., the
suppression of the photon numbers compared to the instantaneous case is
the stronger the more slowly that mass changes are assumed to take
place. Moreover, both $\tilde{m}_{2}(t)$ and $\tilde{m}_{3}(t)$
reproduce the photon spectra for the instantaneous case in the limit
$\tau\rightarrow0$. This is shown in
Fig. \ref{fig:4:photspec_back_tauzero}.
\begin{figure}[htb]
 \begin{center}
  \includegraphics[height=5.0cm]{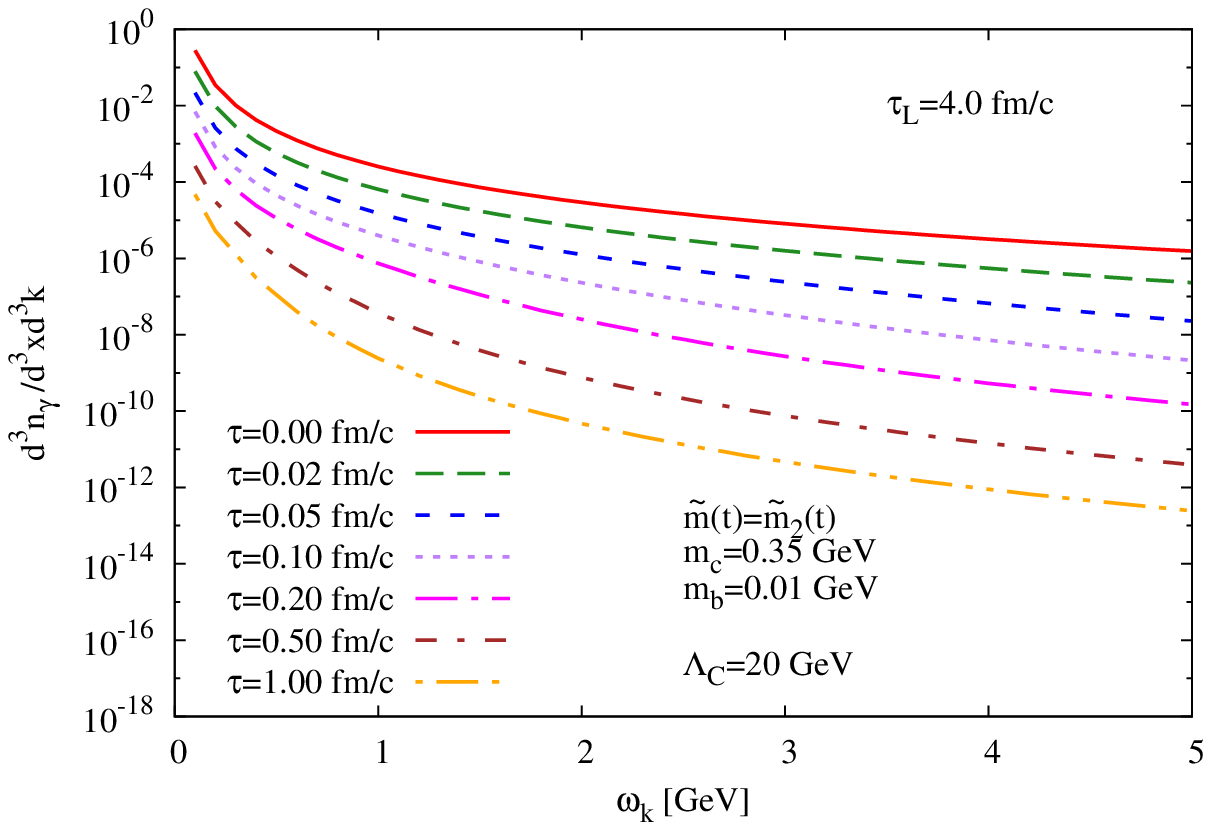}
  \includegraphics[height=5.0cm]{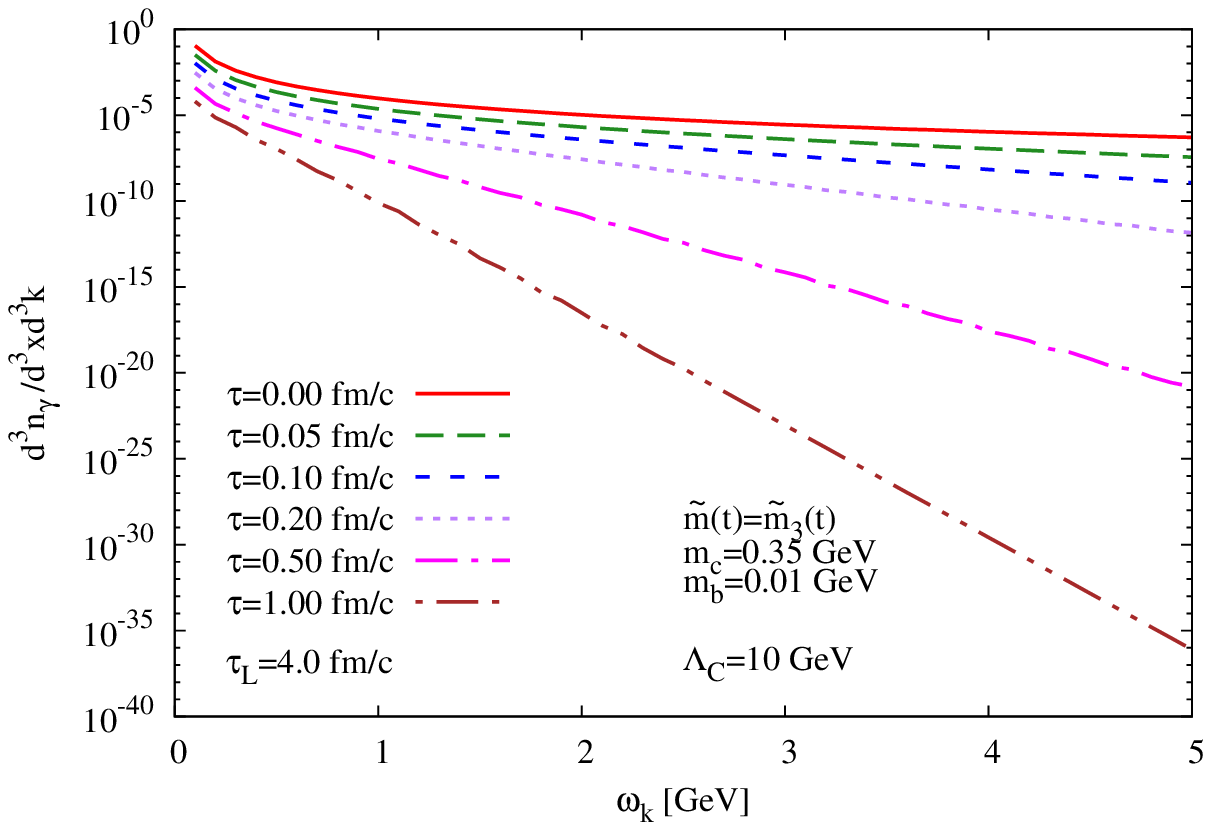}
  \caption{Asymptotic photon numbers for different transition times, $\tau$, for $\tilde{m}_{2}(t)$ (left panel) 
           and $\tilde{m}_{3}(t)$ (right panel). As in the previous scenario, the suppression of the photon 
           numbers with respect to the instantaneous case is the stronger the more slowly ($\tau$ increasing) 
           the mass shifts are assumed to take place.}
  \label{fig:4:photspec_back_tauzero}
 \end{center}
\end{figure}

Hence, we have seen so far that the general dependence of the photon
numbers in the ultraviolet domain on the order of differentiability of
$\tilde{m}(t)$ and the transition time, $\tau$, is the same as for the
first scenario, which one would also expect intuitively. Nevertheless,
there are some differences in the dependence on the magnitude of the
mass shift, $m_{c}-m_{b}$, which can be seen in Fig
\ref{fig:4:massdep_back}.
\begin{figure}[htb]
 \begin{center}
  \includegraphics[height=5.0cm]{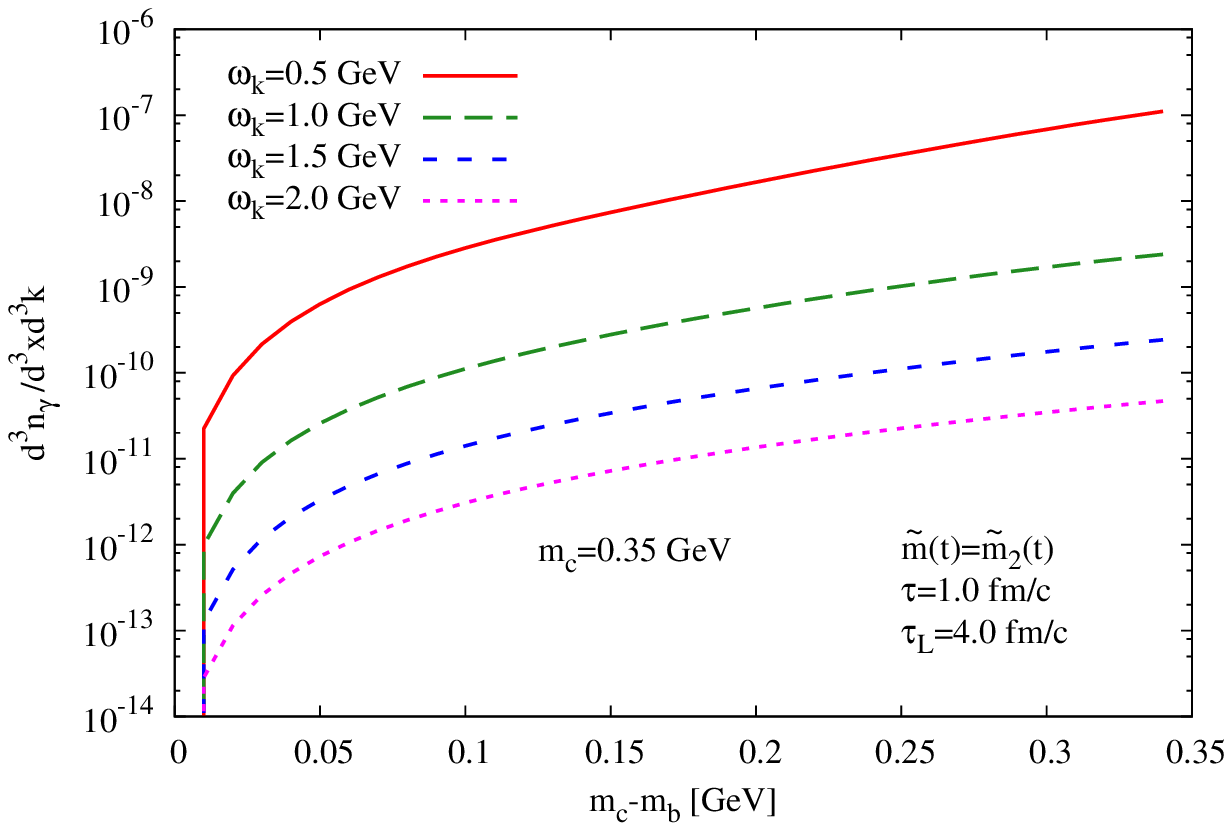}
  \includegraphics[height=5.0cm]{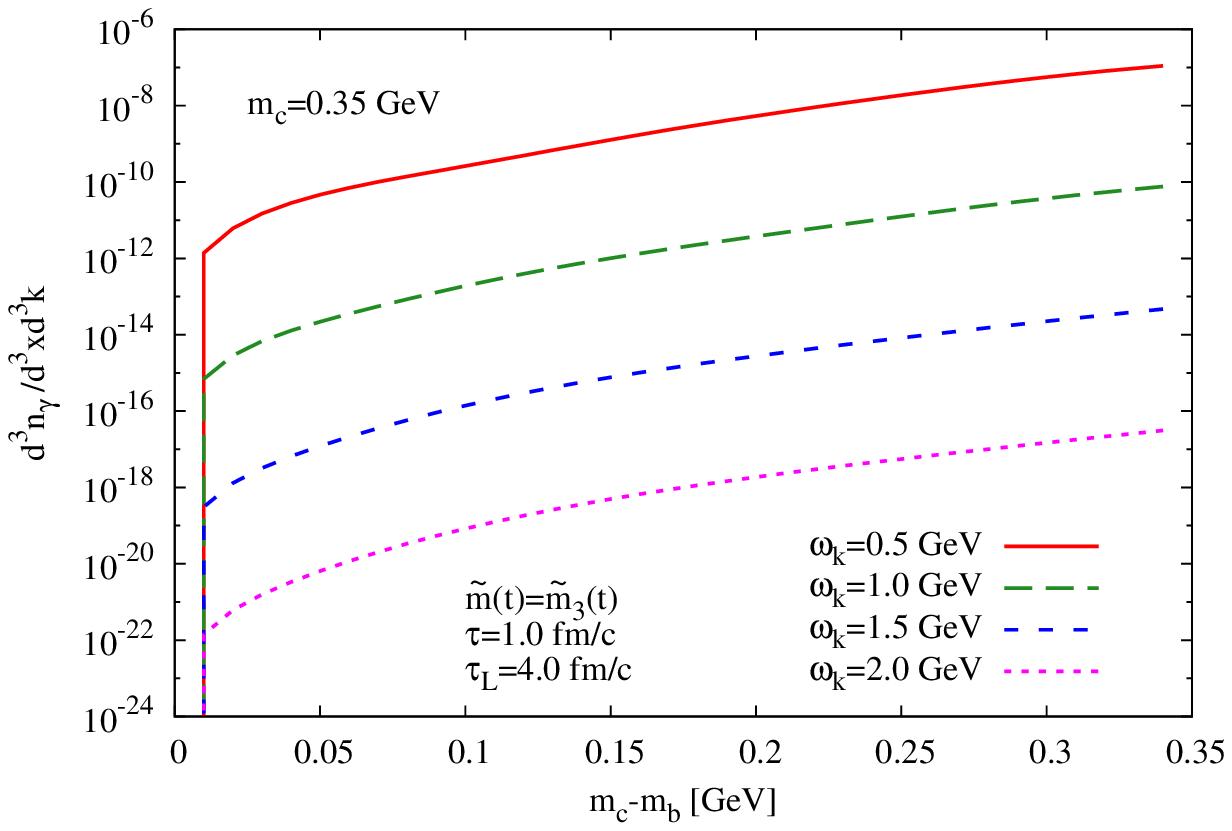}
  \caption{Dependence of the photon numbers on the considered bare mass,
    $m_{b}$, for $\tilde{m}_{2}(t)$ (left panel) and $\tilde{m}_{3}(t)$ (right
    panel). Like for the scenario where the mass is solely changed form 
    $m_{c}$ to $m_{b}$, the photon numbers increase with the magnitude of the
    mass shift, $m_{c}-m_{b}$. Their behavior in the limit $m_{b}\rightarrow0$ 
    does, however, not indicate a divergence.}
  \label{fig:4:massdep_back}
 \end{center}
\end{figure}

As in the previous case, the photon yield increases with the magnitude
of the mass difference for given photon energy, $\omega_{\vec{k}}$. The crucial
difference, however, is that the curvature does not
  change for $m_{b}\rightarrow0$. This indicates that, in contrast to
  the first scenario, the photon numbers converge in that limit and
that the loop integral entering (\ref{eq:2:photon_yield_chiral_aspt})
does not feature a collinear and/or an anticollinear divergence therein.

We have seen in section \ref{sec:pair-production} that the asymptotic
quark and antiquark occupation numbers for $p\gg m_{c},m_{b}$ are
modified by a factor of $4\sin^{2}{p\tau_{L}}$ when turning from
$m_{i}(t)$ to $\tilde{m}_{i}(t)$ ($i=1,2,3$) for any given transition
time, $\tau$. In contrast, the asymptotic photon numbers in the
ultraviolet domain do not exhibit a similar modification by a factor of
$4\sin^{2}{\omega_{\vec{k}}\tau_{L}}$. Solely for $\tilde{m}_{3}(t)$ the photon spectra
exhibit a slightly oscillating behavior (see e.g. right panel of
Fig. \ref{fig:4:photspec_back_tauzero}) in the photon momentum, $k$, for
sufficiently large values of $\tau$. For $m_{3}(t)$ and $\tau=1.0 \;
\fm/c$ this behavior can only be displayed up to $\omega_{\vec{k}}=2.5$ GeV since the
photon numbers for $\omega_{\vec{k}}>2.5$ GeV have again been extrapolated from those
for $1.5\le \omega_{\vec{k}}\le 2.5$ GeV by a linear regression.

Even though one might expect a similar modification as for the
asymptotic quark/antiquark occupation numbers in the first place, there
are two important aspects to be taken into account. On the one hand, the
asymptotic photon numbers incorporate the entire history of the
fermionic wavefunction and hence of the quark and antiquark occupation
numbers, which are extracted from the former. In particular, the
occupation numbers partially coincide with the asymptotic ones of the
first scenario between the two mass shifts. On the other hand, it
follows from (\ref{eq:2:photon_yield_chiral_aspt}) that the dependence
of the wavefunction parameters on the fermion momenta is integrated out
when determining the asymptotic photon numbers. Upon this procedure, a
possible oscillating behavior in the individual contributions to
(\ref{eq:2:photon_yield_chiral_aspt}) from the different momentum modes
can get lost again.

The latter aspect is supported when comparing the asymptotic photon
numbers for both parameterizations $m_{i}(t)$ and $\tilde{m}_{i}(t)$,
which is done in Fig. \ref{fig:4:scencomp} for fixed photon energies,
$\omega_{\vec{k}}$, and different magnitudes of the mass shift, $m_{c}-m_{b}$.
\begin{figure}[htb]
 \begin{center}
  \includegraphics[height=5.0cm]{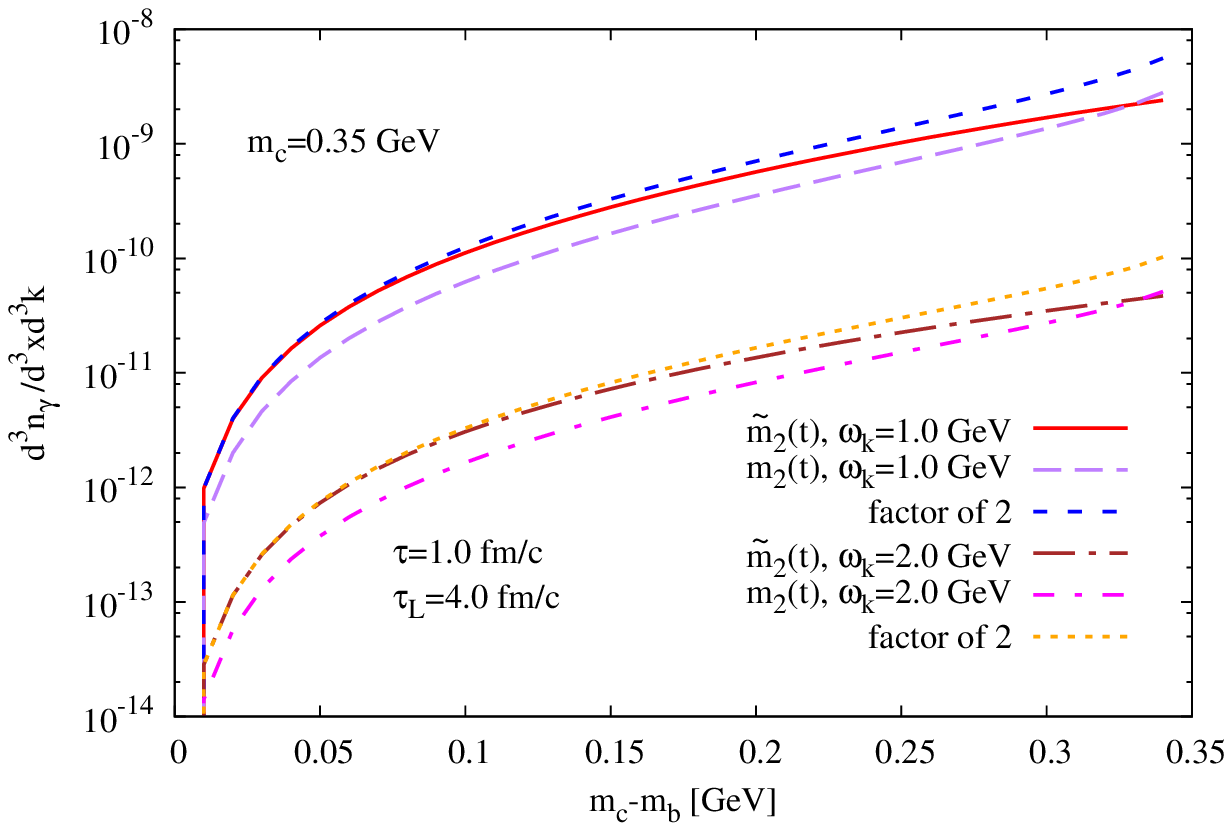}
  \includegraphics[height=5.0cm]{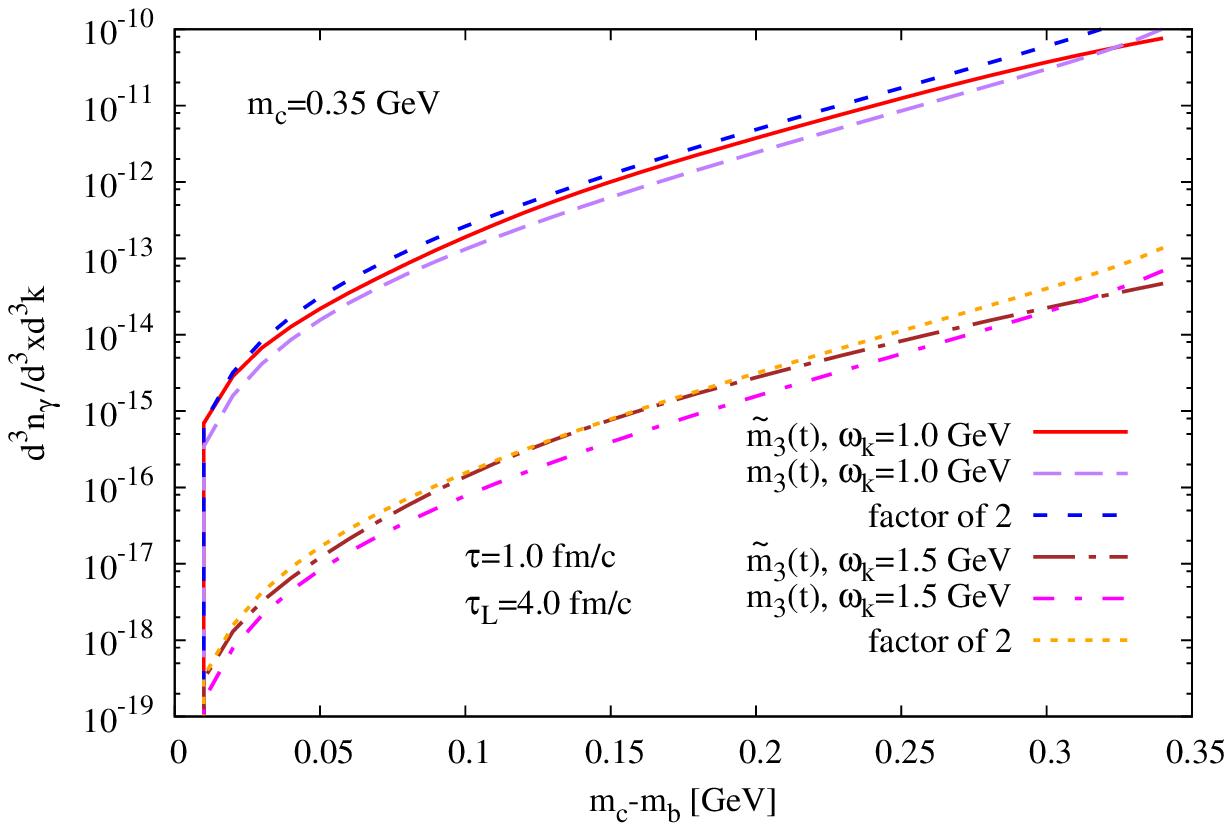}
  \caption{Comparison of photon numbers for $m_{2}(t)$ and $\tilde{m}_{2}(t)$ (left panel) and for 
           $m_{3}(t)$ and $\tilde{m}_{3}(t)$ (right panel). For small mass shifts, $m_{c}-m_{b}$, the 
           photon numbers differ by a factor of roughly $2$, which is particularly distinctive for 
           $m_{2}(t)$ and $\tilde{m}_{2}(t)$, respectively.}
  \label{fig:4:scencomp}
 \end{center}
\end{figure}
There the dotted lines represent the photon spectra of $m_{i}(t)$
multiplied by a factor of $2$ in each case.  Hence, for small magnitudes
of $m_{c}-m_{b}$, the photon numbers roughly double if the
quark/antiquark mass is switched back to its constituent value,
$m_{c}$. The latter feature is particularly distinctive for
$\tilde{m}_{2}(t)$. Such a result is understandable since both mass
shifts are expected to give a comparable contribution to the asymptotic
photon yield (\ref{eq:2:photon_yield_chiral_aspt}). But in particular,
integrating out an additional factor of $4\sin^{2}{p\tau_{L}}$ gives
rise to an overall rescaling by a factor of roughly $2$ if the integrand
does not change significantly over the periodicity interval $\Delta
p=\pi/\tau_{L}$. For increasing $m_{c}-m_{b}$, the asymptotic photon
numbers for $m_i(t)$ and $\tilde{m}_{i}(t)$ start to deviate
from this ratio as the different scaling behavior for
$m_{b}\rightarrow0$ starts to manifest itself.

In contrast to \cite{Wang:2000pv,Wang:2001xh,Boyanovsky:2003qm}, our
asymptotic photon numbers arising from first-order QED processes are UV
finite for both scenarios if the mass shifts are assumed to take place
over a finite time interval, $\tau$. Hence, it is
convenient to compare them to leading-order thermal
contributions. We note again that first-order QED
  contributions vanish in a static thermal equilibrium. There the first
  non-trivial contribution starts at two-loop order. Since a loop
  expansion does not coincide with a coupling-constant expansion,
  resummations are necessary to obtain the thermal rate at quartic order
  in the perturbative coupling constants, i.e. at linear order in
  $\alpha_{e}$ and at linear order in $\alpha_{s}$
  \cite{Arnold:2001ms}. Fig. \ref{fig:4:thermcomp} shows the photon
  numbers from first-order chiral photon production together with
  leading-order thermal contributions. The latter have been obtained by
  integrating the leading order thermal rates taken from
  \cite{Arnold:2001ms} over the lifetime interval of the chirally
  restored phase, $\tau_{L}=4.0$ fm/c, at a temperature of $T=0.2$ GeV.
\begin{figure}[htb]
 \begin{center}
  \includegraphics[height=5.0cm]{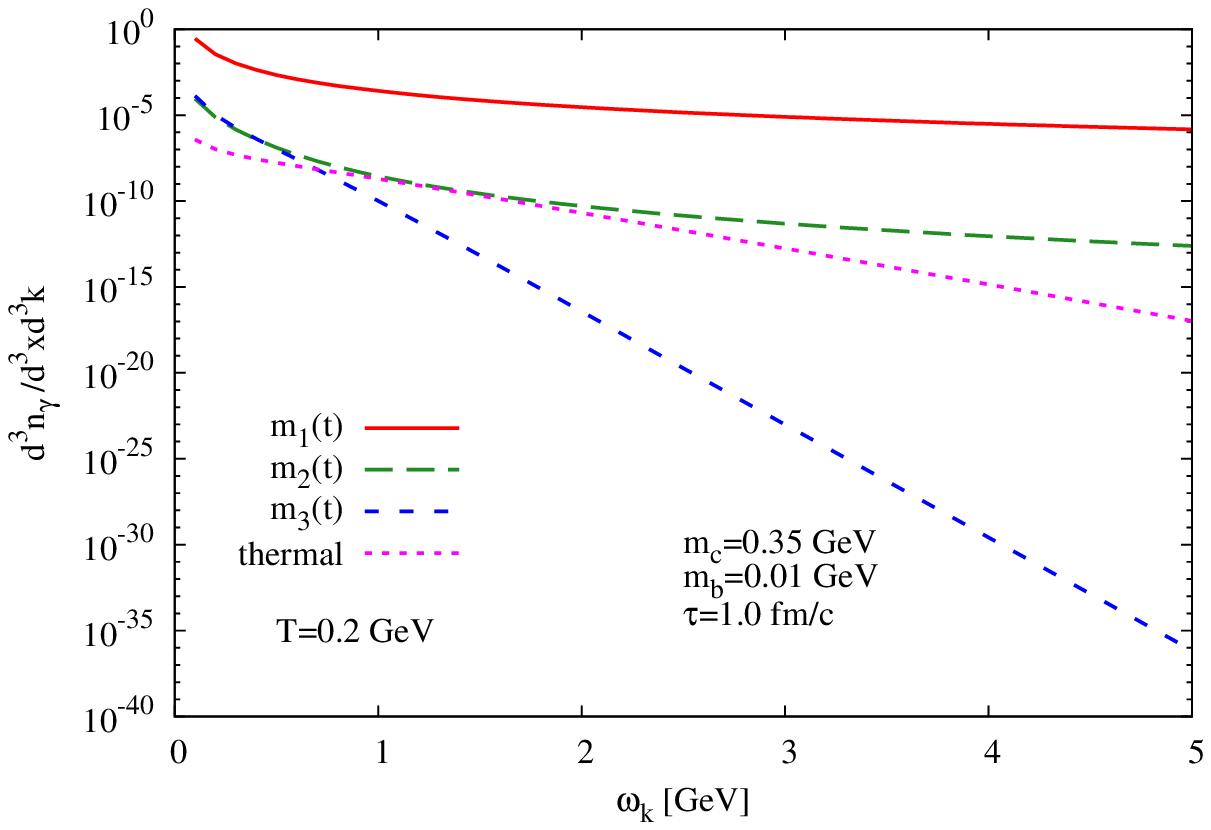}
  \includegraphics[height=5.0cm]{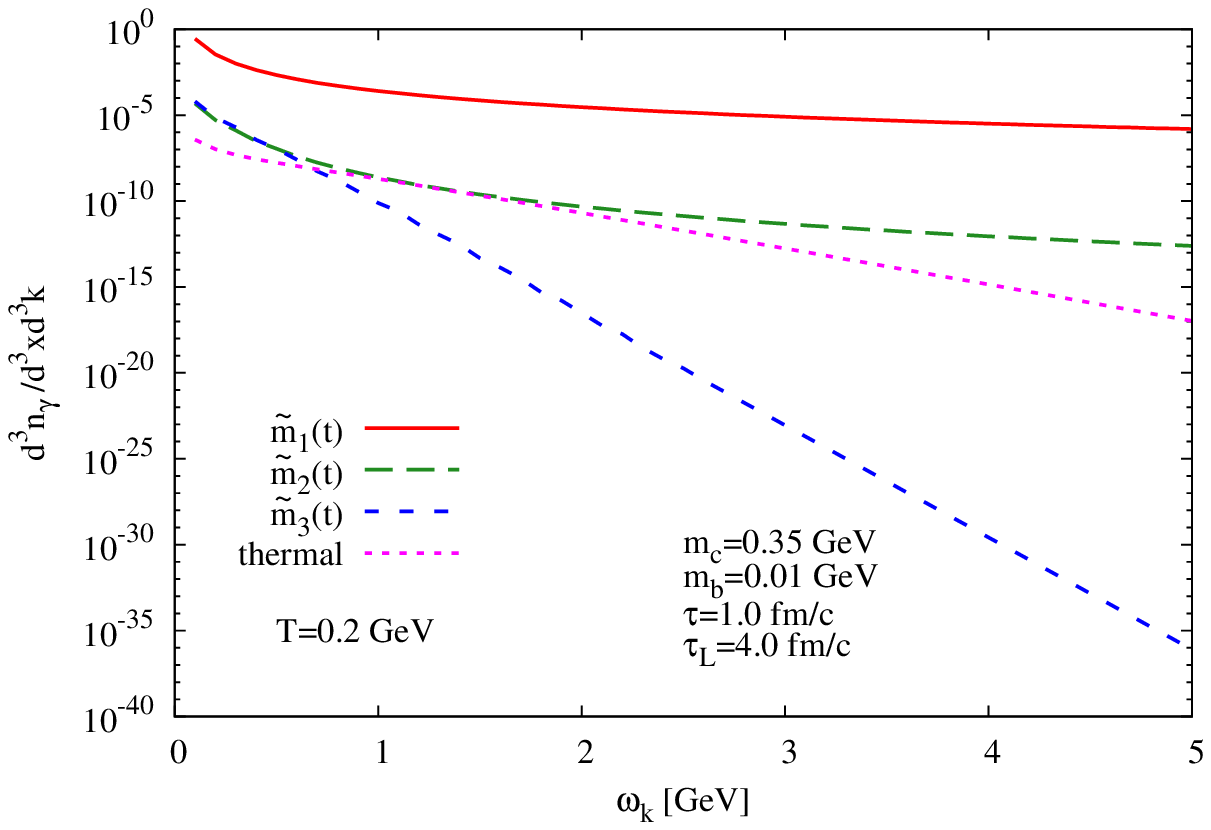}
  \caption{Comparison of first-order mass-shift contributions to time-integrated thermal rates
           for $m_{i}(t)$ (left panel) and $\tilde{m}_{i}(t)$ (right panel). For $i=3$ most likely 
           characterizing a physical scenario, the photon spectra arising from the chiral mass shift are clearly 
           subdominant for $\omega_{\vec{k}}\gtrsim 1.0$ GeV.}
  \label{fig:4:thermcomp}
 \end{center}
\end{figure}

In this context, the photon spectra for $\tilde{m}_{i}(t)$} are the
more meaningful ones, as the underlying scenario considers the finite
lifetime of the chirally restored phase during a heavy-ion collision.
Nevertheless, we see that for both $m_{3}(t)$ and $\tilde{m}_{3}(t)$,
which are continuously differentiable infinitely many times and thus
represent the most physical scenario, the photon
numbers for $\omega_{\vec{k}}\gtrsim1.0$ GeV are subdominant compared to those obtained
from (time-integrated) thermal rates for phenomenologically reasonable
choices of $\tau$ and $T$.

To summarize, we have seen that, if we turn from an instantaneous mass
shift to a mass shift over a finite time interval, $\tau$, the linear
divergence in the loop integral is regulated. Furthermore, the
asymptotic photon spectra are integrable in the ultraviolet domain if
the time evolution of the quark masses is described in a physical
way. Finally, the decay behavior shows a strong sensitivity to the
considered transition time, $\tau$, of the quark mass and we
recover our results for an instantaneous mass shift as
$\tau\rightarrow0$. The dependence on $\tilde{m}(t)$ and $\tau$ is the
same if the quark mass is also restored to its constituent value,
$m_{c}$, as to mimic the finite lifetime of the chirally restored phase.

In particular, for mass parameterizations that are continuously
differentiable infinitely many times and thus represent the most
physical scenario, our photon numbers are subdominant with respect to
those arising from integrated thermal rates in the UV domain for a
physically sensible transition time, $\tau$, and temperature, $T$.

Nevertheless, the dependence of the photon number on the bare mass,
$m_{b}$, indicates that the loop integral entering expression
(\ref{eq:2:photon_yield_chiral_aspt}) for the asymptotic photon numbers
still features a collinear and/or an anticollinear singularity in the
limit $m_{b}\rightarrow0$ for the first scenario (mass solely changed
from $m_{c}$ to $m_{b}$) whereas there is no such indication in the
second one (mass changed back to $m_{c}$). Even though this aspect still
requires further investigations, it is important to point out again that
a possible divergence in the massless limit can be circumvented
leaving $m_{b}$ finite.

\subsection{Summary of results}
We have seen that our prescription of chiral photon production
eliminates possible unphysical vacuum contributions and leads to photon spectra 
being integrable in the UV domain for physical mass-shift scenarios. The crucial
difference compared to
\cite{Wang:2000pv,Wang:2001xh,Boyanovsky:2003qm,Michler:2009hi} has been
the consideration of asymptotic photon numbers. For this purpose,
we have switched the electromagnetic interaction adiabatically according
to (\ref{eq:2:switch}) and determined the photon numbers for
$t\rightarrow\infty$. Only at the end of our calculation, we have taken
$\varepsilon\rightarrow0$. It shall be stressed again that adhering to
the correct order of limits is indeed crucial for two main reasons.

First, the interpretation of
(\ref{eq:2:photon_yield_chiral_eps}) as a photon number is only
justified in the limit $t\rightarrow\pm\infty$ for finite $\varepsilon$
where the electromagnetic field is non-interacting. As a consequence,
taking first $\varepsilon\rightarrow0$ at some finite time, $t$, is
questionable, as we then would have an \textit{interacting} electromagnetic
field such that the interpretation of
(\ref{eq:2:photon_yield_chiral_eps}) as photon number is not
justified. Moreover, such an interpretation remains doubtful for
$t\rightarrow\infty$. Since we would have taken $\varepsilon\rightarrow0$
before, the electromagnetic field would not evolve into a free one for
$t\rightarrow\infty$. A similar problem occurs when only using an
adiabatic switching-on of the interaction for $t\rightarrow-\infty$ but
no adiabatic switching-off for $t\rightarrow \infty$. Such a procedure has been
suggested in \cite{Serreau:2003wr} in order to implement to implement initial correlations 
at some $t=t_{0}$ evolving from an uncorrelated state at $t\rightarrow-\infty.$

Second, interchanging both limits comes along with a
violation of the Ward-Takahashi identities. To see this, we consider the
unphysical scenario, where the electromagnetic interaction is switched on 
an off again instantaneously at $t = \mp1/\varepsilon$, i.e.
 \begin{equation}
 \label{eq:4:switch}
 \hat{H}_{\text{EM}}(t) \rightarrow \theta_{\varepsilon}(t)\hat{H}_{\text{EM}}(t) \ , \quad 
                                    \text{with} \quad \theta_{\varepsilon}(t)=\theta\left(\frac{1}{\varepsilon^{2}}-t^{2}\right) \quad \text{and} \quad \varepsilon>0 \ .
\end{equation}
We shall show in greater detail in appendix \ref{sec:appf} that if we
replace $f_{\varepsilon}(t)$ by $\theta_{\varepsilon}(t)$ in
(\ref{eq:2:photon_yield_chiral_eps}) and consider this expression for
free asymptotic fields, i.e., if we first take $t\rightarrow\infty$
and then $\varepsilon\rightarrow0$, we obtain the same result for
(\ref{eq:4:photnum_mst}) as we would also have obtained if we had adhered
to (\ref{eq:2:switch}) but interchanged the limits
instead. Consequently, interchanging the limits for (\ref{eq:2:switch})
is formally equivalent to switching the electromagnetic interaction on
and off again instantaneously, which is unphysical.

In order to emphasize this aspect, we consider again the Ward-Takahashi
identities (\ref{eq:2:wti}) for the photon self-energy.  We have shown
in section \ref{sec:gaug-inv-phot} and also demonstrated in greater
detail in appendix \ref{sec:appb} that these identities are 
fulfilled when coupling the quark fields to a time-dependent scalar
background field, $\phi(t)$. But it follows from
(\ref{eq:2:photon_yield_chiral_eps}) that a switching of the
electromagnetic interaction leads to a modification of the
one-loop photon self-energy, i.e.,
\begin{equation}
 \ii\Pi^{<}_{\mu\nu}(\vec{k},t_{1},t_{2})\rightarrow\ii\Pi^{<,\varepsilon}_{\mu\nu}(\vec{k},t_{1},t_{2})
                                                   =f_{\varepsilon}(t_{1})f_{\varepsilon}(t_{2})\ii\Pi^{<}_{\mu\nu}(\vec{k},t_{1},t_{2}) \ .
\end{equation}
The electromagnetic interaction is then fully persistent at any given
time, $t$. Because of the additional factor of $f_{\varepsilon}(t_{1})$,
relation (\ref{eq:2:wti}) is not fulfilled in the first place, but we
instead have
\begin{equation}
 \label{eq:4:wti_violation}
 \partial_{t_{1}}\ii\Pi^{<,\varepsilon}_{0\mu}(\vec{k},t_{1},t_{2})-\ii k^{j}\ii\Pi^{<,\varepsilon}_{j\mu}(\vec{k},t_{1},t_{2})
   = -\dot{f}_{\varepsilon}(t_{1})f_{\varepsilon}(t_{2})\ii\Pi^{<,\varepsilon}_{0\mu}(\vec{k},t_{1},t_{2}) \ .
\end{equation}
The time derivative of $f_{\varepsilon}(t_{1})$, however, reads
\begin{equation}
 \dot{f}_{\varepsilon}(t)=\varepsilon\mbox{ sign}(t)\ee^{-\varepsilon|t|} \ .
\end{equation}
As a consequence, the r.h.s. of (\ref{eq:4:wti_violation}) vanishes for
$\varepsilon\rightarrow0$ and, accordingly, (\ref{eq:2:wti}) is restored
again. To the contrary, if we replace $f_{\varepsilon}(t_{1})$ by
$\theta_{\varepsilon}(t_{1})$, we have no such restoration since
\begin{equation}
 \dot{\theta}_{\varepsilon}(t)=\theta\left(\frac{1}{\varepsilon}-t\right)\delta\left(\frac{1}{\varepsilon}+t\right)-
                               \theta\left(\frac{1}{\varepsilon}+t\right)\delta\left(\frac{1}{\varepsilon}-t\right) \ .
\end{equation}
Hence, switching the electromagnetic interaction as in 
(\ref{eq:4:switch}) leads effectively to a violation of the
Ward-Takahashi identities. Since (\ref{eq:4:switch}) leads to the same
asymptotic photon numbers as (\ref{eq:2:switch}) with the
interchanged order of limits, we conclude that the latter
procedure also leads to an effective violation of (\ref{eq:2:wti}).

From this point of view, the Ward-Takahashi identities actually  
imply constraints for a physically reasonable definition of photon numbers. 
Their violation implies an inconsistency of such 
definitions with $U(1)$ gauge invariance which, in turn,  usually leads to unphysical 
artifacts. This issue has also been pointed out in the
context of the Thomas-Reiche-Kuhn sum rules \cite{Knoll:1988dn}, which can
be considered as direct consequences of charge conservation and gauge invariance of QED. 
It has been shown that these sum rules impose severe constraints concerning 
the applicability of transport approaches on photon production from 
non-equilibrated hot hadronic matter presented in \cite{Nifenecker:1985,Ko:1985hq,Nakayama:1986zz,Bauer:1986zz,Neuhauser:1987}.

\section{Summary, conclusions and outlook}
\label{sec:conclusions}

In this work, we have investigated photon emission during the chiral
phase transition in the early stage of a heavy-ion collision. During
this phase transition, the quark mass is changed from its constituent
value, $m_{c}$, to its bare value, $m_{b}$, which leads to a spontaneous
non-perturbative pair production of quarks and antiquarks
\cite{Greiner:1995ac,Greiner:1996wz}. This effectively contributes to
the creation of the QGP, and we have investigated the photon emission
arising from this creation process.

As in \cite{Greiner:1995ac,Greiner:1996wz}, the change of the quark mass
has been modeled by a scalar background field in the QED Lagrangian. We
have restricted our considerations to first-order QED processes. 
Such processes have also been 
investigated in \cite{Blaschke:2011is} for fermions coupled to a 
time-dependent electromagnetic background field \cite{Schmidt:1998vi}.
They are kinematically allowed since the background field
  acts as a source of additional energy. For a proper treatment of the
  non-perturbative nature of the pair-creation process the coupling to
  the background field must be resummed to all
  orders. In order to achieve this, we have constructed an interaction
  picture including the coupling to the source field. The photon
  yield has then been obtained by a standard perturbative QED
  calculation. In the course of our calculations, we have essentially
  pursued an in/out description, where the photon numbers have been
  extracted in the limit $t\rightarrow\infty$ for free asymptotic
  states. In order to obtain such states we have introduced an adiabatic
  switching of the electromagnetic interaction, i.e.,
\begin{equation} 
 \label{eq:5:switch}
 \hat{H}_{\text{EM}}(t) \rightarrow f_{\varepsilon}(t)\hat{H}_{\text{EM}}(t) \ , \quad 
                                    \text{with} \quad f_{\varepsilon}(t)=\ee^{-\varepsilon|t|} \quad \text{and} \quad \varepsilon>0 \ .
\end{equation}
The photon numbers then have been considered for
$t\rightarrow\infty$ and we have taken $\varepsilon\rightarrow0$ at the
very end of our calculation.

Before turning to our investigations on photon production, we have first
provided a digression on pair production of quarks and antiquarks.
There, we have extended the investigations in
\cite{Greiner:1995ac,Greiner:1996wz} to the time evolution of the quark
and antiquark occupation numbers. We have restricted ourselves to the
pair production arising from the chiral mass shift only and not taken
into account radiative corrections. As in
  \cite{Greiner:1995ac,Greiner:1996wz}, we have compared different mass
  parameterizations
\begin{subequations}
 \label{eq:5:chiral_massparam}
 \begin{eqnarray}
   m_{1}(t) & = & \frac{m_{c}+m_{b}}{2}-\frac{m_{c}-m_{b}}{2}\mbox{ }\mbox{sign}(t) \ , \label{eq:5:chiral_massparam_1} \\
   m_{2}(t) & = & \frac{m_{c}+m_{b}}{2}-\frac{m_{c}-m_{b}}{2}\mbox{ }\mbox{sign}(t)\left(1-\ee^{-2|t|/\tau}\right) \ , \label{eq:5:chiral_massparam_2} \\
   m_{3}(t) & = & \frac{m_{c}+m_{b}}{2}-\frac{m_{c}-m_{b}}{2}\mbox{ }\mbox{tanh}\left(\frac{2t}{\tau}\right) \ . \label{eq:5:chiral_massparam_3}
 \end{eqnarray}
\end{subequations}
Here $m_{c}$ and $m_{b}$ denote the initial constituent mass and the
final bare mass, respectively. $m_{1}(t)$ describes an instantaneous mass
shift at $t=0$ whereas $m_{2}(t)$ and $m_{3}(t)$ denote a mass shift over a
finite time interval given by $\tau$.

We have seen that for the case of an instantaneous mass
  shift, the quark and antiquark occupation numbers for $t>0$ scale
  $\propto(m_{c}-m_{b})^{2}/p^{2}$ for $p\gg m_{c},m_{b}$, with $p$
  denoting the (absolute value of) the fermion momentum. This means that
  the total particle number density and the total energy density of the
  fermionic sector are linearly and quadratically divergent,
  respectively.

If we turn from an instantaneous mass shift to a mass shift
  over a finite time interval, $\tau$, the mentioned problems are
  resolved in the asymptotic limit $t\rightarrow\infty$. For $m_{2}(t)$
  which is continuously differentiable once, the occupation numbers for
  $p\gg m_{c},m_{b}$ and $p \gg 1/\tau$ are suppressed to
  $\propto(m_{c}-m_{b})^{2}/p^{6}\tau^{4}$ such that both the total
  particle number density and the total energy density are rendered
  finite. Moreover, if we turn from $m_{2}(t)$ to $m_{3}(t)$ which is 
  continuously differentiable infinitely many times, the occupation
  numbers are suppressed further to an exponential decay.

At finite times, $t$, however, the occupation numbers
  decay $\propto 1/p^{4}$ for large $p$. This means that the total
  particle number density is finite whereas the total energy density
  features a logarithmic divergence. In this context, it is
  nevertheless important to point out that only asymptotic particle
  numbers describe observable quantities and that the analogous
  expressions at finite times do, in general, not allow for a similar
  interpretation \cite{Filatov:2007ha}.

For that reason, we have concentrated our
  investigations on photon production to free asymptotic states. We have
  again compared the different mass parameterizations given by
  (\ref{eq:5:chiral_massparam_1})-(\ref{eq:5:chiral_massparam_3}). As
  one would expect from our investigations on particle production, where
  the total particle number density is linearly divergent for an
  instantaneous mass shift (\ref{eq:5:chiral_massparam_1}), the
  asymptotic photon numbers feature very similar pathologies for such a
  scenario. In particular, the unphysical artifacts we have encountered
  are the following:
\begin{itemize}
\item The loop integral entering our expression for the photon yield
  features a linear divergence. We have shown that this divergence
  does, indeed, arise, from the decay behavior of the
    quark/antiquark occupation numbers $\propto(m_{c}-m_{b})^{2}/p^{2}$
    for $p\gg m_{c},m_{b}$. As this decay behavior and thus the
  mentioned divergence is an artifact of the instantaneous mass shift,
  we have regularized the latter by cutting the loop integral at
  $p=\Lambda_{C}$.
\item After this regularization procedure the asymptotic photon
  spectra decay $\propto 1/\omega^{3}_{\vec{k}}$ for large photon energies, $\omega_{\vec{k}}=|\vec{k}|$ ($\vec{k}$ denotes the three-momentum of the photon), and
  are thus not UV integrable. As in
  \cite{Wang:2000pv,Wang:2001xh,Boyanovsky:2003qm}, the total photon
  numbers density and the total photon energy density are
  logarithmically and linearly divergent, respectively.
\end{itemize}
We then turned from an instantaneous mass shift to a mass shift over a
finite time interval, $\tau$. This renders the loop integral
finite. Moreover, the fermion-momentum range from which
  the main contributions to the photon numbers arise is highly sensitive
  to the order of differentiability, i.e., `the smoothness', of the
  considered mass parametrization, $m_{i}(t)$.  For $m_{2}(t)$
  (continuously differentiable once), we have contributions from quark
  and antiquark momenta up to $p\simeq10 \; \GeV$ whereas we have
  contributions only up to $p\simeq 2.0-3.0 \; \GeV$ for $m_{3}(t)$
  (continuously differentiable infinitely many times). Since the latter
  parametrization describes the physically most realistic scenario,
  chiral photon production can accordingly be considered as a low-energy
  phenomenon.

Furthermore, the resulting photon spectra are rendered integrable in the 
ultraviolet domain if the mass shift is assumed to take place over a 
finite time interval, $\tau$. In particular, their decay behavior in the ultraviolet domain
also crucially depends on the smoothness of the respective $m_{i}(t)$. For
$m_{2}(t)$, they decay as $1/\omega^{6}_{\vec{k}}$ in the ultraviolet domain whereas
they are suppressed even further to an exponential decay for
$m_{3}(t)$. For both parameterizations, the suppression of the photon
numbers with respect to the instantaneous case is the stronger the more
slowly the mass shift is assumed to take place. As expected, both
$m_{2}(t)$ and $m_{3}(t)$ reproduce the photon spectra for an
instantaneous mass shift in the limit $\tau\rightarrow0$.

The asymptotic photon spectra, considered as the only observable ones,
hence do show a very similar sensitivity on the mass parametrization,
$m(t)$, as the asymptotic particle spectra, as to be expected. In
particular, the logarithmic divergence in the energy density of the
fermionic sector at intermediate times does not appear in form of a
similar pathology in the photonic energy density. This could not be
excluded a priori since the asymptotic photon numbers incorporate the
entire history of the fermion-wavefunctions and hence of the
corresponding quark/antiquark occupation numbers.

In order to take into account the finite lifetime of the chirally
restored phase during a heavy-ion collision, we have also considered the
scenario where the quark mass is changed back to its constituent value,
$m_{c}$, after some time interval, $\tau_{L}$. Here we 
have again compared different mass parameterizations
\begin{subequations}
 \label{eq:5:chiral_massparam_back}
 \begin{align}
  \tilde{m}_{1}(t) = \frac{m_{c}+m_{b}}{2}-\frac{m_{c}-m_{b}}{2} & \text{sign}\left(\frac{\tau_{L}^{2}}{4}-t^{2}\right) \ , 
  \label{eq:5:chiral_massparam_back1} \\
  \tilde{m}_{2}(t) = \frac{m_{c}+m_{b}}{2}-\frac{m_{c}-m_{b}}{2} &
  \mbox{sign}\left(t+\frac{\tau_{L}}{2}\right)
  \left(1-\ee^{-|2t+\tau_{L}|/\tau}\right) \nonumber \\
  \times & \mbox{sign}\left(\frac{\tau_{L}}{2}-t\right)
  \left(1-\ee^{-|2t-\tau_{L}|/\tau}\right) \ ,
  \label{eq:5:chiral_massparam_back2} \\
  \tilde{m}_{3}(t) = \frac{m_{c}+m_{b}}{2}-\frac{m_{c}-m_{b}}{2} &
  \mbox{tanh}\left(\frac{2t+\tau_{L}}{\tau}\right)
  \mbox{tanh}\left(\frac{\tau_{L}-2t}{\tau}\right) \ .
  \label{eq:5:chiral_massparam_back3}
 \end{align}
\end{subequations}
The general dependence of the photon numbers on the `smoothness' of the considered 
mass parametrization, $\tilde{m}_{i}(t)$, and
the transition time, $\tau$, has been the same as for the previous
case. In particular, for mass parameterizations being continuously
differentiable infinitely many times, our photon numbers have been
subdominant compared to time-integrated thermal rates in the ultraviolet
domain.

A principal difference between both scenarios, with and without
restoration of the mass to its initial value, has only been observed in
the dependence on the magnitude of the mass shift, $m_{c}-m_{b}$.  In
the first scenario, this dependence indicates a divergence of the photon
numbers in the limit $m_{b}\rightarrow0$ whereas this is not the case
for the second scenario.

In conclusion, we have provided an ansatz for chiral photon production
that eliminates possible unphysical contributions from the vacuum
polarization and, furthermore, renders the resulting photon spectra UV
integrable if the time evolution of the quark mass is modeled in a
physically realistic manner. In contrast to
\cite{Wang:2000pv,Wang:2001xh,Boyanovsky:2003qm,Michler:2009hi}, the
photon numbers have not been considered at finite times, $t$, but for free asymptotic states. In this context,
  we have seen that a consistent definition of photon numbers is actually
  only possible for free asymptotic states. A similar interpretation of
  the respective expression is usually not justified for finite times,
  $t$, as we then do not have free asymptotic states. The same
  problem occurs if the electromagnetic interaction is only 
  switched on adiabatically from $t\rightarrow-\infty$
  but not off again for $t\rightarrow+\infty$. This procedure has been
  suggested in \cite{Serreau:2003wr} such as to implement initial
  correlations at some $t=t_{0}$ evolving from an uncorrelated state at
  $t\rightarrow-\infty.$

Moreover, the consideration of `photon numbers' at 
  finite times again comes along with a 
  violation of the Ward-Takahashi identities. Consequently, our
  investigations support the corresponding concern raised in
  \cite{Fraga:2003sn,Fraga:2004ur} towards
  \cite{Wang:2000pv,Wang:2001xh,Boyanovsky:2003qm}. Our outlook to
future investigations is hence as follows.

Primarily, our results indicate that the problems with the divergent
contribution of the vacuum polarization and/or the decay behavior of the
resulting photon spectra appearing in
\cite{Wang:2000pv,Wang:2001xh,Boyanovsky:2003qm,Michler:2009hi} result
from an inadequate definition of the photon numbers for finite times,
$t$.  This, in turn, gives rise to the question whether they can be
cured if the photon numbers are considered in the realm
  of asymptotic fields. In this context, the actual
role of the Ward-Takahashi identities, which are violated in
\cite{Wang:2000pv,Wang:2001xh,Boyanovsky:2003qm,Michler:2009hi}, but
conserved by our approach on chiral photon production, 
  still requires a more profound consideration.

Furthermore, it is of particular interest whether our asymptotic
definition of photon numbers (\ref{eq:2:photon_yield_chiral_aspt}) can
be extended to finite times in a consistent manner and/or which
alternative quantities can be considered to provide a proper real-time
description of the electromagnetic sector during a heavy-ion
collision. One promising candidate could be the gauge-invariant field
strength tensor,
$$
\hat{F}_{\mu\nu}(x) = \partial_{\mu}\hat{A}_{\nu}(x)-\partial_{\nu}\hat{A}_{\mu}(x) \ ,
$$
and the quantities derived from it. One example is the energy density of
the electromagnetic field, which is given by
\begin{align}
\hat{\varepsilon}(x) = & \left.
                          \hat{F}^{\mu\alpha}(x)\hat{F}_{\alpha}^{\mbox{ }\nu}(x)-
                         \frac{1}{4}\eta^{\mu\nu}\hat{F}^{\alpha\beta}(x)\hat{F}_{\beta\alpha}(x)
                         \right|_{\mu=\nu=0} \nonumber \\
                     = & \hat{F}^{0\alpha}(x)\hat{F}_{\alpha}^{\mbox{ }0}(x)-
                         \frac{1}{4}\hat{F}^{\alpha\beta}(x)\hat{F}_{\beta\alpha}(x) \nonumber \\
                     = & \frac{1}{2}\left(\hat{\vec{E}}^{2}+\hat{\vec{B}}^{2}\right) \ , \nonumber
\end{align}
with $\eta^{\mu\nu}=\text{diag}\left\lbrace1,-1,-1,-1\right\rbrace$ and $\hat{\vec{E}}$ and $\hat{\vec{B}}$ 
denoting the electric and magnetic field operators, respectively.
Investigations on such quantities are, of course, not restricted to
chiral photon production but can also be based on the approach by
Boyanovski et al. \cite{Wang:2000pv,Wang:2001xh,Boyanovsky:2003qm}. To
obtain physically reasonable results in this approach might, however,
require the consideration of correlated initial states
\cite{Moore:2010,Bonitz:2011}. In
\cite{Wang:2000pv,Wang:2001xh,Boyanovsky:2003qm}, the authors assume
that the system is in thermal equilibrium with respect to the strong
interactions, i.e., the initial state is specified by
(\ref{eq:1:initial_condition}). The authors effectively neglected
initial correlations characterized by the interaction part of
$\hat{H}_{\text{QCD}}$ by using the one-loop approximation for the
photon self-energy. It has, however, been shown that the evaluation of
the photon self-energy cannot be performed by simple power counting in
the strong coupling constant, $\alpha_{s}$, but instead requires a
resummation of the so-called ladder diagrams
\cite{Aurenche:1996sh,Aurenche:1998nw,Steffen:2001pv,Arnold:2001ba,Arnold:2001ms}.
Within this procedure, initial correlations are included, which leads to
the question whether this regulates possible problems with the vacuum
polarization and the UV behavior.

As the problem of photon production as well as of the time evolution of
the electromagnetic energy density can be reduced to the calculation of
the photon self-energy, i.e., the current-current correlator, at first
order in $\alpha_{e}$, another alternative is to address the question of
finite-lifetime effects within the 2PI-approximation of the effective
action. This has already been suggested in \cite{Arleo:2004gn}. Even
though the conservation of the Ward-Takahashi identities still turns out
to be difficult, the 2PI approach has the crucial advantage that it does
not make any \textit{ad hoc} assumptions about the two-time dependence
of correlation functions.

\section*{Acknowledgments}
F. M. gratefully acknowledges financial support by the Helmholtz
Research School for Quark Matter Studies (H-QM) and from the Helmholtz
Graduate School for Hadron and Ion Research (HGS-HIRe for FAIR). This
work was (financially) supported by the Helmholtz International Center
for FAIR within the framework of the LOEWE program (Landesoffensive zur
Entwicklung Wissenschaftlich-\"Okonomischer Exzellenz) launched by the
State of Hesse. The authors thank B.\ Schenke, J.\ Knoll, P.\
Danielewicz, and B.\ M\"uller for fruitful discussions.

\appendix

\section{Representation of photon yield as an absolute square}
\label{sec:appa}
In this section, we show that (\ref{eq:2:photon_yield_chiral_eps}) can
be written as an absolute square of a first-order QED
  transition amplitude and is thus positive (semi-)
definite. We also emphasize that keeping the correct
  sequence of limits leading to (\ref{eq:2:photon_yield_chiral_aspt}) is
  crucial to eliminate possible contributions from the vacuum
  contribution (\ref{eq:2:chiral_vacpol}) to the photon self-energy and that an interchange comes
  along with a divergent contribution from it. In order to show that the
  photon yield can be written as an absolute square, we first undo the
contraction
\begin{equation}
 \ii \Pi^{<}_{T}(\vec{k},t_{1},t_{2}) = \gamma^{\mu\nu}(k)\ii \Pi^{<}_{\nu\mu}(\vec{k},t_{1},t_{2}) \ .
\end{equation}
Furthermore, we have 
\begin{eqnarray}
 \label{eq:a:chiral_insert}
 \ii \Pi^{<}_{\nu\mu}(\vec{k},t_{1},t_{2}) & = & \frac{1}{V}\left \langle  0_{q\bar{q}}\right|\hat{j}^{\dagger}_{\mu,\text{J}}(\vec{k},t_{2})
 \hat{j}_{\nu,\text{J}}(\vec{k},t_{1})\left|0_{q\bar{q}}\right \rangle \nonumber \\
 & = & \frac{1}{V}\sum_{f}\left \langle  0_{q\bar{q}}\right|\hat{j}^{\dagger}_{\mu,\text{J}}(\vec{k},t_{2})\left|f\right \rangle
 \left \langle   f\right|\hat{j}_{\nu,\text{J}}(\vec{k},t_{1})\left|0_{q\bar{q}}\right \rangle \ ,
\end{eqnarray}
with $\hat{j}_{\mu,\text{J}}(\vec{k},t)$ given by
\begin{equation}
 \hat{j}_{\mu,\text{J}}(\vec{k},t) = \int \dd^{3}x\mbox{ }\hat{j}_{\mu,\text{J}}(x)\ee^{-\ii \vec{k}\cdot\vec{x}} \ ,
\end{equation}
and $\left|f\right \rangle$ denoting an orthonormal basis of the
fermionic Hilbert subspace. Together with (\ref{eq:2:chiral_polten}) and
(\ref{eq:a:chiral_insert}), we can write
(\ref{eq:2:photon_yield_chiral_eps}) as
\begin{equation}
\begin{split}
 \label{eq:a:chiral_abs_prelim}
 2\omega_{\vec{k}}\frac{\dd^{6}n^{\varepsilon}_{\gamma}(t)}{\dd^{3}x\dd^{3}k} & = 
 \frac{1}{(2\pi)^{3}V}\sum_{\lambda,f}\left|\int_{-\infty}^{t} \dd u f_{\varepsilon}(u)
   \left \langle  f\right|\varepsilon^{\mu}(\vec{k},\lambda)\hat{j}_{\mu,\text{J}}(\vec{k},u)\left|0_{q\bar{q}}\right \rangle
   \ee^{\ii\omega_{\vec{k}}u}\right|^{2}  \\
 & =  \frac{1}{(2\pi)^{3}V}\sum_{\lambda,f}\left|\int\underline{\dd^{4}x}
   \left \langle
     f\right|\varepsilon^{\mu}(\vec{k},\lambda)\hat{j}_{\mu,\text{J}}(x)\left|0_{q\bar{q}}\right
   \rangle
   \ee^{\ii kx}\right|^{2}  \ ,
\end{split} 
\end{equation}
with the underline denoting
$$
\int\underline{\dd^{4}x} = \int\dd^{3}x\int_{-\infty}^{t}\dd u
f_{\varepsilon}(u) \quad \text{with} \quad f_{\varepsilon}(t)=\exp(-\varepsilon |t|) \ .
$$
Using the relations (\ref{eq:2:chiral_phot_decom}) and
(\ref{eq:2:op_phot_a}), expression (\ref{eq:a:chiral_abs_prelim}) can be
rewritten further as
\begin{equation}
\begin{split}
 \label{eq:a:chiral_abs}
 \frac{\dd^{6}n^{\varepsilon}_{\gamma}(t)}{\dd^{3}x\dd^{3}k} & =
 \frac{1}{(2\pi)^{3}V}\sum_{\lambda,f}\left|\int\underline{\dd^{4}x}
   \left \langle f; \vec{k},\lambda
     \left|\hat{A}^{\mu}_{\text{J}}(x)\hat{j}_{\mu,\text{J}}(x)\right|0\right
   \rangle \right|^{2} \\
 & = \frac{1}{(2\pi)^{3}V}\sum_{\lambda,f}\left|\int\underline{\dd^{4}x}
   \left \langle f; \vec{k},\lambda
     \left|\hat{\mathcal{H}}_{\text{J}}(x)\right| 0\right \rangle
 \right|^{2} \\
 & = \frac{1}{(2\pi)^{3}V}\sum_{\lambda,f}\left|\int_{-\infty}^{t} \dd u
   f_{\varepsilon}(u) \left \langle f;\vec{k},\lambda \left|
       \hat{H}_{\text{J}}(u)\right|0\right \rangle \right|^{2} \ .
\end{split}
\end{equation}
To keep the notation a little bit more shorthand, we have introduced
\begin{subequations}
 \begin{eqnarray}
  \left|f;\vec{k},\lambda\right \rangle & = & \left|f\right \rangle \otimes\left|\vec{k},\lambda\right \rangle \ , \\
  \left|0\right \rangle                 & = & \left|0_{q\bar{q}}\right \rangle \otimes\left|0_{\gamma}\right \rangle  \ .
 \end{eqnarray}
\end{subequations}
Hence, (\ref{eq:a:chiral_abs}) shows that
(\ref{eq:2:photon_yield_chiral_eps}) can be written as the absolute
square of the (space-time integrated) transition amplitude at first
order in $e$ between the initial vacuum state and a final state
containing a single photon of momentum, $\vec{k}$, and polarization,
$\lambda$, which is summed over together with the fermionic degrees of
freedom, $\left|f\right \rangle$. Furthermore, we point out that when
taking successively the limits $t\rightarrow\infty$ and then
$\varepsilon\rightarrow 0$, expression
(\ref{eq:2:photon_yield_chiral_eps}) does not contain any unphysical
contributions from the vacuum polarization, i.e.,
Eq. (\ref{eq:2:photon_yield_chiral_eps}) vanishes if the quark mass
stays at its constituent value, $m_{c}$, for all times, $t$. In this
case, we have
\begin{equation}
\begin{split}
 \label{eq:a:pse_vac}
 \ii\Pi^{<}_{T}(\vec{k},t_{1},t_{2})\ee^{\ii\omega_{\vec{k}}(t_{1}-t_{2})}
 & = \ii\Pi^{<}_{T,0}(\vec{k},t_{1}-t_{2})\ee^{\ii\omega_{\vec{k}}(t_{1}-t_{2})} \\
 & = 2e^{2}\int\frac{\dd^{3}p}{(2\pi)^{3}} \left\lbrace
   1+\frac{px(px+\omega_{\vec{k}})+m^{2}_{c}}{E^{c}_{\vec{p}}E^{c}_{\vec{p}+\vec{k}}}
 \right\rbrace
 \ee^{\ii\left(E^{c}_{\vec{p}+\vec{k}}+E^{c}_{\vec{p}}+\omega_{\vec{k}}\right)(t_{1}-t_{2})}
 \ .
\end{split}
\end{equation}
Accordingly, (\ref{eq:2:photon_yield_chiral_eps}) turns into
\begin{align}
 \omega_{\vec{k}}\frac{\dd^{6}n^{\varepsilon}_{\gamma}(t)}{\dd^{3}x\dd^{3}k}
  = \frac{e^{2}}{(2\pi)^{3}}\int\frac{\dd^{3}p}{(2\pi)^{3}}
     \left\lbrace
      1+\frac{px(px+\omega_{\vec{k}})+m^{2}_{c}}{E^{c}_{\vec{p}}E^{c}_{\vec{p}+\vec{k}}}
     \right\rbrace
     \cdot\left|\int_{-\infty}^{t}\dd uf_{\varepsilon}(u)\ee^{\ii\left(E^{c}_{\vec{p}+\vec{k}}+E^{c}_{\vec{p}}+\omega_{\vec{k}}\right)u}\right|^{2} \ .
\end{align}
Letting $t\rightarrow\infty$ first, we obtain
$$
\int_{-\infty}^{\infty}\dd uf_{\varepsilon}(u)\ee^{\ii\left(E^{c}_{\vec{p}+\vec{k}}+E^{c}_{\vec{p}}+\omega_{\vec{k}}\right)u}
 = \frac{2\varepsilon}{\varepsilon^{2}+\left(E^{c}_{\vec{p}+\vec{k}}+E^{c}_{\vec{p}}+\omega_{\vec{k}}\right)^{2}} \ ,
$$
and hence
\begin{equation}
 \omega_{\vec{k}}\frac{\dd^{6}n^{\varepsilon}_{\gamma}}{\dd^{3}x\dd^{3}k}
 = \frac{e^{2}}{(2\pi)^{3}}\int\frac{\dd^{3}p}{(2\pi)^{3}}
     \left\lbrace
      1+\frac{px(px+\omega_{\vec{k}})+m^{2}_{c}}{E^{c}_{\vec{p}}E^{c}_{\vec{p}+\vec{k}}}
     \right\rbrace
     \left\lbrace
      \frac{2\varepsilon}{\varepsilon^{2}+\left(E^{c}_{\vec{p}+\vec{k}}+E^{c}_{\vec{p}}+\omega_{\vec{k}}\right)^{2}}
     \right\rbrace^{2} ß .
\end{equation}
Taking now the second limit $\varepsilon\rightarrow0$, we finally get
\begin{equation}
\begin{split}
 \omega_{\vec{k}}\frac{\dd^{6}n_{\gamma}}{\dd^{3}x\dd^{3}k}
  = & \lim_{\varepsilon\rightarrow0}
      \frac{e^{2}}{(2\pi)^{3}}\int\frac{\dd^{3}p}{(2\pi)^{3}}
       \left\lbrace
       1+\frac{px(px+\omega_{\vec{k}})+m^{2}_{c}}{E^{c}_{\vec{p}}E^{c}_{\vec{p}+\vec{k}}}
       \right\rbrace
       \left\lbrace
        \frac{2\varepsilon}{\varepsilon^{2}+\left(E^{c}_{\vec{p}+\vec{k}}+E^{c}_{\vec{p}}+\omega_{\vec{k}}\right)^{2}}
       \right\rbrace^{2} \\
  < & \lim_{\varepsilon\rightarrow0}
      \frac{4e^{2}}{(2\pi)^{3}}\int\frac{\dd^{3}p}{(2\pi)^{3}}
       \left\lbrace
        1+\frac{px(px+\omega_{\vec{k}})+m^{2}_{c}}{E^{c}_{\vec{p}}E^{c}_{\vec{p}+\vec{k}}}
       \right\rbrace
       \frac{\varepsilon^{2}}{\left(E^{c}_{\vec{p}+\vec{k}}+E^{c}_{\vec{p}}+\omega_{\vec{k}}\right)^{4}} \\
  = & 0 \ ,
\end{split}
\end{equation}
where we have taken into account that
$E^{c}_{\vec{p}+\vec{k}}+E^{c}_{\vec{p}}+\omega_{\vec{k}}$ is positive definite in the
second step.  Thus, we have shown that the vacuum polarization does not
contribute to the asymptotic photon yield.

For completeness, we also emphasize that keeping the correct
order of limits, i.e., taking first $t\rightarrow\infty$ and then
$\varepsilon\rightarrow0$, is indeed crucial to eliminate the
contribution from the vacuum polarization. To clarify this, we show that
taking $\varepsilon\rightarrow0$ first at some finite time, $t$, and
$t\rightarrow\infty$ afterwards comes along with unphysical
artifacts. Such a procedure corresponds to switching on the
electromagnetic interaction solely from $-\infty$ to some time, $t_{0}$,
i.e., with the regulator function
$$
 g_{\varepsilon}(t)=\theta(t_{0}-t)\ee^{\varepsilon(t-t_{0})}+\theta(t-t_{0}) \ ,
$$
which has been suggested in \cite{Serreau:2003wr} to implement initial
correlations at $t=t_{0}$ evolving from an uncorrelated initial state at
$t\rightarrow-\infty$. In this case, the photon yield can still be transformed to
an absolute-square of the form (\ref{eq:a:chiral_abs}) with
$f_{\varepsilon}(t)$ replaced by $g_{\varepsilon}(t)$. If one first
takes $\varepsilon\rightarrow0$ at some finite time, $t$, however, one
encounters a divergent contribution from the vacuum polarization.  In
order to see this, we again split the photon self-energy into the
stationary vacuum polarization and a non-stationary-mass shift
contribution, i.e.,
\begin{equation}
 \ii\Pi^{<}_{T}(\vec{k},t_{1},t_{2}) = \ii\Pi^{<}_{T,0}(\vec{k},t_{1}-t_{2})+\ii\Delta\Pi^{<}_{T}(\vec{k},t_{1},t_{2}) \ ,
\end{equation}
where $\ii\Delta\Pi^{<}_{T}(\vec{k},t_{1},t_{2})$ vanishes for
$t_{1},t_{2}\le t_{0}$. Accordingly, we decompose
(\ref{eq:2:photon_yield_chiral_eps}) as
\begin{equation}
 2\omega_{\vec{k}}\frac{\dd^{6}n^{\varepsilon}_{\gamma}(t)}{\dd^{3}x\dd^{3}k} = \left.2\omega_{\vec{k}}\frac{\dd^{6}n^{\varepsilon}_{\gamma}(t)}{\dd^{3}x\dd^{3}k}\right|_{\text{VAC}}+
                                                                                \left.2\omega_{\vec{k}}\frac{\dd^{6}n^{\varepsilon}_{\gamma}(t)}{\dd^{3}x\dd^{3}k}\right|_{\text{CMS}} \ ,
\end{equation}
with the contributions from the vacuum polarization (VAC) and from
chiral mass shift effects (CMS) reading
\begin{subequations}
 \begin{eqnarray}
  \left.2\omega_{\vec{k}}\frac{\dd^{6}n^{\varepsilon}_{\gamma}(t)}{\dd^{3}x\dd^{3}k}\right|_{\text{VAC}} 
     & = & \frac{1}{(2\pi)^3}\int_{-\infty}^{t}\dd t_{1}\int_{-\infty}^{t}\dd t_{2}\mbox{ }g_{\varepsilon}(t_{1})g_{\varepsilon}(t_{2}) 
           \ii\Pi^{<}_{T,0}(\vec{k},t_{1}-t_{2})\ee^{\ii\omega_{\vec{k}}(t_{1}-t_{2})} \ , \label{eq:a:contr_vac} \\
  \left.2\omega_{\vec{k}}\frac{\dd^{6}n^{\varepsilon}_{\gamma}(t)}{\dd^{3}x\dd^{3}k}\right|_{\text{CMS}}
     & = & \frac{1}{(2\pi)^3}\int_{-\infty}^{t}\dd t_{1}\int_{-\infty}^{t}\dd t_{2}\mbox{ }g_{\varepsilon}(t_{1})g_{\varepsilon}(t_{2}) 
           \ii\Delta\Pi^{<}_{T}(\vec{k},t_{1},t_{2})\ee^{\ii\omega_{\vec{k}}(t_{1}-t_{2})} \ . \label{eq:a:contr_phs}
 \end{eqnarray}
\end{subequations}
After interchanging both time integrations with the loop integral the
contribution from the vacuum polarization at time $t$ is evaluated to
\begin{align}
 \left.\omega_{\vec{k}}\frac{\dd^{6}n^{\varepsilon}_{\gamma}(t)}{\dd^{3}x\dd^{3}k}\right|_{\text{VAC}}
   = \frac{e^{2}}{(2\pi)^{3}}\int\frac{\dd^{3}p}{(2\pi)^{3}}
        & \left\lbrace
           1+\frac{px(px+\omega_{\vec{k}})+m^{2}_{c}}{E^{c}_{\vec{p}}E^{c}_{\vec{p}+\vec{k}}}
          \right\rbrace \nonumber \\
 \times & \left|\int_{-\infty}^{t}\dd u\mbox{ }g_{\varepsilon}(u)\ee^{\ii\left(E^{c}_{\vec{p}+\vec{k}}+E^{c}_{\vec{p}}+\omega_{\vec{k}}\right)u}\right|^{2} \ ,
\end{align}
with the time integral given by
\begin{equation}
\begin{split}
  \int_{-\infty}^{t}\dd u\mbox{
  }g_{\varepsilon}(u)\ee^{ \ii\left(E^{c}_{\vec{p}+\vec{k}}+E^{c}_{\vec{p}}+\omega_{\vec{k}}\right)u}
  = &
  \frac{g_{\varepsilon}(t)\ee^{\ii\left(E^{c}_{\vec{p}+\vec{k}}+E^{c}_{\vec{p}}+\omega_{\vec{k}}\right)t_{0}}}
  {\varepsilon+\ii\left(E^{c}_{\vec{p}+\vec{k}}+E^{c}_{\vec{p}}+\omega_{\vec{k}}\right)}  \\
  & +
  \theta(t-t_{0})\frac{\ee^{\ii\left(E^{c}_{\vec{p}+\vec{k}}+E^{c}_{\vec{p}}+\omega_{\vec{k}}\right)t}-
                       \ee^{\ii\left(E^{c}_{\vec{p}+\vec{k}}+E^{c}_{\vec{p}}+\omega_{\vec{k}}\right)t_{0}}}
  {\ii\left(E^{c}_{\vec{p}+\vec{k}}+E^{c}_{\vec{p}}+\omega_{\vec{k}}\right)} \ .
\end{split}
\end{equation}
Taking now the limit $\varepsilon\rightarrow0$ at this point, the vacuum
contribution does not vanish but instead turns into
\begin{equation}
 \label{eq:a:contr_vac_re}
 \left.\omega_{\vec{k}}\frac{\dd^{6}n_{\gamma}(t)}{\dd^{3}x\dd^{3}k}\right|_{\text{VAC}} 
  =\frac{e^{2}}{(2\pi)^{3}}\int\frac{\dd^{3}p}{(2\pi)^{3}}
    \left\lbrace
     1+\frac{px(px+\omega_{\vec{k}})+m^{2}_{c}}{E^{c}_{\vec{p}}E^{c}_{\vec{p}+\vec{k}}}
    \right\rbrace
    \frac{1}{\left(E^{c}_{\vec{p}+\vec{k}}+E^{c}_{\vec{p}}+\omega_{\vec{k}}\right)^{2}} \ .
\end{equation}
Since the integration measure $\dd^{3}p$ adds another factor of $p^{2}$
to the integrand, the loop integral entering (\ref{eq:a:contr_vac_re})
is linearly divergent. As (\ref{eq:a:contr_vac_re}) is furthermore constant in time, it does
not vanish in the limit $t\rightarrow\infty$. Hence, interchanging the
limits $t\rightarrow\infty$ and $\varepsilon\rightarrow0$ comes along
with a divergent contribution from the vacuum polarization.

Since (\ref{eq:a:contr_vac_re}) is constant in time and hence already
fully present before any mass-shift contributions characterized by
$\ii\Delta\Pi^{<}_{T}(\vec{k},t_{1},t_{2})$ can show up, one might still
argue that (\ref{eq:a:contr_vac_re}) can be uniquely identified with the
virtual cloud of the vacuum and, as a consequence, be subtracted
(renormalized) from the overall photon yield
(\ref{eq:2:photon_yield_chiral_eps}).  This would only leave the
contribution (\ref{eq:a:contr_phs}). The main problem with this
argument, however, is that this contribution on its own is not an
absolute square, which can be seen by rewriting it as follows:
\begin{align}
 \label{eq:a:contr_phs_re}
   \left.2\omega_{\vec{k}}\frac{\dd^{6}n^{\varepsilon}_{\gamma}(t)}{\dd^{3}x\dd^{3}k}\right|_{\text{CMS}}
     = \frac{1}{(2\pi)^3}
        & \left\lbrace
           \int_{t_{0}}^{t}\dd t_{1}\int_{t_{0}}^{t}\dd t_{2}\mbox{ }g_{\varepsilon}(t_{1})g_{\varepsilon}(t_{2})
           \ii\Delta\Pi^{<}_{T}(\vec{k},t_{1},t_{2})\ee^{\ii\omega_{\vec{k}}(t_{1}-t_{2})}
          \right. \nonumber \\
        & \left.
           +2\mbox{Re}\left[
                       \int_{t_{0}}^{t}\dd t_{1}\int_{-\infty}^{t_{0}}\dd t_{2}\mbox{ }
                       g_{\varepsilon}(t_{1})g_{\varepsilon}(t_{2})
                       \ii\Delta\Pi^{<}_{T}(\vec{k},t_{1},t_{2})\ee^{\ii\omega_{\vec{k}}(t_{1}-t_{2})}
                      \right]
          \right\rbrace \ .
\end{align}
Above, we haven taken into account that $\ii\Delta\Pi^{<}_{T}(\vec{k},t_{1},t_{2})=0$ for $t_{1},t_{2}\le t_{0}$. We have, however, 
$\ii\Delta\Pi^{<}_{T}(\vec{k},t_{1},t_{2})\neq0$ if either $t_{1}>t_{0}$ and $t_{2}\le t_{0}$ or
$t_{1}\le t_{0}$ and $t_{2}>t_{0}$. This gives rise to the second term in (\ref{eq:a:contr_phs_re}). For 
this term, we have also made use of
$$
 \ii\Delta\Pi^{<}_{T}(\vec{k},t_{1},t_{2}) = \left[\ii\Delta\Pi^{<}_{T}(\vec{k},t_{2},t_{1})\right]^{*} \ .
$$
Since the overall contribution from $\ii\Pi^{<}_{T,0}(\vec{k},t_{1}-t_{2})$ from the three domains
\begin{itemize}
 \item $t_{1},t_{2}>t_{0}$ \ ,
 \item $t_{1}>t_{0}$ and $t_{2}\le t_{0}$ \ ,
 \item $t_{1}\le t_{0}$ and $t_{2}>t_{0}$ \ ,
\end{itemize}
moreover vanishes in the limit $\varepsilon\rightarrow0$, we can rewrite (\ref{eq:a:contr_phs_re}) as
\begin{align}
 \label{eq:a:contr_phs_rere}
   \left.2\omega_{\vec{k}}\frac{\dd^{6}n^{\varepsilon}_{\gamma}(t)}{\dd^{3}x\dd^{3}k}\right|_{\text{CMS}}
     = \frac{1}{(2\pi)^3}
        & \left\lbrace
           \int_{t_{0}}^{t}\dd t_{1}\int_{t_{0}}^{t}\dd t_{2}\mbox{ }g_{\varepsilon}(t_{1})g_{\varepsilon}(t_{2})
           \ii\Pi^{<}_{T}(\vec{k},t_{1},t_{2})\ee^{\ii\omega_{\vec{k}}(t_{1}-t_{2})}
          \right. \nonumber \\
        & \left.
           +2\mbox{Re}\left[
                       \int_{t_{0}}^{t}\dd t_{1}\int_{-\infty}^{t_{0}}\dd t_{2}\mbox{ }
                       g_{\varepsilon}(t_{1})g_{\varepsilon}(t_{2})
                       \ii\Pi^{<}_{T}(\vec{k},t_{1},t_{2})\ee^{\ii\omega_{\vec{k}}(t_{1}-t_{2})}
                      \right]
          \right\rbrace \ .
\end{align}
In contrast to (\ref{eq:2:photon_yield_chiral_eps}), there is no
contribution to (\ref{eq:a:contr_phs_rere}) from the domain where
$t_{1},t_{2}\le t_{0}$. Therefore, it is not possible to take
(\ref{eq:a:contr_phs_re}) into an absolute-square representation of the
form of (\ref{eq:a:chiral_abs}), but with $f_{\varepsilon}(t)$ replaced
by $g_{\varepsilon}(t)$. Hence, it is not guaranteed that the
(renormalized) photon yield is positive (semi-) definite, and it can, in
principle, acquire unphysical negative values.

\section{Ward-Takahashi identities for $\ii\Pi^{<}_{\mu\nu}(\vec{k},t,u)$}
\label{sec:appb}
In this section, we will provide an explicit verification of the
Ward-Takahashi identities for the photon self-energy, $\ii
\Pi^{<}_{\mu\nu}(\vec{k},t,u)$, which read
\begin{subequations}
 \label{eq:b:chiral_wti}
 \begin{eqnarray}
  \partial_{t}\ii \Pi^{<}_{00}(\vec{k},t,u)-\ii k^{i}\ii \Pi^{<}_{i0}(\vec{k},t,u) & = & 0 \ , \\
  \partial_{t}\ii \Pi^{<}_{0i}(\vec{k},t,u)-\ii k^{j}\ii \Pi^{<}_{ji}(\vec{k},t,u) & = & 0 \ . 
 \end{eqnarray}
\end{subequations}
For this purpose, we first express $\ii\Pi^{<}_{\mu\nu}(\vec{k},t,u)$ in
terms of wavefunction parameters. It follows from
(\ref{eq:2:propagators_momentum}) that
\begin{equation}
\begin{split}
 \label{eq:b:pse_wavef}
 \ii\Pi^{<}_{\mu\nu}(\vec{k},t,u) &=
 e^{2}\sum_{r,s}\int\frac{\dd^{3}p}{(2\pi)^{3}} \left [
   \bar{\psi}^{'}_{\vec{p},r,\uparrow}(t)\gamma_{\mu}
   \psi^{'}_{\vec{p}+\vec{k},s,\downarrow}(t) \right] \left[
   \bar{\psi}^{'}_{\vec{p}+\vec{k},s,\downarrow}(u)
   \gamma_{\nu}\psi^{'}_{\vec{p},r,\uparrow}(u)
 \right] \\
 & = e^{2}\sum_{r,s}\int\frac{\dd^{3}p}{(2\pi)^{3}} \left[
   \bar{\psi}^{'}_{\vec{p},r,\uparrow}(t)\gamma_{\mu}
   \psi^{'}_{\vec{p}+\vec{k},s,\downarrow}(t) \right] \left [
   \bar{\psi}^{'}_{\vec{p},r,\uparrow}(u)\gamma_{\nu}
   \psi^{'}_{\vec{p}+\vec{k},s,\downarrow}(u) \right ]^{*}
\end{split}
\end{equation}
With help of (\ref{eq:2:wave_param}), the matrix elements are evaluated to
\begin{subequations}
 \label{eq:b:matrix}
 \begin{eqnarray}
  \bar{\psi}^{'}_{\vec{p},r,\uparrow}(t)\gamma_{0}\psi^{'}_{\vec{p}+\vec{k},s,\downarrow}(t)
   & = & \alpha^{*}_{\vec{p},\uparrow}(t)\alpha_{\vec{p}+\vec{k},\downarrow}(t)\delta_{rs}+ \nonumber \\
   &   & \beta^{*}_{\vec{p},\uparrow}(t)\beta_{\vec{p}+\vec{k},\downarrow}(t)
         \mbox{ }\bar{\chi}_{r}\left(\vec{\sigma}\cdot\vec{e}_{\vec{p}}\right)
         \left(\vec{\sigma}\cdot\vec{e}_{\vec{p}+\vec{k}}\right)\chi_{s} \ , \label{eq:b:matrix_time} \\
  \bar{\psi}^{'}_{\vec{p},r,\uparrow}(t)\gamma_{i}\psi^{'}_{\vec{p}+\vec{k},s,\downarrow}(t)
   & = & -\alpha^{*}_{\vec{p},\uparrow}(t)\beta_{\vec{p}+\vec{k},\downarrow}(t)
         \mbox{ }\bar{\chi}_{r}\sigma_{i}\left(\vec{\sigma}\cdot\vec{e}_{\vec{p}+\vec{k}}\right)\chi_{s} \nonumber \\
   &   & -\beta^{*}_{\vec{p},\uparrow}(t)\alpha_{\vec{p}+\vec{k},\downarrow}(t)
         \mbox{ }\bar{\chi}_{r}\left(\vec{\sigma}\cdot\vec{e}_{\vec{p}}\right)\sigma_{i}\chi_{s} \ , \label{eq:b:matrix_space} 
 \end{eqnarray}
\end{subequations}
where $\vec{e}_{\vec{p}}$ and $\vec{e}_{\vec{p}+\vec{k}}$ denote the
unit vectors in the directions of $\vec{p}$ and $\vec{p}+\vec{k}$
respectively. Furthermore, we have taken into account that
$$
\gamma_{i} = -\gamma^{i} = \begin{pmatrix}
                            0          & -\sigma^{i} \\
                            \sigma^{i} & 0  
                           \end{pmatrix} \ ,
$$
for spacelike indices, $i$. The representation of
$\ii\Pi^{<}_{\mu\nu}(\vec{k},t,u)$ in terms of wavefunction parameters
is obtained upon insertion of
(\ref{eq:b:matrix_time})-(\ref{eq:b:matrix_space}) into
(\ref{eq:b:pse_wavef}) for the respective indices and carrying out the
summation over the two spin indices, $r$ and $s$. For both indices being
timelike, i.e., $\mu=\nu=0$, we obtain
\begin{equation}
\begin{split}
  \label{eq:b:chiral_pse00}
  \ii \Pi^{<}_{00}(\vec{k},t,u) = 2e^2\int\frac{\dd^{3}p}{(2\pi)^{3}} &
  \left\lbrace
    \alpha^{*}_{\vec{p},\uparrow}(t)\alpha_{\vec{p}+\vec{k},\downarrow}(t)
    \alpha^{*}_{\vec{p}+\vec{k},\downarrow}(u)\alpha_{\vec{p},\uparrow}(u)\right.+\\
  &
  \quad\beta^{*}_{\vec{p},\uparrow}(t)\beta_{\vec{p}+\vec{k},\downarrow}(t)
  \beta^{*}_{\vec{p}+\vec{k},\downarrow}(u)\beta_{\vec{p},\uparrow}(u)+\\
  & \quad\frac{\vec{p}\cdot(\vec{p}+\vec{k})}{p|\vec{p}+\vec{k}|}
  \left[\alpha^{*}_{\vec{p},\uparrow}(t)\alpha_{\vec{p}+\vec{k},\downarrow}(t)
    \beta^{*}_{\vec{p}+\vec{k},\downarrow}(u)\beta_{\vec{p},\uparrow}(u)\right.+\\
  & \quad\quad\quad\quad\quad\quad\left.\left.
      \beta^{*}_{\vec{p},\uparrow}(t)\beta_{\vec{p}+\vec{k},\downarrow}(t)
      \alpha^{*}_{\vec{p}+\vec{k},\downarrow}(u)\alpha_{\vec{p},\uparrow}(u)
    \right]\right\rbrace \ .
\end{split}
\end{equation}
Here we have made use of the identities
\begin{subequations}
 \begin{eqnarray}
  \sum_{r,s}\left|\bar{\chi}_{r}\left(\vec{\sigma}\cdot\vec{u}\right)
  \left(\vec{\sigma}\cdot\vec{v}\right)\chi_{s}\right|^{2}
   = & \mbox{Tr}\left\lbrace\left(\vec{\sigma}\cdot\vec{u}\right)^{2}
         \left(\vec{\sigma}\cdot\vec{v}\right)^{2}\right\rbrace 
 & = 2\vec{u}^{2}\vec{v}^{2} \ , \\
  \sum_{r,s}\bar{\chi}_{s}\left(\vec{\sigma}\cdot\vec{u}\right)
            \left(\vec{\sigma}\cdot\vec{v}\right)\chi_{s}\delta_{rs}
   = & \mbox{Tr}\left\lbrace\left(\vec{\sigma}\cdot\vec{u}\right)
         \left(\vec{\sigma}\cdot\vec{v}\right)\right\rbrace
   & = 2\vec{u}\cdot{\vec{v}} \ ,
 \end{eqnarray}
\end{subequations}
holding for arbitrary vectors, $\vec{u}$ and $\vec{v}$, with real-valued
components. Together with the relations
\begin{equation}
 \label{eq:b:help_a}
   \sum_{r,s}\delta_{rs}\bar{\chi}_{r}\sigma_{i}\left(\vec{\sigma}\cdot\vec{u}\right)\chi_{s}
 = \sum_{r,s}\delta_{rs}\bar{\chi}_{r}\left(\vec{\sigma}\cdot\vec{u}\right)\sigma_{i}\chi_{s}
 = \mbox{Tr}\left\lbrace\sigma_{i}\left(\vec{\sigma}\cdot\vec{u}\right)\right\rbrace
 = 2u_{i}
\end{equation}
and
\begin{subequations}
 \label{eq:b:help_b}
 \begin{eqnarray}
  \sum_{r,s}\left[\bar{\chi}_{r}\left(\vec{\sigma}\cdot\vec{u}\right)\left(\vec{\sigma}\cdot\vec{v}\right)\chi_{s}\right]
            \left[\bar{\chi}_{r}\sigma_{i}\left(\vec{\sigma}\cdot\vec{v}\right)\chi_{s}\right]^{*} 
   & = \mbox{Tr}\left\lbrace
                  \left(\vec{\sigma}\cdot\vec{u}\right)\left(\vec{\sigma}\cdot\vec{v}\right)^{2}\sigma_{i}
                 \right\rbrace
   & = 2\vec{v}^{2}u_{i} \ , \\
  \sum_{r,s}\left[\bar{\chi}_{r}\left(\vec{\sigma}\cdot\vec{u}\right)\left(\vec{\sigma}\cdot\vec{v}\right)\chi_{s}\right]
                  \left[\bar{\chi}_{r}\left(\vec{\sigma}\cdot\vec{u}\right)\sigma_{i}\chi_{s}\right]^{*}
   & = \mbox{Tr}\left\lbrace
                  \left(\vec{\sigma}\cdot\vec{u}\right)^{2}\left(\vec{\sigma}\cdot\vec{v}\right)^{2}\sigma_{i}
                 \right\rbrace
   & = 2\vec{u}^{2}v_{i} \ ,
 \end{eqnarray}
\end{subequations}
the photon self-energy for mixed time- and spacelike indices is
evaluated to
\begin{subequations}
\begin{align}
 \ii \Pi^{<}_{0i}(\vec{k},t,u) = 2e^2\int\frac{\dd^{3}p}{(2\pi)^{3}} &
  \left\lbrace\frac{p_{i}+k_{i}}{|\vec{p}+\vec{k}|}
   \left[\alpha^{*}_{\vec{p},\uparrow}(t)\alpha_{\vec{p}+\vec{k},\downarrow}(t)
    \beta^{*}_{\vec{p}+\vec{k},\downarrow}(u)\alpha_{\vec{p},\uparrow}(u)+
    \right.
  \right. \nonumber \\
 & \quad\quad\quad\quad\mbox{ } \left.
    \beta^{*}_{\vec{p},\uparrow}(t)\beta_{\vec{p}+\vec{k},\downarrow}(t)
    \alpha^{*}_{\vec{p}+\vec{k},\downarrow}(u)\beta_{\vec{p},\uparrow}(u)
  \right] \nonumber \\
 & \quad\quad+\frac{p_{i}}{p}
  \left[\alpha^{*}_{\vec{p},\uparrow}(t)\alpha_{\vec{p}+\vec{k},\downarrow}(t)
   \alpha^{*}_{\vec{p}+\vec{k},\downarrow}(u)\beta_{\vec{p},\uparrow}(u)+
  \right. \nonumber \\
 & \quad\quad\quad\quad\mbox{ } \left.
   \left.\beta^{*}_{\vec{p},\uparrow}(t)\beta_{\vec{p}+\vec{k},\downarrow}(t)
     \beta^{*}_{\vec{p}+\vec{k},\downarrow}(u)\alpha_{\vec{p},\uparrow}(u)
   \right] \right\rbrace \ , \label{eq:b:chiral_pse0i} \\
 \ii \Pi^{<}_{i0}(\vec{k},t,u) = 2e^2\int\frac{\dd^{3}p}{(2\pi)^{3}} &
  \left\lbrace\frac{p_{i}+k_{i}}{|\vec{p}+\vec{k}|}
   \left[\alpha^{*}_{\vec{p},\uparrow}(t)\beta_{\vec{p}+\vec{k},\downarrow}(t)
    \alpha^{*}_{\vec{p}+\vec{k},\downarrow}(u)\alpha_{\vec{p},\uparrow}(u)+
    \right.
   \right. \nonumber \\
  & \quad\quad\quad\quad\mbox{ } \left.
     \beta^{*}_{\vec{p},\uparrow}(t)\alpha_{\vec{p}+\vec{k},\downarrow}(t)
     \beta^{*}_{\vec{p}+\vec{k},\downarrow}(u)\beta_{\vec{p},\uparrow}(u)
   \right] \nonumber \\
   & \quad\quad+\frac{p_{i}}{p}
   \left[\alpha^{*}_{\vec{p},\uparrow}(t)\beta_{\vec{p}+\vec{k},\downarrow}(t)
    \beta^{*}_{\vec{p}+\vec{k},\downarrow}(u)\beta_{\vec{p},\uparrow}(u)+
  \right. \nonumber \\
 & \quad\quad\quad\quad\mbox{ } \left.
  \left.\beta^{*}_{\vec{p},\uparrow}(t)\alpha_{\vec{p}+\vec{k},\downarrow}(t)
    \alpha^{*}_{\vec{p}+\vec{k},\downarrow}(u)\alpha_{\vec{p},\uparrow}(u)
   \right]
 \right\rbrace \ . \label{eq:b:chiral_psei0}
\end{align}
\end{subequations}
Here it is important to note that in contrast to (\ref{eq:b:help_a}) and
(\ref{eq:b:help_b}), the index index $i$ in (\ref{eq:b:chiral_pse0i})
and (\ref{eq:b:chiral_psei0}) is to be taken as covariant and not as
Euclidean, i.e. $\left\lbrace u_{i}\right\rbrace=-\vec{u}$.  Finally, we
take into account that
\begin{subequations}
 \begin{eqnarray}
  \sum_{r,s}\left[\bar{\chi}_{r}\sigma_{i}\left(\vec{\sigma}\cdot\vec{u}\right)\chi_{s}\right]
            \left[\bar{\chi}_{r}\sigma_{j}\left(\vec{\sigma}\cdot\vec{v}\right)\chi_{s}\right]^{*}
   & = & \sum_{r,s}\left[\bar{\chi}_{r}\left(\vec{\sigma}\cdot\vec{u}\right)\sigma_{i}\chi_{s}\right]
                   \left[\bar{\chi}_{r}\left(\vec{\sigma}\cdot\vec{v}\right)\sigma_{j}\chi_{s}\right]^{*} \nonumber \\
   & = & 2\delta_{ij}\vec{u}\cdot\vec{v} \ , \\
  \sum_{r,s}\left[\bar{\chi}_{r}\sigma_{i}\left(\vec{\sigma}\cdot\vec{u}\right)\chi_{s}\right]
            \left[\bar{\chi}_{r}\left(\vec{\sigma}\cdot\vec{v}\right)\sigma_{j}\chi_{s}\right]^{*}
   & = & \sum_{r,s}\left[\bar{\chi}_{r}\left(\vec{\sigma}\cdot\vec{u}\right)\sigma_{i}\chi_{s}\right]
                   \left[\bar{\chi}_{r}\sigma_{j}\left(\vec{\sigma}\cdot\vec{v}\right)\chi_{s}\right]^{*} \nonumber \\
   & = & 2\left(u_{i}v_{j}+u_{j}v_{i}-\delta_{ij}\vec{u}\cdot\vec{v}\right) \ ,
 \end{eqnarray}
\end{subequations}
which allows us to rewrite the photon self-energy for two space-like
indices as
\begin{align}
  \label{eq:b:chiral_pseij}
  \ii \Pi^{<}_{ij}(\vec{k},t,u) = 2e^2 \int\frac{\dd^{3}p}{(2\pi)^{3}} &
  \left\lbrace \ell_{ij} \left[
      \alpha^{*}_{\vec{p},\uparrow}(t)\beta_{\vec{p}+\vec{k},\downarrow}(t)
      \alpha^{*}_{\vec{p}+\vec{k},\downarrow}(u)\beta_{\vec{p},\uparrow}(u)+
    \right.
  \right. \nonumber \\
  & \quad\quad \left.
    \beta^{*}_{\vec{p},\uparrow}(t)\alpha_{\vec{p}+\vec{k},\downarrow}(t)
    \beta^{*}_{\vec{p}+\vec{k},\downarrow}(u)\alpha_{\vec{p},\uparrow}(u)
  \right] \nonumber \\
  & -\eta_{ij} \left[
    \alpha^{*}_{\vec{p},\uparrow}(t)\beta_{\vec{p}+\vec{k},\downarrow}(t)
    \beta^{*}_{\vec{p}+\vec{k},\downarrow}(u)\alpha_{\vec{p},\uparrow}(u)+
  \right. \nonumber \\
  & \quad\quad\mbox{ } \left.  \left.
      \beta^{*}_{\vec{p},\uparrow}(t)\alpha_{\vec{p}+\vec{k},\downarrow}(t)
      \alpha^{*}_{\vec{p}+\vec{k},\downarrow}(u)\beta_{\vec{p},\uparrow}(u)\ell^{b}_{ij}
    \right] \right\rbrace \ .
\end{align}
In order to keep the notation short, we have introduced
\begin{equation}
 \label{eq:b:trace_pauli}
  \ell_{ij} = \frac{p_{i}(p_{j}+k_{j})+p_{j}(p_{i}+k_{i})+\eta_{ij}\vec{p}\cdot(\vec{p}+\vec{k})}{p|\vec{p}+\vec{k}|} \ .
\end{equation}
The verification of (\ref{eq:b:chiral_wti}) in terms of wavefunction
parameters is now straightforward. It follows immediately from
(\ref{eq:2:eom_param}) that
\begin{subequations}
 \label{eq:b:eom}
 \begin{eqnarray}
  \partial_{t}\left(\alpha^{*}_{\vec{p},\uparrow}(t)\alpha_{\vec{p}+\vec{k},\downarrow}(t)\right)
    & = & \ii p\beta^{*}_{\vec{p},\uparrow}(t)\alpha_{\vec{p}+\vec{k},\downarrow}(t)-
          \ii |\vec{p}+\vec{k}|\alpha^{*}_{\vec{p},\uparrow}(t)\beta_{\vec{p}+\vec{k},\downarrow}(t) \ , 
          \label{eq:b:eom_1} \\
          \partial_{t}\left(\beta^{*}_{\vec{p},\uparrow}(t)\beta_{\vec{p}+\vec{k},\downarrow}(t)\right)
          & = & \ii p\alpha^{*}_{\vec{p},\uparrow}(t)\beta_{\vec{p}+\vec{k},\downarrow}(t)-
          \ii |\vec{p}+\vec{k}|\beta^{*}_{\vec{p},\uparrow}(t)\alpha_{\vec{p}+\vec{k},\downarrow}(t) \ .
          \label{eq:b:eom_2} 
 \end{eqnarray}
\end{subequations}
Taking the time derivatives of (\ref{eq:b:chiral_pse00}) and
(\ref{eq:b:chiral_pse0i}) and making use of relations (\ref{eq:b:eom_1})
and (\ref{eq:b:eom_2}), we obtain
\begin{subequations}
 \begin{eqnarray}
  \partial_{t}\ii \Pi^{<}_{00}(\vec{k},t,u) & = & \ii k^{i}\ii \Pi^{<}_{i0}(\vec{k},t,u) \ , \\
  \partial_{t}\ii \Pi^{<}_{0i}(\vec{k},t,u) & = & \ii k^{j}\ii \Pi^{<}_{ji}(\vec{k},t,u) \ ,
 \end{eqnarray}
\end{subequations}
and hence
\begin{subequations}
 \begin{eqnarray}
  \partial_{t}\ii \Pi^{<}_{00}(\vec{k},t,u)-\ii k^{i}\ii \Pi^{<}_{i0}(\vec{k},t,u) & = & 0 \ , \\
  \partial_{t}\ii \Pi^{<}_{0i}(\vec{k},t,u)-\ii k^{j}\ii \Pi^{<}_{ji}(\vec{k},t,u) & = & 0 \ . 
 \end{eqnarray}
\end{subequations}
These are just the Ward-Takahashi identities (\ref{eq:b:chiral_wti}).

\section{Remark on time dependence}
\label{sec:appf}
\begin{figure}[htb]
 \begin{center}
  \includegraphics[height=5.0cm]{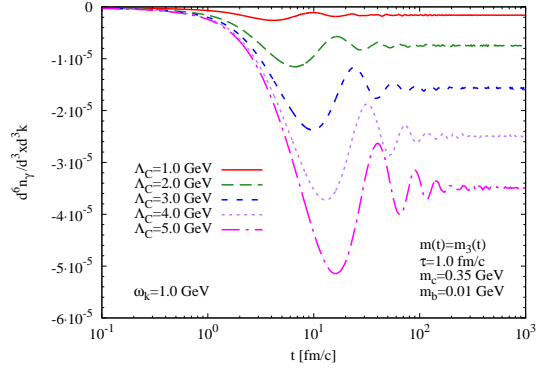}
  \caption{Time evolution of the interference contribution (\ref{eq:f:photnum_int}) for $\omega_{\vec{k}}=1 \; \GeV$ 
           and different values of $\Lambda_{C}$.}
  \label{fig:f:interference}
 \end{center}
\end{figure}

It has been shown in appendix \ref{sec:appa} that keeping the correct
sequence of the limits, \emph{first} $t\rightarrow\infty$ and \emph{then}
$\varepsilon\rightarrow0$, is crucial for eliminating possible
unphysical contributions from the vacuum polarization and that their
interchange comes along with a divergent contribution from the
latter. In this appendix, we consider the time evolution of
(\ref{eq:4:photnum_mst}) and (\ref{eq:4:photnum_int}) in order to
highlight that keeping the correct order of limits is also
essential to obtain physically sensible results from the mass-shift
contribution (\ref{eq:4:pse_mst}) and to eliminate possible
contributions from the interference term (\ref{eq:4:pse_int}). The
latter is essential to obtain photon numbers that can be written as an
absolute square and, as a consequence, are positive (semi-)
definite. For this purpose, we first rewrite (\ref{eq:4:photnum_mst})
and (\ref{eq:4:photnum_int}) as
\begin{subequations}
 \begin{align}
  \left. 2\omega_{\vec{k}}\frac{\dd^{6}n^{\varepsilon}_{\gamma}(t)}{\dd^{3}x\dd^{3}k}\right|_{\text{MST}} 
    = \frac{\gamma^{\mu\nu}(k)}{(2\pi)^{3}}\sum_{r,s}\int\frac{\dd^{3}p}{(2\pi)^{3}} & 
      I^{\varepsilon,*}_{\mu}(\vec{p},\vec{k},r,s,t)I^{\varepsilon}_{\nu}(\vec{p},\vec{k},r,s,t) \ , \label{eq:f:photnum_mst} \\
  \left. 2\omega_{\vec{k}}\frac{\dd^{6}n^{\varepsilon}_{\gamma}(t)}{\dd^{3}x\dd^{3}k}\right|_{\text{INT}}
    = \frac{\gamma^{\mu\nu}(k)}{(2\pi)^{3}}\sum_{r,s}\int\frac{\dd^{3}p}{(2\pi)^{3}} &
       \left[
        I^{\varepsilon,*}_{\mu}(\vec{p},\vec{k},r,s,t)J^{\varepsilon}_{\nu}(\vec{p},\vec{k},r,s,t)
       \right. \nonumber \\
        & +\left.
            J^{\varepsilon,*}_{\mu}(\vec{p},\vec{k},r,s,t)I^{\varepsilon}_{\nu}(\vec{p},\vec{k},r,s,t)
           \right] \ , \label{eq:f:photnum_int} 
 \end{align}
\end{subequations}
where we have introduced
\begin{subequations}
 \label{eq:f:intaspt}
 \begin{eqnarray}
  I^{\varepsilon}_{\mu}(\vec{p},\vec{k},r,s,t) & = & \int_{-\infty}^{t}\dd u f_{\varepsilon}(u)j^{\text{MST}}_{\mu}(\vec{p},\vec{k},r,s,u)\ee^{\ii\omega_{\vec{k}}u} \ , \label{eq:f:intaspt_mst} \\
  J^{\varepsilon}_{\mu}(\vec{p},\vec{k},r,s,t) & = & \int_{-\infty}^{t}\dd u f_{\varepsilon}(u)j^{0}_{\mu}(\vec{p},\vec{k},r,s,u)\ee^{\ii\omega_{\vec{k}}u} \ . \label{eq:f:intaspt_vac}
 \end{eqnarray}
\end{subequations}
The time integral in (\ref{eq:f:intaspt_vac}) is evaluated to
\begin{equation}
 \label{eq:f:intamp_vac}
\begin{split}
 J^{\varepsilon}_{\mu}(\vec{p},\vec{k},r,s,t) = e\bar{u}_{c}(\vec{p},r)\gamma_{\mu}v_{c}(\vec{p}+\vec{k},s)
  & \left\lbrace
     \theta(-t)\frac{e^{\left[\varepsilon+i\omega^{c}_{1}(\vec{p},\vec{k})\right]t}}{\varepsilon+i\omega^{c}_{1}(\vec{p},\vec{k})}
    \right. \\
  & \left.
     +\theta(t)\left[
                \frac{2\varepsilon}{\varepsilon^{2}+\omega^{c,2}_{1}(\vec{p},\vec{k})}-
                \frac{e^{-\left[\varepsilon-i\omega^{c}_{1}(\vec{p},\vec{k})\right]t}}{\varepsilon-i\omega^{c}_{1}(\vec{p},\vec{k})}
               \right] 
    \right\rbrace .
\end{split}
\end{equation}
To handle (\ref{eq:f:intaspt_mst}) for large times $t\ge T\gg\tau$, we
again take into account that the fermionic wavefunctions essentially
have turned into (\ref{eq:4:asymptotics}) in that range. Accordingly, we
obtain
\begin{equation}
\begin{split}
 \label{eq:f:intamp_mst}
 I^{\varepsilon}_{\mu}(\vec{p},\vec{k},r,s,t)
   = & \int_{-\infty}^{T}\dd u f_{\varepsilon}(u)j^{\text{MST}}_{\mu}(\vec{p},\vec{k},r,s,u)\ee^{\ii\omega_{\vec{k}}u}  \\
   + & e\left\lbrace
          \tilde{\alpha}^{*}_{\vec{p}}\tilde{\gamma}_{\vec{p}+\vec{k}}\bar{u}_{b}(\vec{p},r)\gamma_{\mu}v_{b}(\vec{p}+\vec{k},s)
          g_{\varepsilon}\left[\omega^{b}_{1}(\vec{p},\vec{k}),T,t\right]
         \right.  \\
     & +\tilde{\alpha}^{*}_{\vec{p}}\tilde{\delta}_{\vec{p}+\vec{k}}\bar{u}_{b}(\vec{p},r)\gamma_{\mu}u_{b}(\vec{p}+\vec{k},s)
       g^{*}_{\varepsilon}\left[\omega^{b}_{2}(\vec{p},\vec{k}),T,t\right]  \\
     & +\tilde{\beta}^{*}_{\vec{p}}\tilde{\gamma}_{\vec{p}+\vec{k}}\bar{v}_{b}(\vec{p},r)\gamma_{\mu}v_{b}(\vec{p}+\vec{k},s)
       g_{\varepsilon}\left[\omega^{b}_{3}(\vec{p},\vec{k}),T,t\right]  \\
     & +\tilde{\beta}^{*}_{\vec{p}}\tilde{\delta}_{\vec{p}+\vec{k}}\bar{v}_{b}(\vec{p},r)\gamma_{\mu}u_{b}(\vec{p}+\vec{k},s)
       g^{*}_{\varepsilon}\left[\omega^{b}_{4}(\vec{p},\vec{k}),T,t\right]  \\
     & -\left.
         \bar{u}_{c}(\vec{p},r)\gamma_{\mu}v_{c}(\vec{p}+\vec{k},s)
         g_{\varepsilon}\left[\omega^{c}_{1}(\vec{p},\vec{k}),T,t\right]
        \right\rbrace
\end{split}
\end{equation}
for $t\ge T$. To keep the notation short, we have introduced
\begin{equation}
 g_{\varepsilon}(\omega,T,t) = \frac{\ee^{-(\varepsilon-\ii\omega)T}-\ee^{-(\varepsilon-\ii\omega)t}}{\varepsilon-\ii\omega} \ .
\end{equation}
If we now let $\varepsilon\rightarrow0$ at some finite time, $t>T$, we see that (\ref{eq:f:intamp_mst}) turns into
\begin{equation}
\begin{split}
 \label{eq:f:intamp_mst_aspt}
 I^{\varepsilon}_{\mu}(\vec{p},\vec{k},r,s,t)\rightarrow I_{\mu}(\vec{p},\vec{k},r,s,t)
   = & \int_{-\infty}^{T}\dd u\mbox{ }j^{\text{MST}}_{\mu}(\vec{p},\vec{k},r,s,u)\ee^{\ii\omega_{\vec{k}}u}  \\
   + & e\left\lbrace
          \tilde{\alpha}^{*}_{\vec{p}}\tilde{\gamma}_{\vec{p}+\vec{k}}\bar{u}_{b}(\vec{p},r)\gamma_{\mu}v_{b}(\vec{p}+\vec{k},s)
          g\left[\omega^{b}_{1}(\vec{p},\vec{k}),T,t\right]
         \right.  \\
     & +\tilde{\alpha}^{*}_{\vec{p}}\tilde{\delta}_{\vec{p}+\vec{k}}\bar{u}_{b}(\vec{p},r)\gamma_{\mu}u_{b}(\vec{p}+\vec{k},s)
       g^{*}\left[\omega^{b}_{2}(\vec{p},\vec{k}),T,t\right]  \\
     & +\tilde{\beta}^{*}_{\vec{p}}\tilde{\gamma}_{\vec{p}+\vec{k}}\bar{v}_{b}(\vec{p},r)\gamma_{\mu}v_{b}(\vec{p}+\vec{k},s)
       g\left[\omega^{b}_{3}(\vec{p},\vec{k}),T,t\right]  \\
     & +\tilde{\beta}^{*}_{\vec{p}}\tilde{\delta}_{\vec{p}+\vec{k}}\bar{v}_{b}(\vec{p},r)\gamma_{\mu}u_{b}(\vec{p}+\vec{k},s)
       g^{*}\left[\omega^{b}_{4}(\vec{p},\vec{k}),T,t\right]  \\
     & -\left.
         \bar{u}_{c}(\vec{p},r)\gamma_{\mu}v_{c}(\vec{p}+\vec{k},s)
         g \left[\omega^{c}_{1}(\vec{p},\vec{k}),T,t\right]
        \right\rbrace
\end{split}
\end{equation}
with $g(\omega,T,t)$ given by
\begin{equation}
 \label{eq:f:freq_int}
 g(\omega,T,t)=\frac{\ee^{\ii\omega T}}{\ii\omega}\left[\ee^{\ii\omega(t-T)}-1\right] \ .
\end{equation}
On the other hand, (\ref{eq:f:intamp_vac}) does not vanish, but behaves like
\begin{equation}
 J^{\varepsilon}_{\mu}(\vec{p},\vec{k},r,s,t) = e\bar{u}_{c}(\vec{p},r)\gamma_{\mu}v_{c}(\vec{p}+\vec{k},s)
                                                \frac{\ee^{\ii\omega^{c}_{1}(\vec{p},\vec{k})t}}{\ii\omega^{c}_{1}(\vec{p},\vec{k})}+
                                                \mathcal{O}(\varepsilon) \ , \mbox{ for } \varepsilon\rightarrow0 \ .
\end{equation}
Hence, when first taking the limit $\varepsilon\rightarrow0$ at some finite
time, $t$, expression (\ref{eq:f:photnum_int}) does not vanish. Furthermore, we can infer 
from Fig. \ref{fig:f:interference} that it generally persists in the limit $t\rightarrow\infty$.

If one first takes $t\rightarrow\infty$, however, it follows from
(\ref{eq:f:intamp_vac}) and (\ref{eq:f:intamp_mst}) that these
expressions turn into (\ref{eq:4:intaspt_vac_eval}) and
(\ref{eq:4:intaspt_mst_eval}), respectively, so that (\ref{eq:f:photnum_int})
vanishes when taking the subsequent limit $\varepsilon\rightarrow0$.

\begin{figure}[htb]
 \begin{center}
  \includegraphics[height=5.0cm]{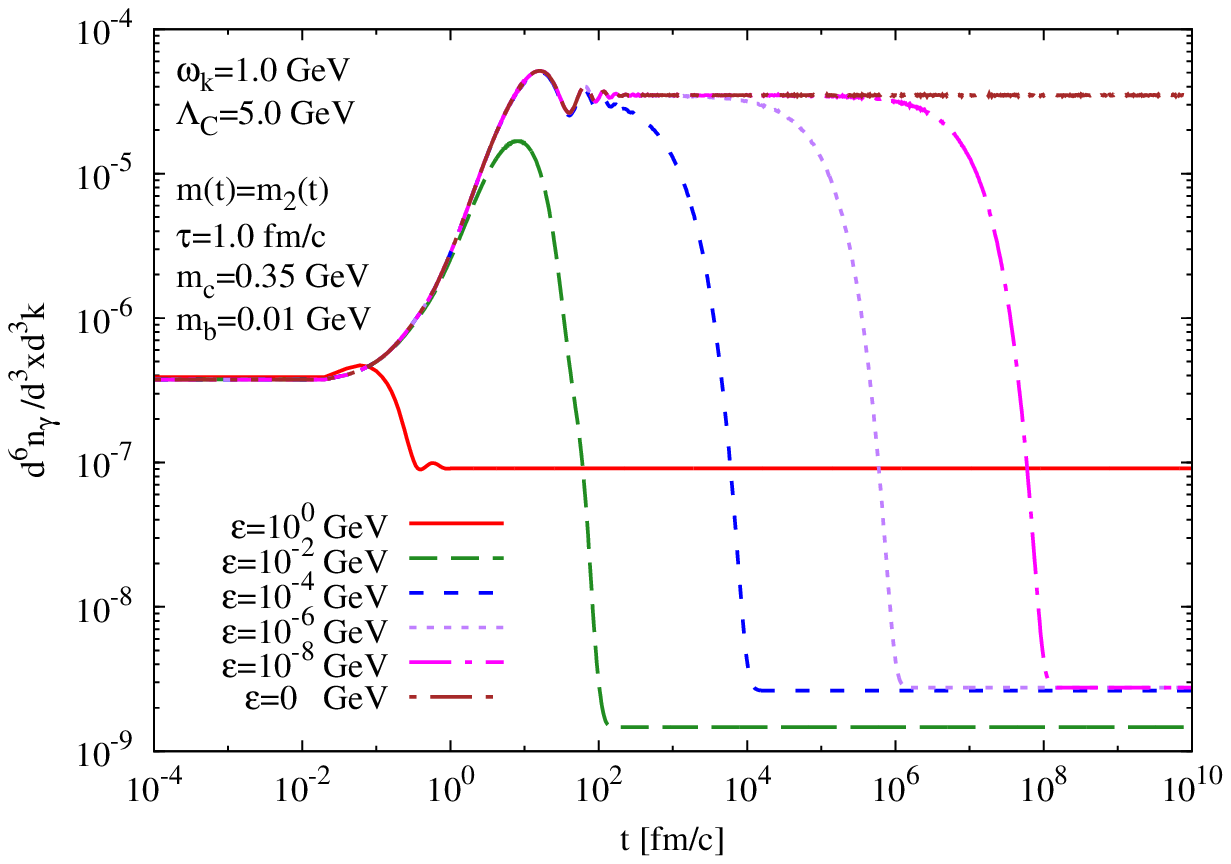}
  \includegraphics[height=5.0cm]{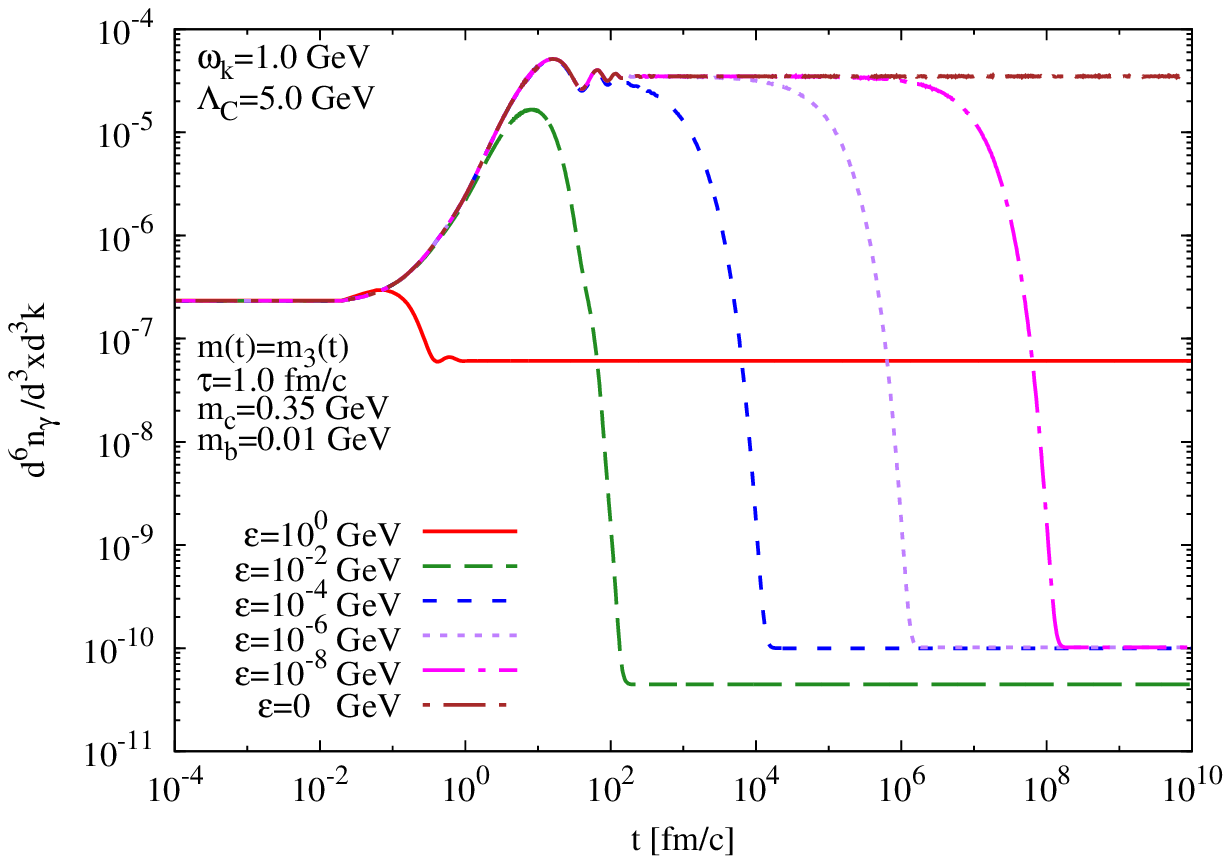}
  \caption{Time evolution of the mass-shift contribution (\ref{eq:f:contr_dir}) for different 
           values of $\varepsilon$ for $m_{2}(t)$ (left panel) and $m_{3}(t)$ (right panel).}
  \label{fig:f:evolution}
 \end{center}
\end{figure}

In addition to the elimination of possible unphysical contributions from
(\ref{eq:4:pse_vac}) and (\ref{eq:4:pse_int}), keeping the correct sequence 
of limits is also crucial to obtain physically reasonable results for the contribution from
(\ref{eq:4:pse_mst}) describing mass-shift effects only. To illustrate
this, we show the time evolution of (\ref{eq:f:photnum_mst}) for different values of 
$\varepsilon$ in Fig. \ref{fig:f:evolution}.

Taking first $\varepsilon\rightarrow0$ at some finite time, $t$,
corresponds to the curve labeled by $\varepsilon=0$ in both panels. If
we then take $t\rightarrow\infty$, we obtain an asymptotic value for
(\ref{eq:f:photnum_mst}), which is by several orders of magnitude larger
than the one against which the asymptotic values of
(\ref{eq:f:photnum_mst}) for finite $\varepsilon$ converge in the limit
$\varepsilon\rightarrow0$ and which accordingly corresponds to
(\ref{eq:2:photon_yield_chiral_aspt}). Hence, we have seen that
interchanging both limits also leads to unphysical results
from the pure mass-shift contribution. Moreover, we see that for sufficiently 
small values of $\varepsilon$, expression (\ref{eq:f:photnum_mst}) for finite 
times $t\ge T\gg\tau$ coincides with its asymptotic value for $\varepsilon=0$. 
Consequently, the interpretation of (\ref{eq:f:photnum_mst}) as a photon number 
at finite times is doubtful as well.

In order to highlight these aspects in particular, we take a closer look at 
(\ref{eq:f:intamp_mst_aspt}) and (\ref{eq:4:intaspt_zero}). We see that the 
integral expressions coincide in both cases whereas the frequency expressions 
entering (\ref{eq:f:intamp_mst_aspt}) feature an additional term of
$\ee^{\ii\omega_{i}t}/\ii\omega_{i}$ ($i=0 \ldots 4$), which has been 
eliminated in the corresponding frequency expressions entering (\ref{eq:4:intaspt_zero}) 
after switching off the electromagnetic interaction adiabatically.

\begin{figure}[htb]
 \begin{center}
  \includegraphics[height=5.0cm]{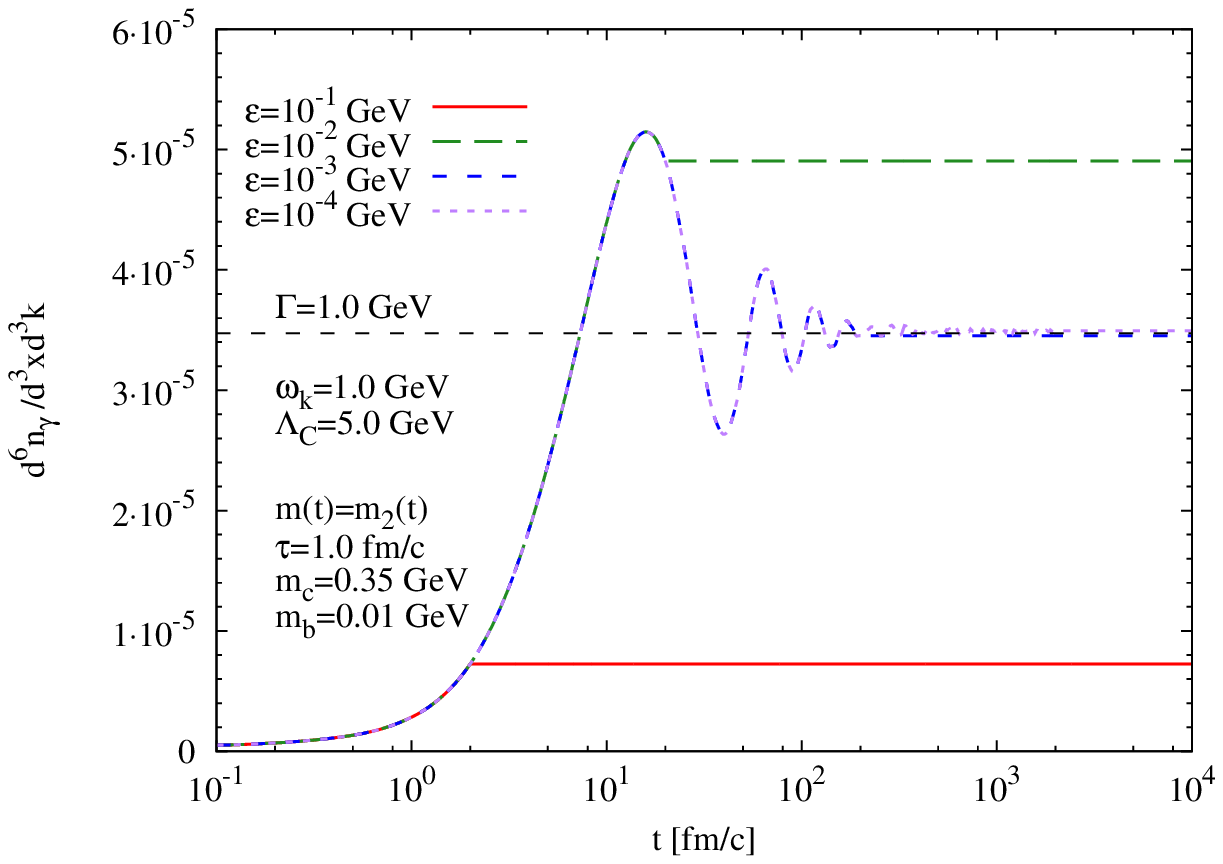}
  \includegraphics[height=5.0cm]{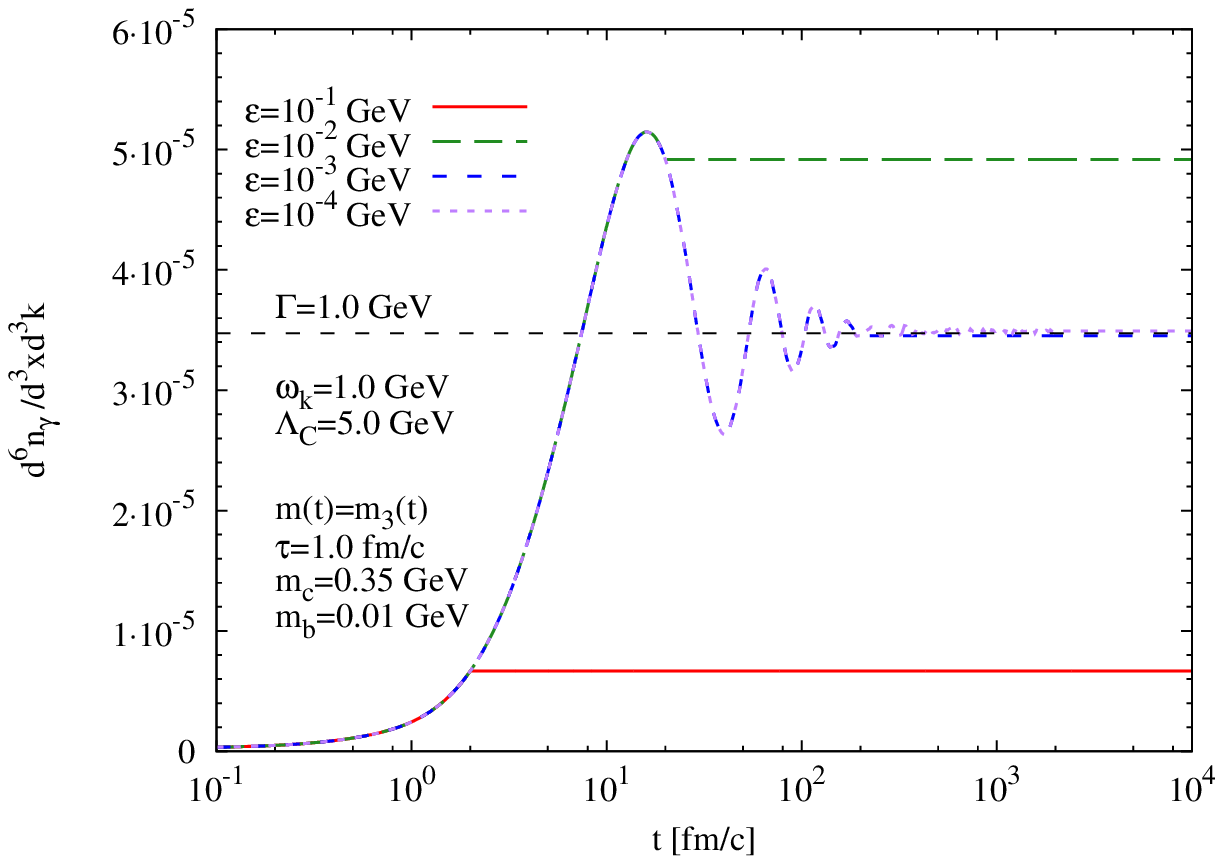}
  \caption{Time evolution of the mass-shift contribution
    (\ref{eq:f:photnum_mst}) for $m_{2}(t)$ (left panel) and $m_{3}(t)$
    (right panel) with the electromagnetic interaction being switched
    in terms of (\ref{eq:f:switch}). If we take the successive limits
    $t\rightarrow\infty$ and then $\varepsilon\rightarrow0$,
    we obtain the same result for (\ref{eq:f:photnum_mst}) as if we had kept the regularization
    (\ref{eq:2:switch}) but interchanged both limits instead. The
    corresponding value is indicated by the dashed horizontal line in each case.}
  \label{fig:f:artificial}
 \end{center}
\end{figure}

In order to see why this additional term is artificial in each case we again consider 
(\ref{eq:f:photnum_mst}) for the correct sequence of limits, but instead perform an instantaneous
switching of the electromagnetic interaction at $t=\mp 1/\varepsilon$, i.e.,
\begin{equation}
 \label{eq:f:switch}
 \hat{H}_{\text{EM}}(t) \rightarrow \theta_{\varepsilon}(t)\hat{H}_{\text{EM}}(t) \ , \quad 
                                    \text{with} \quad \theta_{\varepsilon}(t)=\theta\left(\frac{1}{\varepsilon^{2}}-t^{2}\right) \quad \text{and} \quad \varepsilon>0 \ .
\end{equation}
In this case, (\ref{eq:4:intaspt_mst_eval}) is replaced by
\begin{equation}
\begin{split}
 \label{eq:f:intamp_unphys}
 \tilde{I}^{\varepsilon}_{\mu}(\vec{p},\vec{k},r,s)
  = & \int_{-1/\varepsilon}^{\text{min}(T,1/\varepsilon)}\dd t\mbox{ }j^{\text{MST}}_{\mu}(\vec{p},\vec{k},r,s,t)\ee^{\ii\omega_{\vec{k}}t}  \\
 - & e\left\lbrace
       \tilde{\alpha}^{*}_{\vec{p}}\tilde{\gamma}_{\vec{p}+\vec{k}}\bar{u}_{b}(\vec{p},r)\gamma_{\mu}v_{b}(\vec{p}+\vec{k},s)
       \tilde{g}_{\varepsilon}\left[\omega^{b}_{1}(\vec{p},\vec{k}),T\right] 
      \right.  \\
   & +\tilde{\alpha}^{*}_{\vec{p}}\tilde{\delta}_{\vec{p}+\vec{k}}\bar{u}_{b}(\vec{p},r)\gamma_{\mu}u_{b}(\vec{p}+\vec{k},s)
       \tilde{g}^{*}_{\varepsilon}\left[\omega^{b}_{2}(\vec{p},\vec{k}),T\right]  \\
   & +\tilde{\beta}^{*}_{\vec{p}}\tilde{\gamma}_{\vec{p}+\vec{k}}\bar{v}_{b}(\vec{p},r)\gamma_{\mu}v_{b}(\vec{p}+\vec{k},s)
       \tilde{g}_{\varepsilon}\left[\omega^{b}_{3}(\vec{p},\vec{k}),T\right]  \\
   & +\tilde{\beta}^{*}_{\vec{p}}\tilde{\delta}_{\vec{p}+\vec{k}}\bar{v}_{b}(\vec{p},r)\gamma_{\mu}u_{b}(\vec{p}+\vec{k},s)
       \tilde{g}^{*}_{\varepsilon}\left[\omega^{b}_{4}(\vec{p},\vec{k}),T\right]  \\
   & -\left.
       \bar{u}_{c}(\vec{p},r)\gamma_{\mu}v_{c}(\vec{p}+\vec{k},s)\tilde{g}_{\varepsilon}\left[\omega^{c}_{1}(\vec{p},\vec{k}),T\right]
      \right\rbrace \ .
\end{split}
\end{equation}
In order to keep the notation more shorthand, we have defined
\begin{equation}
 \label{eq:f:g_unphys}
 \tilde{g}_{\varepsilon}(\omega,T)=\theta\left(\frac{1}{\varepsilon}-T\right)
                                   \frac{\ee^{\ii\omega/\varepsilon}-\ee^{\ii\omega T}}{\ii\omega} \ .
\end{equation}
It follows from (\ref{eq:f:freq_int}) and (\ref{eq:f:g_unphys}) that taking the limit $\varepsilon\rightarrow0$ in 
(\ref{eq:f:intamp_unphys}) has the same effect as taking $t\rightarrow\infty$ in (\ref{eq:f:intamp_mst_aspt}). 
This can also be inferred from Fig. \ref{fig:f:artificial} showing the time evolution of (\ref{eq:f:photnum_mst}) 
for different values of $\varepsilon$ with the electromagnetic interaction being switched according to (\ref{eq:f:switch}).

Thus, if we switch the electromagnetic interaction according to (\ref{eq:2:switch}) 
but take the incorrect sequence of limits we obtain the same result for (\ref{eq:f:photnum_mst}) 
as if we took the correct sequence of limits but switched the electromagnetic interaction 
according to (\ref{eq:f:switch}). This shows us that the former case is formally equivalent 
to keeping the exact sequence of limits for an \emph{unphysical} instantaneous switching (\ref{eq:f:switch}).

In order to emphasize why a switching of the electromagnetic interaction 
by means of (\ref{eq:f:switch}) is unphysical, we consider again the Ward-Takahashi
identities (\ref{eq:2:wti}) for the photon self-energy. We have shown in
section \ref{sec:gaug-inv-phot} and also demonstrated in greater detail
in appendix \ref{sec:appb} that these identities are fulfilled when
coupling the quark field to a time-dependent scalar
background field, $\phi(t)$. But one has to keep in mind that a switching 
of the electromagnetic interaction leads to a corresponding modification 
of the photon self-energy, i.e.,
\begin{equation}
 \ii\Pi^{<}_{\mu\nu}(\vec{k},t_{1},t_{2})\rightarrow\ii\Pi^{<,\varepsilon}_{\mu\nu}(\vec{k},t_{1},t_{2})
                                                   =f_{\varepsilon}(t_{1})f_{\varepsilon}(t_{2})\ii\Pi^{<}_{\mu\nu}(\vec{k},t_{1},t_{2}) \ ,
\end{equation}
with the electromagnetic interaction then being fully persistent at any
given time, $t$. Because of the additional factor of
$f_{\varepsilon}(t_{1})$, relation (\ref{eq:2:wti}) is not fulfilled in
the first place, but we instead have
\begin{equation}
 \label{eq:f:wti_violation}
 \partial_{t_{1}}\ii\Pi^{<,\varepsilon}_{0\mu}(\vec{k},t_{1},t_{2})-\ii k^{j}\ii\Pi^{<,\varepsilon}_{j\mu}(\vec{k},t_{1},t_{2})
   = -\dot{f}_{\varepsilon}(t_{1})f_{\varepsilon}(t_{2})\ii\Pi^{<,\varepsilon}_{0\mu}(\vec{k},t_{1},t_{2}) \ .
\end{equation}
However, if the electromagnetic interaction is switched according to
(\ref{eq:2:switch}), the time derivative reads
\begin{equation}
 \dot{f}_{\varepsilon}(t)=\varepsilon\mbox{ sign}(t)\ee^{-\varepsilon|t|} \ ,
\end{equation}
and, as a consequence, the r.h.s. of (\ref{eq:f:wti_violation}) vanishes 
for $\varepsilon\rightarrow0$ such that (\ref{eq:2:wti}) is restored again. 
To the contrary, if we switch the electromagnetic interaction by means of 
(\ref{eq:f:switch}), we have no such restoration since
\begin{equation}
  \dot{\theta}_{\varepsilon}(t)=\theta\left(\frac{1}{\varepsilon}-t\right)
  \delta\left(\frac{1}{\varepsilon}+t\right)-
  \theta\left(\frac{1}{\varepsilon}+t\right)\delta\left(\frac{1}{\varepsilon}-t\right).
\end{equation}
Hence, switching the electromagnetic interaction according to (\ref{eq:f:switch}) effectively leads to
a violation of the Ward-Takahashi identities. Since (\ref{eq:f:switch})
leads to the same asymptotic photon numbers as (\ref{eq:2:switch}) with
the limits being interchanged, we see that the latter procedure also leads 
to an effective violation of (\ref{eq:2:wti}).

\section{Determination of Bogolyubov parameters $\xi_{\vec{p},s}(t)$ and $\eta_{\vec{p},s}(t)$}
\label{sec:appc}

In this appendix, we show explicitly that the system of equations
(\ref{eq:bogolubov_param_sys}) for the Bogolyubov parameters,
$\xi_{\vec{p},s}(t)$ and $\eta_{\vec{p},s}(t)$, is solved by
(\ref{eq:3:bogolyubov_param}). For this purpose, we first rewrite the
fermionic field operator, $\hat{\psi}(x)$, in terms of the transformed
operators (\ref{eq:3:Bogolyubov}), i.e.,
\begin{equation}
 \hat{\psi}(x)=\sum_{s}\int\frac{\dd^{3}p}{(2\pi)^3}
                \left[
                 \hat{\tilde{b}}_{\vec{p},s}(t)\tilde{\psi}_{\vec{p},s,\uparrow}(x)+
                 \hat{\tilde{d}}^{\dagger}_{-\vec{p},s}(t)\tilde{\psi}_{\vec{p},s\downarrow}(x)
                \right] \ .
\end{equation}
Here we have made use of (\ref{eq:3:Bogolyubov_inverse}) and
introduced the transformed positive and negative energy wavefunctions
\begin{subequations}
 \label{eq:c:wavef_trans}
 \begin{eqnarray}
  \tilde{\psi}_{\vec{p},s,\uparrow}(x)   & = & \xi^{*}_{\vec{p},s}(t)\psi_{\vec{p},s\uparrow}(x)+
                                               \eta^{*}_{\vec{p},s}(t)\psi_{\vec{p},s\downarrow}(x) \ , \label{eq:c:wavef_trans_pos} \\
  \tilde{\psi}_{\vec{p},s,\downarrow}(x) & = & \xi_{\vec{p},s}(t)\psi_{\vec{p},s\downarrow}(x)-
                                               \eta_{\vec{p},s}(t)\psi_{\vec{p},s\uparrow}(x) \ . \label{eq:c:wavef_trans_neg}
 \end{eqnarray}
\end{subequations}
It follows from (\ref{eq:3:normalization_param}) that the inverse
transformation relations are given by
\begin{subequations}
  \label{eq:c:wavef_trans_inv}
  \begin{eqnarray}
  \psi_{\vec{p},s,\uparrow}(x)   & = & \xi_{\vec{p},s}(t)\tilde{\psi}_{\vec{p},s\uparrow}(x)-
                                       \eta^{*}_{\vec{p},s}(t)\tilde{\psi}_{\vec{p},s\downarrow}(x) \label{eq:c:wavef_trans_inv_pos} \ , \\
  \psi_{\vec{p},s,\downarrow}(x) & = & \xi^{*}_{\vec{p},s}(t)\tilde{\psi}_{\vec{p},s\downarrow}(x)+
                                       \eta_{\vec{p},s}(t)\tilde{\psi}_{\vec{p},s\uparrow}(x) \label{eq:c:wavef_trans_inv_neg} \ .
 \end{eqnarray} 
\end{subequations}
Now the strategy is to rewrite (\ref{eq:bogolubov_param_sys}) as
matrix-element conditions for
$\tilde{\psi}_{\vec{p},s\uparrow\downarrow}(x)$ from which the latter
are determined. The corresponding Bogolyubov parameters,
$\xi_{\vec{p},s}(t)$ and $\eta_{\vec{p},s}(t)$, are then given by
\begin{subequations}
 \label{eq:c:projection}
 \begin{align}
   \xi_{\vec{p},s}(t) =
   \quad\psi^{\dagger}_{\vec{p},s\downarrow}(x)\tilde{\psi}_{\vec{p},s\downarrow}(x)
   & = \left [
     \psi^{\dagger}_{\vec{p},s\uparrow}(x)\tilde{\psi}_{\vec{p},s\uparrow}(x)
   \right ]^{*}
   = \tilde{\psi}^{\dagger}_{\vec{p},s\uparrow}(x)\psi_{\vec{p},s\uparrow}(x) \ , \\
   \eta_{\vec{p},s}(t) =
   -\psi^{\dagger}_{\vec{p},s\uparrow}(x)\tilde{\psi}_{\vec{p},s\downarrow}(x)
   & = \left [
     \psi^{\dagger}_{\vec{p},s\downarrow}(x)\tilde{\psi}_{\vec{p},s\uparrow}(x)
   \right ]^{*} =
   \tilde{\psi}^{\dagger}_{\vec{p},s\uparrow}(x)\psi_{\vec{p},s\downarrow}(x)
   \ .
 \end{align} 
\end{subequations}
Expressing (\ref{eq:3:hamilton_matrix_a})-(\ref{eq:3:hamilton_matrix_b})
in terms of
(\ref{eq:c:wavef_trans_inv_pos})-(\ref{eq:c:wavef_trans_inv_neg}) and
making use of (\ref{eq:3:normalization_param}), we can rewrite
(\ref{eq:bogolubov_param_sys}) in terms of
$\tilde{\psi}_{\vec{p},s\uparrow\downarrow}(x)$ as
\begin{subequations}
 \label{eq:c:matrix_eq}
 \begin{eqnarray}
  \xi_{\vec{p},s}(t)\mbox{ }\tilde{\psi}^{\dagger}_{\vec{p},s,\uparrow}(x)\hat{h}_{D}(t)\tilde{\psi}_{\vec{p},s,\uparrow}(x)
  -\eta^{*}_{\vec{p},s}(t)\mbox{ }\tilde{\psi}^{\dagger}_{\vec{p},s,\uparrow}(x)\hat{h}_{D}(t)\tilde{\psi}_{\vec{p},s,\downarrow}(x)
  & = E_{\vec{p}}(t)\xi_{\vec{p},s}(t) \ , \label{eq:c:matrix_eq_a} \\
  -\eta_{\vec{p},s}(t)\mbox{ }\tilde{\psi}^{\dagger}_{\vec{p},s,\downarrow}(x)\hat{h}_{D}(t)\tilde{\psi}_{\vec{p},s,\downarrow}(x)
  +\xi^{*}_{\vec{p},s}(t)\mbox{ }\tilde{\psi}^{\dagger}_{\vec{p},s,\uparrow}(x)\hat{h}_{D}(t)\tilde{\psi}_{\vec{p},s,\downarrow}(x)
  & = E_{\vec{p}}(t)\eta_{\vec{p},s}(t) \ , \label{eq:c:matrix_eq_b} 
 \end{eqnarray}
\end{subequations}
To keep the notation short, we have introduced
\begin{equation}
 \label{eq:c:hamilton_density}
 \hat{h}_{D}(t) = -\ii\gamma_{0}\vec{\gamma}\cdot\vec{\nabla}+\gamma_{0}m(t) \ .
\end{equation}
Taking into account that
\begin{equation}
 \tilde{\psi}^{\dagger}_{\vec{p},s,\downarrow}(x)\hat{h}_{D}(t)\tilde{\psi}_{\vec{p},s,\downarrow}(x)
  = -\tilde{\psi}^{\dagger}_{\vec{p},s,\uparrow}(x)\hat{h}_{D}(t)\tilde{\psi}_{\vec{p},s,\uparrow}(x) \ ,
\end{equation}
which follows from (\ref{eq:c:wavef_trans}) and
(\ref{eq:3:hamilton_matrix_a}), one can show that
(\ref{eq:c:matrix_eq_a})-(\ref{eq:c:matrix_eq_b}) are equivalent to
\begin{subequations}
 \label{eq:c:diag_cond_trans}
 \begin{eqnarray}
  \tilde{\psi}^{\dagger}_{\vec{p},s,\uparrow}(x)\hat{h}_{D}(t)\tilde{\psi}_{\vec{p},s,\uparrow}(x) & = & E_{\vec{p}}(t) \ , \label{eq:c:diag_cond_trans_11} \\
  \tilde{\psi}^{\dagger}_{\vec{p},s,\uparrow}(x)\hat{h}_{D}(t)\tilde{\psi}_{\vec{p},s,\downarrow}(x) & = & 0 \ . \label{eq:c:diag_cond_trans_12}
 \end{eqnarray}
\end{subequations}
The Bogolyubov transformation (\ref{eq:3:Bogolyubov}) hence corresponds to a reexpansion of the fermion-field operators in terms of the 
instantaneous eigenstates of the Hamilton-density operator (\ref{eq:c:hamilton_density}). In order to solve
Eqs. (\ref{eq:c:diag_cond_trans_11})-(\ref{eq:c:diag_cond_trans_12}) for
the matrix elements, we parametrize
$\tilde{\psi}_{\vec{p},s\uparrow\downarrow}(x)$ analogously to
(\ref{eq:2:wave_param}) as
\begin{equation}
 \label{eq:c:wave_param_trans}
 \tilde{\psi}_{\vec{p},s,\uparrow\downarrow}(x)=\begin{pmatrix}
                                                 \tilde{\alpha}_{\vec{p},\uparrow\downarrow}(t)\chi_{s} \\
                                                 \tilde{\beta}_{\vec{p},\uparrow\downarrow}(t)
                                                 \frac{\vec{\sigma}\cdot\vec{p}}{p}\chi_{s}
                                                \end{pmatrix}
                                                e^{\ii\vec{p}\cdot\vec{x}}
                                                \text{.}
\end{equation}
Writing (\ref{eq:c:wavef_trans_pos}) and (\ref{eq:c:wavef_trans_neg}) in
terms of wavefunction parameters, it follows from
(\ref{eq:2:param_normalization}) and (\ref{eq:3:normalization_param})
that
\begin{equation}
 \label{eq:c:normalization_trans}
 \left|\tilde{\alpha}_{\vec{p},\uparrow\downarrow}(t)\right|^{2}+\left|\tilde{\beta}_{\vec{p},\uparrow\downarrow}(t)\right|^{2}=1 \ .
\end{equation}
Furthermore, relations (\ref{eq:2:param_link}) imply
\begin{subequations}
 \label{eq:c:param_link_trans}
 \begin{eqnarray}
  \tilde{\alpha}_{\vec{p},\downarrow}(t) & = & \tilde{\beta}^{*}_{\vec{p},\uparrow}(t) \ , \label{eq:c:param_link_trans_a} \\
  \tilde{\beta}_{\vec{p},\downarrow}(t)  & = & -\tilde{\alpha}^{*}_{\vec{p},\uparrow}(t) \ . \label{eq:c:param_link_trans_b}
 \end{eqnarray}
\end{subequations}
Upon insertion of (\ref{eq:c:wave_param_trans}) into
(\ref{eq:c:diag_cond_trans_11})-(\ref{eq:c:diag_cond_trans_12}) and
making use of
(\ref{eq:c:param_link_trans_a})-(\ref{eq:c:param_link_trans_b}) the
diagonalization conditions can be written as
\begin{subequations}
 \label{eq:c:diag_fin}                  
 \begin{eqnarray}
  p\left(
    \tilde{\alpha}^{*}_{\vec{p},\uparrow}(t)\tilde{\beta}_{\vec{p},\uparrow}(t)+
    \tilde{\alpha}_{\vec{p},\uparrow}(t)\tilde{\beta}^{*}_{\vec{p},\uparrow}(t)
    \right)+
  m(t)\left(
       \left|\tilde{\alpha}_{\vec{p},\uparrow}(t)\right|^{2}-
       \left|\tilde{\beta}_{\vec{p},\uparrow}(t)\right|^{2}
      \right) = & E_{\vec{p}}(t) \ , \label{eq:c:diag_fin_a} \\
  p\left(
    \tilde{\beta}^{2}_{\vec{p},\uparrow}(t)-\tilde{\alpha}^{2}_{\vec{p},\uparrow}(t)
   \right)+
  2m(t)\tilde{\alpha}_{\vec{p},\uparrow}(t)\tilde{\beta}_{\vec{p},\uparrow}(t) = & 0 \label{eq:c:diag_fin_b} \ .
 \end{eqnarray}
\end{subequations}
Expressing the wavefunction parameters in terms of polar coordinates,
i.e.,
\begin{subequations}
 \begin{eqnarray}
  \tilde{\alpha}_{\vec{p},\uparrow}(t) & = & \left|\tilde{\alpha}_{\vec{p},\uparrow}(t)\right|e^{\ii\phi_{\alpha}} \ , \\
  \tilde{\beta}_{\vec{p},\uparrow}(t)  & = & \left|\tilde{\beta}_{\vec{p},\uparrow}(t)\right|e^{\ii\phi_{\beta}} \ ,
 \end{eqnarray}
\end{subequations}
and taking into account that the l.h.s.s of both (\ref{eq:c:diag_fin_a})
and (\ref{eq:c:diag_fin_b}) are real valued, one obtains
\begin{eqnarray}
 \sin\left[\phi_{\alpha}(t)-\phi_{\beta}(t)\right] & = & 0 \ , \nonumber \\
 \cos\left[\phi_{\alpha}(t)-\phi_{\beta}(t)\right] & = & (-1)^{n} \ . \nonumber
\end{eqnarray}
Therefore, we can rewrite
(\ref{eq:c:diag_fin_a})-(\ref{eq:c:diag_fin_b}) as
\begin{subequations}
 \begin{eqnarray}
  (-1)^{n}2p\left|\tilde{\alpha}_{\vec{p},\uparrow}(t)\right|\left|\tilde{\beta}_{\vec{p},\uparrow}(t)\right|+
      m(t)\left(
           \left|\tilde{\alpha}_{\vec{p},\uparrow}(t)\right|^{2}-
           \left|\tilde{\beta}_{\vec{p},\uparrow}(t)\right|^{2}
          \right) & = & E_{\vec{p}}(t) \ , \\
  (-1)^{n}p\left(
        \left|\tilde{\alpha}_{\vec{p},\uparrow}(t)\right|^{2}-
        \left|\tilde{\beta}_{\vec{p},\uparrow}(t)\right|^{2}
       \right)+
      2m(t)\left|\tilde{\alpha}_{\vec{p},\uparrow}(t)\right|\left|\tilde{\beta}_{\vec{p},\uparrow}(t)\right| & = & 0 \ ,
 \end{eqnarray}
\end{subequations}
which is equivalent to
\begin{subequations}
 \label{eq:c:diag_fin_abs}
 \begin{eqnarray}
  \left|\tilde{\alpha}_{\vec{p},\uparrow}(t)\right|
  \left|\tilde{\beta}_{\vec{p},\uparrow}(t)\right|       & = & (-1)^{n}\frac{p}{2E_{\vec{p}}(t)} \ , \label{eq:c:diag_fin_abs_a} \\
  \left|\tilde{\alpha}_{\vec{p},\uparrow}(t)\right|^{2}-
  \left|\tilde{\beta}_{\vec{p},\uparrow}(t)\right|^{2}   & = & \frac{m(t)}{E_{\vec{p}}(t)} \ . \label{eq:c:diag_fin_abs_b}
 \end{eqnarray}
\end{subequations}
Relation (\ref{eq:c:diag_fin_abs_a}) implies that $n$ must be
even. Thus, without loss of generality, we can assume that
$\phi_{\alpha}(t) = \phi_{\beta}(t) \equiv \phi(t)$. Together with the
normalization condition (\ref{eq:c:normalization_trans}), it follows
from (\ref{eq:c:diag_fin_abs_a})-(\ref{eq:c:diag_fin_abs_b}) that we
have
\begin{subequations}
 \begin{eqnarray}
  \tilde{\alpha}_{\vec{p},\uparrow}(t) & = & \sqrt{\frac{E_{\vec{p}}(t)+m(t)}{2E_{\vec{p}}(t)}}e^{\ii\phi(t)} \ , \\
  \tilde{\beta}_{\vec{p},\uparrow}(t)  & = & \sqrt{\frac{E_{\vec{p}}(t)-m(t)}{2E_{\vec{p}}(t)}}e^{\ii\phi(t)} \ .
 \end{eqnarray}
\end{subequations}
Since $\psi_{\vec{p},s,\uparrow\downarrow}(x)$ and
$\tilde{\psi}_{\vec{p},s,\uparrow\downarrow}(x)$ coincide with each
other for $t\le t^{'}_{0}$, $\phi(t)$ can, in principle, be chosen
freely as long as it fulfills the asymptotic condition
$\phi(t)\rightarrow E^{c}_{\vec{p}}t$ for $t\rightarrow-\infty$. It is
hence convenient to choose $\phi(t)=E_{\vec{p}}(t)t$ so that our
Bogolyubov transformation corresponds to an expansion of $\hat{\psi}(x)$
in terms of positive and negative energy wavefunctions of the
respective current mass, $m(t)$. As a consequence, we finally have for
the transformed positive and negative energy wavefunction parameters
\begin{subequations}
 \begin{eqnarray}
  \tilde{\alpha}_{\vec{p},\uparrow}(t)   & = & \sqrt{\frac{E_{\vec{p}}(t)+m(t)}{2E_{\vec{p}}(t)}}e^{\ii E_{\vec{p}}(t)t} \ , \\
  \tilde{\beta}_{\vec{p},\uparrow}(t)    & = & \sqrt{\frac{E_{\vec{p}}(t)-m(t)}{2E_{\vec{p}}(t)}}e^{\ii E_{\vec{p}}(t)t} \ , \\
  \tilde{\alpha}_{\vec{p},\downarrow}(t) & = & \sqrt{\frac{E_{\vec{p}}(t)-m(t)}{2E_{\vec{p}}(t)}}e^{-\ii E_{\vec{p}}(t)t} \ , \\
  \tilde{\beta}_{\vec{p},\downarrow}(t)  & = & -\sqrt{\frac{E_{\vec{p}}(t)+m(t)}{2E_{\vec{p}}(t)}}e^{-\ii E_{\vec{p}}(t)t} \ ,
 \end{eqnarray}
\end{subequations}
and hence
\begin{subequations}
 \begin{eqnarray}
  \xi_{\vec{p},s}(t)  & = & e^{\ii E_{\vec{p}}(t)t}\left[
                                                    \sqrt{\frac{E_{\vec{p}}(t)+m(t)}{2E_{\vec{p}}(t)}}\alpha_{\vec{p},\uparrow}(t)+
                                                    \sqrt{\frac{E_{\vec{p}}(t)-m(t)}{2E_{\vec{p}}(t)}}\beta_{\vec{p},\uparrow}(t)
                                                   \right] \ , \\
  \eta_{\vec{p},s}(t) & = & e^{\ii E_{\vec{p}}(t)t}\left[
                                                    \sqrt{\frac{E_{\vec{p}}(t)+m(t)}{2E_{\vec{p}}(t)}}\alpha_{\vec{p},\downarrow}(t)+
                                                    \sqrt{\frac{E_{\vec{p}}(t)-m(t)}{2E_{\vec{p}}(t)}}\beta_{\vec{p},\downarrow}(t)
                                                   \right] \ .
 \end{eqnarray}
\end{subequations}
These are just the Bogolyubov parameters (\ref{eq:3:bogolyubov_param}).

\section{Evaluation of $I_{ij}(\vec{p},\vec{k})$}
\label{sec:appd}

In this appendix, we demonstrate how to evaluate
$I_{ij}(\vec{p},\vec{k})$ ($i,j=1..4$), which
  characterize the direct contributions from first-order QED processes
  (equal indices) and their interference among each other (different
  indices). We will carry out the calculations explicitly for one
direct contribution and one interference contribution, respectively. The
evaluation of the remaining contributions proceeds analogously. In order
to calculate the direct contribution from the spontaneous creation of a
quark-antiquark pair together with a photon, $I_{11}(\vec{p},\vec{k})$, we first
recall that the corresponding transition amplitude is given by
\begin{equation}
 I^{1}_{\mu}(\vec{p},\vec{k},r,s) = \ii e\alpha_{\vec{p}}\alpha_{\vec{p}+\vec{k}}\bar{u}_{b}(\vec{p},r)\gamma_{\mu}
                                    v_{b}(\vec{p}+\vec{k},s)\left(\frac{1}{\omega^{b}_{1}(\vec{p},\vec{k})}-\frac{1}{\omega^{c}_{1}(\vec{p},\vec{k})}\right)^{2} \ .
\end{equation}
In order to carry out the spin summation in $I_{11}(\vec{p},\vec{k})$,
we take into account that
\begin{subequations}
 \label{eq:d:appendix_spinsum}
 \begin{eqnarray}
  \sum_{s}u_{b}(\vec{p},s)\bar{u}_{b}(\vec{p},s) = \frac{\slashed{p}+m_{b}}{2E^{b}_{\vec{p}}}  \ ,     \\
  \sum_{s}v_{b}(\vec{p},s)\bar{v}_{b}(\vec{p},s) = \frac{\slashed{\bar{p}}-m_{b}}{2E^{b}_{\vec{p}}} \ ,
 \end{eqnarray}
\end{subequations}
where $\bar{p}^{\mu}=(E^{b}_{\vec{p}},-\vec{p})$. Hence, 
$I_{11}(\vec{p},\vec{k})$ can be rewritten as
\begin{equation}
 \label{eq:d:contr_11}
 I_{11}(\vec{p},\vec{k}) = e^{2}\frac{\gamma^{\mu\nu}(k)\alpha^{2}_{\vec{p}+\vec{k}}\alpha^{2}_{\vec{p}}}
                           {4E^{b}_{\vec{p}+\vec{k}}E^{b}_{\vec{p}}}\left(\frac{1}{\omega^{b}_{1}(\vec{p},\vec{k})}-\frac{1}{\omega^{c}_{1}(\vec{p},\vec{k})}\right)^{2}
                           \cdot\mbox{Tr}\left\lbrace\gamma_{\nu}(\slashed{\bar{q}}-m_{b})
                           \gamma_{\mu}(\slashed{p}+m_{b})\right\rbrace \ .
\end{equation}

where we have introduced
$q^{\mu}=(E^{b}_{\vec{p}+\vec{k}},\vec{p}+\vec{k})$. Using standard
Dirac trace techniques, we find that
\begin{equation}
\mbox{Tr}\left\lbrace\gamma_{\nu}(\slashed{\bar{q}}-m_{b})\gamma_{\mu}(\slashed{p}+m_{b})\right\rbrace
 = 4\left[\bar{q}_{\nu}p_{\mu}+\bar{q}_{\mu}p_{\nu}-\eta_{\nu\mu}\left(\bar{q}\cdot p+m^{2}_{b}\right)\right] \ .
\end{equation}
Taking into account that $\gamma^{\mu\nu}(k)$ has a purely spacelike
structure given by (\ref{eq:2:chiral_polten}), carrying out the
contraction entering (\ref{eq:d:contr_11}) finally yields
\begin{equation}
   I_{11}(\vec{p},\vec{k},t) = 2e^{2}\alpha^{2}_{\vec{p}+\vec{k}}\alpha^{2}_{\vec{p}}
                               \left(1+\frac{px(px+\omega_{\vec{k}})+m^{2}_{b}}{E^{b}_{\vec{p}+\vec{k}}E^{b}_{\vec{p}}}\right)
                               \left(\frac{1}{\omega^{b}_{1}(\vec{p},\vec{k})}-\frac{1}{\omega^{c}_{1}(\vec{p},\vec{k})}\right)^{2} \ .
\end{equation}
The evaluation of the direct contributions from quark and antiquark
bremsstrahlung and quark-antiquark annihilation into a photon is
performed in the same way. Next, we turn to the evaluation of the
interference contributions. In its course, we encounter spin summations
over mixed tensor products between positive- and negative-energy
spinors. These evaluate to
\begin{subequations}
 \label{eq:d:tenprod}
 \begin{eqnarray}
  \sum_{s}u_{b}(\vec{p},s)\bar{v}_{b}(\vec{p},s) = -\frac{(\slashed{p}+m_{b})\slashed{n}_{\vec{p}}}{2E^{b}_{p}} \ , \\
  \sum_{s}v_{b}(\vec{p},s)\bar{u}_{b}(\vec{p},s) = -\frac{\slashed{n}_{\vec{p}}(\slashed{p}+m_{b})}{2E^{b}_{p}} \ ,
 \end{eqnarray}
\end{subequations}
where we have introduced $n^{\mu}_{\vec{p}}=(0,\vec{p}/p)$ and taken
into account that
\begin{subequations}
 \begin{eqnarray}
  \slashed{n}_{\vec{p}}u_{b}(\vec{p},s) & = & -v_{b}(\vec{p},s) \ , \\
  \slashed{n}_{\vec{p}}v_{b}(\vec{p},s) & = & u_{b}(\vec{p},s) \ .
 \end{eqnarray}
\end{subequations}
With help of these relations, the contribution describing the
interference between the spontaneous creation of a quark/antiquark pair
together with a photon and quark pair annihilation into a photon,
$I_{14}(\vec{p},\vec{k},t)$, can be rewritten as
\begin{align}
 \label{eq:d:contr_14_prelim}
 I_{14}(\vec{p},\vec{k}) = & e^{2}\frac{\alpha_{\vec{p}}\beta_{\vec{p}}\alpha_{\vec{p}+\vec{k}}\beta_{\vec{p}+\vec{k}}}{4E^{b}_{\vec{p}}E^{b}_{\vec{p}+\vec{k}}}
                             \left(\frac{1}{\omega^{b}_{1}(\vec{p},\vec{k})}-\frac{1}{\omega^{c}_{1}(\vec{p},\vec{k})}\right)
                             \left(\frac{1}{\omega^{b}_{4}(\vec{p},\vec{k})}+\frac{1}{\omega^{c}_{1}(\vec{p},\vec{k})}\right) \nonumber \\
                           & \cdot\mbox{Tr}\left\lbrace\gamma_{\nu}(\slashed{p}+\slashed{k}+m_{b})\slashed{n}_{\vec{p}+\vec{k}}
                             \gamma_{\mu}(\slashed{p}+m_{b})\slashed{n}_{\vec{p}}\right\rbrace
\end{align}
Thus, in contrast to the direct contributions, carrying out the spin
summations for the interference contributions gives rise to traces over
products of six Dirac matrices. Making use of the anticommutation
relation
$$
\left\lbrace\gamma^{\mu},\gamma^{\nu}\right\rbrace = 2\eta^{\mu\nu}
$$
and cyclic trace invariance, these traces can be reduced to
\begin{eqnarray}
 \label{eq:d:dirac_reduce}
       \mbox{Tr}\left\lbrace\gamma^{\alpha}\gamma^{\beta}\gamma^{\gamma}\gamma^{\delta}\gamma^{\epsilon}\gamma^{\zeta}\right\rbrace
 & = & \eta^{\alpha\beta}\mbox{Tr}\left\lbrace\gamma^{\gamma}\gamma^{\delta}\gamma^{\epsilon}\gamma^{\zeta}\right\rbrace-
       \eta^{\alpha\gamma}\mbox{Tr}\left\lbrace\gamma^{\beta}\gamma^{\delta}\gamma^{\epsilon}\gamma^{\zeta}\right\rbrace+\nonumber \\
 &   & \eta^{\alpha\delta}\mbox{Tr}\left\lbrace\gamma^{\beta}\gamma^{\gamma}\gamma^{\epsilon}\gamma^{\zeta}\right\rbrace-
       \eta^{\alpha\epsilon}\mbox{Tr}\left\lbrace\gamma^{\beta}\gamma^{\gamma}\gamma^{\delta}\gamma^{\zeta}\right\rbrace+\nonumber \\
 &   & \eta^{\alpha\zeta}\mbox{Tr}\left\lbrace\gamma^{\beta}\gamma^{\gamma}\gamma^{\delta}\gamma^{\epsilon}\right\rbrace \ .
\end{eqnarray}
With the help of (\ref{eq:d:dirac_reduce}), the trace entering
(\ref{eq:d:contr_14_prelim}) is evaluated to
\begin{equation}
\begin{split}
   \mathrm{Tr} &\left\lbrace\gamma_{\nu}(\slashed{p}+\slashed{k}+m_{b})\slashed{n}_{\vec{p}+\vec{k}} 
     \gamma_{\mu}(\slashed{p}+m_{b})\slashed{n}_{\vec{p}}\right\rbrace \\
  &=\frac{E^{b}_{\vec{p}}E^{b}_{\vec{p}+\vec{k}}}{p|\vec{p}+\vec{k}|}
     \left[p_{i}(p_{j}+k_{j})+p_{j}(p_{i}+k_{i})+\eta_{ij}\vec{p}\cdot(\vec{p}+\vec{k})\right]
     +\eta_{ij}p|\vec{p}+\vec{k}| \ .
\end{split}
\end{equation}
We have considered only spacelike indices since the trace expression
entering (\ref{eq:d:contr_14_prelim}) is still contracted with
$\gamma^{\mu\nu}(k)$ which has an exclusively spacelike structure. This
contraction leads to
\begin{align}
   & \gamma^{\mu\nu}(k)\mbox{Tr}\left\lbrace\gamma_{\nu}(\slashed{p}+\slashed{k}+m_{b})\slashed{n}_{\vec{p}+\vec{k}} 
                       \gamma_{\mu}(\slashed{p}+m_{b})\slashed{n}_{\vec{p}}\right\rbrace \nonumber \\
 = & -8p^{2}|\vec{p}+\vec{k}|^{2}\left(1+\frac{px(px+\omega_{\vec{k}})(E^{b}_{\vec{p}}E^{b}_{\vec{p}+\vec{k}}-m^{2}_{b})}{|\vec{p}+\vec{k}|^{2}}\right) \ ,
\end{align}
so that (\ref{eq:d:contr_14_prelim}) finally reads
\begin{equation}
\begin{split}
  I_{14}(\vec{p},\vec{k}) = &
  -2e^{2}\frac{\alpha_{\vec{p}}\beta_{\vec{p}}\alpha_{\vec{p}+\vec{k}}\beta_{\vec{p}+\vec{k}}}
  {4E^{b}_{\vec{p}}E^{b}_{\vec{p}+\vec{k}}}p|\vec{p}+\vec{k}|
  \left(1+\frac{(px(px+\omega_{\vec{k}}))(E^{b}_{\vec{p}}E^{b}_{\vec{p}+\vec{k}}-m^{2}_{b})}{|\vec{p}+\vec{k}|^{2}}\right) \\
  & \cdot \left( \frac{1}{\omega^{b}_{1}(\vec{p},\vec{k})} -
    \frac{1}{\omega^{c}_{1}(\vec{p},\vec{k})}\right)
  \left(\frac{1}{\omega^{b}_{4}(\vec{p},\vec{k})}+\frac{1}{\omega^{c}_{1}(\vec{p},\vec{k})}\right)
  \ .
\end{split}
\end{equation}
The evaluation of the remaining interference contributions follows the
same steps. After evaluating all direct and interference contributions
the substitutions $\vec{p}\rightarrow\vec{p}-\vec{k}$ and
$x\rightarrow-1$ show that the remaining integration over
$\dd^{3}p$ yields the same contribution for
$I_{22}(\vec{p},\vec{k},t)$ and $I_{33}(\vec{p},\vec{k},t)$, for
$I_{12}(\vec{p},\vec{k},t)$ and $I_{13}(\vec{p},\vec{k},t)$ and for
$I_{24}(\vec{p},\vec{k},t)$ and $I_{34}(\vec{p},\vec{k},t)$. For that
reason, theses contributions have been taken together as shown in
(\ref{eq:4:sum_up}) in each case.

\section{Remarks on $m_{b}=0$ for an instantaneous mass shift}
\label{sec:appe}
In this appendix, we discuss the special case of $m_{b}=0$ for an
instantaneous mass shift. It requires some special considerations since
the frequencies describing quark/antiquark bremsstrahlung and pair
annihilation into a photon are no longer negative or positive
definite. We first take into account that we have
\begin{subequations}
 \begin{eqnarray}
   \alpha_{\vec{p}} & = & \frac{1}{\sqrt{2}}\left(\cos\varphi^{c}_{\vec{p}}+\sin\varphi^{c}_{\vec{p}}\right) \ , \\
   \beta_{\vec{p}}  & = & \frac{1}{\sqrt{2}}\left(\cos\varphi^{c}_{\vec{p}}-\sin\varphi^{c}_{\vec{p}}\right) \ ,
 \end{eqnarray}
\end{subequations}
for $m_{b}=0$. The asymptotic direct contributions hence turn into:
\begin{subequations}
 \label{eq:f:contr_dir}
 \begin{align}
  I_{11}(\vec{p},\vec{k})         = \frac{e^{2}}{2} & \left(1+\frac{|\vec{p}+\vec{k}|}{E^{c}_{\vec{p}+\vec{k}}}\right)
                                                      \left(1+\frac{p}{E^{c}_{\vec{p}}}\right)\left(1+\frac{x(px+\omega_{\vec{k}})}
                                                      {|\vec{p}+\vec{k}|}\right) \nonumber \\
                                              \times & \left(\frac{1}{|\vec{p}+\vec{k}|+p+\omega_{\vec{k}}}-\frac{1}{E^{c}_{\vec{p}+\vec{k}}+E^{c}_{\vec{p}}+\omega_{\vec{k}}}\right)^{2} \ , \\
  \tilde{I}_{22}(\vec{p},\vec{k}) = e^{2} & \left(1-\frac{|\vec{p}+\vec{k}|}{E^{c}_{\vec{p}+\vec{k}}}\right)
                                            \left(1+\frac{p}{E^{c}_{\vec{p}}}\right)\left(1-\frac{x(px+\omega_{\vec{k}})}
                                            {|\vec{p}+\vec{k}|}\right) \nonumber \\
                                    \times & \left(\frac{1}{|\vec{p}+\vec{k}|-p-\omega_{\vec{k}}}+\frac{1}{E^{c}_{\vec{p}+\vec{k}}+E^{c}_{\vec{p}}+\omega_{\vec{k}}}\right)^{2} \ , \\
  I_{44}(\vec{p},\vec{k})         = \frac{e^{2}}{2} & \left(1-\frac{|\vec{p}+\vec{k}|}{E^{c}_{\vec{p}+\vec{k}}}\right)
                                                      \left(1-\frac{p}{E^{c}_{\vec{p}}}\right)\left(1+\frac{x(px+\omega_{\vec{k}})}
                                                      {|\vec{p}+\vec{k}|}\right) \nonumber \\
                                                      \times & \left(\frac{1}{|\vec{p}+\vec{k}|+p-\omega_{\vec{k}}}+\frac{1}{E^{c}_{\vec{p}+\vec{k}}+E^{c}_{\vec{p}}+\omega_{\vec{k}}}\right)^{2} \ .                
 \end{align}
\end{subequations}
Furthermore, it follows from (\ref{eq:4:contr_int}) that
$\tilde{I}_{12}(\vec{p},\vec{k})$ and $\tilde{I}_{24}(\vec{p},\vec{k})$
vanish for $m_{b}=0$ and the remaining interference contributions
simplify to
\begin{subequations}
 \label{eq:f:contr_int}
 \begin{align}
  I_{14}(\vec{p},\vec{k}) = -\frac{e^{2}}{2E^{c}_{\vec{p}}E^{c}_{\vec{p}+\vec{k}}}\left(1+\frac{x(px+\omega_{\vec{k}})}{|\vec{p}+\vec{k}|}\right)
                             & \left(\frac{1}{|\vec{p}+\vec{k}|+p+\omega_{\vec{k}}}-\frac{1}{E^{c}_{\vec{p}+\vec{k}}+E^{c}_{\vec{p}}+\omega_{\vec{k}}}\right) \nonumber \\
                       \cdot & \left(\frac{1}{|\vec{p}+\vec{k}|+p-\omega_{\vec{k}}}+\frac{1}{E^{c}_{\vec{p}+\vec{k}}+E^{c}_{\vec{p}}+\omega_{\vec{k}}}\right) \ , \\
  I_{23}(\vec{p},\vec{k}) = -\frac{e^{2}}{2E^{c}_{\vec{p}}E^{c}_{\vec{p}+\vec{k}}}\left(1-\frac{x(px+\omega_{\vec{k}})}{|\vec{p}+\vec{k}|}\right)
                             & \left(\frac{1}{|\vec{p}+\vec{k}|-p-\omega_{\vec{k}}}+\frac{1}{E^{c}_{\vec{p}+\vec{k}}+E^{c}_{\vec{p}}+\omega_{\vec{k}}}\right) \nonumber \\
                       \cdot & \left(\frac{1}{|\vec{p}+\vec{k}|-p+\omega_{\vec{k}}}-\frac{1}{E^{c}_{\vec{p}+\vec{k}}+E^{c}_{\vec{p}}+\omega_{\vec{k}}}\right) \ .
 \end{align}
\end{subequations}
Since the frequencies describing quark/antiquark bremsstrahlung and pair
annihilation into a photon are no longer negative or positive definite,
we have to investigate, whether the loop integral features possible
infrared ($p\rightarrow 0$) and/or (anti-) collinear
($x\rightarrow\pm1$) singularities. In the limit $p\rightarrow 0$, they
behave as
\begin{subequations}
 \begin{eqnarray}
   I_{11}(\vec{p},\vec{k})         & \rightarrow & \frac{e^2(1+x)}{2}\left(1+\frac{\omega_{\vec{k}}}{\sqrt{\omega_{\vec{k}}^{2}+m^{2}_{c}}}\right)
                                                   \left(\frac{1}{2\omega_{\vec{k}}}-\frac{1}{\sqrt{\omega^{2}_{\vec{k}}+m^{2}_{c}}+E^{c}_{\vec{p}}+\omega_{\vec{k}}}\right)^{2}+\mathcal{O}(p) \ , \\
   \tilde{I}_{22}(\vec{p},\vec{k}) & \rightarrow & e^2\left(1-\frac{\omega_{\vec{k}}}{\sqrt{\omega_{\vec{k}}^{2}+m^{2}_{c}}}\right)\frac{1}{p^{2}(1-x)}+\mathcal{O}\left(\frac{1}{p}\right) \ , \\
   I_{44}(\vec{p},\vec{k})         & \rightarrow & \frac{e^2}{2}\left(1-\frac{\omega_{\vec{k}}}{\sqrt{\omega_{\vec{k}}^{2}+m^{2}_{c}}}\right)\frac{1}{p^{2}(1+x)}+\mathcal{O}\left(\frac{1}{p}\right) \ , \\
   I_{14}(\vec{p},\vec{k})         & \rightarrow & -\frac{e^2}{2}\frac{m_{c}}{p\sqrt{\omega_{\vec{k}}^{2}+m^{2}_{c}}}
                                                   \left(\frac{1}{2\omega_{\vec{k}}}-\frac{1}{\sqrt{\omega^{2}_{\vec{k}}+m^{2}_{c}}+E^{c}_{\vec{p}}+\omega_{\vec{k}}}\right)+\mathcal{O}(1) \ , \\
   I_{23}(\vec{p},\vec{k})         & \rightarrow & \frac{e^2}{2}\frac{m_{c}}{p\sqrt{\omega_{\vec{k}}^{2}+m^{2}_{c}}}
                                                   \left(\frac{1}{2\omega_{\vec{k}}}-\frac{1}{\sqrt{\omega^{2}_{\vec{k}}+m^{2}_{c}}+E^{c}_{\vec{p}}+\omega_{\vec{k}}}\right)+\mathcal{O}(1) \ .
 \end{eqnarray}
\end{subequations}
Since the integration measure $\dd^{3}p$ contributes an additional
factor of $p^{2}$ to each contribution, no infrared singularities will
show up in the loop integral for $m_{b}=0$. As a next step, we consider
the collinear limit $x\rightarrow1$. In this limit, the direct
contribution from quark/antiquark bremsstrahlung turns into
\begin{equation}
  \tilde{I}_{22}(\vec{p},\vec{k}) \rightarrow -e^2\left.\left(1-\frac{|\vec{p}+\vec{k}|}{E^{c}_{\vec{p}+\vec{k}}}\right)\right|_{x=1} 
  \left(1+\frac{p}{E^{c}_{\vec{p}}}\right)\frac{(p+\omega_{\vec{k}})^{2}+p^{2}}{p^{2}\omega_{\vec{k}}^{2}(x-1)}
  +\mathcal{O}(1)
\end{equation}
with all other contributions staying finite. Furthermore, the direct
contribution from pair annihilation into a photon features an
anti-collinear singularity in the limit $x\rightarrow\-1$ if $p<\omega_{\vec{k}}$,
i.e.,
\begin{equation}
  I_{44}(\vec{p},\vec{k}) \rightarrow \frac{e^2}{2}\left.\left(1-\frac{|\vec{p}+\vec{k}|}{E^{c}_{\vec{p}+\vec{k}}}\right)\right|_{x=-1}
  \left(1-\frac{p}{E^{c}_{\vec{p}}}\right)\frac{(p-\omega_{\vec{k}})^{2}+p^{2}}{p^{2}\omega_{\vec{k}}^{2}(x+1)} 
  +\mathcal{O}(1) \ .
\end{equation}
For $p>\omega_{\vec{k}}$, $I_{44}(\vec{p},\vec{k},t)$ stays finite for
$x\rightarrow-1$. All other contributions generally stay finite in that limit. 
Therefore, in the limit $m_{b}\rightarrow0$, the loop integral
develops a collinear and an anticollinear singularity for the
contributions from quark/antiquark Bremsstrahlung and quark pair
annihilation into a photon, respectively.


\end{document}